Quantum Calculations of a Large Section of the Voltage Sensing Domain of the K$_v$1.2 Channel Show that Proton Transfer, Not S4 Motion, Provides the Gating Current

Alisher. M. Kariev and Michael. E. Green[(1)]

Department of Chemistry and Biochemistry

City College of New York

New York NY 10031

and

PhD Program in Chemistry

City University of New York

New York NY 10016

to whom correspondence should be addressed: mgreen@ccny.cuny.edu

ABSTRACT: Quantum calculations on the voltage sensing domain (VSD) of the $K_v1.2$ potassium channel (pdb: 3Lut) have been carried out on a 904 atom subset of the VSD, plus 24 water molecules. Side chains pointing away from the center of the VSD were truncated; S1,S2,S3 end atoms were fixed (all calculations); S4 end atoms could be fixed or free. Open conformations (membrane potentials ≥ 0) closely match the known X-ray structure of the open state with salt bridges in the VSD not ionized ($H^+$ on the acid), whether S4 end atoms were fixed or free (slightly closer fixed than free). The S4 segment backbone, free or not, moves less than 2.5 Å for positive to negative membrane potential switches, not entirely in the expected direction, leaving $H^+$ motion as the principal component of gating current. Groups of 3-5 side chains are important for proton transport, based on the calculations. A proton transfers from tyrosine, (Y266), through arginine (R300), to glutamate (E183), accounting for approximately 20-25% of the gating charge. Clusters of amino acids that can transfer protons (acids, bases, tyrosine, histidine) are the main paths for proton transport. A group of five amino acids, bounded by the conserved aromatic F233, appears to exchange a proton. Dipole rotations may also contribute. A proton path (calculations still in progress) is proposed for the remainder of the VSD, suggesting a hypothesis for a complete gating mechanism.

INTRODUCTION:

The mechanism of gating of voltage gated ion channels has been the subject of study ever since their existence was proposed by Hodgkin and Huxley[1-3]. They pointed out that the response to depolarization of the membrane would have to be preceded by a capacitative current, called gating current, as charge rearranged in response to the changing electric field. This was first measured in 1974 by two groups, Keynes and Rojas[4], and Armstrong and Bezanilla[5]. The channel structure turned out to be tetrameric, exactly so for $K^+$ channels, not exactly for $Na^+$ channels; in this paper we will be concerned with a $K^+$ channel. Each channel contains four domains, each with a voltage sensing domain (VSD). Each VSD is composed of four transmembrane (TM) segments; each domain also has 2 TM segments that contribute to the 8 TM segments forming the pore through which the ion traverses the membrane. The way in which the VSDs bring about the opening of the pore as the membrane depolarizes has long been an active field of investigation. Horn and coworkers adapted the substituted cysteine accessibility method (SCAM)[6] to the VSD; S4, the VSD TM segment attached, via a linker, to the pore segments, has arginines (and usually one lysine, typically near the intracellular end of the S4 segment) in every third position; for ion channels, the SCAM method was originally applied, by Horn and coworkers, to R→C mutations in that segment, plus reaction of the cysteine with methanethiosulfonate (MTS) reagents that kill the channel; the method assumes that the cysteine must reach the surface of the membrane to react, because the cysteine must ionize in order to react, with ionization considered impossible in the putatively low dielectric constant medium in the interior of the membrane. By determining whether the membrane reacted when the MTS reagent was added to the intracellular or extracellular side of the membrane, or reacted in both cases, or neither, Horn and coworkers concluded that the cysteine, hence the entire S4 segment, had moved toward the intracellular side of the membrane when polarized, the extracellular side when depolarized[7]. The latter would open the channel by pulling on the linker between the S4 segment and the pore S5 segment. X-ray determination of the $K^+$ ion channel structure by MacKinnon and coworkers[8] (first KcsA, a channel without a VSD, but then the open state of channels with VSD) made possible a specific discussion of mechanism. Only the open state was determined, because the channel could only be crystallized in the absence of the field. The extent of the displacement of the VSD has been the subject of extensive research since, and the SCAM method has been supplemented by other experiments, especially fluorescence quenching, plus molecular dynamics (MD), all of which have been interpreted in terms of what has become the standard paradigm. This may be summarized as: 1) membrane polarized: the S4 TM segment of the VSD is "down" (intracellular), pushing, through the S4-S5 linker (the S5 TM segment being part of the pore), to close the intracellular gate; 2) membrane depolarized: the S4 rises (moves extracellularly), pulls on the S4-S5 linker, separating the residues at the gate, making it open for the ions to pass through the pore, giving the observed ion current. There have been a number of proposals for the details of the motion involved, which are not necessarily consistent with each other. We avoid discussing these details, in favor of presenting an alternative that does not move the S4 to either surface. The SCAM assumption that the cysteine must move to a membrane surface to react is not necessarily correct in this instance. The cysteine side chain is essentially a single sulfur atom when ionized. If it replaces an arginine, as in the original experiments, there is a large cavity; the arginine side chain, guanidinium, is much larger than a single atom; consequences are discussed below. The cysteine can ionize in the cavity that is left by the mutation, between the outer and inner membrane surfaces, without the backbone moving to the surface. The standard models all require that the arginines remain charged throughout the entire gating cycle, that they exchange partners with alternate charged glutamates or aspartates, thus limiting the energy cost of breaking the salt bridge. However, a $Q_{10}$ value (increase in gating probability with 10°C rise in temperature) of 1.5, (which is roughly in the neighborhood of values for some channels[9-11]) corresponds to about 30 kJ for gating current. The literature suggests that there may be non-Arrhenius behavior in gating, however, so this may not be a well-defined value. The $H_v1$ channel, which resembles a VSD through much of the transmembrane sections, shows this behavior[12] too. Since this channel transmits protons, it shows that protons can traverse the VSD. Another set of experiments, in which the end arginines are mutated to histidine, allow the VSD to transmit protons as well[13-14]. Our calculations show that proton transfers do occur under the influence of the field, leaving the arginines not always charged, and never completely charged, so that the fundamental assumption of the standard model is called into question. In addition, the S4 segment does not move more than about 2.5 Å relative to the S1,S2,S3 block, and not in the direction expected with standard models, so that even if the arginines remain ionized, they do not contribute to gating current.

Green long ago proposed an alternate gating mechanism in which the gating current was the motion of protons through the VSD, rather than the motion of positively charged arginines [15-18]. Even earlier, Green suggested that water played a key role at the gate[19]. Kariev and Green have discussed a possible way to understand the evidence supporting the standard paradigm that is compatible with the model we propose. The effects of R→C mutation can be understood by looking at the difference in size of cysteine and arginine [20-21]. There will be enough space that the cys could transfer a proton to another amino acid, as there are acid residues that were originally in salt bridges with the arginine, and would be missing their partner in the cys mutant; this would account for the required ionization of the cys. Also, the cavity remaining after cys replaces the much larger arg leaves room for a MTS headgroup, which could then react with the ionized cys. A fuller discussion of the reinterpretation of the evidence for the standard models is given in previous work[20-21]. Here we show a step in which proton motion can contribute to gating current, while the S4 backbone remains stationary.

The TASK 2-pore channel is a potassium channel somewhat analogous to the $K_v1.2$ channel, but pH gated. Niemeyer et al found a single arginine residue was responsible for the pH sensitivity [22]; the pK of the relevant arginine had shifted to 8.0, over four units compared to the arginine solution pK. They proposed that this was a consequence of the hydrophobic environment of the arginine. While we do not find the environment of all the arginines in the VSD to be so hydrophobic, the fact that the gating of a $K^+$ channel depends upon a large pK shift is in accord with our proposal; the environment is different for different arginines, but a similar effect is possible where the proton needs to transit. The protonation states in our model are necessarily transient. (Also, recall that standard models use SCAM, including the assumption of a hydrophobic environment for S4, as the reason the S4 must move to the membrane surface to react.), In addition to the arginines and glutamates, other residues, including a tyrosine, play an important role in the part we have calculated so far. Calculations of the structures and energies of the VSD with $H^+$ in several possible positions leads to a possible $H^+$ path through the VSD, which includes changes in ionization states of several residues, and groupings of three to five amino acids that participate in $H^+$ transfer. The path appears to be modulated by a phenylalanine which is well conserved. As this phenylalanine is near the edge of the section that was calculated, further work will be needed to determine exactly how it plays its role.

Several other matters appear to be important for the gating mechanism:

One involves the sharing/delocalization of a proton. In earlier work, we have examined, by use of quantum calculations, the effects of water with salt bridges [23-24]. This showed, in much smaller systems, that a proton could be part of a ring of hydrogen bonds that included water, which in turn affected the state of ionization. It is even possible to have what is essentially a resonance hybrid of hydrogen bonded structures. Proton delocalization is only possible over about 1 Å, much less than electron delocalization.

Second, an arginine, if it moves into the headgroup region of the membrane, will be held in a complex with the phosphates of the membrane lipids. This makes optimizations with S4 fixed more appropriate than with S4 free; we consider this further in Results, section b, and Discussion, section c.

We have also examined the way the pore responds to opening; this requires that an incoming $K^+$ be held, in effect complexed at the entrance to the pore, to allow the incoming ion to push an ion in the cavity of the pore toward the selectivity filter, rather than have that ion push the incoming ion back out [25]. It appears that the location where the ion is held may be at the top of the T1 moiety that hangs below the membrane, below the PVPV conserved sequence at the membrane surface.

Proton tunneling [26], which would correspond to the "piquito", the rapid rise time initial component of gating current [27] may need to be considered; to account for the piquito it has been proposed that there is discontinuous movement of side chains in a complex energy landscape [28-29]. However, the tunneling proposal provides a mechanism that directly leads to the observation. Such an initial step in the gating current also suggests the crossing of a threshold as the membrane depolarizes, allowing the energy levels of the initial and final proton positions to match, as needed to allow tunneling. If there are multiple channels, with the threshold energy (potential) distributed (almost exactly a Gaussian distribution) over channels with a width in energy of about thermal energy ($k_BT$, $k_B$=Boltzmann's constant, T=temperature), one can reproduce the observed current-voltage curve as well as the "Boltzmann" curve does [30].

Experimental indications that proton transport in the VSD should be considered include:

It is known that substituting histidine for one of the end arginines allows a proton current to pass completely through the VSD [13-14, 31-32]. This tells us that proton passage is possible, although it remains to work out the path.

The $H_v1$ channel, which transmits protons, has a structure very similar to that of the VSD in about half the channel, and with a structure that follows a similar pattern in the remainder, except for a section in which the protons appear to be redirected, toward the pore for the VSD and to the membrane surface for $H_v1$. A recent review by deCoursey et al [33] discusses the evidence pointing to the proton path; they conclude that there is a small shift of S4 ("one click"), corresponding to a single exchange of partners, to produce gating. However, their evidence could be understood even without this shift. $H_3O^+$ may also be involved[34-35]. A substantial controversy as to the proton transport mechanism of $H_v1$ has grown over the last two years, which we do not have space to review; we believe a Grotthuss mechanism is not necessary for proton transport, and see an alternate path that does not require water. For the purposes of our paper, it is only necessary to note that the $H_v1$ contains groupings of amino acids similar to those in the $K_v1$ channel, and these groupings appear adequate to provide proton transport without invoking water or a Grotthuss mechanism.

Zhao and Blunck have recently shown that a truncated VSD forms a cation channel with a strong preference for protons [36]. While not a surprise in itself, and not entirely incompatible with standard models, this finding adds to the evidence that protons can, and should, move through the VSD.

Proton transfer has also been reported in other membrane proteins, including cytochrome c (cyt c)[37], where a mechanism related to what we suggest here appears possible. (Proton transport through M2 of the influenza virus may use a somewhat different mechanism[38-40], and we do not discuss it here).

Perez-Gonzalez et al showed that substituting charges of either sign reduced voltage sensitivity, which is more easily understood if it affects proton transport than with electric field remodeling in standard models[41] (however, the authors interpreted their results in terms of standard models).

The majority of the path is not a water wire. It is necessary to determine how the $H^+$ moves, but the evidence cited above does appear to show that it does move (there may be a need to explain separately how certain R→H mutants turn the VSD into a complete $H^+$ channel, but for purposes of understanding how the VSD transmits protons internally (capacitative gating current), it is not necessary to understand how the $H^+$ passes the intracellular and extracellular ends of the VSD.

There have been extensive MD studies on VSDs; only a sample of recent studies is referred to here [42-44]. A very recent example uses the Drude polarizable potential on a model that more nearly resembles a $Na^+$ channel[45]. Papers with MD calculations may also include experimental data. However, classical force fields cannot usually describe proton transfer, nor do they get hydrogen bonds consistently correct, when different length hydrogen bonds are involved. The new Sun and Gong paper with Drude potentials proposes a partial trajectory for S4, but does not agree with other MD simulations. Including polarizability should improve its accuracy to some extent, but it still could not allow for proton transfer, hence, like almost all MD simulations, cannot speak to the question of whether it exists. There is another problem with MD calculations: gating transitions must eventually return the channel to its original state, albeit not necessarily by the same path; there is evidence of hysteresis[46], so there must be alternate paths. One expects channels to live for tens of thousands of cycles of opening and closing before the channel protein is replaced (e.g., if there is a ten minute minimum lifetime, with the channel opening at about 10 Hz on average—and some channels operate more rapidly than this—with 4 domains, this means 24,000 refoldings). Whatever transitions are made in gating, it must be possible to return with essential certainty to the original. The standard models, with S4 moving more than 10 Å, effectively unfold the protein, or at last a substantial section of it. Refolding with the high accuracy required, without the protein being trapped in a local minimum in any cycle, seems very difficult. Known membrane protein refolding appears to require chaperonins or equivalent [47-50]. Enzymes and receptors that do undergo conformational transitions and refold tend to show hinge-like motion, rarely destroying secondary structure. Ion channels like $K_v1.2$ do not appear to have any long-term chaperonins; an initial insertion chaperonin would not help here. At this point this is only a plausibility argument. However, the idea of a large scale unfolding/refolding event to open the channel does look somewhat implausible; the versions that have been proposed require major unfolding of the protein, internally, not hinge-like changes. This would have to be followed repeatedly by refolding, possibly along a different path, without the protein becoming trapped in a local energy minimum. So far, we have not seen MD simulations, even with a polarizable potential, that show that they can return the protein reliably, in multiple repeats, to the original state. This said, it is not impossible to conceive of the channel refolding accurately; possibly the stationary parts of the VSD somehow also act as chaperonins, or there is some other mechanism for a mobile section to be guided back to its original position. At the moment, possible difficulty in refolding must be accounted for (*e.g.,* if the neighboring segments also act as chaperonins, how do they do so?), but is not definitive in ruling out the standard model. It is surprising that the question does not appear to have been addressed in previous work.

Here we present the results of quantum calculations on larger sections of the VSD than any previously reported. These calculations are consistent with the idea that proton motion constitutes the gating current. We propose a specific path for part of this motion, based directly on the calculation, with determination of the energies of the states of the VSD with the proton in the proposed sequence of positions. The conformation of the VSD is optimized in the quantum calculation, and negligible motion of arginine backbone atoms is found. One proton transfer studied here in detail is sufficient to account for 20-25% of the gating current. In contrast, the motion of the S4 backbone is of the order of 1-2 Å, and not entirely in the correct direction for it to contribute to gating current. Furthermore, the conventional attribution of the ionization states of the salt bridges leads, in our calculations, to positions of the salt bridge components that differ with the X-ray structure at 0 V; however, switching proton positions, hence ionization states, from ionized to neutral salt bridges, leads to interatomic distances that are a good match for those determined by X-ray structures. Since the X-ray structures do not show the protons, probably the most accurate way to get these ionization states is to use quantum calculations. We determine whether the energies of the proton-transferred cases are lower or higher than the energies of the corresponding ionized salt bridge structures, which depends in part on the applied external field. Furthermore, the structures that transfer protons often must be considered in groups of three to five residues; one member in a group we have calculated is a phenol, tyrosine. A millisecond rate of proton exchange for a tyrosine has been shown to be possible by Takeda and coworkers[51]. These structures appear in the results of the calculations for the upper (extracellular) part of the VSD. Three calculated cases are shown in Fig. 1, and in a different form, Fig. 2. The remainder of the path leading to the gate remains to be calculated, and these calculations are in progress.

CALCULATIONS: We have carried out optimizations (i.e., energy minimizations) with 976 atoms (24 water molecules, 904 protein atoms from 43 amino acids, 3700 electrons, and 8,178 basis functions; of the protein atoms, 426 are hydrogen, 478 are "heavy"), varying four parameters: *i)* field; *ii)* S4 free/fixed; *iii)* proton in the salt bridge involving R297 shifted, and in one case with a proton shifted from R300; *iv*) initial direction of the average dipole of the water cluster (this turned out to make relatively little difference, so it will not be discussed further). In each calculation, 42 atoms from S1, S2, and S3 are fixed, all from the ends of the segments. In the *S4 free* case these were the only fixed atoms. In *S4 fixed* cases, 19 atoms from the ends of the S4 segment were also frozen, so that S4 could not move vertically. While it might seem that this defeats the purpose of the computation, comparison of S4 free, S4 fixed, and the X-ray structure shows that the S4 fixed open state reproduced the local X-ray structure slightly better than S4 free. Even when S4 was free, it failed to move in a manner that could produce gating current; for the backbone atoms, the fixed-free differences were essentially always <2Å. We will discuss this below, but it appears that S4 fixed is justified. Water molecules were always free. The fixed-free comparison shows the interactions with S1, S2, and S3 were sufficient to keep the S4 in place, or nearly so. One additional case could be of interest: S4 free and the field set at +$10^7$ Vm$^{-1}$, equivalent to +70 mV across the membrane. (For all calculations, $10^7$ V m$^{-1}$ is considered equivalent to 70 mV across the channel; the field is applied as either 0.5 or 1 x $10^7$ V m$^{-1}$.) The +70 mV should drag S4 in the opposite direction, not only open but pulled even further outward, to see whether in fact S4 could respond to such a field, or whether it would be trapped in its normal position. This large a positive field is on the edge of what is possible in many experimental preparations; still, the S4 backbone moved <2.5 Å (mostly not in a manner that would produce gating current). Hydrogens were added to the initial file using gOpenMole, and then optimized with the rest of the structure.

All optimizations were done at HF/6-31G* level, using Gaussian09 (versions D or E) [52]. The 976 atom calculations included the full amino acids shown in Table 1. Coordinates of 14 representative optimized cases out of 30 total are in the supplementary material.

**Table 1:** Amino acids with full side chain included in the calculation[+]

| S1 | S2 | S3 | S4 |
|---|---|---|---|
| A166 | T219 | M255 | I292 |
| S169 | F222 | N256 | I296 |
| V170 | V225 | I258 | R297* |
| V172 | E226* | D259** | V299 |
| I173 | T227 | A262 | R300** |
| S176 | C229 | I263 | R303 |
| I177 | I230 | P265 | K306*** |
| S179 | F233 | Y266 | L307 |
| F180 |  | T269 | S308 |
| C181 |  | L270 | R309*** |
| E183* |  | G271 | H310 |
| T184 |  |  |  |

*salt bridge;  **,***apparent triads, or larger groups

+all other amino acid backbone atoms are included; side chains are truncated for amino acids in the section calculated, but not listed in the Table.

All side chains that point in, and can interact, and possibly contribute to a proton pathway, are included. 976 atoms stretched the limits of this type of calculation with the computers available to us. An earlier computation included only 672 atoms, truncating more side chains, allowing side chains and some backbone atoms to bulge into vacated space, and producing physically extremely improbable conformations (for one thing, it would be impossible to re-fold these conformations), invalidating the results. The 976 atom case did not allow this; its structure appears to be stable. Voltage made small structural differences, principally to side chains, that turned out to be within the range that could reasonably be expected to return to the initial state with only proton transfers. F233 is included; for comparison, six of the seven hydrophobic amino acids discussed in the paper by Schwaiger et al [53], emphasizing the role of F233 in gating, are included here, the only exception being I231 (We note that Schwaiger et al had the phenylalanine ring rotate after a translation of only 1 Å). The result in our calculation has the phenyl group motion limited to a very small translation of the side chain. A smaller, 311 atom calculation centered on F233 showed very little motion of this side chain. While a $H^+$ - phenyl π-electron complex should exist, it would be relatively weak; however, if it exists, the complex should not be too strong or the proton could not transfer to the next residue in the chain. It does appears that such complexes are of about the right strength for the necessary transfer [54-57]. However, it is possible to see a path around F233,

The structures from HF calculations are accurate enough to use, judging from comparisons with the X-ray structures. The errors do not affect interatomic distances by enough to change any conclusions. However, energy values provided by HF calculations may be inaccurate by multiples of $k_BT$, even for differences in energy in which most errors cancel. Therefore, single point calculations on the HF optimized structures were carried out at B3LYP/6-31G** level; these required approximately one day each, and are much more accurate. They include the exchange and correlation energy, quantum mechanical effects; the correlation energy is also missing in the HF calculation, and accounts for energy differences greater than those that are crucial for the proton transfer. See Table 6 for the values of these quantum contributions to the energy. The differences in the sums of the quantum terms are comparable in magnitude to the total differences between open and closed (*i.e.,* several $k_BT$), and they reverse sign when going from open to closed, so that omitting them would produce results with approximately 100% error in the magnitude of the open-closed transition energy.

Proton positions were mostly inserted in the initial coordinates, as there are multiple local minima; there is usually a local minimum at each acid, base, and salt bridge, so all those that were tested were selected in the initial file. However, in the smaller calculations surrounding the F233 residue, a proton transferred during the optimization from K309 to D259. This happened in all of the several versions of the calculation, indicating that not all local minima were so deep as to trap the proton. See the section on local vs global minima, below.

Absolute values of energy are not used to draw conclusions in this work; only energy differences are compared, so there will be cancellation of systematic error for almost all the system, only a very small part of which changes in single proton steps, or with application of voltage. With quantum terms included, differences are reliable enough to use in forming a proton path. These energy values are used in the comparison of the structures, in particular in answering the question of whether the proton transfer to a neutral structure is more or less stable than the ionized salt bridge structure. Interatomic distances, for some specific pairs of atoms, show how much, or little, the backbone moved. The distances are compared with those in the X-ray structures for certain key pairs, in Table 3, showing the X-ray structure is closer to the S4 *fixed* than the S4 *free* cases (for a more detailed discussion of this point, see Results, part (b), and Discussion, part b).

It was also possible to examine the contributions to the energy from classical and exchange and correlation terms, the latter two being strictly quantum mechanical. While they are small compared to the classical terms, they are much larger than $k_BT$, hence important in deciding which configurations of the proton are of lowest energy, and by how much. With these terms included, it is possible to see a reasonable proton pathway. The energy contributions are considered in the Results, part (j).

The charges on individual atoms were determined by NBO calculations on the optimized structures (using B3LYP/6-31G**). The results for atoms of particular importance are shown in Results, part (f).

**Local minima vs. a global minimum**: The calculations presented here do not allow us to conclude that a global minimum has been achieved in any single calculation. What is more, this may not even be a meaningful question. We need the positions of the protons in a configuration space defined by the possible proton positions and potentials. In order to know which is most likely to be occupied at any given voltage we must do the energy minimization with relaxation of the protein; all the protein side chains are free to relax, so that any local changes are included in the computation. Each plausible configuration of protons on acids and bases was chosen in turn, and at different voltages. It is not the case that any particular proton position choice was the global minimum at all voltages. This had to be tested in order to show that the position of the proton minimum was in fact voltage dependent. Once we had explored a sufficient fraction of the configuration space, and found the potential at which proton transfer occurred, we could be reasonably sure that we had the step in the path successfully determined. We will see that the transfer steps almost all involve three or more amino acids. We have included so much of the VSD (including all of the local side chains, fully relaxed) that we can resolve relatively small differences. In this work we can determine at which points a proton must cross to the next step, and at what voltage/field. Experiment can tell that protons have passed through the VSD, but not the path taken. To understand how the proton passes through, and what possibilities exist for mutations to test a path, a determination of the configurational space, and the effects of voltage within it, is required. This can be done by systematically exploring the plausible low-lying (in energy) parts of this space. The global minimum is not relevant when the steps along the path are determined.

A related question appears precisely because these are energy minima, thus necessarily states that exist at T = 0 K. For comparison with X-ray results this is not a problem, as the X-ray structure determination is done at a temperature below the lowest phase transition of water, hence effectively at zero temperature (even though the actual temperature is higher, it is not enough higher to alter structure). The transitions between states that we are attempting to determine take place at temperatures of biological organisms; these structures may show some temperature dependence. However, we believe this will not seriously change any conclusions. For one thing, the X-ray structures appear to be a rather good representation of the actual structures, so the T= 0 K structures we determine, starting from the X-ray structures, should be comparably good. Second, by obtaining multiple minima, we are able to find the *differences* in energy rather accurately; given that the differences in structure on moving a proton are not large, the vibrational spectrum should not differ greatly from one minimum to another. It is the vibrational spectrum of the complete system that determines almost all the temperature dependences; the local contribution is small. As a result, the energy differences, on which the interpretation depends, should be fairly reliable. The added error due to finite temperature is expected to be of the order of $k_BT$, while the energy differences on proton transfer, and with field, are of the order of 10 $k_BT$.

The local/global minima and temperature problems that arise in this type of calculation must be considered; running multiple cases, as we have done, is expected to be not merely necessary, but sufficient, for deciding the order of energy minima, and the possibility of proton transfer. The expected errors are about an order of magnitude smaller than the size of the effect we calculate.

**RESULTS**:

We have calculated the optimized structures and energies for those cases listed in Table 2. A key result is the S4 free, negative voltage (closed state), with all salt bridges ionized; this is the state that the standard models assume must produce a large downward shift of the S4 domain. No such shift is found in these calculations. Fig. 1 shows the X-ray and optimized structures superimposed; there is almost no motion of S4, except for the side chain of I292, which moves in the wrong direction. Table 3 gives quantitative results (part b below).

In all cases the protein starts from the X-ray structure (pdb: 3Lut) which has protons added by gOpenMole. In the cases labeled "proton shift" in Table 2 at least one salt bridge proton has been transferred "by hand" (*i.e*, in the initial structure), as labeled. As many cases were done as were needed to suitably investigate proton transfers in the plausible energy range in the configurational space defined by voltage and proton positions (see Table 2). Enough cases were optimized and their energy determined to allow us to conclude that we could determine a proton pathway through the upper part of the VSD. The side chains throughout the VSD are free to relax, as are all the backbone atoms save those at the ends of S1, S2, and S3. Therefore, the immediate neighborhood of all the transitions considered are as realistic as possible.

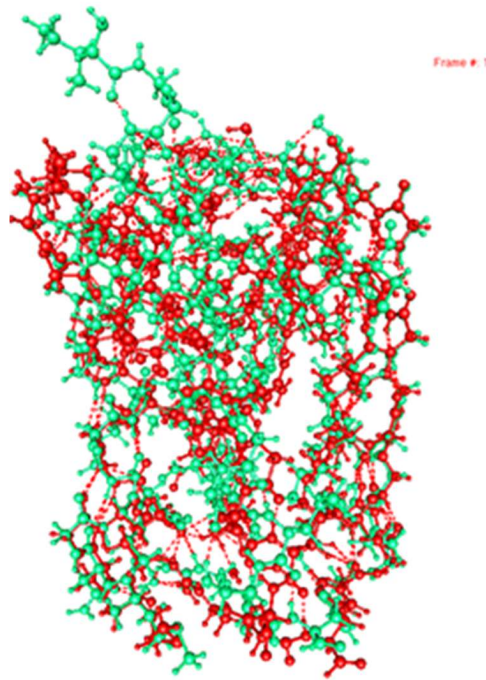

*Fig. 1: Comparison of X-ray structure (considered open) (red) and optimized V= -70 mV (therefore presumably closed) structure (green). Optimized structure differs from X-ray structure by ≈1 Å, generally less (see Table 3). The only exception is the I292 side chain, which, absent other extracellular atoms, pops into open space. S4 shows no intracellular motion at all. The top arginine, R294, and surrounding residues are missing.*

The key question for the comparison with standard models is whether there is an S4 shift. We ran enough cases with S4 free to allow comparison with standard models to be possible. Having found that the S4 motion was nowhere near what standard models would require (see Table 3A), we switched to S4 fixed, which is almost certainly a better representation of the actual situation (see Section (b) below, and Discussion, section (b), for a more complete justification of fixing S4.) The results in Table 3 make it clear that the backbone motion was not different enough with S4 free for it to matter; side chain motion, not necessarily perpendicular to the membrane, was larger, especially for the acids in S2 and S3, whether fixed or free, but, as just noted, the fixed results are likely to be slightly more accurate. Comparison with S4 fixed (not shown) shows that there is hardly any difference in the motion, so that free vs fixed differences can be ignored.

**Table 2:** Optimized Cases

| S4 | Electric potential* | Proton shift** |
|---|---|---|
| Free | -70 | R297 – E183 |
| Free | -70 | None |
| Free | 0 | None |
| Free | +70 | R297 – E183 |
| Fixed | 0 | None |
| Fixed | -70 | None |
| Fixed | -70, -35 | R300 – E183 |
| Fixed | -70, -35 | R297 – E183 |
| Fixed | -70, -35 | Y266 -- R300 |
| Fixed | -70 | R297 – E183/Y266 – E226 |
| Fixed | +70 | None |
| Fixed | +70, +35 | R297 – E183 |
| Fixed | +70, +35 | R300 – E183 |
| Fixed | +70, +35 | Y266 - R300 |
| Fixed | +70 | R297 – E183/Y266 – E226 |
| Fixed | 0 | R297-E183 |
| Fixed | -70 | Y266-E183/R300-E226 |
| Fixed | +70 | Y266-E183/R300-E226 |
| Fixed | +70, +35 | Y266 – E183 |
| Fixed | -70, -35 | Y266 – E183 |
| Fixed | 0 | R300 – E183 |
| Fixed | 0 | Y266 – E183 |

*The electric field magnitude is 1 x $10^7$ V $m^{-1}$ along the z-axis, inside negative (closed), or positive (open), corresponding to 70 mV across the membrane, or else 0.5 x $10^7$ V $m^{-1}$ for 35 mV across the membrane; this is the average field, while the local field depends on local charges. If the field is zero, this should reproduce the X-ray structure, presumed to be open. The positive field would correspond to about a potential of +70 mV across the membrane, near the limit of what is possible, but should keep the open conformation.*

*\*\*Proton shifts are from the residue on the left of the pair to the residue on the right. If 2 pairs are shown, two protons were shifted. "None" means that all arginines are positive, all glutamates negative, and Y266 is neutral. For example, R297 – E183 means that the proton shifts from R297 to E183, such that both arginine and glutamate become neutral; if Y266 is shown as the proton source, that residue becomes negative.*

**Table 3 shows the results of calculations with S4 free to move** against the S1,S2,S3 bundle in terms of distances moved compared to the X-ray structure. The results show essentially no motion where "motion" is defined as differences in coordinates of two optimized configurations. If the field is negative, in standard models the backbone of S4 must be displaced inward to reach the closed state. The results in Table 3A show that this does not happen; there is almost no difference among field negative, positive, or zero. Positive voltage optimizations (equivalent to +70mV across the membrane) were calculated, and are included for completeness. With positive voltage, in principle, the results should be the same as for zero volts, since both should correspond to the same open state. In fact, within a few tenths of an Angstrom, they do agree, and both agree within <2 Å with the X-ray structure.

**Table 3:** *Distances* (Å) *moved by backbone* N *atoms with various field and salt bridge ionization states, from the X-ray structure positions.*

| Displacement/ amino acid- N atom | V < 0 (SB I) | V < 0 (SB NI) | V = 0 (SB I) | V > 0 (SB I) | V > 0 (SB NI) |
|---|---|---|---|---|---|
| R297 | 5.1 | 1.9 | 1.8 | 2.4 | 1.8 |
| R300 | 1.7 | 1.4 | 1.2 | 1.4 | 1.3 |
| R303 | 1.8 | 1.4 | 1.2 | 1.6 | 1.3 |
| K306 | 2.1 | 1.1 | 1.0 | 1.6 | 1.0 |
| R309 | 2.1 | 1.0 | 0.6 | 1.4 | 0.9 |
| H310 | 2.2 | 1.5 | 1.2 | 1.4 | 1.4 |

*V < 0 corresponds to the closed state, intracellular 70mV negative; V > 0, intracellular 70 mV positive. Two cases have V = 0. Results with 35 mV were smaller than with 70 mV, as expected.*
*SB = R297–E183 salt bridge: if I, H$^+$ is on R297, making the salt bridge ionized; if NI, H$^+$ is on E183, making the salt bridge neutral (uncharged).*

When we go to negative voltage, with S4 free, the motion that corresponds to the gating current (intracellular direction, as the channel closes) in standard models is not found; very little motion of any kind is found. The only exception is R297, which moves ≈ 5 Å (and there is appreciable motion of the side chains especially of nearby hydrophobic residue I292); while adding nothing to gating current (the motion is not in the right direction) it does show that when S4 is not constrained, the conditions of the calculation do allow motion. Nothing like the motion required by standard models is found. Motion in earlier 672 atom calculations was substantial, but into the empty space beyond the edge of the limited system. Because the 672 atom calculations gave results that were obviously erroneous, they are not further discussed here (what motion there was did not agree with any standard model either). With 976 atoms, the structure is maintained. An arginine-phospholipid complex would also prevent S4 from moving. (See Discussion, section b). Some side chains do rotate, affecting local interatomic distances, and thus affecting the probability of proton transfer. We cannot predict that an R294 mutation would necessarily make a huge difference to the channel, as the S4 may also be tied down by another means, such as salt bridge/triad/pentad interactions with S1/S2/S3. Fig. 1 shows S4 free with negative voltage, compared to the X-ray structure, showing little change in structure, almost none for the backbone; Fig. 2 shows results for the lowest energy structures with no voltage (A), and with negative voltage (B), both with S4 fixed. The shifts in proton position among Y266, R300, and E183 define the low energy cases. Fig. 3 shows the main energy relations among the V=0 and V= -70 mV cases, with the proton shifts labeled.

**Table 4:** *Energy values of cases in which Y266 does/does not transfer a proton*

| Charge: Y266,R300,E183** | Energy*: V < 0 (closed) | Energy*: V = 0 |
|---|---|---|
| 0,+,- | -700 | -700 |
| 0,0,0 | **-727** | -691 |
| -,+,0 | -695 | **-711** |

*+25111 Ha (*i.e., E=-.700 means total energy = -25111.700 Ha*)

** The three charge states shown are those of Y266, R300, and E183 respectively. R297 was always +. This shows the initial position of the proton, which was free to shift during the calculation. In no case did the proton transfer back completely. (0,+,-) means Y266 is not ionized (charge=0), R300 and E183 are ionized (+,-). Boldface: low energy states for closed and open cases, showing position of the protons in each case, and therefore the proton transfer associated with opening the channel. Units of energy are mHa, 1 mHa≈$k_B$T at 35°C; only the differences are significant (the differences are more reliable as the errors almost entirely cancel in very similar systems).

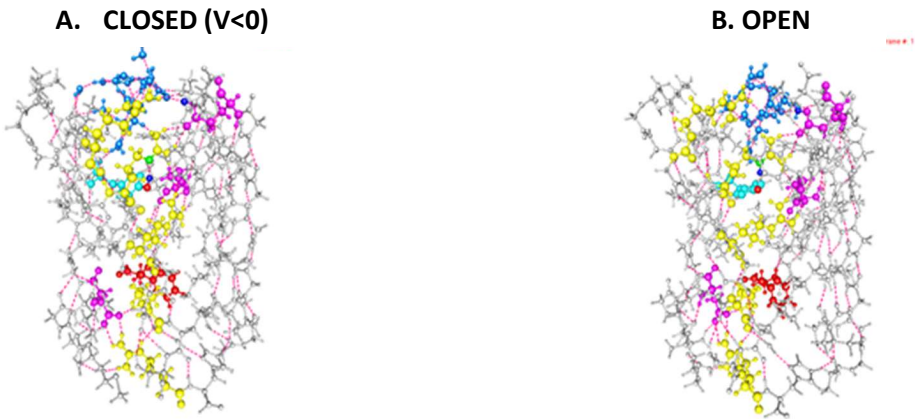

**A. CLOSED (V<0)**   **B. OPEN**

***Fig. 2:*** *Optimized structures, S4 **fixed**: A) E183 protonated, V<0 (closed) B) E183 protonated, V=0 (open); **A)** proton neutralizing E183 comes from R300, which is neutral, while Y266 is uncharged; **B)** the proton comes from Y266, which becomes an anion, and reaches E183 by way of R300, which remains positively charged. The proton giving R300 charge resides on nitrogen NE, not on the $NH_2$ amines of the guanidinium. Colors: yellow, arg; magenta, acids; red, F233; light blue, Y266, deep blue, water, with end atoms of S1/S2/S3/S4 all fixed. The energy values, given in Table 4 (see also Fig. 3), show these to be the minimum energy cases in the open and closed paths. There is appreciable rotation of F233 between open and closed cases, and the relative arrangements of Y266 (pale blue) and E183 also change between open and closed cases*

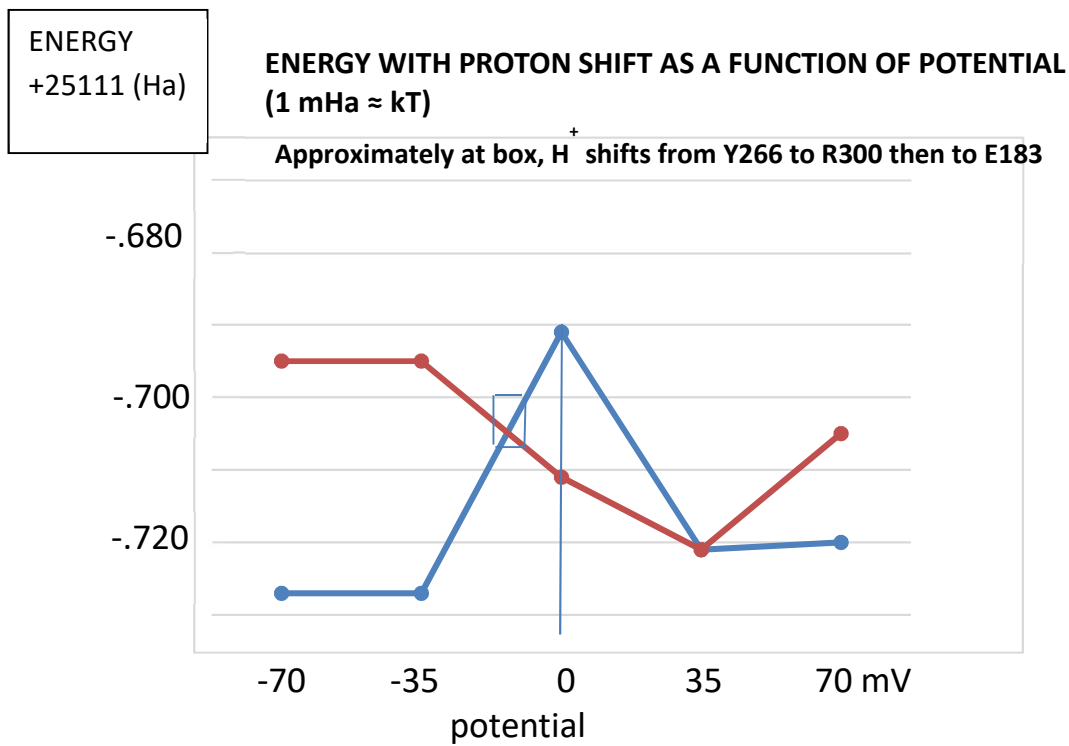

**Fig. 3:** *Energy shifts corresponding to proton transitions. As in Table 4, E+25111 Ha. The box shows approximately where the proton of this part of the VSD transfers R300 → Y266; Table 4 gives the energy values for three closed and three open states (Notation: as in Table 4, **ORANGE CURVE: -,+,0 = Y266 negative, R297 positive, E183 neutral, BLUE CURVE: 0,0,0 = all neutral**) See Table 6 for the classical and quantum contributions to the energy.*

**Figs. 3 and 4 illustrate a key shift by the protons**. See also section (h), below. Several other transfers were calculated, but led to transitions to states that were higher in energy. The lowest energy in Table 4 is for the closed state, with the proton on the glutamate from R300, making it neutral. The lowest energy open state is approximately 17 $k_BT$ above this state, suggesting that for this state at least there is a less than 40 kJ activation energy, possibly appreciably less if the 35 mV results are considered. This is not surprising, given the experimental range of temperature dependence.

**Salt bridge ionization:** The fact that some of the salt bridges need not be ionized is not a surprise: water is required to ionize a salt bridge (other polar solvents are not relevant here, and hydrophilic residues are not placed appropriately to help with the ionization). One water is not enough, two may be sufficient, and three certainly are[23]. In the center of the VSD, there is a hydrophobic region in which there is no water, and we do not expect ionization, so shifting a proton to neutralize the two components of the salt bridge at that point is expected. Furthermore, it is possible to find structures that suggest the type of proton resonance that appears in much simpler calculations of isolated salt bridges with 2 or 3 water molecules[24]. Such structures help to stabilize the charge transfer, and the seemingly anomalous pK values. These are quantum effects, and would not appear with any form of classical potentials, as far as we can determine.

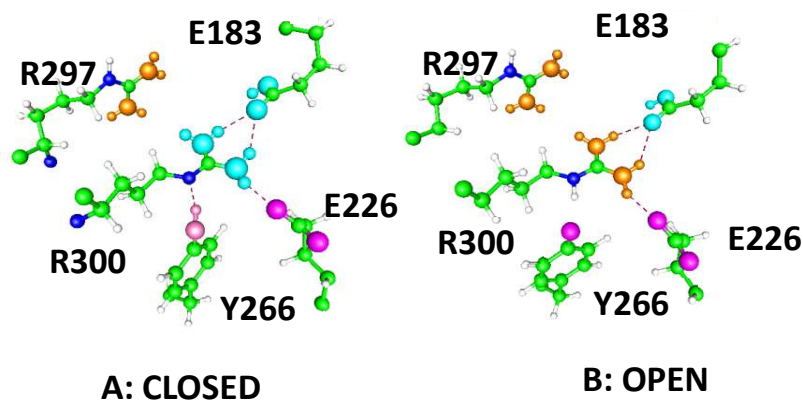

**A: CLOSED**  **B: OPEN**

*Fig. 4: The key proton transfer involving Y266 for the energy diagram in Fig. 3; the proton moves from the tyrosine in the closed state to the nitrogen on the R300 in the open state. The guanidinium group is shown as blue when R300 is neutral (A) and gold when it is charged (B).The shift also influences the position of the side chain of E183, so that a two step shift (Y266 →R300→ E183) can occur, with R300 serving as pathway; however, E183 appears to be neutral in both open and closed states. In addition, when the proton is on R297 (as shown for open) the Y266 sidechain rotates downward toward R303 at almost the same energy, so that the position of Y266 is not as well defined as in the closed state, where Y266 is locked, in the position shown in the figure above. This also suggests another possible step in the chain of proton transfers, which however is just outside the range for which the calculation has been completed.*

**Charges**: Table 5 gives atomic charges under various conditions. Charges on key groups: arginines, glutamates, and tyrosine, are non-integral; the electron wave functions produce a charge distribution among the atoms that shows the fraction of charge on the various groups. There is shared charge.

**Table 5a:** Charges from NBO on Y266 and glutamates*

|      | 0 field; R300(+) | V>0, R300(+) | V>0, R300(0) | V<0, R300(+) | V<0, R300(0) |
|------|------------------|--------------|--------------|--------------|--------------|
| Y266 | -0.241           | -0.843       | -0.247       | -0.843       | -0.248       |
| E183 | -0.742           | -0.014       | -0.016       | -0.009       | -0.013       |
| E226 | -0.720           | -0.733       | -0.729       | -0.732       | -0.728       |

*For Y266, the charge is on the O if ionized, (so R300(+)); if it is not ionized (so R300 (0)), the H is left mainly on E183; for glutamates, charge is the sum of that on the carboxyl oxygens plus the carbon to which they are bound, plus the proton if H is on the carboxyl (not ionized case, which applies to all except the first column; the E183 proton comes from Y266 in columns 2 and 4, and from R300 in columns 3 and 5).

**TABLE 5b:** Charges from NBO on arginines*

|         | 0 field; R300(+) | V>0, R300(+) | V>0, R300(0) | V<0, R300(+) | V<0, R300(0) |
|---------|------------------|--------------|--------------|--------------|--------------|
| R297(6) | 0.144            | 0.158        | 0.162        | 0.147        | 0.160        |
| R297(7) | 0.829            | 0.838        | 0.842        | 0.835        | 0.840        |
| R297(9) | 0.635            | 0.651        | 0.653        | 0.655        | 0.653        |
| R300(6) | 0.126            | 0.091        | 0.070        | 0.093        | -0.071       |
| R300(7) | 0.813            | 0.778        | 0.572        | 0.781        | 0.570        |
| R300(9) | 0.620            | 0.622        | -0.094       | 0.624        | -0.095       |
| R303(6) | 0.142            | 0.116        | 0.135        | 0.111        | 0.134        |
| R303(7) | 0.838            | 0.798        | 0.817        | 0.793        | 0.816        |
| R303(9) | 0.627            | 0.577        | 0.599        | 0.577        | 0.598        |

*Charges on each arginine are summed over 6, 7, or 9 atoms: 6 sums over both $NH_2$ groups, or $NH+NH_2$ if one proton has transferred to another group; 7, add the central carbon; 9, add the preceding (nearer backbone) NH. i.e., if the $H^+$ from arginine has transferred to another group, the sum over two $NH_2$ groups becomes the sum over $NH + NH_2$ and the number of atoms becomes 5, 6, or 8.

We see first that the charges are not particularly close to integral, meaning that the electrons are spread out over several atoms, and the charges likewise. This is not a surprise, as comparison with a particle in a box would suggest that a bigger box means a lower energy. Concentrating charge should increase electrostatic repulsion. As a result, we ought to expect *a priori* a result like the one that is found, rather than integral charge on groups. This does have implications for proton transfer, as the electronic wave function forms a partial bridge between groups. There may be some effect of delocalization error[58] (charges are calculated with B3LYP, a DFT method), but this is not expected to be large enough to affect the overall result.

It is not required that the path followed by the protons is the same for the open→closed and the closed→open transitions. Instead, it appears more probable, based on the charges that must exist at intermediate states, combined with the field, that two different paths are required. It is known that there is hysteresis in the gating transitions [46]. The existence of multiple states from closed to open has been suggested on the basis of kinetic data on channel opening by many workers, which supports the idea that more than one path may exist.

**Role for Y266**: The shifts in pK values that appear to exist, based on these calculations, are not simply a consequence of local electric fields. We have been discussing Y266, which, in the open state, moves so as to place its –OH group near an R300 nitrogen of the guanidinium that has lost its proton. The hydrogen from the Y266 –OH group is partially shared with this nitrogen, making the R300 deprotonation incomplete, and allowing a short chain of partially bonded protons to form. Thus the R300 deprotonation does not require so large a pK shift as it would in solution. Fig. 3 and Table 5 show how this affects energy and charge, and Fig. 4 shows the local structural relation. This step is itself an intermediate step in gating. States with high energy are not discussed. In the following, the nominal charge is in parentheses, corresponding to proton position. The overall system is unchanged except for the proton transfers. The summary of these steps is that a charge has moved from Y266 up to E183 upon depolarization. In the closed state the ionized Y266 corresponds to high energy, and in the open state to low energy, so the proton must transfer in opening the channel. The transfer is in the correct direction (proton moves upward).

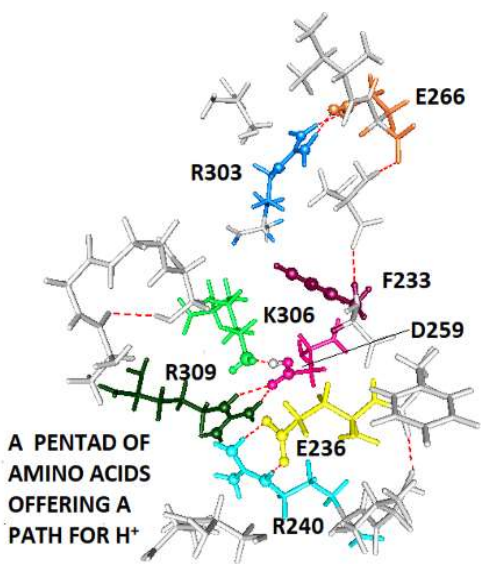

**Fig. 5:** *This figure shows a possible next step in proton transfer. The proton that is between K306 and D259 always moves to D259 in the calculation. This section also illustrates a pentad of amino acids that is predicted to provide a path for a proton transfer. The role of F233 is not entirely clear from this result. We need a path for a proton, and a reasonable hypothesis is the path R303-K306-D259, going around F233. Further work will be needed to test this suggestion. It is conceivable that F233 is required to block a path for a water molecule. This section is being included as part of a larger calculation, in work now under way.*

**Charge shifts on opening the channel:** Combining the results shown in Table 5 with those in the preceding paragraph, we have the change in charge on the relevant residues on opening the channel. This shows approximately 0.5-0.6 charges having transferred, in moving from Y266 to R300, so that this step provides about 20-25% of the gating current for the entire VSD. This is not unreasonable for a single (set of) step(s). It is consistent with the need for at least one more H$^+$ transfer (probably 2), and a possible contribution from the dipoles, but these latter transitions remain for future calculation.

**Role for F233**: Several side chains were fairly mobile. F233 should be particularly important. To further examine the role of this residue, a series of smaller calculations (311 atoms) with F233 and neighboring amino acids, with and without applied negative voltage, were carried out, and the motion of the F233, as well as the proton transfer from K309 to D259, was determined. There were no major changes other than this. In particular, there is only limited rotation of the F233 phenyl ring itself in these results. The proton transfer is the only obvious major response, although some minor other changes also occurred. Figure 5 shows the amino acids that were calculated, and illustrates the pentad that is expected to transfer a proton to another group further along the path towards the gate. The K309-D259 proton transfer is within this pentad.

**Exchange and Correlation Energy**

Classical, exchange and correlation energy terms for the two lowest energy cases are shown in Table 6. The quantum terms make a difference of the order of 0.010 Ha ≈ 10 k$_B$T (equal at 35$^O$C), hence are significant on the scale of deciding which states are most stable at room temperature. The upper line in both parts of Table 6 corresponds to the orange curve in Fig 3, the lower line to the blue curve.

**Table 6a:** Energy (Ha) of the two low energy states with channel closed (-70mV)

|  | Total + 25111# | Sum classical terms +22421 | Exchange (x) +2526 | Correlation (c) +164 | Total xc +2690 |
|---|---|---|---|---|---|
| Y266→E183* | -.695 | -.538 | -.077 | -.080 | -.157 |
| R300→E183** | **-.727** | -.567 | -.073 | -.087 | -.160 |

**Table 6b:** Energy (Ha) of the two low energy states with channel open (0mV)

|  | Total + 25111# | Sum classical terms +22421 | Exchange (x) +2526 | Correlation (c) +164 | Total xc +2690 |
|---|---|---|---|---|---|
| Y266→E183* | **-.711** | **-.523** | -.098 | -.090 | -.188 |
| R300→E183** | -.691 | -.518 | -.088 | -.086 | -.174 |

*Y266 is deprotonated   **Y266 is protonated, R300 is neutral  #For the total energy, add 25111 Ha, (e.g., the total energy for V=-70mV, Y266→E183 is -25111.695 Ha). For each column, add the amount of energy shown in the heading. Energy units can be understood as 0.001 Ha ≈ $k_BT$

The differences between V= 0 mV and V = -70 mV are significant, the sums of the two purely quantum terms (exchange + correlation) being different (ionized (0mV) – unionized (0mV)) for open channel by 14 $k_BT$, and by 3 $k_BT$ (two closed states, favoring unionized). The relaxed form (0 mv, unionized) has the lowest exchange plus correlation energy, and presumably corresponds to the normal open state. These differences are comparable, or even larger than, the total energy differences, so that ignoring the quantum contributions to the energy would lead to an error approaching a factor of two in the energy differences.

The exchange and correlation energy contributions are small compared to the one electron, electron-electron repulsion, and nuclear repulsion energy contributions, but not small enough to be neglected; the *differences* are tens of $k_BT$, hence of considerable importance in determining which states are lowest in energy, and the size of the activation barriers. The energy contributions are shown in Table 6 for the relevant states (*i.e.,* those that are the lowest energy with V = -70 mV and with V = 0). The largest single term is the one electron energy, and it about 95 % canceled by the repulsion terms. The *total* energy crosses from proton on Y266 to proton on R297 at around -20 mV. This is consistent with the open probability for the channel passing the 50% mark around there (since the resolution of the calculation is only 35 mV, it is fortuitous that it happens to come so close to that point, but it is certainly consistent). The two electron and nuclear terms are less sensitive to voltage. In other words, the sum of these three classical terms leads to too large a difference in energy, over 40 $k_BT$, which would make transitions impossible, but the two quantum terms shown in Table 6 compensate sufficiently to make the difference 16 $k_BT$. The error in this sum could be several $k_BT$, but it is at least a reasonable value. Labeling the terms "quantum" and "classical", although formally correct, ignores the point that the energy landscape for the structure also changes with exchange and correlation terms. Therefore, these will affect the "classical" terms. It is clear that the quantum effects must be included to obtain an accurate energy, reasonable charges on the atoms, and even a correct structure, especially as regards putting the protons in the right places.

**Tests of the accuracy of the calculation**: There are two tests, one offering comparison with experiment, the other a plausibility argument. **1)** structures, especially key interatomic distances, in the proton-shifted (neutral salt bridge) cases, with zero voltage, S4 fixed, came out close to the X-ray structure, as they must. This also suggests that the structures resulting from these calculations are realistic. The S4 free structures are not quite so close. Had the S4 free structures come out closer, we would have concluded that S4 free was the better representation. If neither had, we would have concluded that the calculation itself was at fault. However, the fact that the structures that gave the closest replica of the X-ray structures were S4 fixed, field = 0, and were quite close, strongly suggests that the calculation is correct, and the S4 fixed structures can be trusted to within a fraction of an Angstrom for interatomic distances, especially for key salt bridges and bond lengths. By extension, we trust the negative voltage results, although no X-ray structures exist. The key R297 – E183 salt bridge, with applied negative potential, which should have broken if S4 were to move down in order to reach a closed position, instead tightened from approximately 6 Å (guanidinium N to carboxyl O) to approximately 4 Å, when the salt bridge was charged, with S4 free, as assumed in standard models. For the salt bridge not charged (*i.e.,* $H^+$ transferred from arginine to glutamate), there should be almost no response to the field (only a dipole remains, and the interaction energy with the field would be too small to show any structural effect); nothing in the calculation contradicts this point, and the same result would be expected on any model. However, the uncharged closed case has lower energy, strongly suggesting that the proton has transferred from arginine to glutamate. Standard models that require physical motion at constant charge to open the channel are not consistent with this result; these models assume that all salt bridges remain charged. **2)** The difference in energy of the salt bridge ionized and unionized structures is generally of the order of 10 $k_BT$, making it possible for the charge to shift back and forth with the field. The one large barrier is the energy difference between ionized and unionized Y266 in open and closed states; a large barrier appears to lock down the closed state, while the reversal of this barrier allows the proton transfer required for the channel to open. The calculation suggests that the proton has already continued on to E183 in the closed state, with the opening transfer only from Y266 to R297. This implies that the gating charge accompanying this step is fairly close to 0.5 – 0.6e, as suggested earlier. An erroneous calculation would be certainly expected to produce much larger differences in energy, as the independent calculations would then fail to be closely related.

**Dipoles**: We earlier discussed the dipole of the water group alone. However, the dipole of the entire system leads to another question that must be considered. Table 7 gives the total dipole for the two low energy states, plus the fully ionized model that is assumed in all standard models. The dipole moments on the left correspond to the case of the closed channel, energy+25111= -.727 Ha, the middle case is the open channel, -.711 Ha, and the channel as assumed in standard models,

**Table 7:** Dipole moments of three low energy states[+]

| V = -70 | | V = 0 | | V = 0 | |
|---|---|---|---|---|---|
| H shift: R300→E183 | | H shift: Y266→E183 | | No H shift, salt bridges ionized | |
| μx | μ* | μx | μ* | μx | μ* |
| 27.1 | 31.6 | 24.8 | 28.0 | 52.3 | 56.3 |

[+] *μx is the component along the channel axis, μ\* the total dipole of the 976 atom system*

with all H on arginine, and tyrosine neutral, has E =-.700 Ha, with channel open but about 11k$_B$T higher energy than the proton shift shown. The changes in the dipole moment might contribute to the classical part of the energy, but the total **μ\*E** term, taking everything as homogeneous, is an order of magnitude too small to account for the effects observed. If the field is sufficiently inhomogeneous, and the dipole is large where the field is large, the contribution may become substantial. We cannot yet say that that can be ruled out. If that is the case, the rotation of the dipole, not evident in the overall value shown here, could contribute substantially to the gating current. The water dipole (Fig. 6, Table 8) is about 2/3 of **μx**, and does rotate, in the calculation; more work is needed to understand the complete system; however, results so far suggest that the effect should be too subtle to be significant

The dipole moments for these droplets are given in Table 8, showing the rotation of the dipole from parallel to field, meaning approximately parallel to channel axis (closed, with field) to orthogonal to the axis (open, no field).

**Water Dipoles**

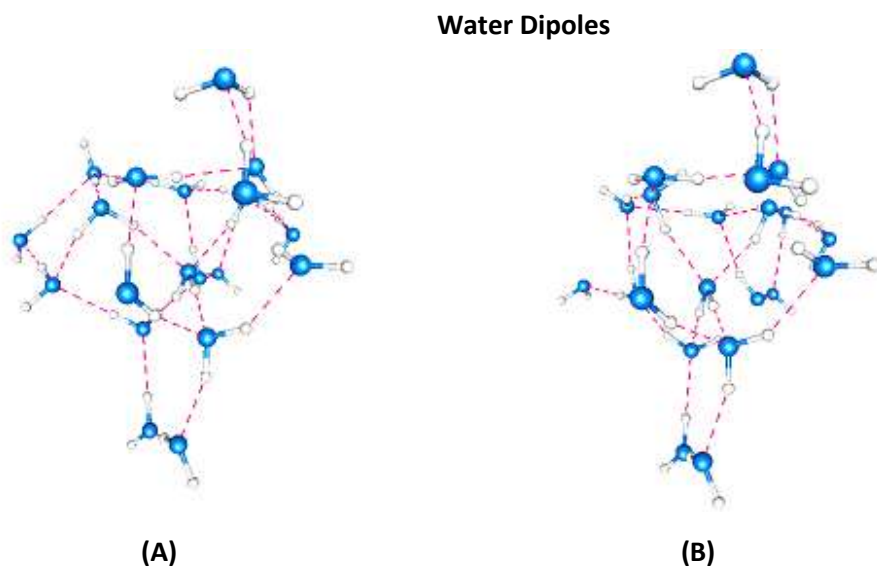

(A)　　　　　　　　　　(B)

*Fig. 6: Water cluster for open (A) and closed (B) cases.*

**TABLE 8:** Properties of the water cluster: energy and dipole moment

|  | Energy** | Total dipole (D) | X component | Y component | Z component |
|---|---|---|---|---|---|
| Zero field | -.203 | 16.9 | 16.0 | 5.1 | -1.4 |
| $-10^7$ Vm$^{-1}$ along X* | -.199 | 17.4 | -1.3 | 4.7 | 16.7 |

*Field negative (closed), corresponding to -70 mV along X. 1 Debye (D) is 3.33 x 10$^{-30}$ C m.

**Hartrees + 1605. (i.e. total energy -1605.203, -1605.199; closed conformation has about 4$k_B$T higher energy, with 1mHa ≈ 1 $k_B$T.)

The droplet shown in Fig. 6 consists of the water molecules as optimized in the complete 976 atom system; these coordinates are used for a single point B3LYP/6-31G++ calculation of energy and dipole of the water molecules alone. We cannot draw quantitative conclusions from the water droplet; the number of water molecules is approximately what fit in the space with the VSD, hence reasonable, but the exact number would be affected by lipids, and the neighboring solution, both omitted from the calculation. These water molecules form a reasonable boundary between the VSD and the missing external neighboring molecules in the calculation, at the high-dielectric extracellular end. We draw conclusions primarily from the interior of the remainder of the protein; the vacuum boundary in the remaining region would be relatively low dielectric constant in the complete system, but not as low as is effectively the case in this calculation. The vacuum boundary limits the accuracy of the calculation for some groups near the boundary, like I292. However, the water makes the extracellular end more realistic, avoiding a vacuum termination as nearly as is practicable. Table 8 shows the rotation of the dipole from parallel to field, meaning approximately parallel to channel axis (closed, with field) to orthogonal to the axis (open, no field). The majority of the total water dipole rotates from alignment along the X coordinate, through the VSD, when the field is off, to mostly orthogonal to that direction, along the Z coordinate parallel to the plane of the membrane, with field negative (closed). The difference would correspond to an appreciable contribution to the gating charge, but because it cannot be taken as quantitative, we only note that the water *could* make a significant contribution to the gating charge. The total dipole of the system, which is only about 70% greater than the water dipole, does not appear to rotate like the water dipole, and there is good reason to not assume that the dipole is even approximately homogeneous through the system. So far we cannot estimate the contribution of rotating dipoles to the gating current. Standard models generally do not consider this possible contribution explicitly, although an MD calculation would include some contribution.

**DISCUSSION**: The results point to a mechanism for gating current that depends on the shift of protons rather than a shift of the physical location of the backbone. We can summarize what we have found so far, with consideration of the implications for a gating mechanism. The remainder of the proton path is only a plausible hypothesis so far. Calculations of this part of the path are now under way. Since specific proton transfers involve specific amino acids, calculations are also under way for possible mutations of these amino acids. For example, we are looking at Y266F, an obvious mutation. Some mutations of other residues in the proposed proton path are also being studied. We expect to determine plausible steps in the path in this manner. This also suggests possible experimental mutations. Since the mutations that are required to test these calculations involve principally the arginines, the acids, and occasionally a less obvious residue, such as Y266, it is not difficult to choose appropriate mutations to make.

*a)* **Gating steps:** So far, we have hypothesized that **A**) the gating current is carried primarily by protons instead of S4 motion, possibly with a contribution from water at the extracellular end, and **B**) the ionization state of the salt bridges can be influenced by neighboring residues, allowing proton transfer. We must understand B in order to understand A; furthermore, we extend "salt bridges" to include larger groupings of amino acids that can join in transferring a proton: triads, or even quartets or pentads. A similar triad (tyr—arg—glu) has been reported by Chowdhury et al[59] as significant in relation to potassium channel gating. In this work we see an example of a triad which is part of a pentad (Fig. 4), and another pentad (Fig. 5). Similar groupings are found, for example, in cyt-c and in $H_v1$, proteins known to transmit protons. The existence of these clusters suggests a path for the protons to follow as their transfers create a gating current. However, our calculations show that the salt bridges are not merely influenced by third amino acids, but that there are transfers of a proton through an intermediate residue between two neighboring sets of residues; the final position of the proton allows it to move to the next grouping of three or more amino acids. A single residue can participate in two neighboring triads, something we see in Fig. 4; alternatively, the grouping can be considered a single grouping of five, as in Fig. 5. An initial ionization/proton transfer shifts the local potential so that the next transfer in the sequence occurs, and the remainder of the shifts that generate gating current follow. The initial pulse of gating current in the type of channel considered here (the "piquito") [28-29] is too fast to measure, and seems consistent, in speed, and size of the effect, with proton tunneling [20-21, 26] although an alternate possibility has been suggested[28]. A proton tunneling event fits well with our model; we cannot yet state which amino acids constitute the key pair. However, there is a large field drop, very possibly constituting most of the voltage drop across the membrane, near R297, so the transfer of a proton there is a candidate for the possible key initial step [60].

For the remainder of the VSD additional work will be required, but the principal types of transitions that are possible, and the participation of three or sometimes four or even five amino acids in a single transfer, is established. When similar groupings are found in the range of the path that we hypothesize the proton might take, we can regard these as plausibly parts of the path. Obviously, those parts that are now hypothesized to be parts of the path must still be calculated before they can be considered established parts. However, the present calculation not only establishes the initial steps, but shows the groupings that can reasonably be expected to be parts of the path, and thus the calculations that ought to be attempted to complete the path. Since alternate paths for gating charges are expected based on hysteresis[61], as well as on the many examples of multi-state models based on kinetics that have been published, we expect to find groups that can allow parallel paths for the protons; a single mutation would alter kinetics, but not destroy the channel.

The remainder of this section constitutes a hypothesis, with computations on these sections just beginning: The $H^+$ on opening comes from (R303,E226), with the path beginning at the lower end of the VSD: the path includes 3) K306,D259,R309,R240,E236; 4) extends through S169,H310, and 2 water molecules. The step from E226, R303 to K306, D259 is short enough that there should not be a major barrier (the role of F233 remains to be determined in this regard, however). Calculations already completed suggest that the proton transfer from K306 to D259 is inevitable. The system continues on, via a path that includes the upper end of the T1 segment, with any water molecules that are between T1 and the VSD, toward H418, which is hypothesized to have an added proton in the closed state, and the conserved PVPV sequence at the gate. The proton, on moving up away from H418, allows the gate to open. Both classical and quantum terms depend on the location of the proton. The PVPV sequences of the four domains are spaced far enough apart that the actual gate appears to be at the junction of the PVPV sequence with the upper part of the T1 moiety that hangs below it, where the opening is smaller. Since it is sometimes possible to get a functioning channel with T1 truncated, it may be that a path through the water at the inner surface of the membrane can organize to transmit a proton, although exactly how the gate would close in this case is not yet clear; a move of 2 – 3 Å should be sufficient. The central part of the $K_v1.2$ sequence is similar to $H_v1$; the R300-E226-R303 triad for $K_v1.2$, is comparable to R201, D108, R204 for $H_v1$: the distance of the closest N atom of R303 to the nearest carboxyl of E226 is 4.26 Å (X-ray distance), the corresponding distance for $H_v1$ is 4.34 Å (R201 – D108); corresponding distances for R300 to E226 = 2.71 Å ($K_v1.2$) and R204 to E108 = 2.71 Å also ($H_v1$). The two channels diverge below this. The $H_v1$ proton pathway continues toward the cytoplasmic side through several sets of residues that, taken together, produce a pathway that is somewhat similar to that in the VSD of $K_v1.2$, albeit differently oriented, and with different residues that are able to transfer a proton. While everything in this paragraph requires further computation, (which has begun) it is possible to see a specific path with reasonable distances over essentially the entire required distance; only the arrangement of water above T1 has unknown distances, and a water wire for transmitting protons is not a new or implausible proposal.

*b)* **S4 free vs S4 fixed**: Section (b) of the Results section gave reasons based on the calculations for taking S4 fixed. In addition, there is a physical mechanism for keeping it fixed; it is likely that the end of the S4 segment is held by the membrane. The first arginine (R294) is in the membrane region, near phosphate groups. At least some negative charges must exist in the membrane to have a functioning channel[62], and the mechanical state of the membrane also matters [63]; this seems unlikely if S4 simply went through it. Arginines form fairly strong complexes with phosphates [64-67], which would anchor R294, making a fixed position of S4 unavoidable. R294 is omitted in this calculation, as its side chain is unattached to the rest of the system. This side chain would therefore be incorrectly treated, so it must be omitted. Its effect on the remainder of the chain is best represented by fixing the last residue. If a phosphate-R294 complex were to form and if S4 were to move, the complex would have to be pushed apart by the force of the S4 breaking out of the central part of the membrane, which appears to require far more energy than is available. Furthermore, it is well known that phosphatidylinositol (4,5) bisphosphate ($PIP_2$) affects gating, making the channel more stable, especially in the open state[68-69]. This is further evidence that S4 fixed cases appear to be the correct ones for the interpretation of the results, and for understanding the gating current. (The preference for the open state would occur in our model if the $PIP_2$ is in the inner leaflet of the membrane, holding protons there; if the protons left on closing, the positive charge on R294 would be held in their place) Table 3 shows how small the motions are of the S4 segment with the S4 free, V<0.

*c)* **The role of Y266**: We have several times mentioned that Y266 is a key residue in the sequence of proton transfer steps. This residue can provide a proton to R300, which in turn passes it to E183 and R297. R300 is oriented slightly differently than the other arginines, which makes the transfer of a proton from Y266 easier. The proton donation from Y266 to R300 is one of two paths that can provide gating charge (section b above). A deprotonated tyrosine also can act as a valve to insure unidirectional proton transfer in the K-channel analog of B-type cyt-c oxidase[37]. That system has a somewhat different chain than the one we observe in the VSD, but still uses a tyrosine as a valve for proton transfer. Another voltage sensor based on tyrosine has recently been reported for the M2 muscarinic receptor[70], albeit with a different interpretation. Our calculation predicts that Y266 mutations would make an appreciable change in gating.

*d)* **Charges on several amino acids** (see Table 5). The NBO charge calculations give us a more accurate value for the charge transfer accompanying proton shifts. In some cases the proton final position is in the center of a short bond, and it is important to know where the charge resides. Charges are assigned to each atom by the calculation. The charges are not close to integral for the guanidinium or for the acid. It is too simple to say that the charge has transferred as a unit; partial charge transfer, with electron density shared among atoms, and groups, is found. This must be taken into account in any calculations done on the system. Protons can delocalize over about a bond length.

*e)* **Our hypothesis agrees with previous experimental results on H418**. The final (hypothesized) steps include protonation of H418, which is close to the gate at the C-terminal end of the channel, and is expected to play a key role in opening and closing the channel. Deleting this residue has been reported to kill the channel[71].

*f)* The individual optimizations provide 30 energy minima (cases in Table 2), presumably mostly local minima, which include those we consider relevant to the possible trajectory of the proton, and to allow for the possibility of motion of S4. Because the side chains are free to move, as is the backbone in the relevant S4 free cases, the space of possible local minima has been fairly thoroughly explored; the local minima found include all those with plausible positions for the protons, and thus include cases that cannot be studied in most MD simulations, as these cannot move the protons across a salt bridge, or otherwise alter their bonding. The low energy cases among those found are most likely to include the correct path of the proton. It is apparent that the proton motion is such as to allow for the existence of gating current, when three to four more steps are included; further work on the role of dipole rotation may also be needed. While the entire path cannot be found from calculations on half the VSD, it is possible to find a detailed description of one of the steps, and observe that the remainder of the path can follow similar sets of three to five polar amino acids through the VSD. It is hard to understand the presence of such residues if they are not required for such a path. However, combined with the questions that must be raised concerning the standard model, it is at least plausible, and we believe very likely, that protons form the gating current, and S4 is nearly immobile on the scale required to produce gating current.

**Summary:** In this paper, we present quantum calculations that include a large fraction of the VSD. *The key point: our results show a gating current composed, at least in large part, of proton motion, with an essentially immobile S4 transmembrane segment*. Quantum calculations are generally limited by the size of the system that can be calculated. We here present a set of calculations larger, to the best of our knowledge, than any previously offered. These represent much of the VSD, and they include an electric field of appropriate magnitude applied to the VSD. Previous work has generally been understood in terms of gating models with mobile S4 segments of the VSD providing the gating current. In earlier reviews, we have considered reinterpretation of the evidence that has hitherto been used to support the idea of S4 motion, particularly SCAM results and MD calculations [21, 72]. Our new calculations enable us to propose specific proton transfers in part of the channel, and hypothesize additional transfers that would continue the chain all the way to the gate, thus providing the gating current and opening and closing the channel. Further work is needed to complete the proton transfer chain, so as to fully understand gating in the channel.

Author contributions: A. M. Kariev and M. E. Green designed and carried out research, and analyzed results. M.E. Green wrote the paper

Conflicts of interest: Both authors declare no conflicts of interest

Supplementary information: Optimized coordinates for 14 of the 30 optimized cases are given in the supplementary material.

ACKNOWLEDGMENTS: This research was supported, in part, by a grant of computer time from the City University of New York High Performance Computing Center under NSF Grants CNS-0855217, CNS-0958379 and ACI-1126113, and this research used resources of the Center for Functional Nanomaterials, which is a U.S. DOE Office of Science Facility, at Brookhaven National Laboratory under Contract No. DE-SC0012704.

**REFERENCES:**

1. Hodgkin, A. L.; Huxley, A. F., Currents carried by sodium and potassium ions through the giant membrane of *Loligo*. *J. Physiol. (Lond)* **1952,** *116*, 449-472.
2. Hodgkin, A. L.; Huxley, A. F., The components of membrane conductance in the giant axon of *Loligo*. *J. Physiol. (Lond)* **1952,** *116*, 473-496.
3. Hodgkin, A. L.; Huxley, A. F., The dual effect of membrane potential on sodium conductance in the giant axon of *Loligo*. *J. Physiol. (Lond)* **1952,** *116*, 497-506.
4. Keynes, R. D.; Rojas, E., Kinetics and steady-state properties of the charged system controlling sodium conductance in the squid giant axon. *J. Physiol. (Lond.)* **1974,** *239*, 393-434.
5. Armstrong, C. M.; Bezanilla, F., Charge movement associated with the opening and closing of of the activation gates of the Na channels. *J. Gen. Physiol.* **1974,** *63*, 533-552.
6. Karlin, A.; Akabas, M. A., Substituted Cysteine Accessibility Method. In *Methods in Enzymology*, Conn, P. M., Ed. Academic Press: New York, 1998; pp 123-145.
7. Horn, R., Cysteine Scanning. In *Methods in Enzymology*, Conn, P. M., Ed. Academic Press: New York, 1998; pp 145-155.
8. Doyle, D. A.; Cabral, J. M.; Pfuetzner, R. A.; Kuo, A.; Gulbis, J. M.; Cohen, S. L.; Chait, B. T.; MacKinnon, R., The structure of the potassium channel: molecular basis of K$^+$ conduction and selectivity. *Science* **1998,** *280*, 69-77.
9. Kirsch, G. E.; Sykes, J. S., Temperature dependence of sodium currents in rabbit and frog muscle membranes. *J. Gen. Physiol.* **1987,** *89* (2), 239-51.

10. Kuno, M.; Ando, H.; Morihata, H.; Sakai, H.; Mori, H.; Sawada, M.; Oiki, S., Temperature dependence of proton permeation through a voltage-gated proton channel. *J. Gen. Physiol.* **2009,** *134* (3), 191-205.
11. Milburn, T.; Saint, D. A.; Chung, S.-H., The temperature dependence of conductance of the sodium channel: implications for mechanisms of ion permeation. *Recept. Channels* **1995,** *3* (3), 201-11.
12. Cherny, V. V.; Morgan, D.; DeCoursey, T. E.; Musset, B.; Chaves, G.; Smith, S. M. E., Tryptophan 207 is crucial to the unique properties of the human voltage-gated proton channel, hHV1. *J Gen Physiol* **2015,** *146* (5), 343-56.
13. Starace, D.; Stefani, E.; Bezanilla, F., Histidine scanning mutagenesis indicates full translocation of two charges of the Shaker K channel voltage sensor. *Biophys. J.* **1998,** *74*, A215.
14. Starace, D. M.; Bezanilla, F., Histidine scanning mutagenesis of uncharged residues of the *Shaker* $K^+$ channel S4 segment. *Biophys. J.* **2001,** *80*, 217a.
15. Kariev, A. M.; Znamenskiy, V. S.; Green, M. E., Quantum mechanical calculations of charge effects on gating the KcsA channel. *Biochim. Biophys. Acta, Biomembr.* **2007,** *1768* (5), 1218-1229.
16. Lu, J.; Yin, J.; Green, M. E., A model for ion channel voltage gating with static S4 segments. *Ferroelectrics* **1999,** *220* (3-4), 249-271.
17. Sapronova, A.; Bystrov, V.; Green, M. E., Ion channel gating and proton transport. *J. Mol. Struct.: THEOCHEM* **2003,** *630*, 297-307.
18. Sapronova, A.; Bystrov, V. S.; Green, M. E., Water, proton transfer, and hydrogen bonding in ion channel gating. *Front Biosci* **2003,** *8*, s1356-70.
19. Green, M. E., Electrorheological effects and gating of membrane channels. *J.Theor. Biol.* **1989,** *138*, 413-428.
20. Kariev, A. M.; Green, M. E., Voltage gated ion channel function: gating, conduction, and the role of water and protons. *Int. J. Mol. Sci.* **2012,** *13*, 1680-1709.
21. Kariev, A. M.; Green, M. E., Caution is required in interpretation of mutations in the voltage sensing domain of voltage gated channels as evidence for gating mechanisms. *Int'l J. Molec. Sci.* **2015,** *16*, 1627-1643.
22. Niemeyer, M. I.; Gonzalez-Nilo, F. D.; Zuniga, L.; Gonzalez, W.; Cid, L. P.; Sepulveda, F., Neutralization of a single arginine residue gates open a two-pore domain, alkali-activated $K^+$ channel. *Proc Natl Acad Sci U S A* **2007,** *104*, 666-671.
23. Liao, S.; Green, M. E., Quantum calculations on salt bridges with water: Potentials, structure, and properties. *Comput. Theo. Chem* **2011,** *963*, 207-214.
24. Kariev, A. M.; Green, M. E., Quantum Effects in a Simple Ring with Hydrogen Bonds *J. Phys. Chem. B* **2015,** *119*, 5962-5969.
25. Kariev, A. M.; Njau, P.; Green, M. E., The Open Gate of the Kv1.2 Channel: Quantum Calculations Show The Key Role Of Hydration. *Biophys J* **2014,** *106*, 548-555.
26. Yin, J.; Green, M. E., Intermolecular proton transfer between two methylamine molecules with an external electric field in the gas phase. *J. Phys. Chem. A* **1998,** *102*, 7181-7190.
27. Stefani, E.; Bezanilla, F., Voltage dependence of the early events in voltage gating. *Biophys. J.* **1997,** *72*, A131.
28. Sigg, D.; Bezanilla, F.; Stefani, E., Fast gating in the Shaker $K^+$ channel and the energy landscape of activation. *Proc. Natl Acad. Sci USA* **2003,** *100*, 7611-7615.

29. Stefani, E.; Sigg, D.; Bezanilla, F., Correlation between the early component of gating current and total gating current in Shaker K channels. *Biophysical Journal* **2000,** *78*, 7A.

30. Fatade, A.; Snowhite, J.; Green, M. E., A Resonance Model gives the Response to Membrane Potential for an Ion Channel: II. Simplification of the Calculation, and Prediction of Stochastic Resonance. *J. Theor. Biol.* **2000,** *206* (3), 387-393.

31. Starace, D. M.; Bezanilla, F., Histidine scanning mutagenesis of basic residues of the S4 segment of the *Shaker* K$^+$ channel. *J. Gen'l. Physiol.* **2001,** *117*, 469-490.

32. Starace, D. M.; Bezanilla, F., A proton pore in a potassium channel voltage sensor reveals a focused electric field. *Nature* **2004,** *427*, 548-553.

33. DeCoursey, T. E.; Morgan, D.; Musset, B.; Cherny, V. V., Insights into the structure and function of H$_v$1 from a meta-analysis of mutation studies. *J. Gen'l Physiol.* **2016,** *148*, 97-118.

34. van Keulen, S. C.; Gianti, E.; Carnevale, V.; Klein, M. L.; Rothlisberger, U.; Delemotte, L., Does proton conduction in the voltage-gated H+ channel hHv1 involve Grotthuss-like hopping via acidic residues? *J. Phys. Chem. B* **2016**, Ahead of Print.

35. van Keulen, S. C.; Gianti, E.; Carnevale, V.; Klein, M. L.; Rothlisberger, U.; Delemotte, L., Does Proton Conduction in the Voltage-Gated H$^+$ Channel hHv1 Involve Grotthuss-Like Hopping via Acidic Residues? *J. Phys. Chem. B* **2017,** *121* (15), 3340-3351.

36. Zhao, J.; Blunck, R., The isolated voltage sensing domain of the Shaker potassium channel forms a voltage-gated cation channel. *Elife* **2016,** *5*.

37. Woelke, A. L.; Wagner, A.; Galstyan, G.; Meyer, T.; Knapp, E.-W., Proton Transfer in the K-Channel Analog of B-Type Cytochrome c Oxidase from Thermus thermophilus. *Biophys. J.* **2014,** *107* (9), 2177-2184.

38. Hong, M.; DeGrado, W. F., Structural basis for proton conduction and inhibition by the influenza M2 protein. *Protein Sci.* **2012,** *21* (11), 1620-1633.

39. Hong, M.; Fritzsching, K. J.; Williams, J. K., Hydrogen-bonding partner of the proton-conducting histidine in the influenza M2 proton channel revealed from 1H chemical shifts. *J. Am. Chem. Soc.* **2012,** *134* (36), 14753-14755.

40. Williams, J. K.; Zhang, Y.; Schmidt-Rohr, K.; Hong, M., pH-Dependent Conformation, Dynamics, and Aromatic Interaction of the Gating Tryptophan Residue of the Influenza M2 Proton Channel from Solid-State NMR. *Biophys. J.* **2013,** *104* (8), 1698-1708.

41. Gonzalez-Perez, V.; Stack, K.; Boric, K.; Naranjo, D., Reduced voltage sensitivity in a K$^+$-channel voltage sensor by electric field remodeling. *Proc. Natl Acad Sci.* **2010,** *107*, 5178-5183.

42. Tronin, A. Y.; Nordgren, C. E.; Strzalka, J. W.; Kuzmenko, I.; Worcester, D. L.; Lauter, V.; Freites, J. A.; Tobias, D. J.; Blasie, J. K., Direct Evidence of Conformational Changes Associated with Voltage Gating in a Voltage Sensor Protein by Time-Resolved X-ray/Neutron Interferometry. *Langmuir* **2014,** *30* (16), 4784-4796.

43. Delemotte, L.; Klein, M. L.; Tarek, M., Molecular dynamics simulations of voltage-gated cation channels: insights on voltage-sensor domain function and modulation. *Front. Pharmacol. Ion Channels Channelopathies* **2012,** *3* (May), 97.

44. Jensen, M. O.; Borhani, D. W.; Lindorff-Larsen, K.; Maragakis, P.; Jogini, V.; Eastwood, M. P.; Dror, R. O.; Shaw, D. E., Principles of conduction and hydrophobic gating in K$^+$ channels *Proc. Nat'l Acad. US* **2010,** *107*, 5833-5838.

45. Sun, R.-N.; Gong, H., Simulating the Activation of Voltage Sensing Domain for a Voltage Gated Sodium Channel Using Polarizable Force Field. *J. Phys. Chem. Letters* **2017,** *8*, 901-908.
46. Villalba-Galea, C. A., Hysteresis in voltage-gated channels. *Channels (Austin)* **2016**, 1-16.
47. Plummer, A. M.; Fleming, K. G., From Chaperones to the Membrane with a BAM! *Trends Biochem. Sci.* **2016,** *41* (10), 872-882.
48. Schiffrin, B.; Calabrese, A. N.; Devine, P. W. A.; Harris, S. A.; Ashcroft, A. E.; Brockwell, D. J.; Radford, S. E., Skp is a multivalent chaperone of outer-membrane proteins. *Nat. Struct. Mol. Biol.* **2016,** *23* (9), 786-793.
49. Serdiuk, T.; Balasubramaniam, D.; Sugihara, J.; Mari, S. A.; Kaback, H. R.; Muller, D. J., YidC assists the stepwise and stochastic folding of membrane proteins. *Nat. Chem. Biol.* **2016,** *12* (11), 911-917.
50. Thoma, J.; Burmann, B. M.; Hiller, S.; Muller, D. J., Impact of holdase chaperones Skp and SurA on the folding of β-barrel outer-membrane proteins. *Nat. Struct. Mol. Biol* **2015,** *22* (10), 795-802.
51. Takeda, M.; Jee, J. G.; Ono, A. M.; Terauchi, T.; Kainosho, M., Hydrogen Exchange Rate of Tyrosine Hydroxyl Groups in Proteins As Studied by the Deuterium Isotope Effect on Cζ Chemical Shifts. *J. Am. Chem. Soc.* **2009,** *131* (51), 18556-18562.
52. Frisch, M. J.; Trucks, G. W.; Schlegel, H. B.; Scuseria, G. E.; Robb, M. A.; Cheeseman, J. R.; Scalmani, G.; Barone, V.; Mennucci, B.; Petersson, G. A.; Nakatsuji; H.; Caricato, M.; Li, X.; Hratchian, H. P.; Izmaylov, A. F.; Bloino, J.; Zheng, G.; Sonnenberg, J. L.; Hada, M.; Ehara, M.; Toyota, K.; Fukuda, R.; Hasegawa, J.; Ishida, M.; Nakajima, T.; Honda, Y.; Kitao, O.; Nakai, H. V., T.; Montgomery, J.; Peralta, J. A.; J. E.; Ogliaro, F.; Bearpark, M.; Heyd, J. J.; Brothers, E.; Kudin, K. N.; Staroverov, V. N.; Kobayashi, R. N., J.; Raghavachari, K.; Rendell, A.; Burant, J. C.; Iyengar, S. S.; Tomasi, J.; Cossi, M.; Rega, N.; Millam, J. M.; Klene, M.; Knox, J. E.; Cross, J. B.; Bakken, V.; Adamo, C.; Jaramillo, J.; Gomperts, R.; Stratmann, R. E.; Yazyev, O.; Austin, A. J.; Cammi, R.; Pomelli, C.; Ochterski, J. W.; Martin, R. L.; Morokuma, K.; Zakrzewski, V. G.; Voth, G. A.; Salvador, P.; Dannenenberg, J. J.; Dapprich, S.; Daniels, A. D.; Farkas, Ö.; Foresman, J. B.; Ortiz, J. V.; Cioslowski, J.; Fox, D. J. *Gaussian 09, Revision D.01*, Gaussian, Inc.: Wallingford CT, 2009.
53. Schwaiger, C. S.; Liin, S. I.; Elinder, F.; Lindahl, E., The Conserved Phenylalanine in the K+ Channel Voltage-Sensor Domain Creates a Barrier with Unidirectional Effects. *Biophys. J.* **2013,** *104* (1), 75-84.
54. Brown, H. C.; Melchiore, J. J., Complexes of hydrogen chloride and hydrogen bromide with aromatic hydrocarbons in n-heptane solution. *J. Am. Chem. Soc.* **1965,** *87* (23), 5269-75.
55. Cockcroft, V. B.; Osguthorpe, D. J.; Barnard, E. A.; Lung, G. G., Modeling of agonist binding to the ligand-gated ion channel superfamily of receptors. *Proteins: Struct., Funct., Genet.* **1990,** *8* (4), 386-97.
56. Grimmer, M.; Heidrich, D., Quantum-mechanical studies of the mechanism of electrophilic substitution. 2. Role of σ- and π-complex structures of protonated aromatics of different reactivity. *Z. Chem.* **1974,** *14* (12), 481-2.
57. Mons, M.; Robertson, E. G.; Simons, J. P., Intra- and Intermolecular π-Type Hydrogen Bonding in Aryl Alcohols: UV and IR-UV Ion Dip Spectroscopy. *J. Phys. Chem. A* **2000,** *104* (7), 1430-1437.
58. Otero-de-la-Roza, A.; DiLabio, G. A.; Johnson, E. R., Exchange-Correlation Effects for Noncovalent Interactions in Density Functional Theory. *J. Chem. Theory Comput.* **2016,** *12* (7), 3160-3175.
59. Chowdhury, S.; Haehnel, B. M.; Chanda, B., Interfacial gating triad is crucial for electromechanical transduction in voltage-activated potassium channels. *J. Gen. Physiol.* **2014,** *144* (5), 457-467.
60. Asamoah, O. K.; Wuskell, J. P.; Loew, L. M.; Bezanilla, F., A Fluorometric Approach to Local Electric Field Measurements in a Voltage-Gated Ion Channel. *Neuron* **2003,** *37*, 85-97.

61. Villalba-Galea, C. A., Hysteresis in voltage-gated channels. *Channels (Austin)* **2017,** *11* (2), 140-155.
62. Schmidt, D.; Jiang, Q.-X.; MacKinnon, R., Phospholipids and the origin of cationic gating charges in voltage sensors. *Nature (London, U. K.)* **2006,** *444* (7120), 775-779.
63. Schmidt, D.; MacKinnon, R., Voltage-dependent $K^+$ channel gating and voltage sensor toxin sensitivity depend on the mechanical state of the lipid membrane. *Proc. Nat'l Academy of Sciences* **2008,** *105*, 19276-19281.
64. Green, M. E., A possible role for phosphate in complexing the arginines of S4 in voltage gated channels. *J. Theor. Biol.* **2005,** *233*, 337-341.
65. Green, M. E., Consequences of Phosphate-Arginine Complexes in Voltage Gated Ion Channels. *Channels* **2008,** *2*, 395-400.
66. Pradhan, P.; Ghose, R.; Green, M. E., Voltage gating and anions, especially phosphate: a model system. *Biochem Biophys. Acta (Membranes)* **2005,** *1717*, 97-103.
67. Freites, J. A.; Tobias, D. J.; von, H. G.; White, S. H., Interface connections of a transmembrane voltage sensor. *Proc. Natl. Acad. Sci. U. S. A.* **2005,** *102*
68. Falkenburger, B. H.; Jensen, J. B.; Dickson, E. J.; Suh, B.-C.; Hille, B., Phosphoinositides: lipid regulators of membrane proteins. *J. Physiol. (Oxford, U. K.)* **2010,** *588* (17), 3179-3185.
69. Suh, B.-C.; Hille, B., PIP2 is a necessary cofactor for ion channel function: how and why? *Annu. Rev. Biophys.* **2008,** *37*, 175-195.
70. Barchad-Avitzur, O.; Priest, M. F.; N., D.; Bezanilla, F.; Parnas, H.; Ben-Chaim, Y., A Novel Voltage Sensor in the Orthosteric Binding Site of the M2 Muscarinic Receptor. *Biophys J* **2016,** *111*, 1396-1408.
71. Zhao, L.-L.; Qi, Z.; Zhang, X.-E.; Bi, L.-J.; Jin, G., Regulatory role of the extreme C-terminal end of the S6 inner helix in C-terminal-truncated Kv1.2 channel activation *Cell Biol. International* **2010,** *34*, 433-439.
72. Kariev, A. M.; Green, M. E., Voltage Gated Ion Channel Function: Gating, Conduction, and the Role of Water and Protons. *Int'l J. Molec. Sci.* **2012,** *13*, 1680-1709.

SUPPLEMENTARY MATERIAL: COORDINATES OF SELECTED CASES;
Notation: the positions of protons are indicated as charges on amino acids; if an R (either 300 or 297) is listed as "neutral" it has transferred its proton to another amino acid, generally E183; if E183 is listed as "neutral" it has accepted a proton, whether from an arginine or from Y266. Similarly, if Y266 is shown as negative, it has transferred its ionizable proton to another amino acid. Except as noted, R297 is positive and E183 is the only carboxylic acid that can accept a proton.

S4 is fixed at the ends, except where "S4 Free" is noted, in which case the S4 is allowed to move freely with respect to the other three transmembrane segments. The voltage for each case is noted.

The final file (Case 14) has S4 free, -70mV negative voltage, and all amino acids in their "native" charge state--all arginines are positive, all acids negative, and tyrosine (Y266) neutral. This is the case that would have been expected to have S4 move, if it were going to move. It did not move.

Case 1:
voltage=0, Y266 neutral, R300 neutral, E183 neutral

| | | | |
|---|---|---|---|
| N | 15.802748 | -6.133741 | 5.261069 |
| H | 15.000474 | -5.906329 | 5.815964 |
| C | 15.709624 | -5.529811 | 3.929295 |
| C | 15.001670 | -6.466409 | 2.931375 |
| O | 14.254140 | -6.041260 | 2.065796 |
| C | 17.109416 | -5.209681 | 3.391311 |
| H | 16.607142 | -5.796305 | 5.751991 |
| H | 15.121687 | -4.621509 | 3.935920 |
| H | 17.063954 | -4.816033 | 2.382904 |
| H | 17.577240 | -4.457711 | 4.016567 |
| H | 17.739550 | -6.092207 | 3.389811 |
| N | 15.285550 | -7.763112 | 3.044991 |
| H | 15.877215 | -8.030363 | 3.802303 |
| C | 14.673853 | -8.778914 | 2.186456 |
| H | 14.746006 | -8.500238 | 1.147858 |
| H | 15.190942 | -9.717305 | 2.336458 |
| C | 13.193214 | -8.946854 | 2.530571 |
| O | 12.362027 | -9.161640 | 1.687330 |
| N | 12.866158 | -8.803500 | 3.829326 |
| H | 13.587727 | -8.723517 | 4.508537 |
| C | 11.484286 | -8.958399 | 4.289736 |
| H | 11.482590 | -8.954602 | 5.371574 |
| H | 11.055077 | -9.884466 | 3.940086 |
| C | 10.593529 | -7.816666 | 3.783179 |
| O | 9.450421 | -8.019874 | 3.471227 |
| N | 11.170827 | -6.604632 | 3.745066 |
| C | 10.507953 | -5.436464 | 3.221573 |
| C | 10.309540 | -5.486674 | 1.704078 |
| O | 9.264435 | -5.084642 | 1.249788 |
| C | 11.252668 | -4.174356 | 3.686533 |
| O | 12.642423 | -4.322692 | 3.673183 |

```
H          12.127772     -6.510127      3.994548
H           9.504595     -5.385282      3.611648
H          10.937622     -3.329139      3.084844
H          10.985755     -3.974887      4.713991
H          12.971644     -4.488421      2.798266
N          11.292688     -5.962586      0.920590
C          11.037128     -6.153732     -0.504232
C           9.910206     -7.168917     -0.707747
O           9.151731     -7.071253     -1.644702
C          12.331963     -6.504276     -1.299021
C          12.565671     -8.002012     -1.535208
C          12.339033     -5.762081     -2.636815
H          12.130428     -6.325788      1.320764
H          10.642682     -5.226036     -0.887172
H          13.157179     -6.123167     -0.707998
H          13.520521     -8.137387     -2.032441
H          12.581601     -8.579208     -0.621618
H          11.805032     -8.418528     -2.188472
H          13.238657     -5.997828     -3.194163
H          11.484320     -6.047105     -3.240534
H          12.313833     -4.688852     -2.490505
N           9.817331     -8.136373      0.212020
H          10.569257     -8.257922      0.851718
C           8.868022     -9.241609      0.120497
H           9.237689    -10.037783      0.754129
H           8.809266     -9.613071     -0.890983
C           7.451044     -8.911795      0.540059
O           6.533445     -9.381790     -0.092114
N           7.258889     -8.114639      1.597322
C           5.914032     -7.670855      1.882615
C           5.458354     -6.661318      0.834886
O           4.274943     -6.583092      0.580055
C           5.655596     -7.187100      3.323779
C           5.841147     -8.335144      4.321046
C           6.461916     -5.955941      3.742489
H           8.032219     -7.837947      2.162579
H           5.253503     -8.507948      1.717107
H           4.605128     -6.912388      3.327921
H           5.548252     -8.015614      5.315475
H           5.227594     -9.189776      4.055306
H           6.874357     -8.660447      4.365866
H           6.110686     -5.597222      4.704005
H           7.511554     -6.196499      3.848301
H           6.368667     -5.140781      3.032580
N           6.354726     -5.914430      0.181730
C           5.931727     -5.102530     -0.947122
C           5.527744     -5.988931     -2.133641
O           4.492212     -5.764840     -2.716587
C           6.987312     -4.041396     -1.317312
C           7.147864     -3.028089     -0.170286
C           6.594155     -3.335295     -2.619478
C           8.408132     -2.169551     -0.276491
H           7.320806     -5.958333      0.425317
H           5.013931     -4.602217     -0.677309
```

```
H         7.936597    -4.546701    -1.467448
H         6.267486    -2.384975    -0.148062
H         7.177721    -3.549449     0.777112
H         7.269413    -2.518046    -2.832373
H         6.620301    -4.003373    -3.470330
H         5.588856    -2.930740    -2.553669
H         8.482391    -1.500994     0.574242
H         9.293835    -2.793589    -0.281833
H         8.420624    -1.561022    -1.173564
N         6.331194    -7.000322    -2.485061
H         7.231876    -7.098295    -2.065280
C         6.006648    -7.842851    -3.616850
H         6.851172    -8.498178    -3.789603
H         5.839077    -7.256977    -4.508291
C         4.755158    -8.697411    -3.415597
O         3.916274    -8.770019    -4.282039
N         4.658416    -9.353786    -2.252550
H         5.399685    -9.303325    -1.586413
C         3.490006   -10.137156    -1.926906
H         3.264766   -10.851309    -2.704675
H         3.694665   -10.679454    -1.012301
C         2.238767    -9.284259    -1.731649
O         1.159870    -9.694837    -2.091310
N         2.423893    -8.083818    -1.184805
C         1.273961    -7.190680    -1.156699
C         0.821010    -6.740694    -2.548666
O        -0.363806    -6.622665    -2.764319
C         1.484533    -5.917358    -0.353069
O         1.626429    -6.121778     1.018217
H         3.320025    -7.766764    -0.892777
H         0.434883    -7.723825    -0.740868
H         2.328249    -5.376915    -0.756689
H         0.592172    -5.322698    -0.482288
H         2.493379    -6.471572     1.180093
N         1.741980    -6.433439    -3.470856
C         1.337628    -6.029174    -4.806418
C         0.583618    -7.161829    -5.495852
O        -0.414651    -6.924998    -6.142244
C         2.542054    -5.518012    -5.630954
C         2.941030    -4.121456    -5.125238
C         2.240675    -5.499481    -7.135097
C         4.263526    -3.607619    -5.693153
H         2.713252    -6.475322    -3.248263
H         0.609644    -5.234427    -4.726866
H         3.368439    -6.201764    -5.461649
H         2.145325    -3.420994    -5.372929
H         3.016607    -4.143456    -4.045455
H         3.093572    -5.123035    -7.685782
H         2.028242    -6.489108    -7.520646
H         1.390108    -4.863974    -7.360555
H         4.553952    -2.689064    -5.193861
H         5.059399    -4.329581    -5.539671
H         4.204307    -3.396494    -6.754782
N         1.052900    -8.403395    -5.341579
```

```
H         1.906718    -8.561590    -4.848372
C         0.292540    -9.526508    -5.833032
H         0.178789    -9.493322    -6.906357
H         0.820525   -10.433212    -5.566498
C        -1.111732    -9.575076    -5.239009
O        -2.067570    -9.804500    -5.945007
N        -1.217901    -9.367827    -3.927608
C        -2.522875    -9.413595    -3.306249
C        -3.454420    -8.284898    -3.755120
O        -4.649155    -8.508635    -3.796956
C        -2.424017    -9.396615    -1.773672
O        -3.663373    -9.680314    -1.186800
H        -0.402368    -9.276158    -3.358981
H        -3.030589   -10.321873    -3.598442
H        -1.733119   -10.164896    -1.454967
H        -2.043728    -8.437411    -1.438209
H        -4.360911    -9.288971    -1.698526
N        -2.922992    -7.092240    -4.019939
C        -3.699120    -5.963151    -4.507090
C        -4.266581    -6.254021    -5.894860
O        -5.431084    -6.023322    -6.151709
C        -2.797331    -4.703446    -4.499222
C        -3.209253    -3.587394    -5.442368
C        -4.494906    -3.057994    -5.451093
C        -2.278404    -3.070113    -6.334911
C        -4.840068    -2.054184    -6.339080
C        -2.613330    -2.051179    -7.212352
C        -3.900316    -1.543837    -7.220504
H        -1.947778    -6.960565    -3.851005
H        -4.557863    -5.807294    -3.872755
H        -2.750007    -4.341042    -3.477091
H        -1.797726    -5.011419    -4.768673
H        -5.237413    -3.445645    -4.779343
H        -1.281404    -3.473773    -6.353442
H        -5.848912    -1.682233    -6.350090
H        -1.874235    -1.666058    -7.891564
H        -4.168910    -0.763907    -7.911241
N        -3.424976    -6.765524    -6.794730
C        -3.901721    -7.102960    -8.117293
C        -5.015665    -8.146642    -8.051272
O        -5.986586    -8.038891    -8.755275
C        -2.770058    -7.620611    -9.008468
S        -1.576341    -6.355968    -9.547156
H        -2.484376    -6.973635    -6.530493
H        -4.350060    -6.233736    -8.573782
H        -2.246139    -8.436966    -8.529932
H        -3.206103    -7.999536    -9.923341
H        -0.946915    -6.192580    -8.393968
N        -4.846012    -9.182107    -7.204948
H        -3.988537    -9.313012    -6.711513
C        -5.828424   -10.255158    -7.239410
H        -6.001956   -10.584056    -8.252976
H        -5.430288   -11.085676    -6.668447
C        -7.207602    -9.905697    -6.656110
```

```
O         -8.221584    -10.297350    -7.164233
N         -7.176471     -9.204614    -5.516534
C         -8.405087     -8.888694    -4.813583
C         -9.295053     -7.892144    -5.549413
O        -10.487679     -7.899805    -5.403543
C         -8.123801     -8.374777    -3.392059
C         -7.335396     -7.061709    -3.261420
C         -8.076275     -5.748480    -3.272464
O         -9.376603     -5.830778    -3.312950
H         -9.777446     -4.958579    -3.223018
O         -7.492178     -4.695781    -3.211041
H         -6.297958     -8.952920    -5.115385
H         -9.002589     -9.785822    -4.732765
H         -9.073941     -8.290667    -2.885504
H         -7.556829     -9.148504    -2.885423
H         -6.809518     -7.062844    -2.312834
H         -6.568258     -6.978572    -4.012532
N         -8.659483     -6.956499    -6.285211
C         -9.413283     -5.938871    -6.996586
C        -10.435645     -6.574016    -7.933625
H        -10.226589     -7.596577    -8.250185
O        -11.392722     -5.979786    -8.314266
C         -8.412166     -5.020115    -7.727878
O         -7.712074     -4.259095    -6.766165
C         -9.057410     -4.040272    -8.697726
H         -7.703276     -7.105189    -6.523395
H         -9.976016     -5.342482    -6.289748
H         -7.706396     -5.637513    -8.273658
H         -6.962397     -4.744461    -6.429734
H         -8.292053     -3.357331    -9.048877
H         -9.491093     -4.546015    -9.549108
H         -9.844202     -3.472736    -8.213192
N        -12.664563     -9.429889     3.164222
H        -12.922766     -9.946953     2.349065
C        -11.516487     -8.546511     2.893191
C        -10.519175     -8.500792     4.031525
O         -9.392645     -8.085782     3.829688
C        -11.932060     -7.082274     2.633439
O        -12.474845     -6.633573     3.848899
C        -12.935884     -6.926080     1.497481
H        -13.458516     -8.892653     3.454407
H        -10.971693     -8.917900     2.037330
H        -11.033097     -6.527056     2.397600
H        -12.541560     -5.680343     3.845233
H        -13.128231     -5.872036     1.327427
H        -12.569422     -7.352180     0.570999
H        -13.881137     -7.394637     1.743802
N        -10.912409     -8.912490     5.227241
H        -11.836108     -9.282675     5.278305
C        -10.117759     -8.874740     6.429074
H        -10.065494     -7.874086     6.829758
H        -10.611782     -9.494111     7.168415
C         -8.667437     -9.345701     6.335156
O         -7.838228     -8.745009     6.973389
```

```
N         -8.330303    -10.424052     5.604750
H         -9.032680    -10.929901     5.114057
C         -6.931824    -10.817710     5.495285
H         -6.890673    -11.799038     5.039984
H         -6.481718    -10.878653     6.473277
C         -6.079123     -9.857234     4.644362
O         -4.909138     -9.699485     4.884645
N         -6.757552     -9.275458     3.661449
C         -6.038043     -8.278086     2.880352
C         -6.032951     -6.932366     3.598651
O         -5.071207     -6.205153     3.495938
C         -6.365409     -8.211815     1.375404
C         -7.563889     -7.414372     0.907779
C         -7.554388     -6.022942     0.960824
C         -8.650173     -8.045961     0.317469
C         -8.615467     -5.285316     0.463498
C         -9.703604     -7.310186    -0.200852
C         -9.695539     -5.927888    -0.119682
H         -7.751975     -9.304257     3.637764
H         -5.007752     -8.584803     2.922330
H         -5.479883     -7.798417     0.905878
H         -6.438147     -9.235389     1.027308
H         -6.703622     -5.508566     1.370048
H         -8.660485     -9.119306     0.232669
H         -8.582834     -4.211204     0.502953
H        -10.517229     -7.813391    -0.690783
H        -10.505638     -5.358031    -0.535137
N         -7.068614     -6.639245     4.386842
C         -7.049148     -5.501546     5.271473
C         -5.961075     -5.671111     6.331199
O         -5.372862     -4.692340     6.740093
C         -8.450381     -5.333199     5.899359
C         -8.751341     -3.960707     6.461252
C         -8.965609     -2.885350     5.600933
C         -8.874792     -3.754368     7.827316
C         -9.286704     -1.632532     6.093501
C         -9.195603     -2.500391     8.327972
C         -9.398100     -1.436530     7.465149
H         -7.903950     -7.178532     4.319669
H         -6.779097     -4.608454     4.727083
H         -9.172151     -5.551465     5.119998
H         -8.582433     -6.083469     6.668561
H         -8.872206     -3.025792     4.536719
H         -8.718599     -4.572081     8.508817
H         -9.448415     -0.817091     5.412546
H         -9.275206     -2.356397     9.390682
H         -9.617167     -0.457767     7.853658
N         -5.685488     -6.900498     6.761219
H         -6.315177     -7.639717     6.543351
C         -4.631720     -7.174764     7.711953
H         -4.686018     -6.524319     8.571651
H         -4.754942     -8.196190     8.048555
C         -3.227715     -7.014292     7.133762
O         -2.367537     -6.481107     7.793480
```

```
N         -3.003625     -7.484309      5.900788
C         -1.725803     -7.275557      5.240277
C         -1.513059     -5.794499      4.908379
O         -0.427999     -5.287363      5.091746
C         -1.611442     -8.184485      3.996883
C         -0.441793     -7.783753      3.093626
C         -1.478461     -9.650672      4.424243
H         -3.713061     -8.010858      5.437440
H         -0.929713     -7.514148      5.930938
H         -2.527989     -8.070681      3.423080
H         -0.359090     -8.489264      2.273441
H         -0.559254     -6.798317      2.663451
H          0.496694     -7.796616      3.639845
H         -1.469878    -10.294669      3.550747
H         -0.544786     -9.805130      4.959131
H         -2.292382     -9.972110      5.060412
N         -2.558118     -5.116075      4.420471
C         -2.426588     -3.732439      4.010585
C         -2.324387     -2.803144      5.221773
O         -1.475246     -1.932802      5.235110
C         -3.575238     -3.292506      3.091832
C         -3.628136     -3.979870      1.713639
C         -2.479503     -3.590899      0.790552
O         -1.342591     -3.954180      1.143503
O         -2.739628     -2.938473     -0.234984
H         -3.410076     -5.596120      4.225320
H         -1.497605     -3.622503      3.481084
H         -4.519995     -3.463047      3.596899
H         -3.484559     -2.219278      2.954851
H         -3.614135     -5.053802      1.840203
H         -4.563888     -3.711327      1.237959
N         -3.177610     -2.969248      6.241614
C         -2.974605     -2.214912      7.461881
C         -1.640742     -2.582382      8.119778
O         -0.980839     -1.722762      8.662099
C         -4.109964     -2.244122      8.495206
O         -4.313550     -3.515362      9.037236
C         -5.394365     -1.610222      7.972251
H         -3.839910     -3.713084      6.219782
H         -2.847728     -1.180019      7.184399
H         -3.742723     -1.638615      9.314637
H         -4.785192     -4.049063      8.409040
H         -6.133761     -1.601223      8.763995
H         -5.216438     -0.585765      7.660682
H         -5.803610     -2.157134      7.133035
N         -1.210544     -3.848625      8.020389
H         -1.843776     -4.589353      7.811082
C          0.050815     -4.201498      8.622318
H          0.083492     -3.965337      9.676576
H          0.189537     -5.267439      8.496148
C          1.221575     -3.477235      7.972298
O          2.158276     -3.080226      8.626546
N          1.159934     -3.329648      6.637167
C          2.250933     -2.710311      5.932569
```

```
C        2.307638   -1.197139    6.114518
O        3.391932   -0.651233    6.088007
C        2.246399   -3.076065    4.437518
S        1.378895   -1.930579    3.318776
H        0.431044   -3.787567    6.128394
H        3.178921   -3.066031    6.357491
H        3.270579   -3.062558    4.087529
H        1.859735   -4.077278    4.311634
H        0.164234   -1.986988    3.842867
N        1.164542   -0.526618    6.278802
C        1.235250    0.904035    6.497816
C        1.821366    1.192585    7.867627
O        2.444194    2.205491    8.080769
C       -0.076753    1.680972    6.250513
C       -1.140346    1.495259    7.344063
C       -0.598007    1.407534    4.835329
C       -2.418168    2.302508    7.103229
H        0.290780   -1.009638    6.259376
H        1.960746    1.295620    5.801259
H        0.231373    2.719880    6.294890
H       -1.387546    0.448886    7.459256
H       -0.718301    1.808024    8.294147
H       -1.340271    2.142554    4.552689
H        0.207234    1.461563    4.108774
H       -1.053415    0.429760    4.751920
H       -3.057469    2.263117    7.979825
H       -2.199018    3.347843    6.901398
H       -2.989606    1.916613    6.265996
N        1.617625    0.243785    8.787960
H        0.981755   -0.499819    8.600173
C        2.050768    0.429130   10.154569
H        1.549636   -0.311413   10.765252
H        1.802070    1.414861   10.518231
C        3.555361    0.267358   10.306953
O        4.214056    1.036300   10.953169
N        4.098247   -0.786619    9.658822
H        3.511996   -1.446043    9.192314
C        5.527090   -0.933177    9.619740
H        5.763342   -1.888623    9.166057
H        5.955740   -0.908823   10.610386
C        6.228689    0.159508    8.816932
O        7.355392    0.482952    9.101689
N        5.555629    0.653155    7.778697
C        6.120025    1.811755    7.133439
C        5.515616    3.112176    7.655833
H        5.026430    3.053767    8.626533
O        5.611046    4.137864    7.055268
C        6.008296    1.737836    5.608566
C        6.784554    0.591938    4.994027
C        8.162248    0.463588    5.177200
C        6.135465   -0.354819    4.207908
C        8.871836   -0.563028    4.569974
C        6.843548   -1.379535    3.599952
C        8.214360   -1.484291    3.769247
```

```
H            4.630602      0.352148     7.562127
H            7.158632      1.845458     7.426406
H            4.964569      1.648830     5.333407
H            6.358299      2.683683     5.209822
H            8.682240      1.149650     5.824100
H            5.068106     -0.303880     4.103691
H            9.931830     -0.647230     4.736986
H            6.321910     -2.105105     3.001245
H            8.752085     -2.283556     3.292809
N           12.383857      9.886082    -1.227636
H           11.630001     10.191693    -1.814265
C           12.112715      8.542590    -0.708150
C           11.085237      8.693381     0.406429
O           10.095792      7.994430     0.438015
C           13.391383      7.886848    -0.167975
C           14.475049      7.620190    -1.218230
S           13.868045      6.540985    -2.547125
C           15.347148      6.442829    -3.583754
H           13.198078      9.888251    -1.809547
H           11.654092      7.888713    -1.438492
H           13.818114      8.515339     0.607837
H           13.111438      6.947613     0.295098
H           14.847990      8.546716    -1.640023
H           15.316364      7.145977    -0.725587
H           15.107796      5.789869    -4.409240
H           15.617264      7.418448    -3.966705
H           16.180919      6.022870    -3.036382
N           11.295982      9.655610     1.320992
C           10.388487      9.820145     2.439510
C            9.050916     10.452758     2.053155
O            8.065149     10.198791     2.699204
C           11.057152     10.594026     3.582483
C           12.252313      9.817485     4.104239
O           13.377804     10.107592     3.756372
N           11.979720      8.806247     4.938560
H           12.103934     10.232935     1.227910
H           10.118612      8.840609     2.799267
H           11.411283     11.562274     3.251529
H           10.327490     10.740272     4.368871
H           12.739835      8.228462     5.226151
H           11.068087      8.398111     4.980888
N            9.009229     11.252354     0.970760
H            9.848677     11.443606     0.473489
C            7.733763     11.672740     0.429119
H            7.906541     12.427355    -0.327496
H            7.112401     12.098598     1.199931
C            6.943222     10.524008    -0.199576
O            5.741296     10.577846    -0.230249
N            7.645152      9.495791    -0.720961
C            6.983405      8.293529    -1.197890
C            6.479952      7.422750    -0.035888
O            5.395673      6.896153    -0.116478
C            7.877311      7.526978    -2.201847
C            8.140504      8.400451    -3.443380
```

```
C         7.242399      6.187890     -2.595638
C         9.233707      7.867386     -4.372701
H         8.609818      9.417835     -0.499519
H         6.080261      8.599108     -1.703230
H         8.825439      7.321574     -1.712961
H         7.211700      8.507628     -3.999226
H         8.417932      9.400511     -3.127307
H         7.870034      5.672220     -3.311277
H         7.114159      5.534008     -1.741643
H         6.265695      6.329547     -3.046508
H         9.404524      8.562101     -5.188613
H        10.175992      7.740289     -3.846574
H         8.975894      6.911343     -4.813183
N         7.241240      7.312631      1.071146
C         6.758910      6.575094      2.227283
C         5.394786      7.116569      2.675935
O         4.464116      6.382429      2.904781
C         7.717131      6.680898      3.424162
C         9.152789      6.174196      3.253817
O        10.028205      6.826612      3.815162
O         9.354669      5.124504      2.592663
H         8.192235      7.612284      1.048095
H         6.588370      5.543074      1.962990
H         7.778016      7.705052      3.761116
H         7.279348      6.108991      4.239792
N         5.320965      8.452235      2.829043
H         6.136771      9.010091      2.691104
C         4.116095      9.086744      3.304149
H         4.329186     10.135870      3.467203
H         3.790535      8.665205      4.241985
C         2.949105      9.005195      2.329710
O         1.807928      9.098765      2.718544
N         3.277772      8.856085      1.043021
H         4.236666      8.887923      0.786687
C         2.315110      8.922066     -0.012995
H         2.845449      9.132165     -0.933090
H         1.621128      9.734982      0.150817
C         1.437959      7.682495     -0.250768
O         0.387946      7.788238     -0.819563
N         1.951160      6.542334      0.253240
C         1.171912      5.351145      0.482825
C         0.361371      5.446887      1.779559
O        -0.831956      5.257420      1.757791
C         2.073747      4.117060      0.486749
H         2.855588      6.600485      0.670987
H         0.438235      5.266224     -0.301549
H         1.494563      3.225435      0.701115
H         2.537735      4.009248     -0.486920
H         2.858441      4.202792      1.231184
N         0.998212      5.717252      2.934108
C         0.275445      5.525141      4.183080
C        -0.868864      6.522241      4.349713
O        -1.890221      6.183620      4.908404
C         1.199852      5.466570      5.423842
```

```
C         1.765598      6.822965      5.871129
C         2.308303      4.425232      5.206004
C         2.607500      6.740673      7.146901
H         1.984395      5.864619      2.948433
H        -0.225898      4.569811      4.123881
H         0.553566      5.110669      6.220525
H         2.361547      7.246916      5.069795
H         0.948472      7.515283      6.045784
H         2.693298      4.073751      6.152106
H         1.930509      3.563554      4.664756
H         3.138550      4.836578      4.642139
H         2.895374      7.736762      7.469958
H         2.044140      6.284611      7.955656
H         3.513073      6.162905      7.009565
N        -0.692923      7.764629      3.879136
H         0.180600      8.032433      3.475516
C        -1.694154      8.774355      4.102975
H        -1.314008      9.716363      3.728426
H        -1.904515      8.893212      5.155959
C        -3.041855      8.466288      3.455299
O        -4.047221      8.649548      4.101027
N        -3.070667      7.966250      2.206026
C        -4.406233      7.836078      1.623995
C        -5.094550      6.531640      1.984415
O        -6.304689      6.488564      1.983499
C        -4.208045      8.094939      0.123981
C        -2.974220      9.003829      0.124771
C        -2.092246      8.357384      1.180601
H        -5.048940      8.599905      2.033488
H        -4.012536      7.172667     -0.407075
H        -5.088412      8.550702     -0.310665
H        -2.471857      9.079763     -0.821686
H        -3.257158     10.005150      0.440630
H        -1.588773      7.486191      0.802424
H        -1.347523      9.028405      1.576536
N        -4.330088      5.480246      2.318004
C        -4.939773      4.287578      2.879019
C        -5.662807      4.598426      4.194002
O        -6.734970      4.091019      4.436276
C        -3.908856      3.172631      3.151819
C        -3.498995      2.288368      1.988751
C        -4.070440      1.031025      1.821627
C        -2.471099      2.643734      1.124587
C        -3.608548      0.142446      0.863177
C        -1.993049      1.765001      0.169450
C        -2.533348      0.493957      0.054352
O        -1.980588     -0.355925     -0.831761
H        -3.339965      5.599635      2.365121
H        -5.702994      3.930468      2.205395
H        -3.030555      3.626387      3.598977
H        -4.338682      2.532664      3.914101
H        -4.883486      0.726258      2.458948
H        -2.011482      3.605910      1.209833
H        -4.057421     -0.825899      0.753641
```

```
H          -1.186879    2.052809   -0.482149
H          -2.410248   -1.214754   -0.789572
N          -5.041506    5.405707    5.070182
H          -4.118312    5.738167    4.882226
C          -5.658300    5.744123    6.333700
H          -4.888942    6.133471    6.989071
H          -6.095539    4.871295    6.791254
C          -6.764388    6.791694    6.209783
O          -7.810298    6.680503    6.806277
N          -6.491976    7.831950    5.415786
H          -5.600377    7.910776    4.975696
C          -7.437641    8.938719    5.201742
H          -6.977888    9.611436    4.489175
H          -7.611958    9.479724    6.120432
C          -8.800673    8.534415    4.682785
O          -9.791069    9.104447    5.066729
N          -8.835835    7.546389    3.772404
C         -10.092268    7.121937    3.209379
C         -10.929987    6.249957    4.136488
O         -12.007398    5.854152    3.729266
C          -9.867420    6.401904    1.860625
O         -11.079957    6.338429    1.139119
C          -9.255115    5.015759    2.015190
H          -7.982515    7.141712    3.446454
H         -10.703100    7.995526    3.023360
H          -9.200640    7.024912    1.279211
H         -11.780557    6.166429    1.764307
H          -9.038669    4.594942    1.039850
H          -8.328106    5.042938    2.566355
H          -9.940372    4.350556    2.524124
N         -10.456258    5.965807    5.348047
C         -11.302903    5.322569    6.337304
C         -12.343941    6.267083    6.945964
O         -13.316478    5.802174    7.481217
C         -10.474461    4.700681    7.469339
C          -9.476803    3.616299    7.037939
C          -8.687130    3.140127    8.261023
C         -10.151012    2.434062    6.338097
H          -9.553545    6.284560    5.629687
H         -11.883924    4.561948    5.842119
H          -9.932887    5.490667    7.979045
H         -11.180362    4.285920    8.181983
H          -8.769630    4.051118    6.342279
H          -7.940700    2.402325    7.980096
H          -8.173974    3.968077    8.739364
H          -9.342523    2.685460    9.001295
H          -9.412913    1.679941    6.082584
H         -10.901186    1.971142    6.977240
H         -10.637981    2.725219    5.414616
N         -12.075339    7.588921    6.914073
C         -13.044269    8.545836    7.358486
C         -13.963462    9.038987    6.261640
H         -14.736248    9.737014    6.603628
O         -13.879623    8.735855    5.118199
```

```
H           -11.334800      7.927312      6.339070
H           -13.647074      8.109472      8.143917
H           -12.541807      9.412364      7.780728
N           -11.774269     10.301570     -2.867425
H           -11.550479     10.511687     -1.915085
C           -12.057381      8.873136     -3.031538
C           -11.637701      8.561088     -4.461752
O           -10.573975      8.047280     -4.711819
C           -13.477255      8.411683     -2.638576
C           -13.738860      8.720923     -1.152848
C           -13.669542      6.921041     -2.946100
C           -15.150582      8.382236     -0.670443
H           -12.569189     10.862537     -3.109816
H           -11.346835      8.335175     -2.419433
H           -14.202226      8.982667     -3.223557
H           -13.012665      8.183499     -0.547093
H           -13.568194      9.778381     -0.975725
H           -14.695271      6.617766     -2.780854
H           -13.428081      6.677733     -3.976698
H           -13.037964      6.313969     -2.305421
H           -15.290557      8.734224      0.345229
H           -15.905575      8.856387     -1.292304
H           -15.336959      7.314989     -0.667488
N           -12.454351      8.971177     -5.463497
H           -13.386990      9.230152     -5.234506
C           -12.203840      8.540113     -6.822271
H           -12.239052      7.462990     -6.914349
H           -12.972240      8.963496     -7.457590
C           -10.843941      8.968308     -7.379885
O           -10.218054      8.251090     -8.106106
N           -10.441202     10.224953     -7.045037
H           -11.000653     10.735764     -6.401013
C            -9.186693     10.805485     -7.496849
H            -9.082878     10.689732     -8.564582
H            -9.192905     11.859428     -7.251726
C            -7.950004     10.154849     -6.859773
O            -7.030748      9.742790     -7.510898
N            -7.924642     10.148202     -5.491991
H            -8.750879     10.403576     -4.999946
C            -6.746635      9.782074     -4.713273
H            -5.864645     10.225397     -5.147240
H            -6.874295     10.170116     -3.710183
C            -6.476702      8.292089     -4.610371
O            -5.330600      7.909733     -4.664048
N            -7.492467      7.436523     -4.399787
C            -7.153435      6.077348     -3.999693
C            -6.570537      5.265334     -5.149732
O            -5.743984      4.414192     -4.923375
C            -8.275357      5.318294     -3.252192
C            -9.452254      4.883170     -4.142003
C            -8.705067      6.123713     -2.020850
C           -10.516487      4.070834     -3.400982
H            -8.439857      7.756500     -4.382877
H            -6.326213      6.147364     -3.309479
```

```
H         -7.789055    4.415569   -2.893135
H         -9.915572    5.746170   -4.604973
H         -9.065322    4.267012   -4.949600
H         -9.336642    5.531938   -1.373288
H         -7.842089    6.427344   -1.436729
H         -9.259547    7.013582   -2.296123
H        -11.254875    3.692881   -4.100414
H        -10.083257    3.214659   -2.897101
H        -11.040818    4.667261   -2.664049
N         -6.998006    5.503105   -6.407942
C         -6.362224    4.816119   -7.512847
C         -4.962937    5.348416   -7.824774
O         -4.110131    4.591822   -8.224643
C         -7.258994    4.807640   -8.759840
C         -8.375610    3.758623   -8.670076
C         -7.828880    2.341058   -8.860548
N         -8.821775    1.285910   -8.671057
C         -9.140480    0.701623   -7.518305
N         -8.653572    1.117996   -6.365612
N         -9.980252   -0.325030   -7.514862
H         -7.655215    6.233623   -6.576107
H         -6.167801    3.806302   -7.192415
H         -7.691418    5.791967   -8.909186
H         -6.641402    4.606136   -9.629449
H         -8.889923    3.841983   -7.717773
H         -9.118264    3.951122   -9.436925
H         -7.456656    2.226713   -9.869675
H         -6.987218    2.139202   -8.211451
H         -9.195402    0.868814   -9.493519
H         -8.081304    1.928561   -6.314791
H         -9.071247    0.799664   -5.496124
H        -10.263431   -0.722632   -8.382475
H         -9.927012   -0.971018   -6.738523
N         -4.754067    6.653678   -7.639028
H         -5.497497    7.241151   -7.336808
C         -3.458658    7.285624   -7.930073
H         -3.131748    7.045217   -8.929764
H         -3.589236    8.356009   -7.845070
C         -2.379194    6.827494   -6.960669
O         -1.247966    6.625809   -7.326991
N         -2.790477    6.693948   -5.694638
C         -1.893827    6.400219   -4.603203
C         -1.572135    4.921459   -4.490809
O         -0.430339    4.569941   -4.297964
C         -2.493377    6.963548   -3.308186
C         -1.882206    6.271030   -2.091065
C         -2.303139    8.486730   -3.327439
H         -3.721010    6.968701   -5.464970
H         -0.939124    6.870919   -4.789568
H         -3.555952    6.750370   -3.321073
H         -2.113193    6.798492   -1.184326
H         -2.261283    5.259152   -1.992765
H         -0.808109    6.233310   -2.164805
H         -3.049481    8.984000   -2.724442
```

```
H         -1.319034    8.756132   -2.958611
H         -2.409773    8.878835   -4.332153
N         -2.575259    4.037305   -4.605964
C         -2.222951    2.639263   -4.694383
C         -1.324197    2.410477   -5.902598
O         -0.418234    1.611602   -5.856302
C         -3.442005    1.700715   -4.705460
C         -4.272908    1.753024   -3.415302
C         -5.392390    0.695236   -3.355790
N         -4.964371   -0.487506   -2.638029
C         -5.736553   -1.483163   -2.534551
N         -7.015059   -1.592161   -3.091903
N         -5.426364   -2.549541   -1.735825
H         -3.511859    4.333369   -4.781694
H         -1.608930    2.380591   -3.844315
H         -4.073469    1.938191   -5.559538
H         -3.068722    0.693524   -4.856035
H         -3.627602    1.600034   -2.559519
H         -4.713334    2.737240   -3.320084
H         -6.258872    1.141773   -2.868512
H         -5.707339    0.465385   -4.378258
H         -7.156105   -1.056987   -3.921307
H         -7.328802   -2.535277   -3.219982
H         -4.495864   -2.589562   -1.369950
H         -5.831305   -3.430454   -1.968264
N         -1.546899    3.166383   -6.996490
H         -2.384008    3.695807   -7.104179
C         -0.684621    2.988588   -8.146855
H         -1.048786    3.629593   -8.939611
H         -0.677718    1.965414   -8.494111
C          0.761414    3.366436   -7.833196
O          1.704631    2.751445   -8.266111
N          0.886570    4.434447   -7.030068
H          0.077947    4.970822   -6.811251
C          2.150711    4.947380   -6.607804
H          2.774349    5.237305   -7.444073
H          1.969938    5.828493   -6.005802
C          2.988861    4.006517   -5.765766
O          4.192980    4.151955   -5.695960
N          2.354379    3.042206   -5.086466
C          3.081572    2.233818   -4.149153
C          4.274256    1.480442   -4.743854
O          5.224074    1.228848   -4.027510
C          2.091308    1.277985   -3.483307
C          2.701082    0.445943   -2.366101
C          1.630090   -0.376493   -1.677118
N          2.267514   -1.276559   -0.727735
C          1.605599   -2.275340   -0.135155
N          0.319277   -2.373148   -0.211845
N          2.349633   -3.204568    0.493186
H          1.359993    2.965328   -5.140961
H          3.535366    2.865429   -3.394631
H          1.287387    1.888614   -3.085004
H          1.642934    0.639218   -4.236031
```

```
H            3.460178    -0.218313    -2.766251
H            3.194403     1.086214    -1.640124
H            0.929690     0.260970    -1.154057
H            1.071727    -0.947326    -2.412459
H            3.242728    -1.428775    -0.853500
H           -0.224512    -1.676438    -0.678853
H           -0.238034    -3.061896     0.324159
H            3.185036    -2.867870     0.917627
H            1.859640    -3.863432     1.066685
N            4.255160     1.062184    -6.017655
H            3.495761     1.315850    -6.616364
C            5.356056     0.264885    -6.527831
H            5.134253     0.006589    -7.555780
H            5.460038    -0.651013    -5.967442
C            6.711559     0.940334    -6.498776
O            7.730666     0.298281    -6.395134
N            6.706442     2.270224    -6.621776
H            5.839990     2.768362    -6.583816
C            7.927348     3.039685    -6.568360
H            8.635467     2.679711    -7.301300
H            7.693344     4.066636    -6.815194
C            8.681407     3.052067    -5.236507
O            9.813889     3.485641    -5.192020
N            8.055645     2.544411    -4.166993
C            8.650716     2.387740    -2.850913
C            9.934176     1.546812    -2.871461
O           10.756469     1.643148    -1.974849
C            7.568167     1.691852    -1.995590
C            7.779636     1.560365    -0.487081
C            7.404784     2.811082     0.317200
C            7.314321     2.526708     1.812828
N            8.659764     2.532590     2.459248
O           11.041439     1.863703     0.850900
H           10.883284     2.079267    -0.068494
H           11.538746     1.048805     0.791572
H            7.115177     2.222033    -4.274544
H            8.931345     3.349542    -2.442288
H            6.639722     2.219138    -2.168699
H            7.417572     0.704827    -2.417773
H            7.125707     0.758072    -0.158624
H            8.786470     1.234814    -0.261518
H            8.091564     3.634946     0.144614
H            6.426425     3.156445    -0.001087
H            6.732013     3.275340     2.327000
H            6.865679     1.563523     2.002905
H            9.380389     2.094429     1.894021
H            8.951158     3.512073     2.617746
H            8.627893     2.047850     3.343990
N           10.058650     0.696606    -3.881387
C           11.129472    -0.259667    -4.125770
C           12.489143     0.339165    -4.489112
O           13.465331    -0.366932    -4.376726
C           10.735989    -1.171172    -5.309690
C            9.624889    -2.198394    -5.049916
```

```
C         9.198826    -2.829031    -6.379392
C        10.086508    -3.281167    -4.074183
H         9.318266     0.683740    -4.552079
H        11.306775    -0.846992    -3.236697
H        10.447111    -0.529551    -6.136758
H        11.630117    -1.700232    -5.616872
H         8.758645    -1.689694    -4.637881
H         8.416909    -3.565845    -6.221833
H         8.817637    -2.078794    -7.060355
H        10.035197    -3.334490    -6.855692
H         9.317612    -4.033170    -3.933316
H        10.968253    -3.782358    -4.459495
H        10.334401    -2.881562    -3.095488
N        12.542524     1.584484    -4.968577
C        13.765787     2.202097    -5.444716
C        14.910696     2.409630    -4.434670
O        15.954544     2.815639    -4.870377
C        13.411961     3.526644    -6.124218
O        12.813856     4.419940    -5.228417
H        11.710868     2.135565    -4.957977
H        14.211829     1.552525    -6.188459
H        14.322623     3.972013    -6.493133
H        12.762343     3.324321    -6.969834
H        11.866893     4.390685    -5.292755
N        14.757554     2.087271    -3.130783
C        15.898589     2.034411    -2.238236
C        16.805271     0.821235    -2.476649
O        17.935779     0.830173    -2.063942
C        15.472907     1.943926    -0.767740
C        14.595386     3.079522    -0.245368
C        14.258499     2.817660     1.223026
N        13.216154     3.685176     1.757003
C        13.370553     4.832917     2.397392
N        14.577967     5.334231     2.632910
O        14.887347     7.785931     4.204727
H        14.578776     8.665622     3.971311
H        15.604143     7.908461     4.814163
N        12.306995     5.491581     2.807899
H        13.873120     1.768387    -2.811523
H        16.510192     2.908101    -2.403639
H        14.952133     1.003743    -0.617587
H        16.388419     1.888623    -0.189083
H        15.091918     4.038693    -0.360045
H        13.670505     3.144588    -0.806631
H        13.886981     1.808507     1.328228
H        15.141340     2.878477     1.848384
H        12.282184     3.369420     1.588875
H        15.395644     4.865869     2.322748
H        14.695290     6.187779     3.151491
H        11.357379     5.190863     2.652281
H        12.383061     6.331278     3.336222
N        16.227348    -0.247290    -3.050912
C        16.929294    -1.490185    -3.266240
C        17.718560    -1.436230    -4.570917
```

```
H         18.090134   -0.445400   -4.841683
O         17.940201   -2.387623   -5.246250
C         15.957168   -2.674677   -3.251207
C         15.430677   -3.029401   -1.893381
N         14.492872   -2.287890   -1.202385
O         12.530502   -0.160862   -0.526063
H         13.189211   -0.736126   -0.919235
H         12.046009    0.237110   -1.241214
C         15.759939   -4.117039   -1.161279
C         14.285604   -2.926542   -0.097458
N         15.027110   -4.037674   -0.003582
H         15.309545   -0.168152   -3.433238
H         17.662417   -1.596689   -2.474392
H         15.130048   -2.462611   -3.919868
H         16.477984   -3.531995   -3.655863
H         16.436989   -4.920770   -1.356654
H         13.603393   -2.621376    0.670151
H         14.950024   -4.728121    0.710725
O        -14.438086   -2.383573    5.091980
H        -14.767973   -1.787825    4.422158
H        -14.099199   -1.827430    5.786474
O        -12.890300   -0.584917    6.951697
H        -12.132422   -1.045627    7.292992
H        -13.246819   -0.080253    7.672366
O        -12.135746    0.209585    4.158463
H        -12.314066    0.221434    5.092147
H        -12.957235   -0.047264    3.739165
O        -12.837848   -2.314602    1.292744
H        -12.098380   -1.737748    1.093041
H        -12.582524   -2.869011    2.028398
O         -7.036657    1.814962   -0.038654
H         -7.225215    1.880872    0.900906
H         -6.141776    1.502057   -0.098539
O         -7.717176    1.874269    2.715620
H         -8.670871    1.824102    2.688340
H         -7.496990    2.582716    3.315511
O         -8.793847    0.063963   -1.356228
H         -8.229756    0.699752   -0.902017
H         -8.202043   -0.592126   -1.723694
O        -13.028528   -2.718408   -1.533118
H        -13.194345   -2.680124   -0.593613
H        -12.154913   -3.082244   -1.625007
O        -13.061449    1.290371   -0.778209
H        -12.441591    0.650155   -0.422362
H        -12.639838    2.134677   -0.618408
O        -14.098216    4.428300    2.099492
H        -13.767483    4.776763    2.922080
H        -14.582774    3.631437    2.285102
O        -15.111912    1.660826    1.203236
H        -15.948448    1.770159    0.769981
H        -14.468988    1.521245    0.502477
O        -12.668751   -0.443425   -3.150057
H        -12.982382    0.332983   -2.692212
H        -12.818264   -1.162455   -2.528201
```

```
O        -12.299681   -2.511599   -5.076945
H        -12.624584   -1.709556   -4.666812
H        -13.058730   -3.041389   -5.287997
O        -10.592271   -3.361633   -2.937916
H         -9.930690   -2.692983   -3.079320
H        -11.253597   -3.197991   -3.610429
O        -11.005167   -0.364898    0.396741
H        -10.270494   -0.380491   -0.219277
H        -10.747130    0.248030    1.089289
O        -10.582475    1.634128    2.322356
H        -11.130577    1.312913    3.041652
H        -11.058205    2.324770    1.862869
O        -11.974884    3.550962    0.543654
H        -12.747957    3.850999    1.031936
H        -11.464680    4.343137    0.405926
O         -9.971490    0.540979   -3.903732
H        -10.851526    0.198577   -3.758063
H         -9.549248    0.514014   -3.044481
O        -14.525230   -0.619052    2.842211
H        -14.890310    0.126340    2.366871
H        -14.119313   -1.189423    2.188264
O        -12.385611   -3.733866    3.807224
H        -13.142009   -3.345255    4.266564
H        -11.623186   -3.381574    4.250543
O         -9.420324   -2.321268   -5.567122
H        -10.278427   -2.728276   -5.512558
H         -8.803969   -2.991078   -5.871417
```

Case 2:
Voltage = 0; Y266 negative, R300 positive, E183 neutral
```
N          15.815911   -6.171507    5.162723
H          15.012115   -5.956338    5.720342
C          15.718932   -5.550103    3.839266
C          15.012365   -6.475093    2.829898
O          14.264709   -6.039831    1.969796
C          17.116994   -5.218940    3.303618
H          16.617975   -5.835149    5.658171
H          15.128406   -4.643710    3.858853
H          17.068657   -4.811800    2.300728
H          17.583876   -4.474228    3.938259
H          17.749422   -6.099719    3.289140
N          15.298602   -7.772676    2.926189
H          15.890371   -8.048875    3.680154
C          14.687521   -8.778222    2.055239
H          14.756666   -8.484825    1.020508
H          15.207847   -9.717029    2.190850
C          13.208188   -8.956166    2.399416
O          12.377761   -9.166801    1.554734
N          12.882012   -8.827389    3.699958
H          13.603979   -8.750648    4.379008
```

```
C        11.501007   -8.990668    4.160085
H        11.500864   -9.001261    5.241886
H        11.073110   -9.912783    3.798654
C        10.607647   -7.843954    3.669621
O         9.465056   -8.044802    3.355847
N        11.183937   -6.630558    3.646549
C        10.519402   -5.456882    3.137645
C        10.320011   -5.486917    1.619588
O         9.275433   -5.077566    1.172390
C        11.261767   -4.199859    3.618586
O        12.652266   -4.343916    3.599424
H        12.141561   -6.538812    3.893979
H         9.516113   -5.412177    3.528535
H        10.943084   -3.347172    3.029667
H        10.997165   -4.015588    4.649510
H        12.978203   -4.499861    2.721504
N        11.302309   -5.952837    0.828231
C        11.043830   -6.125794   -0.598581
C         9.920193   -7.142080   -0.815398
O         9.166260   -7.039064   -1.754875
C        12.337943   -6.461504   -1.400249
C        12.579909   -7.955836   -1.649226
C        12.336292   -5.708423   -2.731942
H        12.139273   -6.323653    1.222652
H        10.645791   -5.194130   -0.967833
H        13.162736   -6.080200   -0.808697
H        13.535225   -8.081946   -2.148189
H        12.599157   -8.540735   -0.740568
H        11.820913   -8.370919   -2.305224
H        13.234740   -5.935553   -3.294916
H        11.480231   -5.991949   -3.334339
H        12.306844   -4.636609   -2.576471
N         9.827345   -8.118301    0.095592
H        10.577542   -8.244162    0.736091
C         8.879798   -9.223877   -0.009149
H         9.249095  -10.026356    0.616674
H         8.822607   -9.584183   -1.024712
C         7.461395   -8.899175    0.411818
O         6.546433   -9.371842   -0.222127
N         7.267304   -8.106214    1.471719
C         5.922028   -7.662010    1.759189
C         5.467797   -6.643661    0.719015
O         4.288010   -6.579464    0.447490
C         5.664027   -7.190546    3.204285
C         5.843202   -8.349467    4.190133
C         6.476248   -5.967681    3.634998
H         8.041084   -7.827455    2.034821
H         5.261471   -8.497403    1.586382
H         4.615521   -6.908950    3.209784
H         5.551122   -8.038826    5.187782
H         5.225578   -9.198378    3.915493
H         6.874807   -8.680547    4.232569
H         6.133470   -5.624046    4.605080
H         7.526348   -6.211817    3.728883
```

| | | | |
|---|---:|---:|---:|
| H | 6.375877 | -5.141572 | 2.939627 |
| N | 6.363071 | -5.877460 | 0.087204 |
| C | 5.936976 | -5.048159 | -1.027868 |
| C | 5.525631 | -5.920726 | -2.221910 |
| O | 4.493851 | -5.686874 | -2.806180 |
| C | 6.996928 | -3.989225 | -1.391640 |
| C | 7.172004 | -2.991739 | -0.232712 |
| C | 6.598023 | -3.262041 | -2.680452 |
| C | 8.441738 | -2.146489 | -0.335961 |
| H | 7.326475 | -5.909064 | 0.342282 |
| H | 5.024579 | -4.544371 | -0.746201 |
| H | 7.941423 | -4.499968 | -1.555842 |
| H | 6.298333 | -2.342640 | -0.198191 |
| H | 7.201641 | -3.525092 | 0.708021 |
| H | 7.279065 | -2.447758 | -2.886933 |
| H | 6.612884 | -3.918794 | -3.540651 |
| H | 5.598150 | -2.848743 | -2.599071 |
| H | 8.528653 | -1.488305 | 0.521968 |
| H | 9.321457 | -2.779131 | -0.354172 |
| H | 8.455705 | -1.527074 | -1.225707 |
| N | 6.325141 | -6.934911 | -2.580644 |
| H | 7.225240 | -7.038912 | -2.161767 |
| C | 6.001247 | -7.762011 | -3.723308 |
| H | 6.848899 | -8.409883 | -3.909055 |
| H | 5.826972 | -7.163637 | -4.605121 |
| C | 4.755989 | -8.628406 | -3.532713 |
| O | 3.916813 | -8.694495 | -4.399884 |
| N | 4.667477 | -9.305886 | -2.381526 |
| H | 5.405486 | -9.252845 | -1.712100 |
| C | 3.503229 | -10.098282 | -2.062547 |
| H | 3.279708 | -10.803426 | -2.849034 |
| H | 3.713572 | -10.651382 | -1.155703 |
| C | 2.248189 | -9.253726 | -1.853224 |
| O | 1.171058 | -9.664212 | -2.221428 |
| N | 2.432037 | -8.060222 | -1.290929 |
| C | 1.278730 | -7.174304 | -1.222456 |
| C | 0.821087 | -6.680836 | -2.597439 |
| O | -0.365668 | -6.562916 | -2.809270 |
| C | 1.450110 | -5.958168 | -0.316435 |
| O | 1.609903 | -6.302678 | 1.019995 |
| H | 3.322655 | -7.755572 | -0.969563 |
| H | 0.444643 | -7.739153 | -0.839360 |
| H | 2.267481 | -5.338639 | -0.668220 |
| H | 0.533492 | -5.397167 | -0.393286 |
| H | 2.521394 | -6.515636 | 1.166869 |
| N | 1.735690 | -6.364452 | -3.523254 |
| C | 1.322748 | -5.943973 | -4.850174 |
| C | 0.580855 | -7.071626 | -5.563151 |
| O | -0.413139 | -6.829197 | -6.216064 |
| C | 2.517962 | -5.398415 | -5.666596 |
| C | 2.889693 | -4.003825 | -5.135603 |
| C | 2.218321 | -5.360277 | -7.170647 |
| C | 4.199134 | -3.450508 | -5.695892 |
| H | 2.708044 | -6.396674 | -3.303160 |
| H | 0.584750 | -5.160708 | -4.754566 |
| H | 3.356524 | -6.069514 | -5.507594 |
| H | 2.079121 | -3.315016 | -5.367377 |
| H | 2.968073 | -4.043740 | -4.056935 |
| H | 3.064376 | -4.956828 | -7.712833 |
| H | 2.027063 | -6.347486 | -7.573536 |
| H | 1.354974 | -4.738850 | -7.386983 |
| H | 4.471026 | -2.538718 | -5.174713 |
| H | 5.009933 | -4.159439 | -5.561594 |
| H | 4.131718 | -3.215916 | -6.752424 |
| N | 1.051637 | -8.314455 | -5.427477 |
| H | 1.903118 | -8.479454 | -4.931906 |
| C | 0.299633 | -9.430045 | -5.949333 |
| H | 0.188985 | -9.370557 | -7.021905 |
| H | 0.833390 | -10.339442 | -5.704314 |
| C | -1.106360 | -9.504557 | -5.359776 |
| O | -2.060501 | -9.732088 | -6.069224 |
| N | -1.211208 | -9.314421 | -4.045751 |
| C | -2.512865 | -9.372129 | -3.423385 |
| C | -3.439843 | -8.244078 | -3.877452 |
| O | -4.629907 | -8.477155 | -3.965478 |
| C | -2.417153 | -9.359004 | -1.890916 |
| O | -3.658129 | -9.655855 | -1.311407 |
| H | -0.393945 | -9.226632 | -3.477870 |
| H | -3.017615 | -10.280884 | -3.719320 |
| H | -1.723131 | -10.124238 | -1.572312 |
| H | -2.045592 | -8.399100 | -1.549005 |
| H | -4.354114 | -9.250711 | -1.813396 |
| N | -2.907816 | -7.043418 | -4.102035 |
| C | -3.672128 | -5.903629 | -4.584533 |
| C | -4.250155 | -6.189476 | -5.969087 |
| O | -5.419612 | -5.966818 | -6.216103 |
| C | -2.740833 | -4.667344 | -4.583016 |
| C | -3.140595 | -3.505856 | -5.477645 |
| C | -4.452640 | -3.068153 | -5.613869 |
| C | -2.151970 | -2.841425 | -6.196663 |
| C | -4.768087 | -2.019308 | -6.463105 |
| C | -2.457839 | -1.778325 | -7.029332 |
| C | -3.773179 | -1.369202 | -7.173458 |
| H | -1.940464 | -6.907594 | -3.891406 |
| H | -4.523322 | -5.730873 | -3.944494 |
| H | -2.642331 | -4.339305 | -3.553063 |
| H | -1.762702 | -4.996428 | -4.899528 |
| H | -5.238883 | -3.562430 | -5.077952 |
| H | -1.129356 | -3.164030 | -6.111914 |
| H | -5.797626 | -1.730485 | -6.582119 |
| H | -1.672976 | -1.280946 | -7.570224 |
| H | -4.017539 | -0.559636 | -7.838649 |
| N | -3.413087 | -6.685176 | -6.880158 |
| C | -3.897966 | -7.005665 | -8.204042 |
| C | -5.017455 | -8.043767 | -8.143195 |
| O | -5.995957 | -7.920163 | -8.835136 |
| C | -2.771445 | -7.517283 | -9.105095 |
| S | -1.573702 | -6.249553 | -9.627714 |

```
H     -2.467960    -6.887333    -6.625516
H     -4.344116    -6.129229    -8.649026
H     -2.250491    -8.342440    -8.638619
H     -3.211211    -7.881628   -10.024132
H     -0.934757    -6.111133    -8.476292
N     -4.844098    -9.092525    -7.315832
H     -3.986326    -9.228331    -6.823724
C     -5.823744   -10.166269    -7.361259
H     -5.998837   -10.483794    -8.378229
H     -5.423212   -11.002472    -6.800525
C     -7.201995    -9.825979    -6.773518
O     -8.216432   -10.208508    -7.288888
N     -7.173193    -9.142581    -5.624140
C     -8.391642    -8.823245    -4.901199
C     -9.283781    -7.797061    -5.591752
O    -10.466389    -7.765069    -5.375929
C     -8.096228    -8.332563    -3.471062
C     -7.205135    -7.086206    -3.329419
C     -7.800066    -5.706614    -3.216403
O     -9.097387    -5.610557    -3.211020
H     -9.355896    -4.698419    -3.016040
O     -7.078197    -4.742800    -3.109370
H     -6.290394    -8.899911    -5.226338
H     -9.000226    -9.713712    -4.828684
H     -9.047620    -8.180115    -2.980848
H     -7.596964    -9.147277    -2.957528
H     -6.608600    -7.182011    -2.429863
H     -6.487896    -7.019283    -4.128150
N     -8.660391    -6.884845    -6.361752
C     -9.415942    -5.859912    -7.059435
C    -10.431904    -6.481777    -8.011729
H    -10.231784    -7.508021    -8.323554
O    -11.380705    -5.879493    -8.400540
C     -8.417634    -4.915254    -7.759585
O     -7.705796    -4.201511    -6.767873
C     -9.062589    -3.889129    -8.679890
H     -7.723338    -7.068918    -6.648105
H     -9.984924    -5.285577    -6.339541
H     -7.716845    -5.510514    -8.335664
H     -6.960535    -4.713374    -6.461190
H     -8.291656    -3.209303    -9.024310
H     -9.524748    -4.355751    -9.538420
H     -9.826674    -3.318814    -8.162469
N    -12.648687    -9.488839     3.058955
H    -12.898815   -10.006846     2.242047
C    -11.501708    -8.600016     2.799741
C    -10.502043    -8.566087     3.937168
O     -9.378989    -8.139091     3.740185
C    -11.938870    -7.137400     2.575382
O    -12.528045    -6.745581     3.788926
C    -12.915009    -6.968443     1.416922
H    -13.445335    -8.955729     3.348272
H    -10.957018    -8.950087     1.934190
H    -11.048317    -6.553161     2.384466
```

```
H            -12.575610   -5.792938    3.839620
H            -13.114900   -5.913419    1.263050
H            -12.519116   -7.371986    0.492343
H            -13.859641   -7.454439    1.629614
N            -10.894750   -8.995872    5.126307
H            -11.816196   -9.373014    5.168304
C            -10.099020   -8.986222    6.329155
H            -10.040548   -7.993589    6.747950
H            -10.599214   -9.614499    7.056662
C             -8.649982   -9.465316    6.235949
O             -7.827061   -8.892465    6.906533
N             -8.309789  -10.518610    5.470870
H             -9.009768  -11.003230    4.955995
C             -6.910425  -10.909733    5.351390
H             -6.869764  -11.878907    4.870802
H             -6.461186  -10.995747    6.327774
C             -6.060457   -9.924029    4.525509
O             -4.895128   -9.755453    4.776420
N             -6.740983   -9.331139    3.550898
C             -6.033225   -8.278275    2.833869
C             -6.007952   -7.002819    3.669651
O             -5.030775   -6.291139    3.650802
C             -6.396261   -8.084692    1.346360
C             -7.568891   -7.207952    0.955169
C             -7.543264   -5.834533    1.182797
C             -8.642130   -7.737862    0.251122
C             -8.578963   -5.022811    0.752165
C             -9.666735   -6.923897   -0.205355
C             -9.647577   -5.564417    0.056905
H             -7.735512   -9.354337    3.536299
H             -5.003903   -8.591058    2.825407
H             -5.505074   -7.672503    0.885742
H             -6.517106   -9.076108    0.926010
H             -6.700272   -5.388422    1.677479
H             -8.665919   -8.791570    0.031549
H             -8.537277   -3.963588    0.934797
H            -10.466721   -7.347268   -0.784215
H            -10.442741   -4.930731   -0.290120
N             -7.049127   -6.748929    4.464945
C             -7.024620   -5.641532    5.388471
C             -5.905281   -5.827862    6.411844
O             -5.303885   -4.855587    6.816474
C             -8.412228   -5.508044    6.057831
C             -8.770988   -4.119843    6.543022
C             -9.060277   -3.113744    5.622108
C             -8.880458   -3.833445    7.895282
C             -9.444169   -1.852038    6.042019
C             -9.264792   -2.569863    8.323131
C             -9.545565   -1.577011    7.400791
H             -7.898873   -7.253945    4.337922
H             -6.778110   -4.727777    4.867053
H             -9.148563   -5.804355    5.319014
H             -8.483927   -6.216503    6.873131
H             -8.976501   -3.317597    4.567203
```

```
H         -8.664250    -4.595398     8.623286
H         -9.661181    -1.090200     5.315373
H         -9.334441    -2.363724     9.376175
H         -9.816783    -0.590633     7.732615
N         -5.619961    -7.062417     6.820643
H         -6.245690    -7.803215     6.594606
C         -4.545704    -7.337510     7.748180
H         -4.601650    -6.706879     8.622582
H         -4.643967    -8.368186     8.064548
C         -3.148347    -7.140437     7.163329
O         -2.289709    -6.625997     7.839962
N         -2.927161    -7.568375     5.915304
C         -1.644480    -7.363947     5.259385
C         -1.417029    -5.885553     4.915951
O         -0.325947    -5.386531     5.090879
C         -1.534355    -8.285086     4.024036
C         -0.365100    -7.899908     3.114967
C         -1.412866    -9.747757     4.466376
H         -3.647561    -8.058372     5.429539
H         -0.854950    -7.604622     5.956511
H         -2.451610    -8.172182     3.450721
H         -0.282651    -8.616523     2.304696
H         -0.483529    -6.922214     2.669145
H          0.574127    -7.904811     3.659889
H         -1.406955   -10.400258     3.599250
H         -0.481722    -9.904223     5.005130
H         -2.231111   -10.057025     5.103644
N         -2.462416    -5.209014     4.428393
C         -2.351452    -3.823587     4.014602
C         -2.255259    -2.904640     5.230666
O         -1.384995    -2.064571     5.292789
C         -3.523300    -3.394152     3.120749
C         -3.698811    -4.161346     1.800074
C         -2.559594    -4.000107     0.806157
O         -1.425530    -4.298592     1.205748
O         -2.839077    -3.594287    -0.336331
H         -3.319862    -5.689625     4.266155
H         -1.431504    -3.693230     3.476674
H         -4.447910    -3.478920     3.683518
H         -3.385731    -2.337629     2.908573
H         -3.805050    -5.217214     2.001896
H         -4.617668    -3.823819     1.333817
N         -3.159420    -3.058434     6.216537
C         -2.965211    -2.314132     7.438570
C         -1.638603    -2.690243     8.107663
O         -0.994708    -1.847714     8.694142
C         -4.101043    -2.331440     8.472183
O         -4.279417    -3.583414     9.065579
C         -5.398045    -1.740572     7.929362
H         -3.822660    -3.800511     6.181485
H         -2.832323    -1.278631     7.166347
H         -3.744373    -1.689655     9.268072
H         -4.713583    -4.161683     8.450169
H         -6.135168    -1.712113     8.723086
```

```
H    -5.238320   -0.726688    7.575824
H    -5.800833   -2.327256    7.114479
N    -1.197653   -3.951979    7.973464
H    -1.830960   -4.682641    7.734735
C     0.003080   -4.342738    8.667209
H    -0.059057   -4.157051    9.730255
H     0.146161   -5.402676    8.503573
C     1.237435   -3.607536    8.171552
O     2.102829   -3.265860    8.945185
N     1.325146   -3.403013    6.841039
C     2.486714   -2.759176    6.293230
C     2.387884   -1.240921    6.125658
O     3.434717   -0.641171    5.966455
C     3.049315   -3.443415    5.048091
S     2.128357   -3.210935    3.503743
H     0.602257   -3.757359    6.246744
H     3.257834   -2.853466    7.042792
H     4.030048   -3.030993    4.859988
H     3.158248   -4.500369    5.248178
H     1.129924   -4.055273    3.717552
N     1.217735   -0.611864    6.259419
C     1.238052    0.824396    6.510019
C     1.794097    1.110658    7.890819
O     2.380680    2.139145    8.125306
C    -0.090012    1.560084    6.229855
C    -1.170786    1.337480    7.298077
C    -0.558518    1.263495    4.800381
C    -2.468403    2.103498    7.031414
H     0.367182   -1.136075    6.287271
H     1.965612    1.245472    5.835012
H     0.180966    2.609235    6.280873
H    -1.385652    0.283748    7.404624
H    -0.779411    1.661131    8.258154
H    -1.339599    1.949919    4.499413
H     0.261485    1.378762    4.096987
H    -0.939427    0.256355    4.697867
H    -3.126271    2.039165    7.892814
H    -2.278363    3.156373    6.839105
H    -3.006759    1.703073    6.179312
N     1.624535    0.134949    8.790803
H     0.978993   -0.601056    8.604012
C     2.056048    0.307676   10.161042
H     1.574375   -0.455910   10.758663
H     1.788502    1.281710   10.542231
C     3.565620    0.174490   10.291442
O     4.230222    0.975923   10.888992
N     4.109447   -0.890960    9.658440
H     3.517903   -1.620984    9.320683
C     5.539001   -1.035248    9.619347
H     5.776951   -1.990977    9.166794
H     5.975315   -1.007508   10.607224
C     6.235779    0.058236    8.812790
O     7.365000    0.374501    9.095208
N     5.560517    0.564232    7.781950
```

```
C        6.122095    1.732150    7.151180
C        5.506677    3.022334    7.686752
H        5.007124    2.947621    8.651217
O        5.603153    4.057193    7.102411
C        6.018820    1.678819    5.624689
C        6.800787    0.546473    4.991476
C        8.173076    0.399184    5.197993
C        6.161161   -0.364379    4.155921
C        8.885539   -0.612384    4.568227
C        6.871254   -1.373758    3.524991
C        8.237285   -1.497815    3.721170
H        4.638390    0.260910    7.558519
H        7.158684    1.768431    7.450170
H        4.977120    1.590720    5.342930
H        6.367648    2.632409    5.243407
H        8.686791    1.057202    5.877802
H        5.097511   -0.297577    4.028059
H        9.941252   -0.711803    4.753095
H        6.354816   -2.068333    2.886670
H        8.777869   -2.284269    3.227130
N       12.361248    9.925824   -1.111669
H       11.607165   10.236416   -1.695397
C       12.093075    8.575192   -0.609438
C       11.064833    8.708373    0.506884
O       10.076589    8.007466    0.528494
C       13.372990    7.915314   -0.077231
C       14.457403    7.663971   -1.130498
S       13.852162    6.602173   -2.474096
C       15.332976    6.516953   -3.509331
H       13.175987    9.936922   -1.692771
H       11.636210    7.929764   -1.348324
H       13.798365    8.534853    0.706508
H       13.094939    6.969780    0.374010
H       14.829618    8.596577   -1.539453
H       15.298874    7.183920   -0.643812
H       15.095814    5.872993   -4.342489
H       15.603069    7.497037   -3.880799
H       16.166126    6.091458   -2.965280
N       11.273753    9.657624    1.435601
C       10.366455    9.803216    2.556976
C        9.028867   10.442990    2.182404
O        8.044381   10.182399    2.827618
C       11.035541   10.557243    3.712854
C       12.231983    9.773158    4.220211
O       13.357221   10.073305    3.879898
N       11.960963    8.744733    5.033726
H       12.081004   10.237273    1.351557
H       10.096464    8.817616    2.899843
H       11.388413   11.531527    3.398664
H       10.306347   10.688797    4.502252
H       12.722525    8.163167    5.309693
H       11.050654    8.332288    5.065346
N        8.986275   11.257116    1.110692
H        9.824694   11.449963    0.612326
```

```
C         7.710276    11.682666     0.574253
H         7.882141    12.446668    -0.173097
H         7.088658    12.098225     1.350430
C         6.920876    10.540892    -0.068393
O         5.719031    10.595894    -0.100007
N         7.623050     9.519392    -0.602366
C         6.960358     8.323386    -1.093860
C         6.455159     7.440744     0.058105
O         5.367844     6.921273    -0.026341
C         7.853574     7.566587    -2.105710
C         8.118482     8.453259    -3.337450
C         7.216500     6.232894    -2.514202
C         9.211617     7.929022    -4.271868
H         8.587225     9.437195    -0.380283
H         6.057923     8.635863    -1.596173
H         8.801229     7.354414    -1.618956
H         7.190092     8.567423    -3.892641
H         8.396868     9.449567    -3.010333
H         7.844015     5.723455    -3.234362
H         7.086215     5.570229    -1.667335
H         6.240564     6.380765    -2.964738
H         9.382972     8.631781    -5.080816
H        10.153718     7.796296    -3.746844
H         8.953630     6.977455    -4.721733
N         7.218120     7.313118     1.162330
C         6.737683     6.556151     2.306925
C         5.372651     7.086484     2.765458
O         4.444387     6.346364     2.983614
C         7.696646     6.645266     3.504442
C         9.131690     6.140408     3.323947
O        10.009432     6.788324     3.887800
O         9.330287     5.097239     2.652589
H         8.170779     7.607712     1.140359
H         6.569782     5.527992     2.026480
H         7.757910     7.664687     3.855564
H         7.259832     6.061540     4.312146
N         5.294712     8.419499     2.941340
H         6.108013     8.982643     2.810320
C         4.087638     9.040679     3.428277
H         4.298045    10.086107     3.616558
H         3.761212     8.596211     4.355205
C         2.922311     8.979805     2.450004
O         1.780343     9.070771     2.838296
N         3.253967     8.852155     1.162305
H         4.213960     8.872983     0.909639
C         2.295757     8.928673     0.103162
H         2.829052     9.148479    -0.812753
H         1.600012     9.738683     0.273361
C         1.420403     7.690813    -0.149664
O         0.380838     7.805684    -0.738045
N         1.912066     6.546181     0.362212
C         1.123506     5.352439     0.553946
C         0.313178     5.409139     1.853539
O        -0.877361     5.209508     1.833230
```

```
C         2.012984    4.110251    0.518352
H         2.813662    6.590612    0.787695
H         0.390839    5.297549   -0.233497
H         1.419564    3.220840    0.699178
H         2.479393    4.029755   -0.456658
H         2.795442    4.161629    1.268589
N         0.955799    5.660374    3.011342
C         0.239445    5.450441    4.260131
C        -0.898248    6.451699    4.443906
O        -1.921349    6.116603    5.001188
C         1.170294    5.375023    5.495596
C         1.729503    6.726861    5.965015
C         2.284753    4.345685    5.253594
C         2.583608    6.626235    7.231326
H         1.941700    5.807866    3.023563
H        -0.265507    4.497618    4.190218
H         0.531177    5.000233    6.289362
H         2.314368    7.172563    5.167081
H         0.908238    7.408459    6.161388
H         2.676532    3.981320    6.191941
H         1.909816    3.491205    4.699318
H         3.109279    4.772504    4.692781
H         2.864117    7.618261    7.573014
H         2.032374    6.146875    8.034925
H         3.493833    6.061115    7.073924
N        -0.713871    7.698838    3.986985
H         0.160725    7.965963    3.585406
C        -1.707708    8.711512    4.224758
H        -1.322197    9.656200    3.862518
H        -1.918129    8.817721    5.279115
C        -3.055908    8.418387    3.572063
O        -4.061986    8.598362    4.218495
N        -3.085487    7.934665    2.316397
C        -4.424396    7.823301    1.734004
C        -5.116728    6.512006    2.064702
O        -6.328728    6.465419    2.052662
C        -4.227832    8.114139    0.239506
C        -2.986630    9.013024    0.255735
C        -2.106725    8.338901    1.296390
H        -5.062018    8.581303    2.161601
H        -4.041579    7.202242   -0.312361
H        -5.105584    8.586759   -0.182585
H        -2.486164    9.104999   -0.690577
H        -3.260891   10.009845    0.592859
H        -1.610347    7.471344    0.900543
H        -1.355957    8.996173    1.703517
N        -4.347909    5.461800    2.378965
C        -4.936672    4.237281    2.892134
C        -5.664171    4.491604    4.215738
O        -6.726167    3.954287    4.441223
C        -3.879793    3.126534    3.083612
C        -3.508391    2.339492    1.842038
C        -3.840413    0.993666    1.727777
C        -2.778502    2.895067    0.792263
```

```
C         -3.488422    0.236764    0.619653
C         -2.410634    2.155176   -0.311044
C         -2.756568    0.794211   -0.450244
O         -2.435971    0.116882   -1.507373
H         -3.358657    5.587871    2.429389
H         -5.699459    3.898649    2.208395
H         -3.000335    3.572773    3.539111
H         -4.281092    2.432575    3.814075
H         -4.389685    0.516311    2.523769
H         -2.469836    3.919371    0.842249
H         -3.741596   -0.808619    0.575727
H         -1.840033    2.616058   -1.099800
N         -5.067567    5.302063    5.107633
H         -4.145253    5.644315    4.932713
C         -5.693050    5.616465    6.372271
H         -4.926728    5.981968    7.044812
H         -6.143195    4.737061    6.803590
C         -6.790404    6.676691    6.267850
O         -7.850558    6.542823    6.836045
N         -6.502383    7.752530    5.528492
H         -5.606659    7.844328    5.098052
C         -7.450200    8.860410    5.330187
H         -6.985408    9.553054    4.640519
H         -7.641875    9.376803    6.259295
C         -8.807183    8.459359    4.784921
O         -9.801618    9.021967    5.170285
N         -8.848541    7.481791    3.858232
C        -10.122203    7.081631    3.312035
C        -10.996107    6.294821    4.288441
O        -12.127110    6.014457    3.922102
C         -9.957857    6.295381    1.992081
O        -11.202947    6.239333    1.319743
C         -9.383135    4.898576    2.172762
H         -8.000713    7.085166    3.506572
H        -10.697469    7.970397    3.088210
H         -9.295231    6.875273    1.363152
H        -11.886911    6.182756    1.983503
H         -9.240611    4.434163    1.203771
H         -8.424859    4.917573    2.670240
H        -10.059779    4.274645    2.744698
N        -10.493119    5.929455    5.458921
C        -11.290888    5.249506    6.455670
C        -12.227379    6.182969    7.228948
O        -13.145923    5.710381    7.846659
C        -10.417291    4.481507    7.457125
C         -9.518761    3.392849    6.853049
C         -8.677461    2.758808    7.965061
C        -10.308433    2.319055    6.099498
H         -9.557277    6.181496    5.702259
H        -11.953027    4.565534    5.948913
H         -9.795483    5.191083    7.991836
H        -11.093433    4.038694    8.180951
H         -8.833475    3.854093    6.152016
H         -7.992728    2.018660    7.561887
```

```
H            -8.089825     3.508385     8.485244
H            -9.307298     2.265549     8.702828
H            -9.636406     1.559004     5.713106
H           -11.032148     1.831378     6.750495
H           -10.850622     2.722898     5.251637
N           -11.934518     7.498613     7.234464
C           -12.824643     8.434965     7.854322
C           -13.986224     8.853713     6.977627
H           -14.699319     9.525147     7.469433
O           -14.140171     8.521188     5.850014
H           -11.256254     7.858362     6.599361
H           -13.222525     8.011962     8.767909
H           -12.274908     9.331477     8.126763
N           -11.802826    10.320195    -2.707646
H           -11.574041    10.515984    -1.753568
C           -12.080551     8.894437    -2.897354
C           -11.661085     8.609736    -4.333310
O           -10.593715     8.108261    -4.591929
C           -13.499574     8.421126    -2.515250
C           -13.765554     8.699174    -1.024442
C           -13.687547     6.936446    -2.853503
C           -15.187413     8.376057    -0.561416
H           -12.598744    10.884176    -2.939163
H           -11.367682     8.347624    -2.295805
H           -14.224940     9.002043    -3.089443
H           -13.052606     8.134577    -0.427428
H           -13.577772     9.748484    -0.819679
H           -14.711016     6.624705    -2.689583
H           -13.451348     6.716628    -3.890397
H           -13.048797     6.318288    -2.230495
H           -15.326440     8.694612     0.465557
H           -15.927280     8.890201    -1.168929
H           -15.400687     7.314264    -0.601578
N           -12.482064     9.026362    -5.328160
H           -13.415170     9.280236    -5.095505
C           -12.231399     8.610726    -6.691936
H           -12.263572     7.534536    -6.795283
H           -13.001719     9.038466    -7.321946
C           -10.872964     9.048477    -7.245658
O           -10.244980     8.339256    -7.978283
N           -10.472282    10.301224    -6.896290
H           -11.035067    10.805894    -6.250380
C            -9.219019    10.889862    -7.340682
H            -9.115474    10.788962    -8.409984
H            -9.227674    11.940526    -7.081967
C            -7.980405    10.233171    -6.713787
O            -7.066038     9.820172    -7.371600
N            -7.951248    10.211886    -5.346525
H            -8.771970    10.475474    -4.849700
C            -6.773204     9.834574    -4.573715
H            -5.890666    10.284242    -5.000350
H            -6.899976    10.209176    -3.565441
C            -6.501152     8.343717    -4.490810
O            -5.355181     7.961999    -4.529533
```

```
N         -7.525544     7.484685    -4.328211
C         -7.212550     6.104827    -3.984309
C         -6.697719     5.300123    -5.171359
O         -5.964196     4.357729    -4.977792
C         -8.325959     5.343452    -3.223307
C         -9.530425     4.945661    -4.093763
C         -8.711969     6.118577    -1.958772
C        -10.556737     4.074592    -3.365414
H         -8.470676     7.811758    -4.318455
H         -6.360554     6.132083    -3.321863
H         -7.840997     4.425181    -2.905434
H        -10.018557     5.828581    -4.486623
H         -9.168986     4.382434    -4.950137
H         -9.337015     5.516951    -1.312643
H         -7.830679     6.392646    -1.387496
H         -9.258152     7.024593    -2.195671
H        -11.336698     3.762157    -4.053138
H        -10.099766     3.177811    -2.963036
H        -11.037854     4.604080    -2.551566
N         -7.073816     5.638270    -6.420812
C         -6.450987     4.974479    -7.548094
C         -5.026674     5.464849    -7.812071
O         -4.194572     4.698191    -8.233710
C         -7.341763     5.034464    -8.797238
C         -8.478254     4.005581    -8.715805
C         -7.972126     2.586652    -9.006672
N         -8.874871     1.528173    -8.557812
C         -8.929258     1.019825    -7.329007
N         -8.167173     1.472639    -6.346679
N         -9.763834     0.035226    -7.059737
H         -7.671243     6.424048    -6.562250
H         -6.294403     3.947677    -7.266612
H         -7.750978     6.032684    -8.914300
H         -6.731258     4.842944    -9.674116
H         -8.934965     4.045107    -7.732555
H         -9.258920     4.255105    -9.425698
H         -7.851080     2.449718   -10.072283
H         -6.993488     2.411991    -8.578136
H         -9.394778     1.045896    -9.256042
H         -7.534446     2.227690    -6.471548
H         -8.420750     1.224286    -5.400433
H        -10.408646    -0.319764    -7.729813
H         -9.630508    -0.532613    -6.242181
N         -4.779992     6.748345    -7.543284
H         -5.522997     7.341551    -7.251500
C         -3.485316     7.385726    -7.827284
H         -3.157246     7.158327    -8.829714
H         -3.616712     8.454655    -7.727422
C         -2.411303     6.912320    -6.860550
O         -1.281726     6.704121    -7.226618
N         -2.828457     6.767479    -5.597011
C         -1.933862     6.465321    -4.505927
C         -1.619439     4.984875    -4.401593
O         -0.482104     4.629507    -4.203840
```

```
C         -2.524255    7.029328   -3.207201
C         -1.917662    6.322582   -1.996582
C         -2.315398    8.550100   -3.217773
H         -3.754329    7.056494   -5.366174
H         -0.975713    6.928771   -4.692138
H         -3.589324    6.829660   -3.217866
H         -2.138346    6.848850   -1.087086
H         -2.310547    5.315859   -1.902646
H         -0.844625    6.271233   -2.073744
H         -3.050147    9.052153   -2.604354
H         -1.325181    8.804270   -2.854846
H         -2.425205    8.951212   -4.218869
N         -2.622060    4.096465   -4.535400
C         -2.249510    2.701983   -4.666890
C         -1.300735    2.556267   -5.857580
O         -0.352381    1.806056   -5.797896
C         -3.452172    1.756963   -4.766294
C         -4.371062    1.844521   -3.540722
C         -5.377987    0.695911   -3.437060
N         -4.779198   -0.493092   -2.836825
C         -5.420980   -1.630103   -2.679624
N         -6.662734   -1.800277   -3.160596
N         -4.893685   -2.637278   -2.008631
H         -3.547101    4.396667   -4.754613
H         -1.669259    2.409970   -3.805424
H         -4.021020    1.966291   -5.670195
H         -3.053329    0.754143   -4.865686
H         -3.778806    1.873021   -2.639593
H         -4.937775    2.762648   -3.580318
H         -6.218814    1.015974   -2.831531
H         -5.751777    0.452053   -4.427706
H         -3.873939   -0.374788   -2.370380
H         -7.032973   -1.165159   -3.828895
H         -7.042367   -2.728999   -3.173415
H         -4.032180   -2.598623   -1.492245
H         -5.384573   -3.504658   -1.972268
N         -1.566245    3.258454   -6.951797
H         -2.408528    3.780791   -7.050437
C         -0.705504    3.110063   -8.109480
H         -1.081093    3.759599   -8.889565
H         -0.694360    2.092078   -8.471540
C          0.738936    3.490488   -7.797955
O          1.678902    2.890814   -8.255833
N          0.863994    4.538401   -6.969587
H          0.057483    5.074073   -6.741814
C          2.130051    5.040619   -6.542744
H          2.753786    5.334352   -7.377563
H          1.954385    5.918720   -5.934853
C          2.971154    4.092986   -5.704450
O          4.172380    4.256194   -5.621505
N          2.338355    3.115574   -5.046633
C          3.044667    2.277830   -4.112784
C          4.231133    1.506149   -4.702015
O          5.158942    1.213471   -3.974725
```

```
C         1.993220    1.369435   -3.474783
C         2.410727    0.551241   -2.258610
C         1.124748    0.122338   -1.560007
N         1.348493   -0.933938   -0.594893
C         0.468646   -1.918857   -0.361199
N        -0.683847   -1.954598   -0.981799
N         0.812819   -2.890773    0.461268
H         1.351437    3.003230   -5.150594
H         3.503519    2.894102   -3.349131
H         1.186624    2.031845   -3.178985
H         1.573820    0.717178   -4.231932
H         2.985510   -0.321632   -2.548607
H         3.035032    1.137878   -1.588586
H         0.645021    0.975608   -1.092811
H         0.436720   -0.255535   -2.299305
H         2.060229   -0.803538    0.088182
H        -1.169330   -1.107744   -1.267857
H        -1.315144   -2.722055   -0.791504
H         1.725740   -2.942064    0.847983
H         0.092020   -3.526417    0.788710
N         4.240680    1.146385   -5.997086
H         3.489390    1.422198   -6.595078
C         5.342997    0.362579   -6.528185
H         5.116615    0.124591   -7.560112
H         5.451333   -0.563932   -5.986890
C         6.697633    1.039094   -6.494589
O         7.719103    0.398703   -6.406783
N         6.689816    2.371358   -6.597599
H         5.821817    2.865356   -6.540036
C         7.908114    3.142744   -6.529888
H         8.618848    2.794830   -7.266098
H         7.672690    4.172779   -6.762091
C         8.658375    3.137022   -5.196083
O         9.787596    3.577964   -5.140840
N         8.033768    2.605874   -4.137287
C         8.626709    2.428875   -2.822709
C         9.916197    1.597634   -2.856545
O        10.736975    1.683746   -1.957497
C         7.546867    1.710575   -1.982732
C         7.753961    1.563742   -0.475517
C         7.378842    2.807633    0.339139
C         7.283515    2.509382    1.831527
N         8.627504    2.502213    2.482073
O        11.015629    1.860772    0.870565
H        10.855107    2.085997   -0.046208
H        11.515347    1.048271    0.801235
H         7.096247    2.278780   -4.252134
H         8.900012    3.385226   -2.396740
H         6.613581    2.230408   -2.152548
H         7.406374    0.727525   -2.417021
H         7.098581    0.758652   -0.157753
H         8.760160    1.235532   -0.250773
H         8.067175    3.632258    0.176046
H         6.401623    3.156858    0.021479
```

```
H         6.703350    3.255265    2.352169
H         6.830952    1.546206    2.010013
H         9.347976    2.068717    1.913160
H         8.922083    3.477566    2.653864
H         8.592148    2.006337    3.360750
N        10.048263    0.767818   -3.882324
C        11.120660   -0.183210   -4.140789
C        12.479071    0.423004   -4.495817
O        13.457480   -0.281435   -4.392270
C        10.729413   -1.076474   -5.339706
C         9.621079   -2.111009   -5.097777
C         9.196927   -2.719634   -6.438086
C        10.085480   -3.209472   -4.141253
H         9.306122    0.759165   -4.551083
H        11.298615   -0.784169   -3.261008
H        10.439224   -0.421711   -6.156008
H        11.625048   -1.597996   -5.655393
H         8.753347   -1.612402   -4.677127
H         8.416907   -3.460847   -6.292979
H         8.813745   -1.958866   -7.106148
H        10.034753   -3.214775   -6.922789
H         9.317113   -3.963745   -4.011255
H        10.966963   -3.703454   -4.536675
H        10.334560   -2.826091   -3.156495
N        12.529197    1.674109   -4.960993
C        13.749777    2.301832   -5.429800
C        14.895161    2.499829   -4.418460
O        15.936125    2.918497   -4.849509
C        13.390804    3.634240   -6.091133
O        12.792738    4.515035   -5.183061
H        11.696047    2.222647   -4.940441
H        14.197106    1.663837   -6.182794
H        14.299390    4.086863   -6.456238
H        12.739799    3.441029   -6.937874
H        11.845476    4.479890   -5.240670
N        14.745759    2.156649   -3.119625
C        15.888817    2.094079   -2.230321
C        16.797862    0.886888   -2.488267
O        17.930596    0.893702   -2.081380
C        15.466628    1.980130   -0.760393
C        14.580191    3.100712   -0.221206
C        14.240604    2.810898    1.241228
N        13.194226    3.665261    1.788343
C        13.344071    4.807400    2.439160
N        14.549949    5.311373    2.678313
O        14.869775    7.744209    4.281061
H        14.559394    8.627585    4.064773
H        15.588619    7.856532    4.889987
N        12.278327    5.458053    2.856558
H        13.864843    1.823854   -2.804898
H        16.497298    2.971953   -2.384381
H        14.953772    1.033928   -0.622501
H        16.383573    1.924418   -0.183972
H        15.070472    4.065122   -0.318019
```

```
H         13.656352    3.168884   -0.783708
H         13.871479    1.799097    1.326258
H         15.121645    2.862748    1.870126
H         12.261500    3.346651    1.617495
H         15.368948    4.847276    2.365265
H         14.665354    6.158838    3.206931
H         11.329638    5.155213    2.698915
H         12.350647    6.292512    3.393565
N         16.219636   -0.176222   -3.072156
C         16.923677   -1.414983   -3.303777
C         17.713392   -1.343009   -4.607284
H         18.076203   -0.346683   -4.869724
O         17.944470   -2.286793   -5.290166
C         15.953638   -2.601351   -3.303997
C         15.427756   -2.974136   -1.950792
N         14.489769   -2.242001   -1.250289
O         12.538275   -0.116070   -0.541504
H         13.188485   -0.693087   -0.945935
H         12.040395    0.281344   -1.247765
C         15.757560   -4.070970   -1.232846
C         14.282588   -2.895072   -0.153798
N         15.024810   -4.006840   -0.074175
H         15.299861   -0.095232   -3.449370
H         17.657020   -1.530904   -2.513430
H         15.126055   -2.382251   -3.969829
H         16.476123   -3.452464   -3.719471
H         16.434523   -4.872121   -1.438818
H         13.599794   -2.600431    0.617371
H         14.947314   -4.707149    0.630519
O        -14.615809   -2.482410    4.768728
H        -14.852064   -1.958458    4.005649
H        -14.381115   -1.864405    5.453289
O        -13.191662   -0.594781    6.670794
H        -12.500574   -1.061104    7.126518
H        -13.560371    0.016906    7.296437
O        -12.192924   -0.130395    3.899330
H        -12.475726   -0.050048    4.803963
H        -12.977667   -0.344870    3.394555
O        -12.405320   -2.798951    1.211586
H        -11.728545   -2.129098    1.172760
H        -12.301413   -3.242307    2.053422
O         -6.815806    1.307713    0.056255
H         -7.122601    1.524476    0.940636
H         -5.886832    1.122256    0.145264
O         -7.842742    1.787559    2.655195
H         -8.795719    1.803286    2.617068
H         -7.570302    2.464721    3.269371
O         -8.640067   -0.227926   -1.283258
H         -7.940030    0.247482   -0.813982
H         -8.308093   -1.090753   -1.508414
O        -12.290583   -2.121506   -1.641122
H        -12.795314   -2.785666   -1.184193
H        -11.987773   -1.539874   -0.945389
O        -12.763189    1.432073   -1.313844
```

```
H                    -12.139541    0.861008   -0.866891
H                    -12.664237    2.276669   -0.873701
O                    -13.848145    3.968839    2.991320
H                    -13.378146    4.625144    3.502738
H                    -14.765878    4.071430    3.203923
O                    -15.159707    0.586181    0.082183
H                    -15.883469    0.443557   -0.511663
H                    -14.418114    0.868655   -0.452221
O                    -12.217179   -0.192514   -3.771701
H                    -12.546238    0.566212   -3.293630
H                    -12.356182   -0.923388   -3.170533
O                    -12.037246   -1.698877   -6.309050
H                    -12.302049   -1.130879   -5.586897
H                    -12.774237   -2.265124   -6.500076
O                     -9.712164   -2.958409   -2.642603
H                     -9.649003   -2.605256   -3.529519
H                    -10.590330   -2.757793   -2.325358
O                    -10.989194   -0.379079    0.185945
H                    -10.137952   -0.359434   -0.261297
H                    -10.902620    0.223367    0.933169
O                    -10.757759    1.531259    2.195718
H                    -11.261943    1.136567    2.911674
H                    -11.281357    2.236936    1.820157
O                    -12.331550    3.511267    0.661366
H                    -13.019697    3.628153    1.320285
H                    -11.863933    4.342652    0.640087
O                     -9.337734    0.612626   -3.890324
H                    -10.257950    0.355419   -3.911572
H                     -9.049652    0.439643   -2.992154
O                    -14.401300   -1.015400    2.311194
H                    -14.793411   -0.426865    1.665242
H                    -13.901486   -1.656320    1.807755
O                    -12.391809   -3.876001    3.886340
H                    -13.195693   -3.458100    4.222416
H                    -11.684883   -3.529042    4.417399
O                     -9.406459   -2.362849   -5.388474
H                    -10.279235   -2.531589   -5.736960
H                     -8.807301   -2.996491   -5.791534

Case 3:
Voltage -70 mV;   Y266 neutral; R300, neutral; E183 neutral
N           15.580944    -6.451469     5.391416
H           14.780731    -6.200923     5.939305
C           15.509643    -5.856520     4.054221
C           14.783016    -6.782962     3.060452
O           14.052228    -6.345797     2.186827
C           16.919741    -5.576790     3.520757
H           16.389845    -6.128186     5.884447
H           14.945059    -4.933534     4.050349
H           16.889045    -5.190425     2.508988
H           17.403781    -4.831921     4.142207
H           17.526994    -6.475176     3.529510
```

```
N      15.032378    -8.085604     3.186404
H      15.612369    -8.361765     3.949488
C      14.398661    -9.092386     2.333235
H      14.482431    -8.823798     1.292859
H      14.891368   -10.042265     2.492758
C      12.912802    -9.220816     2.671363
O      12.081430    -9.424712     1.825901
N      12.582804    -9.055884     3.966877
H      13.302707    -8.986374     4.648923
C      11.195108    -9.170814     4.421376
H      11.188268    -9.157965     5.503131
H      10.743743   -10.088315     4.077111
C      10.336944    -8.010595     3.900734
O       9.190715    -8.185973     3.584565
N      10.947166    -6.814584     3.855271
C      10.318595    -5.634057     3.317570
C      10.126801    -5.690990     1.799352
O       9.096244    -5.262541     1.337380
C      11.093003    -4.388030     3.776268
O      12.478911    -4.570920     3.769522
H      11.905594    -6.744371     4.106682
H       9.314756    -5.553195     3.701226
H      10.801687    -3.539751     3.167074
H      10.827406    -4.174248     4.801233
H      12.805998    -4.755413     2.897553
N      11.099962    -6.200428     1.023953
C      10.844256    -6.396651    -0.400382
C       9.693378    -7.385344    -0.601380
O       8.944764    -7.279524    -1.544861
C      12.132565    -6.785513    -1.187311
C      12.333415    -8.290910    -1.404097
C      12.159258    -6.060439    -2.534205
H      11.924166    -6.585903     1.431072
H      10.474583    -5.462318    -0.791691
H      12.964681    -6.415407    -0.598905
H      13.287085    -8.454107    -1.895336
H      12.332057    -8.857111    -0.483479
H      11.566150    -8.698123    -2.055406
H      13.053903    -6.324343    -3.087057
H      11.298808    -6.332248    -3.135779
H      12.159266    -4.985220    -2.401000
N       9.570431    -8.340344     0.328319
H      10.315059    -8.474836     0.973641
C       8.593019    -9.421180     0.241223
H       8.934634   -10.218590     0.888835
H       8.533193    -9.803631    -0.766085
C       7.181698    -9.045745     0.642909
O       6.257483    -9.502080     0.010630
N       7.003520    -8.226545     1.685802
C       5.669684    -7.739286     1.954979
C       5.251021    -6.727652     0.893553
O       4.074248    -6.630710     0.617445
C       5.415518    -7.234086     3.389334
C       5.557535    -8.378469     4.397949
```

```
C      6.257448    -6.025280     3.802561
H      7.780930    -7.960683     2.250279
H      4.986652    -8.558610     1.792682
H      4.374715    -6.925106     3.382500
H      5.267463    -8.041397     5.387567
H      4.919466    -9.215853     4.134954
H      6.579689    -8.735715     4.453711
H      5.917068    -5.653359     4.763052
H      7.299760    -6.295845     3.909491
H      6.185275    -5.210641     3.090221
N      6.171801    -6.003043     0.249776
C      5.776269    -5.184213    -0.884109
C      5.343447    -6.067678    -2.062536
O      4.319744    -5.815739    -2.653842
C      6.868947    -4.165014    -1.264115
C      7.066958    -3.148300    -0.125802
C      6.498212    -3.454662    -2.570627
C      8.360686    -2.341733    -0.239822
H      7.133074    -6.061030     0.508390
H      4.878561    -4.647265    -0.616972
H      7.798814    -4.706502    -1.411874
H      6.211977    -2.473862    -0.109879
H      7.076657    -3.662085     0.826136
H      7.203684    -2.665133    -2.790646
H      6.498225    -4.129893    -3.416516
H      5.510152    -3.011194    -2.503906
H      8.461932    -1.668112     0.604440
H      9.222078    -2.999285    -0.239564
H      8.396893    -1.742148    -1.142435
N      6.115624    -7.109718    -2.399307
H      7.009706    -7.232401    -1.972530
C      5.773400    -7.949754    -3.527335
H      6.603987    -8.623480    -3.697355
H      5.618446    -7.363666    -4.420900
C      4.504403    -8.778232    -3.325492
O      3.666321    -8.836197    -4.193943
N      4.393375    -9.431857    -2.162281
H      5.131093    -9.388629    -1.491888
C      3.207303   -10.187596    -1.834320
H      2.967719   -10.900050    -2.609395
H      3.399564   -10.730343    -0.917280
C      1.974183    -9.306851    -1.644814
O      0.887188    -9.696141    -2.004685
N      2.187877    -8.107003    -1.107171
C      1.062466    -7.184491    -1.071036
C      0.616982    -6.720560    -2.460413
O     -0.565457    -6.584496    -2.680891
C      1.278265    -5.930431    -0.230694
O      1.413888    -6.189100     1.126776
H      3.087713    -7.817102    -0.798466
H      0.211115    -7.703913    -0.662271
H      2.121256    -5.365061    -0.613768
H      0.384351    -5.341622    -0.351912
H      2.307761    -6.456344     1.293756
```

```
N         1.544191    -6.431998    -3.383126
C         1.145560    -6.024187    -4.718096
C         0.378786    -7.146591    -5.411039
O        -0.608559    -6.891422    -6.067885
C         2.355968    -5.521342    -5.538583
C         2.762317    -4.129358    -5.026355
C         2.060425    -5.496360    -7.043553
C         4.087807    -3.618829    -5.589442
H         2.514587    -6.487425    -3.159323
H         0.424503    -5.223000    -4.638776
H         3.176906    -6.211370    -5.367887
H         1.970489    -3.423667    -5.272036
H         2.836027    -4.155457    -3.946763
H         2.918349    -5.124936    -7.589997
H         1.841543    -6.483228    -7.432707
H         1.215561    -4.853593    -7.270059
H         4.381029    -2.705502    -5.082374
H         4.879679    -4.345737    -5.439446
H         4.030768    -3.400397    -6.649947
N         0.820049    -8.397338    -5.251532
H         1.667134    -8.572954    -4.752448
C         0.043277    -9.504951    -5.753923
H        -0.064069    -9.462747    -6.827588
H         0.556407   -10.421435    -5.491701
C        -1.365276    -9.540103    -5.166934
O        -2.319135    -9.771082    -5.874320
N        -1.472639    -9.318869    -3.858022
C        -2.776408    -9.352422    -3.235418
C        -3.692485    -8.216308    -3.692213
O        -4.890821    -8.424043    -3.732787
C        -2.679029    -9.329546    -1.702831
O        -3.923334    -9.599034    -1.118463
H        -0.656825    -9.231529    -3.289008
H        -3.292832   -10.256969    -3.524128
H        -1.996629   -10.103273    -1.379346
H        -2.290277    -8.372770    -1.370431
H        -4.613441    -9.195680    -1.630922
N        -3.143422    -7.035403    -3.971064
C        -3.907948    -5.914822    -4.493113
C        -4.449216    -6.233753    -5.886715
O        -5.613112    -6.041171    -6.155960
C        -3.013316    -4.655286    -4.488139
C        -3.472078    -3.518231    -5.385380
C        -4.792824    -3.082106    -5.426259
C        -2.544447    -2.883423    -6.203112
C        -5.168659    -2.055609    -6.276261
C        -2.913664    -1.844149    -7.041965
C        -4.233507    -1.431282    -7.085985
H        -2.166724    -6.912808    -3.802062
H        -4.776065    -5.749068    -3.875653
H        -2.930369    -4.319418    -3.458829
H        -2.021582    -4.947355    -4.803627
H        -5.537506    -3.560182    -4.818737
H        -1.519848    -3.211204    -6.194444
```

```
H         -6.201301    -1.752629    -6.312157
H         -2.174840    -1.370521    -7.663061
H         -4.528535    -0.637244    -7.749953
N         -3.576116    -6.724283    -6.775034
C         -4.019462    -7.062119    -8.108086
C         -5.145266    -8.094859    -8.085430
O         -6.033051    -8.031072    -8.895558
C         -2.870758    -7.593788    -8.968672
S         -1.652356    -6.342773    -9.485001
H         -2.634887    -6.908353    -6.497413
H         -4.443881    -6.190243    -8.582972
H         -2.366820    -8.412362    -8.472689
H         -3.288935    -7.972050    -9.891765
H         -1.047772    -6.179428    -8.318771
N         -5.078558    -9.066913    -7.155619
H         -4.254248    -9.192308    -6.607817
C         -6.085907   -10.115215    -7.183050
H         -6.258813   -10.452766    -8.193234
H         -5.714550   -10.947558    -6.596550
C         -7.459140    -9.722570    -6.613183
O         -8.481540   -10.086905    -7.126858
N         -7.417939    -9.016209    -5.478356
C         -8.641654    -8.653168    -4.788765
C         -9.504927    -7.658196    -5.556657
O        -10.702666    -7.658824    -5.450443
C         -8.361797    -8.103652    -3.378663
C         -7.519565    -6.820350    -3.271137
C         -8.193200    -5.469023    -3.256946
O         -9.495268    -5.469727    -3.327708
H         -9.844193    -4.571910    -3.238798
O         -7.547134    -4.455075    -3.162024
H         -6.535785    -8.774572    -5.078978
H         -9.260681    -9.533426    -4.686412
H         -9.317927    -7.967214    -2.894271
H         -7.836469    -8.883813    -2.838322
H         -6.956763    -6.846764    -2.345164
H         -6.778024    -6.766612    -4.050597
N         -8.837301    -6.733962    -6.273855
C         -9.555196    -5.702617    -7.000840
C        -10.386302    -6.273665    -8.141462
H        -10.409108    -7.356394    -8.248911
O        -10.987947    -5.559205    -8.887919
C         -8.631217    -4.564540    -7.490669
O         -9.434347    -3.442192    -7.832385
C         -7.723201    -4.928742    -8.657701
H         -7.856541    -6.841277    -6.421677
H        -10.268224    -5.253622    -6.319020
H         -8.030048    -4.257022    -6.646351
H        -10.110641    -3.751079    -8.429613
H         -7.045481    -4.104314    -8.851716
H         -7.131392    -5.809814    -8.458451
H         -8.304290    -5.105593    -9.556372
N        -12.952122    -9.022524     3.174171
H        -13.214938    -9.550528     2.367628
```

```
C    -11.773531   -8.180020    2.898781
C    -10.789152   -8.123178    4.050099
O     -9.659878   -7.714257    3.857081
C    -12.146694   -6.722637    2.563055
O    -12.716169   -6.204126    3.737580
C    -13.114369   -6.605222    1.390979
H    -13.731500   -8.453468    3.442359
H    -11.222755   -8.599619    2.068278
H    -11.230811   -6.196532    2.328527
H    -12.850275   -5.263293    3.637750
H    -13.281620   -5.557270    1.166354
H    -12.728750   -7.082400    0.497945
H    -14.074828   -7.045224    1.631540
N    -11.197257   -8.533635    5.241900
H    -12.119322   -8.910812    5.275372
C    -10.412532   -8.529004    6.451303
H    -10.336968   -7.533916    6.860733
H    -10.933725   -9.138450    7.180182
C     -8.974764   -9.043519    6.375023
O     -8.140210   -8.478406    7.038097
N     -8.658727  -10.119661    5.632233
H     -9.369329  -10.599092    5.127070
C     -7.270140  -10.552705    5.532084
H     -7.253507  -11.532129    5.071308
H     -6.830742  -10.631426    6.513558
C     -6.387460   -9.608364    4.692694
O     -5.219627   -9.468420    4.949188
N     -7.045394   -9.018012    3.701158
C     -6.302000   -8.005885    2.963248
C     -6.261921   -6.695390    3.744988
O     -5.267278   -6.009548    3.699390
C     -6.611046   -7.876391    1.459456
C     -7.789973   -7.050307    0.991656
C     -7.776294   -5.663150    1.107344
C     -8.850003   -7.647187    0.322326
C     -8.803632   -4.895981    0.584954
C     -9.869113   -6.881098   -0.220160
C     -9.853576   -5.503358   -0.083978
H     -8.039939   -9.013977    3.679635
H     -5.279281   -8.337534    2.997281
H     -5.714131   -7.456725    1.017445
H     -6.688005   -8.886111    1.073921
H     -6.944102   -5.173253    1.579317
H     -8.862559   -8.715641    0.190486
H     -8.766571   -3.824847    0.672310
H    -10.658922   -7.355970   -0.772726
H    -10.635584   -4.909949   -0.520432
N     -7.292893   -6.385747    4.534672
C     -7.221569   -5.273489    5.451656
C     -6.102284   -5.491033    6.469414
O     -5.455956   -4.539151    6.852734
C     -8.598656   -5.101583    6.124751
C     -8.810243   -3.808954    6.883919
C     -8.979045   -2.606893    6.201181
```

```
C           -8.906046    -3.804699     8.268743
C           -9.226379    -1.430441     6.886055
C           -9.163887    -2.629068     8.960916
C           -9.323736    -1.438445     8.271779
H           -8.158167    -6.869402     4.428727
H           -6.951315    -4.368588     4.926161
H           -9.342075    -5.168088     5.336903
H           -8.761766    -5.940556     6.787251
H           -8.925242    -2.589088     5.126356
H           -8.781945    -4.722924     8.815886
H           -9.345858    -0.512282     6.340225
H           -9.227190    -2.645511    10.034243
H           -9.506599    -0.522727     8.805758
N           -5.861756    -6.728179     6.898228
H           -6.516417    -7.449843     6.693430
C           -4.785610    -7.023263     7.817870
H           -4.824644    -6.393718     8.694135
H           -4.898019    -8.053223     8.131547
C           -3.389823    -6.846971     7.222572
O           -2.515980    -6.352256     7.894544
N           -3.189413    -7.272080     5.970444
C           -1.911491    -7.092406     5.298841
C           -1.665164    -5.620520     4.938846
O           -0.566782    -5.133818     5.107238
C           -1.829952    -8.027673     4.072110
C           -0.664512    -7.670389     3.147846
C           -1.728546    -9.487857     4.527528
H           -3.922668    -7.753219     5.495583
H           -1.118126    -7.338226     5.989758
H           -2.750581    -7.903937     3.506544
H           -0.600417    -8.397138     2.345039
H           -0.774998    -6.696017     2.693021
H            0.280109    -7.682140     3.682962
H           -1.741719   -10.148433     3.666644
H           -0.794904    -9.654675     5.058852
H           -2.545387    -9.777676     5.175684
N           -2.700639    -4.933584     4.446697
C           -2.569829    -3.551378     4.028841
C           -2.420113    -2.639002     5.244019
O           -1.528605    -1.820135     5.290629
C           -3.758286    -3.090364     3.171672
C           -3.994034    -3.840510     1.850012
C           -2.860081    -3.730762     0.845346
O           -1.747217    -4.135874     1.218613
O           -3.112647    -3.254404    -0.273866
H           -3.564207    -5.403285     4.285141
H           -1.662752    -3.440843     3.465207
H           -4.667153    -3.158040     3.761700
H           -3.604929    -2.035165     2.961915
H           -4.149292    -4.890266     2.048879
H           -4.899444    -3.450472     1.399192
N           -3.309568    -2.774884     6.244298
C           -3.093978    -2.024901     7.456985
C           -1.778252    -2.422295     8.133804
```

```
O        -1.106968    -1.582694     8.694212
C        -4.226302    -2.004940     8.493910
O        -4.436303    -3.246513     9.098559
C        -5.507646    -1.387323     7.948154
H        -3.988197    -3.503289     6.218708
H        -2.937574    -0.995415     7.174780
H        -3.851323    -1.365655     9.283504
H        -4.889420    -3.818315     8.490671
H        -6.240042    -1.327303     8.743421
H        -5.322003    -0.384111     7.576784
H        -5.928460    -1.976911     7.144841
N        -1.380397    -3.700457     8.043328
H        -2.031177    -4.418219     7.811975
C        -0.187160    -4.107493     8.739874
H        -0.237332    -3.891849     9.797887
H        -0.075804    -5.175185     8.604000
C         1.064036    -3.420003     8.218010
O         1.940520    -3.076040     8.977882
N         1.153963    -3.258734     6.881772
C         2.332594    -2.660245     6.317782
C         2.291096    -1.138359     6.164142
O         3.359963    -0.574312     6.022983
C         2.830820    -3.345975     5.045590
S         1.922866    -2.976896     3.520328
H         0.422815    -3.612332     6.296775
H         3.116491    -2.797974     7.047090
H         3.840685    -3.007650     4.863293
H         2.853489    -4.415035     5.205088
H         0.859339    -3.741312     3.719541
N         1.140119    -0.470673     6.273050
C         1.204068     0.968497     6.491343
C         1.755726     1.271283     7.870550
O         2.368197     2.289064     8.085588
C        -0.095730     1.741716     6.182264
C        -1.196980     1.568389     7.238506
C        -0.554707     1.440122     4.750694
C        -2.466369     2.370643     6.944548
H         0.271804    -0.965355     6.291883
H         1.951513     1.351097     5.814811
H         0.208520     2.782230     6.222188
H        -1.446000     0.523604     7.355738
H        -0.807789     1.892646     8.199120
H        -1.310124     2.147457     4.431579
H         0.277973     1.520390     4.057207
H        -0.966271     0.444389     4.656851
H        -3.136044     2.340272     7.798665
H        -2.241293     3.414113     6.738921
H        -3.007065     1.973406     6.092389
N         1.549788     0.323258     8.791266
H         0.889828    -0.402217     8.613495
C         1.978276     0.512612    10.160076
H         1.470069    -0.222732    10.771000
H         1.737711     1.502298    10.518179
C         3.482342     0.337369    10.305033
```

```
O         4.164480    1.128924   10.896138
N         3.999955   -0.753863    9.695360
H         3.390737   -1.466241    9.351675
C         5.425061   -0.938911    9.667405
H         5.638287   -1.909420    9.234458
H         5.855763   -0.904784   10.657461
C         6.156882    0.119117    8.844832
O         7.291893    0.411950    9.129239
N         5.502078    0.621285    7.799030
C         6.098617    1.759274    7.146315
C         5.514803    3.076764    7.649619
H         5.013135    3.037215    8.615064
O         5.636470    4.094801    7.040848
C         6.003008    1.674981    5.620777
C         6.755411    0.506853    5.017547
C         8.121035    0.322829    5.238389
C         6.095144   -0.401675    4.195769
C         8.807450   -0.722633    4.635688
C         6.779320   -1.444771    3.591585
C         8.139375   -1.605749    3.801627
H         4.575274    0.334619    7.572613
H         7.134062    1.774610    7.451350
H         4.960818    1.610760    5.334266
H         6.381340    2.609398    5.220643
H         8.649059    0.978934    5.909006
H         5.034983   -0.305689    4.057496
H         9.858141   -0.849996    4.831622
H         6.247195   -2.136641    2.963271
H         8.659788   -2.418340    3.328667
N        12.613550    9.596234   -1.248802
H        11.873463    9.913870   -1.846454
C        12.304852    8.264740   -0.719487
C        11.274100    8.450699    0.386872
O        10.265804    7.779082    0.416374
C        13.562726    7.580985   -0.165336
C        14.645038    7.276039   -1.206574
S        14.017038    6.202095   -2.529928
C        15.499712    6.053551   -3.555389
H        13.433688    9.573142   -1.821868
H        11.834462    7.616523   -1.447350
H        14.001313    8.205395    0.607189
H        13.255653    6.653816    0.304804
H        15.045481    8.188485   -1.633895
H        15.470190    6.783597   -0.704576
H        15.248366    5.398907   -4.375947
H        15.798746    7.017618   -3.946054
H        16.318070    5.616646   -2.998124
N        11.504281    9.412877    1.296997
C        10.594285    9.607741    2.408568
C         9.277001   10.274697    2.010348
O         8.280895   10.053125    2.652572
C        11.276205   10.369076    3.552095
C        12.447964    9.564527    4.084498
O        13.582765    9.826353    3.744622
```

```
N        12.144410     8.562238     4.918949
H        12.328099     9.967716     1.206347
H        10.295508     8.637691     2.771304
H        11.657189    11.326324     3.218975
H        10.546022    10.537651     4.333516
H        12.888282     7.966933     5.213303
H        11.222912     8.176002     4.954870
N         9.263811    11.067697     0.922328
H        10.110810    11.229654     0.427419
C         8.003502    11.517470     0.368859
H         8.201251    12.260241    -0.393337
H         7.389096    11.966669     1.131932
C         7.186384    10.384439    -0.254324
O         5.986288    10.470076    -0.291498
N         7.862643     9.333319    -0.763272
C         7.170445     8.145634    -1.233957
C         6.634883     7.300728    -0.067335
O         5.534494     6.808351    -0.148247
C         8.048947     7.344296    -2.224251
C         8.343686     8.197252    -3.472933
C         7.379505     6.019216    -2.608174
C         9.428319     7.625157    -4.388995
H         8.823869     9.230816    -0.536812
H         6.279720     8.471711    -1.748418
H         8.987987     7.117876    -1.727335
H         7.421774     8.323284    -4.036274
H         8.645788     9.192886    -3.165432
H         7.996920     5.478892    -3.314375
H         7.228376     5.377920    -1.748363
H         6.409824     6.182873    -3.066713
H         9.622358     8.305747    -5.211606
H        10.363628     7.479590    -3.855310
H         9.148450     6.671115    -4.820162
N         7.387601     7.175215     1.043839
C         6.879055     6.458277     2.201910
C         5.527970     7.039470     2.639634
O         4.575502     6.332974     2.866901
C         7.833208     6.546127     3.403444
C         9.254686     5.997273     3.244152
O        10.145886     6.631293     3.802334
O         9.428294     4.935727     2.595041
H         8.347661     7.444953     1.022551
H         6.682440     5.429463     1.943291
H         7.921267     7.570736     3.733003
H         7.375097     5.992729     4.220654
N         5.489851     8.377405     2.786069
H         6.320917     8.912597     2.648950
C         4.299225     9.045885     3.249968
H         4.538258    10.090241     3.407632
H         3.957292     8.638800     4.188424
C         3.136629     8.987513     2.268583
O         1.995144     9.109701     2.649473
N         3.470176     8.824775     0.985192
H         4.431315     8.829449     0.735279
```

```
C         2.517191    8.905260   -0.078305
H         3.059863    9.091715   -0.996305
H         1.843943    9.738046    0.071914
C         1.609133    7.687020   -0.310194
O         0.561833    7.815066   -0.879403
N         2.094474    6.538092    0.201332
C         1.289982    5.364375    0.431021
C         0.472387    5.479549    1.721794
O        -0.722708    5.301804    1.692203
C         2.166758    4.112402    0.445792
H         3.001528    6.575688    0.615931
H         0.559552    5.290356   -0.357506
H         1.566520    3.234004    0.657550
H         2.636496    3.991476   -0.523448
H         2.946380    4.184428    1.196930
N         1.102514    5.751803    2.879294
C         0.364292    5.585250    4.123057
C        -0.756432    6.611697    4.274013
O        -1.792392    6.302325    4.823364
C         1.276011    5.511847    5.372732
C         1.867809    6.857295    5.819745
C         2.362801    4.445288    5.169506
C         2.698994    6.761005    7.101577
H         2.090092    5.889616    2.900944
H        -0.159830    4.642184    4.064825
H         0.615200    5.173522    6.165183
H         2.478118    7.266410    5.021408
H         1.064424    7.567408    5.986930
H         2.732516    4.092368    6.121037
H         1.969615    3.588567    4.631398
H         3.205855    4.834276    4.608937
H         3.004293    7.751938    7.424490
H         2.121237    6.317646    7.907208
H         3.593843    6.165183    6.971697
N        -0.542960    7.847390    3.800966
H         0.341472    8.089930    3.404868
C        -1.513857    8.887768    4.018661
H        -1.103669    9.816493    3.642539
H        -1.724196    9.016468    5.070588
C        -2.866972    8.618762    3.365891
O        -3.871073    8.837212    4.003256
N        -2.902689    8.108966    2.120757
C        -4.239977    8.013753    1.535301
C        -4.984715    6.749522    1.927679
O        -6.195865    6.757018    1.919701
C        -4.021961    8.221377    0.030789
C        -2.766351    9.099777    0.020009
C        -1.907226    8.456184    1.096092
H        -4.854248    8.814662    1.917299
H        -3.842397    7.278855   -0.469760
H        -4.886376    8.684666   -0.427231
H        -2.255711    9.142462   -0.923535
H        -3.027230   10.114461    0.311403
H        -1.420606    7.566072    0.740457
```

```
H          -1.151144     9.119009     1.484459
N          -4.269046     5.673779     2.291121
C          -4.938190     4.512574     2.850520
C          -5.638483     4.853566     4.168965
O          -6.741401     4.413645     4.408446
C          -3.973826     3.331470     3.086429
C          -3.635815     2.505644     1.861076
C          -4.332918     1.334795     1.576597
C          -2.597742     2.864590     1.013744
C          -4.007257     0.547987     0.482013
C          -2.270778     2.094060    -0.087910
C          -2.970648     0.931871    -0.361397
O          -2.617894     0.205846    -1.447897
H          -3.414708    -0.117636    -1.907891
H          -3.275163     5.750834     2.345475
H          -5.725584     4.209370     2.178854
H          -3.068141     3.711073     3.548038
H          -4.443505     2.682244     3.816470
H          -5.141527     1.028394     2.216937
H          -2.032818     3.752495     1.204511
H          -4.536059    -0.366907     0.289642
H          -1.479397     2.397067    -0.750378
N          -4.970841     5.618292     5.048293
H          -4.034958     5.911381     4.856514
C          -5.582761     6.005336     6.299177
H          -4.805951     6.385456     6.951028
H          -6.052296     5.158934     6.774233
C          -6.651500     7.084436     6.135928
O          -7.706439     7.027961     6.726190
N          -6.342147     8.091247     5.313906
H          -5.445011     8.139362     4.880103
C          -7.257687     9.220267     5.085304
H          -6.781273     9.871884     4.364233
H          -7.419732     9.775731     5.997427
C          -8.632474     8.840758     4.569496
O          -9.611456     9.429009     4.959389
N          -8.690353     7.854323     3.658142
C          -9.953936     7.451763     3.088772
C         -10.766367     6.542276     3.995474
O         -11.807718     6.069059     3.559312
C          -9.761225     6.816107     1.689871
O         -10.954474     6.903389     0.953235
C          -9.239336     5.382290     1.729344
H          -7.846104     7.427544     3.334380
H         -10.571605     8.331054     2.962637
H          -9.046138     7.439464     1.169803
H         -11.641616     6.500567     1.473253
H          -8.968511     5.061323     0.731363
H          -8.362386     5.288810     2.354441
H         -10.001049     4.705882     2.098831
N         -10.353337     6.327943     5.240900
C         -11.259067     5.732432     6.212920
C         -12.445055     6.655970     6.501978
O         -13.529692     6.188238     6.758346
```

```
C         -10.538004    5.397364    7.526365
C          -9.534250    4.235087    7.447170
C          -8.716902    4.184513    8.741571
C         -10.219887    2.887550    7.196651
H          -9.462967    6.657426    5.550927
H         -11.696982    4.842273    5.794089
H         -10.028467    6.287234    7.878934
H         -11.302465    5.156738    8.260029
H          -8.847363    4.423679    6.629000
H          -7.988780    3.378765    8.709922
H          -8.181943    5.114221    8.898199
H          -9.358073    4.014626    9.603746
H          -9.479693    2.092166    7.178789
H         -10.931085    2.662089    7.989148
H         -10.750009    2.858102    6.251782
N         -12.200522    7.978949    6.518744
C         -13.273039    8.917896    6.674532
C         -14.049258    9.175384    5.400083
H         -14.889087    9.868465    5.516222
O         -13.795693    8.696578    4.344326
H         -11.339629    8.323361    6.149409
H         -13.965271    8.567023    7.429623
H         -12.873459    9.866487    7.019929
N         -11.399969   10.603551   -3.013216
H         -11.150848   10.793386   -2.062913
C         -11.816457    9.206991   -3.172692
C         -11.420851    8.859935   -4.600828
O         -10.381233    8.297434   -4.846936
C         -13.274734    8.886640   -2.781030
C         -13.514063    9.244538   -1.302422
C         -13.600995    7.414363   -3.063721
C         -14.953793    9.046013   -0.824864
H         -12.140699   11.235339   -3.252888
H         -11.162961    8.606955   -2.555415
H         -13.942021    9.511372   -3.379459
H         -12.841528    8.656467   -0.682305
H         -13.250014   10.285347   -1.142897
H         -14.652333    7.209527   -2.907474
H         -13.370472    7.129681   -4.086153
H         -13.034591    6.765561   -2.402933
H         -15.063023    9.412864    0.189451
H         -15.657185    9.588738   -1.450951
H         -15.244933    8.002168   -0.821506
N         -12.217787    9.294727   -5.606272
H         -13.139288    9.595964   -5.384742
C         -11.972872    8.843880   -6.959258
H         -12.033869    7.767144   -7.039983
H         -12.729040    9.278162   -7.601904
C         -10.601635    9.233244   -7.517693
O          -9.996993    8.496745   -8.241578
N         -10.162849   10.477782   -7.182627
H         -10.710967   11.007726   -6.544640
C          -8.894341   11.023967   -7.637100
H          -8.789701   10.891253   -8.702753
```

```
H           -8.878587       12.081461       -7.407588
C           -7.672564       10.359017       -6.988311
O           -6.753288        9.935059       -7.630848
N           -7.659101       10.352152       -5.619641
H           -8.482401       10.628144       -5.133559
C           -6.495253        9.961264       -4.832348
H           -5.599190       10.377760       -5.264327
H           -6.616272       10.359353       -3.832338
C           -6.268292        8.463912       -4.714277
O           -5.133741        8.047355       -4.711552
N           -7.321035        7.642772       -4.555883
C           -7.043209        6.284748       -4.116210
C           -6.395979        5.449108       -5.213283
O           -5.518963        4.674302       -4.918169
C           -8.232873        5.569489       -3.436346
C           -9.368536        5.183245       -4.399222
C           -8.697051        6.392810       -2.229037
C          -10.549916        4.494220       -3.715501
H           -8.255734        7.997394       -4.570135
H           -6.261740        6.342785       -3.373841
H           -7.807333        4.647372       -3.048883
H           -9.726245        6.052590       -4.935411
H           -8.960384        4.506207       -5.145609
H           -9.400969        5.838945       -1.623718
H           -7.854514        6.643390       -1.592420
H           -9.177968        7.313947       -2.533284
H          -11.255488        4.128223       -4.455611
H          -10.227310        3.646433       -3.122056
H          -11.086931        5.173571       -3.065645
N           -6.814696        5.564734       -6.493518
C           -6.106090        4.822427       -7.516751
C           -4.760271        5.444493       -7.899241
O           -3.890185        4.715921       -8.316589
C           -6.966879        4.499655       -8.746711
C           -8.036858        3.417976       -8.510429
C           -7.450679        2.076806       -8.041102
N           -8.388348        0.950372       -8.074819
C           -9.198500        0.573081       -7.082223
N           -9.461714        1.417389       -6.096291
N           -9.735184       -0.627041       -7.081494
H           -7.489226        6.258287       -6.735688
H           -5.795827        3.898635       -7.061643
H           -7.450821        5.399941       -9.113901
H           -6.295934        4.164635       -9.530188
H           -8.777741        3.779653       -7.803319
H           -8.564457        3.255563       -9.444503
H           -6.625621        1.798488       -8.683815
H           -7.050857        2.146512       -7.038324
H           -8.318670        0.320537       -8.843177
H           -9.209212        2.374113       -6.170856
H           -9.975771        1.152008       -5.272487
H           -9.386628       -1.392606       -7.629608
H          -10.455373       -0.868109       -6.419059
N           -4.571383        6.757521       -7.721942
```

```
H           -5.312315      7.350933     -7.425900
C           -3.257811      7.352384     -8.011367
H           -2.933574      7.093301     -9.006671
H           -3.359217      8.427153     -7.938226
C           -2.194224      6.878592     -7.032969
O           -1.067347      6.644049     -7.392535
N           -2.610525      6.772889     -5.765796
C           -1.723015      6.464254     -4.671163
C           -1.444922      4.978007     -4.544804
O           -0.316679      4.592094     -4.340327
C           -2.307517      7.057597     -3.382763
C           -1.724815      6.352660     -2.158849
C           -2.065393      8.573596     -3.416044
H           -3.535125      7.069357     -5.539459
H           -0.755120      6.905332     -4.861635
H           -3.376238      6.879910     -3.397953
H           -1.939327      6.893010     -1.256063
H           -2.140827      5.355759     -2.055908
H           -0.652629      6.275441     -2.228523
H           -2.802863      9.103968     -2.830568
H           -1.077849      8.814563     -3.037015
H           -2.144576      8.956417     -4.426975
N           -2.474279      4.124764     -4.663732
C           -2.164153      2.717524     -4.734152
C           -1.265570      2.444041     -5.932826
O           -0.396160      1.606643     -5.875480
C           -3.408746      1.806689     -4.774910
C           -4.360548      1.970537     -3.579979
C           -5.448860      0.884498     -3.485581
N           -5.007994     -0.292803     -2.757323
C           -5.828029     -1.254631     -2.564076
N           -7.132262     -1.285043     -3.042009
N           -5.529317     -2.307063     -1.768386
H           -3.401452      4.451776     -4.832759
H           -1.576330      2.442308     -3.871014
H           -3.956468      1.985261     -5.698705
H           -3.044992      0.786285     -4.822080
H           -3.794850      1.985348     -2.660536
H           -4.853445      2.927338     -3.654639
H           -6.318442      1.325235     -2.997340
H           -5.765388      0.631232     -4.501235
H           -7.269977     -0.782391     -3.890775
H           -7.547087     -2.198151     -3.082629
H           -4.606193     -2.417209     -1.394936
H           -6.033447     -3.156481     -1.906178
N           -1.459759      3.193337     -7.035104
H           -2.264461      3.769652     -7.144117
C           -0.595841      2.983570     -8.179101
H           -0.940101      3.626229     -8.979236
H           -0.614301      1.957456     -8.516738
C            0.858326      3.326035     -7.861317
O            1.786260      2.682485     -8.284293
N            1.007445      4.395989     -7.064151
H            0.213483      4.959514     -6.860698
```

```
C         2.282309    4.880648   -6.640293
H         2.918466    5.143119   -7.476200
H         2.121619    5.773634   -6.050398
C         3.094242    3.926519   -5.780065
O         4.300265    4.045735   -5.701433
N         2.428061    2.983367   -5.101940
C         3.101620    2.134935   -4.154869
C         4.282205    1.343779   -4.722365
O         5.201482    1.052542   -3.983650
C         2.031171    1.219401   -3.563977
C         2.428250    0.354637   -2.373360
C         1.133384   -0.209485   -1.811071
N         1.349158   -1.211023   -0.786673
C         0.367638   -2.053383   -0.419188
N        -0.791056   -2.021281   -1.034301
N         0.596532   -2.937674    0.523852
H         1.433630    2.938723   -5.174896
H         3.548660    2.735576   -3.371638
H         1.222627    1.874467   -3.257090
H         1.623761    0.598106   -4.353093
H         3.090803   -0.451472   -2.668464
H         2.949613    0.944398   -1.622858
H         0.514109    0.599037   -1.434189
H         0.590668   -0.683203   -2.616428
H         2.061201   -1.044576   -0.110880
H        -1.129890   -1.169180   -1.431149
H        -1.541493   -2.653972   -0.762209
H         1.513554   -3.109966    0.865333
H        -0.183117   -3.502400    0.855326
N         4.281487    0.944899   -6.004135
H         3.540425    1.225017   -6.613436
C         5.363433    0.116651   -6.504612
H         5.138940   -0.142416   -7.531832
H         5.441318   -0.797906   -5.938283
C         6.736189    0.756137   -6.475992
O         7.738501    0.089706   -6.364641
N         6.766107    2.085332   -6.608300
H         5.912164    2.604987   -6.573201
C         8.005490    2.823550   -6.551058
H         8.709874    2.439443   -7.275292
H         7.800317    3.854257   -6.807692
C         8.748767    2.826982   -5.213265
O         9.889900    3.236396   -5.161819
N         8.104020    2.338761   -4.145791
C         8.686068    2.174994   -2.824665
C         9.950624    1.305640   -2.832450
O        10.769444    1.388760   -1.931350
C         7.582249    1.508360   -1.973451
C         7.781315    1.384520   -0.462949
C         7.442009    2.654382    0.326956
C         7.332442    2.387275    1.824347
N         8.672831    2.353850    2.481175
O        11.043759    1.617662    0.892575
H        10.893507    1.828893   -0.029265
```

```
H         11.522402      0.791091      0.841458
H          7.158111      2.035733     -4.258492
H          8.985609      3.132478     -2.419373
H          6.666397      2.054216     -2.156485
H          7.412186      0.521668     -2.388594
H          7.101435      0.605609     -0.131487
H          8.776633      1.030649     -0.229089
H          8.155686      3.454433      0.151271
H          6.477032      3.027453     -0.000996
H          6.772474      3.159663      2.328139
H          6.851243      1.441279      2.019126
H          9.383412      1.890053      1.923632
H          8.992797      3.324044      2.635486
H          8.619521      1.875958      3.368834
N         10.062466      0.448215     -3.837530
C         11.109254     -0.537626     -4.069489
C         12.486198      0.023151     -4.427404
O         13.443814     -0.705721     -4.301297
C         10.700747     -1.444010     -5.252697
C          9.562911     -2.442684     -4.997425
C          9.130917     -3.067132     -6.327922
C          9.990732     -3.533414     -4.015314
H          9.324095      0.446353     -4.510534
H         11.264786     -1.125192     -3.176512
H         10.433793     -0.798369     -6.084028
H         11.583581     -1.996252     -5.551753
H          8.706525     -1.911855     -4.593193
H          8.329833     -3.783547     -6.173252
H          8.773306     -2.310222     -7.014278
H          9.957889     -3.595023     -6.796351
H          9.201125     -4.263650     -3.875431
H         10.860862     -4.059408     -4.394208
H         10.243855     -3.136731     -3.036861
N         12.574050      1.261829     -4.919244
C         13.814631      1.844748     -5.393664
C         14.958652      2.033747     -4.379133
O         16.013009      2.415248     -4.811960
C         13.497155      3.171312     -6.087132
O         12.917856      4.088294     -5.202925
H         11.756474      1.833638     -4.917124
H         14.248647      1.177688     -6.129059
H         14.420302      3.590289     -6.455810
H         12.846475      2.977264     -6.933908
H         11.970501      4.079249     -5.267949
N         14.792295      1.722326     -3.074279
C         15.928389      1.647509     -2.176950
C         16.804567      0.409888     -2.402431
O         17.934351      0.393805     -1.987788
C         15.495714      1.578526     -0.707339
C         14.639621      2.736041     -0.198122
C         14.286898      2.490214      1.269214
N         13.261669      3.385117      1.791185
C         13.439234      4.535908      2.419308
N         14.656891      5.011309      2.656558
```

```
O        15.028868     7.467212     4.211320
H        14.745472     8.353707     3.972084
H        15.747678     7.573264     4.821449
N        12.389203     5.224410     2.815850
H        13.901122     1.419326    -2.758405
H        16.562104     2.504313    -2.346960
H        14.954936     0.650825    -0.551362
H        16.407795     1.509207    -0.124733
H        15.157465     3.683640    -0.315118
H        13.720191     2.817208    -0.766111
H        13.890572     1.490897     1.376863
H        15.166566     2.533534     1.900576
H        12.321594     3.089418     1.619820
H        15.464852     4.518742     2.358635
H        14.792087     5.866171     3.168376
H        11.433766     4.944541     2.657797
H        12.481544     6.067381     3.336224
N        16.200238    -0.648980    -2.967241
C        16.870746    -1.911547    -3.168273
C        17.669182    -1.889188    -4.468289
H        18.064101    -0.909799    -4.747558
O        17.874238    -2.852596    -5.131737
C        15.867974    -3.070182    -3.149443
C        15.324792    -3.399919    -1.791907
N        14.403950    -2.627679    -1.111941
O        12.508282    -0.434546    -0.465242
H        13.146230    -1.037812    -0.850717
H        12.028284    -0.039090    -1.184792
C        15.620109    -4.490418    -1.049700
C        14.173035    -3.251908    -0.003455
N        14.883490    -4.381979     0.103203
H        15.285820    -0.550151    -3.353081
H        17.596166    -2.030815    -2.371151
H        15.050388    -2.842211    -3.824572
H        16.368621    -3.944015    -3.544099
H        16.275589    -5.314100    -1.235269
H        13.494850    -2.922513     0.757658
H        14.783037    -5.064865     0.821930
O       -14.929722    -2.440555     5.284115
H       -15.070645    -1.570308     4.920278
H       -14.426611    -2.308294     6.080648
O       -12.772073    -1.570060     7.335180
H       -12.082833    -2.061380     7.768270
H       -12.982013    -0.850239     7.917504
O       -11.882781    -0.958346     4.587550
H       -11.986054    -1.109474     5.523663
H       -12.698956    -0.533300     4.317258
O       -13.470918    -1.497777     1.153079
H       -12.611455    -1.069906     1.129696
H       -13.392863    -2.234513     1.762452
O        -7.423101     2.196895    -0.103248
H        -7.555141     2.263076     0.846437
H        -6.505869     1.972041    -0.203877
O        -7.633194     2.149865     2.721403
```

| | | | |
|---|---|---|---|
| H | -8.411248 | 1.691243 | 3.031058 |
| H | -7.483551 | 2.881695 | 3.315408 |
| O | -9.180003 | 0.348640 | -1.275893 |
| H | -8.589395 | 1.014137 | -0.904100 |
| H | -8.610225 | -0.340588 | -1.612839 |
| O | -13.000615 | -1.922159 | -1.654690 |
| H | -13.314073 | -1.992348 | -0.752737 |
| H | -13.713766 | -2.168422 | -2.234912 |
| O | -13.813123 | 1.513256 | 0.356362 |
| H | -14.241874 | 0.686036 | 0.545542 |
| H | -12.888966 | 1.336987 | 0.542750 |
| O | -14.680102 | 5.246503 | 4.124659 |
| H | -14.014772 | 5.838141 | 3.788336 |
| H | -14.614055 | 5.349535 | 5.069799 |
| O | -14.703140 | 2.662480 | 2.986863 |
| H | -14.903586 | 3.540852 | 3.322938 |
| H | -14.603806 | 2.715437 | 2.041412 |
| O | -13.336314 | 0.833938 | -2.396199 |
| H | -13.632472 | 1.332355 | -1.635709 |
| H | -13.077163 | -0.016978 | -2.039838 |
| O | -14.474249 | -1.109944 | -4.141616 |
| H | -14.414024 | -0.260518 | -3.708060 |
| H | -15.298233 | -1.129403 | -4.610488 |
| O | -10.572576 | -2.940057 | -3.030987 |
| H | -10.892165 | -2.579067 | -3.854251 |
| H | -11.254346 | -2.730956 | -2.398086 |
| O | -11.266363 | 0.240360 | 0.789619 |
| H | -10.601802 | 0.163027 | 0.104301 |
| H | -10.806630 | 0.507530 | 1.590718 |
| O | -10.209266 | 0.992222 | 3.285589 |
| H | -10.577648 | 0.309838 | 3.844682 |
| H | -10.727928 | 1.781921 | 3.462077 |
| O | -11.957294 | 3.113818 | 3.705388 |
| H | -11.912013 | 4.064500 | 3.654371 |
| H | -12.872601 | 2.889280 | 3.553491 |
| O | -10.746130 | 0.954414 | -3.524643 |
| H | -11.593027 | 1.223341 | -3.167919 |
| H | -10.181390 | 0.783597 | -2.763551 |
| O | -14.430978 | -0.068486 | 3.680545 |
| H | -14.660585 | 0.857060 | 3.580157 |
| H | -14.387286 | -0.433740 | 2.802022 |
| O | -13.081754 | -3.334161 | 3.367518 |
| H | -13.830590 | -3.189757 | 3.954727 |
| H | -12.406096 | -2.767751 | 3.737159 |
| O | -11.696982 | -1.220164 | -5.030317 |
| H | -12.648753 | -1.298531 | -4.990898 |
| H | -11.457902 | -0.551984 | -4.388114 |

Case 2:

```
* F-976-E-MINUS-35mv-H-Tyr-266
N         15.580944    -6.451469     5.391416
H         14.780605    -6.201176     5.939239
C         15.509567    -5.856494     4.054239
C         14.783016    -6.782962     3.060452
O         14.052292    -6.345799     2.186782
C         16.919631    -5.576623     3.520772
H         16.389693    -6.127924     5.884523
H         14.944896    -4.933563     4.050387
H         16.888874    -5.190224     2.509018
H         17.403616    -4.831732     4.142242
H         17.526957    -6.474962     3.529482
N         15.032378    -8.085604     3.186404
H         15.612322    -8.361768     3.949519
C         14.398661    -9.092386     2.333235
H         14.482433    -8.823803     1.292858
H         14.891371   -10.042262     2.492770
C         12.912807    -9.220830     2.671366
O         12.081463    -9.424755     1.825889
N         12.582804    -9.055884     3.966877
H         13.302703    -8.986414     4.648929
C         11.195108    -9.170814     4.421376
H         11.188270    -9.157957     5.503131
H         10.743739   -10.088316     4.077120
C         10.336944    -8.010595     3.900734
O          9.190736    -8.186000     3.584535
N         10.947160    -6.814579     3.855252
C         10.318570    -5.634049     3.317564
C         10.126840    -5.690931     1.799336
O          9.096371    -5.262336     1.337324
C         11.092959    -4.388035     3.776306
O         12.478878    -4.570872     3.769343
H         11.905570    -6.744345     4.106716
H          9.314721    -5.553204     3.701198
H         10.801517    -3.539712     3.167236
H         10.827475    -4.174374     4.801325
H         12.805795    -4.755324     2.897303
N         11.099999    -6.200434     1.023965
C         10.844305    -6.396689    -0.400370
C          9.693436    -7.385377    -0.601404
O          8.944937    -7.279603    -1.544975
C         12.132613    -6.785588    -1.187278
C         12.333425    -8.290993    -1.404044
C         12.159325    -6.060542    -2.534189
H         11.924065    -6.586137     1.431140
H         10.474649    -5.462356    -0.791691
H         12.964734    -6.415498    -0.598871
H         13.287101    -8.454227    -1.895265
H         12.332025    -8.857181    -0.483417
H         11.566158    -8.698188    -2.055359
H         13.053968    -6.324471    -3.087038
H         11.298871    -6.332352    -3.135753
H         12.159341    -4.985320    -2.401010
N          9.570431    -8.340344     0.328319
```

```
H           10.314998    -8.474790     0.973714
C            8.593019    -9.421180     0.241223
H            8.934621   -10.218583     0.888853
H            8.533220    -9.803641    -0.766081
C            7.181685    -9.045756     0.642870
O            6.257522    -9.502155     0.010573
N            7.003461    -8.226499     1.685712
C            5.669615    -7.739212     1.954813
C            5.251037    -6.727566     0.893366
O            4.074289    -6.630604     0.617199
C            5.415370    -7.234001     3.389149
C            5.557241    -8.378393     4.397775
C            6.257332    -6.025242     3.802446
H            7.780838    -7.960621     2.250218
H            4.986577    -8.558522     1.792481
H            4.374586    -6.924964     3.382236
H            5.267122    -8.041301     5.387372
H            4.919129    -9.215730     4.134739
H            6.579369    -8.735711     4.453607
H            5.916866    -5.653281     4.762890
H            7.299619    -6.295871     3.909499
H            6.185287    -5.210612     3.090082
N            6.171856    -6.002978     0.249613
C            5.776344    -5.184142    -0.884278
C            5.343491    -6.067588    -2.062706
O            4.319796    -5.815624    -2.653994
C            6.869047    -4.164972    -1.264289
C            7.067139    -3.148304    -0.125951
C            6.498301    -3.454558    -2.570765
C            8.360934    -2.341843    -0.239949
H            7.133119    -6.060987     0.508240
H            4.878652    -4.647170    -0.617137
H            7.798892    -4.706491    -1.412092
H            6.212216    -2.473797    -0.110016
H            7.076791    -3.662111     0.825975
H            7.203790    -2.665038    -2.790771
H            6.498266    -4.129757    -3.416679
H            5.510258    -3.011068    -2.504004
H            8.462235    -1.668249     0.604329
H            9.222273    -2.999469    -0.239707
H            8.397192    -1.742241    -1.142551
N            6.115623    -7.109673    -2.399467
H            7.009719    -7.232356    -1.972736
C            5.773357    -7.949703    -3.527488
H            6.603932    -8.623441    -3.697534
H            5.618384    -7.363603    -4.421041
C            4.504357    -8.778168    -3.325618
O            3.666280    -8.836135    -4.194066
N            4.393359    -9.431802    -2.162407
H            5.131091    -9.388570    -1.492038
C            3.207309   -10.187565    -1.834412
H            2.967702   -10.899995    -2.609501
H            3.399615   -10.730343    -0.917398
C            1.974183    -9.306851    -1.644814
```

```
O         0.887188    -9.696194   -2.004603
N         2.187877    -8.107003   -1.107171
C         1.062409    -7.184556   -1.070943
C         0.616899    -6.720548   -2.460282
O        -0.565535    -6.584413   -2.680697
C         1.278111    -5.930572   -0.230466
O         1.413950    -6.189414    1.126952
H         3.087725    -7.817064   -0.798549
H         0.211098    -7.704080   -0.662224
H         2.120966    -5.365010   -0.613564
H         0.384092    -5.341882   -0.351483
H         2.307864    -6.456596    1.293773
N         1.544084    -6.431961   -3.383021
C         1.145401    -6.024110   -4.717962
C         0.378713    -7.146528   -5.410977
O        -0.608617    -6.891359   -6.067831
C         2.355745    -5.521072   -5.538423
C         2.761868    -4.129042   -5.026137
C         2.060227    -5.496091   -7.043398
C         4.087304    -3.618286   -5.589143
H         2.514483    -6.487418   -3.159265
H         0.424266    -5.223003   -4.638574
H         3.176787    -6.210979   -5.367731
H         1.969949    -3.423461   -5.271828
H         2.835530    -4.155166   -3.946542
H         2.918116    -5.124544   -7.589820
H         1.841475    -6.482977   -7.432575
H         1.215284    -4.853434   -7.269908
H         4.380365    -2.704936   -5.082017
H         4.879282    -4.345086   -5.439155
H         4.030270    -3.399810   -6.649639
N         0.820021    -8.397267   -5.251521
H         1.667088    -8.572880   -4.752418
C         0.043272    -9.504886   -5.753937
H        -0.064086    -9.462635   -6.827599
H         0.556435   -10.421367   -5.491762
C        -1.365276    -9.540103   -5.166934
O        -2.319108    -9.771152   -5.874321
N        -1.472639    -9.318869   -3.858022
C        -2.776401    -9.352477   -3.235399
C        -3.692486    -8.216370   -3.692178
O        -4.890811    -8.424118   -3.732703
C        -2.679003    -9.329616   -1.702814
O        -3.923262    -9.599234   -1.118426
H        -0.656826    -9.231548   -3.289012
H        -3.292805   -10.257033   -3.524111
H        -1.996532   -10.103288   -1.379346
H        -2.290336    -8.372808   -1.370401
H        -4.613427    -9.195876   -1.630806
N        -3.143423    -7.035461   -3.971032
C        -3.907940    -5.914860   -4.493056
C        -4.449187    -6.233746   -5.886671
O        -5.613068    -6.041126   -6.155912
C        -3.013296    -4.655333   -4.488046
```

```
C         -3.472029    -3.518247    -5.385264
C         -4.792743    -3.082017    -5.426058
C         -2.544404    -2.883530    -6.203072
C         -5.168555    -2.055510    -6.276057
C         -2.913599    -1.844251    -7.041929
C         -4.233412    -1.431280    -7.085866
H         -2.166731    -6.912867    -3.802026
H         -4.776058    -5.749113    -3.875597
H         -2.930343    -4.319494    -3.458727
H         -2.021567    -4.947411    -4.803548
H         -5.537422    -3.560018    -4.818474
H         -1.519831    -3.211402    -6.194476
H         -6.201174    -1.752456    -6.311892
H         -2.174785    -1.370705    -7.663102
H         -4.528428    -0.637244    -7.749839
N         -3.576093    -6.724267    -6.775011
C         -4.019466    -7.062103    -8.108057
C         -5.145297    -8.094816    -8.085388
O         -6.033114    -8.030967    -8.895470
C         -2.870780    -7.593811    -8.968643
S         -1.652311    -6.342848    -9.484955
H         -2.634890    -6.908431    -6.497377
H         -4.443858    -6.190215    -8.582945
H         -2.366882    -8.412419    -8.472675
H         -3.288966    -7.972037    -9.891749
H         -1.047846    -6.179409    -8.318677
N         -5.078558    -9.066913    -7.155619
H         -4.254248    -9.192320    -6.607830
C         -6.085910   -10.115215    -7.183049
H         -6.258812   -10.452778    -8.193230
H         -5.714556   -10.947552    -6.596537
C         -7.459140    -9.722570    -6.613183
O         -8.481524   -10.086878    -7.126898
N         -7.417939    -9.016209    -5.478356
C         -8.641654    -8.653167    -4.788754
C         -9.504928    -7.658159    -5.556595
O        -10.702650    -7.658734    -5.450267
C         -8.361790    -8.103681    -3.378645
C         -7.519477    -6.820431    -3.271096
C         -8.193038    -5.469071    -3.256679
O         -9.495079    -5.469695    -3.327671
H         -9.844036    -4.571891    -3.238586
O         -7.546903    -4.455184    -3.161451
H         -6.535789    -8.774594    -5.078968
H         -9.260690    -9.533420    -4.686410
H         -9.317922    -7.967172    -2.894287
H         -7.836541    -8.883885    -2.838288
H         -6.956541    -6.846961    -2.345210
H         -6.778057    -6.766643    -4.050670
N         -8.837301    -6.733962    -6.273855
C         -9.555196    -5.702617    -7.000840
C        -10.386046    -6.273602    -8.141689
H        -10.409089    -7.356342    -8.248993
O        -10.987293    -5.559103    -8.888423
```

```
C         -8.631312   -4.564298   -7.490284
O         -9.434533   -3.441998   -7.831927
C         -7.723015   -4.928132   -8.657220
H         -7.856607   -6.841481   -6.421929
H        -10.268410   -5.253865   -6.319060
H         -8.030343   -4.256844   -6.645801
H        -10.110596   -3.750863   -8.429428
H         -7.045359   -4.103576   -8.850925
H         -7.131131   -5.809168   -8.458024
H         -8.303897   -5.104881   -9.556039
N        -12.952122   -9.022524    3.174171
H        -13.215000   -9.550467    2.367604
C        -11.773531   -8.180020    2.898781
C        -10.789122   -8.123249    4.050087
O         -9.659793   -7.714465    3.857064
C        -12.146619   -6.722601    2.563120
O        -12.715837   -6.204029    3.737738
C        -13.114462   -6.605062    1.391196
H        -13.731488   -8.453499    3.442454
H        -11.222783   -8.599613    2.068258
H        -11.230715   -6.196590    2.328454
H        -12.850019   -5.263217    3.637899
H        -13.281661   -5.557085    1.166643
H        -12.729025   -7.082233    0.498078
H        -14.074921   -7.044986    1.631883
N        -11.197257   -8.533635    5.241900
H        -12.119355   -8.910736    5.275381
C        -10.412545   -8.529115    6.451311
H        -10.337037   -7.534078    6.860877
H        -10.933713   -9.138687    7.180104
C         -8.974757   -9.043554    6.374947
O         -8.140154   -8.478376    7.037913
N         -8.658722  -10.119724    5.632203
H         -9.369346  -10.599264    5.127171
C         -7.270130  -10.552738    5.532057
H         -7.253485  -11.532149    5.071252
H         -6.830728  -10.631478    6.513529
C         -6.387460   -9.608364    4.692694
O         -5.219619   -9.468422    4.949191
N         -7.045394   -9.018012    3.701158
C         -6.301973   -8.005930    2.963216
C         -6.261971   -6.695345    3.744811
O         -5.267301   -6.009523    3.699222
C         -6.610841   -7.876580    1.459383
C         -7.789996   -7.050923    0.991428
C         -7.776654   -5.663737    1.106805
C         -8.849933   -7.648230    0.322340
C         -8.804228   -4.896947    0.584329
C         -9.869293   -6.882523   -0.220215
C         -9.854080   -5.504749   -0.084365
H         -8.039943   -9.014006    3.679598
H         -5.279248   -8.337545    2.997382
H         -5.714001   -7.456606    1.017512
H         -6.687392   -8.886335    1.073854
```

```
H          -6.944552    -5.173537     1.578631
H          -8.862239    -8.716719     0.190761
H          -8.767446    -3.825784     0.671445
H         -10.659064    -7.357730    -0.772552
H         -10.636287    -4.911642    -0.520867
N          -7.293002    -6.385610     4.534364
C          -7.221730    -5.273353     5.451352
C          -6.102461    -5.490852     6.469145
O          -5.456099    -4.538968     6.852411
C          -8.598843    -5.101510     6.124410
C          -8.810397    -3.809065     6.883898
C          -8.979288    -2.606845     6.201470
C          -8.906107    -3.805157     8.268734
C          -9.226647    -1.430576     6.886655
C          -9.163955    -2.629711     8.961213
C          -9.323917    -1.438928     8.272379
H          -8.158220    -6.869398     4.428525
H          -6.951490    -4.368444     4.925868
H          -9.342225    -5.167801     5.336512
H          -8.762024    -5.940633     6.786700
H          -8.925588    -2.588776     5.126645
H          -8.781967    -4.723519     8.815638
H          -9.346256    -0.512296     6.341059
H          -9.227231    -2.646428    10.034539
H          -9.506863    -0.523364     8.806590
N          -5.861975    -6.727968     6.898076
H          -6.516693    -7.449620     6.693395
C          -4.785924    -7.022971     7.817853
H          -4.825001    -6.393313     8.694036
H          -4.898403    -8.052884     8.131654
C          -3.390082    -6.846814     7.222645
O          -2.516229    -6.352123     7.894630
N          -3.189628    -7.271995     5.970549
C          -1.911667    -7.092411     5.299011
C          -1.665261    -5.620551     4.938950
O          -0.566846    -5.133893     5.107319
C          -1.830104    -8.027759     4.072347
C          -0.664591    -7.670603     3.148131
C          -1.728812    -9.487918     4.527864
H          -3.922885    -7.753130     5.495678
H          -1.118350    -7.338224     5.989987
H          -2.750689    -7.904009     3.506718
H          -0.600501    -8.397395     2.345361
H          -0.774988    -6.696245     2.693253
H           0.280001    -7.682386     3.683300
H          -1.741975   -10.148550     3.667022
H          -0.795210    -9.654755     5.059259
H          -2.545713    -9.777642     5.175983
N          -2.700683    -4.933602     4.446723
C          -2.569791    -3.551413     4.028848
C          -2.420055    -2.639035     5.244026
O          -1.528454    -1.820255     5.290671
C          -3.758179    -3.090370     3.171592
C          -3.993855    -3.840527     1.849922
```

```
C         -2.859849    -3.730828     0.845315
O         -1.746949    -4.135755     1.218717
O         -3.112387    -3.254706    -0.274000
H         -3.564290    -5.403253     4.285196
H         -1.662684    -3.440938     3.465257
H         -4.667090    -3.158025     3.761551
H         -3.604787    -2.035175     2.961839
H         -4.149166    -4.890275     2.048787
H         -4.899225    -3.450481     1.399036
N         -3.309527    -2.774871     6.244281
C         -3.093978    -2.024901     7.456985
C         -1.778227    -2.422283     8.133769
O         -1.106870    -1.582682     8.694100
C         -4.226306    -2.005092     8.493911
O         -4.436264    -3.246772     9.098351
C         -5.507665    -1.387418     7.948248
H         -3.988319    -3.503125     6.218592
H         -2.937615    -0.995394     7.174834
H         -3.851340    -1.365921     9.283604
H         -4.889516    -3.818426     8.490419
H         -6.240075    -1.327579     8.743512
H         -5.322056    -0.384126     7.577075
H         -5.928451    -1.976867     7.144820
N         -1.380397    -3.700457     8.043328
H         -2.031219    -4.418216     7.812066
C         -0.187162    -4.107473     8.739880
H         -0.237315    -3.891790     9.797886
H         -0.075815    -5.175172     8.604040
C          1.064024    -3.420008     8.217958
O          1.940562    -3.076028     8.977771
N          1.153879    -3.258765     6.881711
C          2.332480    -2.660282     6.317666
C          2.291031    -1.138383     6.164108
O          3.359924    -0.574352     6.023012
C          2.830599    -3.345982     5.045415
S          1.922615    -2.976798     3.520199
H          0.422681    -3.612342     6.296759
H          3.116420    -2.798068     7.046915
H          3.840481    -3.007717     4.863090
H          2.853201    -4.415051     5.204870
H          0.859018    -3.741107     3.719458
N          1.140073    -0.470671     6.273042
C          1.204068     0.968497     6.491343
C          1.755867     1.271210     7.870506
O          2.368584     2.288860     8.085489
C         -0.095735     1.741771     6.182377
C         -1.196941     1.568454     7.238669
C         -0.554806     1.440244     4.750822
C         -2.466310     2.370770     6.944793
H          0.271734    -0.965324     6.291801
H          1.951475     1.351091     5.814764
H          0.208555     2.782274     6.222312
H         -1.446001     0.523677     7.355885
H         -0.807694     1.892671     8.199275
```

```
H            -1.310205      2.147625      4.431774
H             0.277840      1.520488      4.057287
H            -0.966433      0.444537      4.656977
H            -3.135946      2.340396      7.798938
H            -2.241207      3.414239      6.739191
H            -3.007062      1.973583      6.092649
N             1.549788      0.323258      8.791266
H             0.889674     -0.402097      8.613548
C             1.978259      0.512664     10.160074
H             1.470031     -0.222645     10.771021
H             1.737710      1.502369     10.518137
C             3.482320      0.337408     10.305053
O             4.164481      1.128932     10.896182
N             3.999933     -0.753835      9.695408
H             3.390718     -1.466103      9.351485
C             5.425034     -0.938857      9.667453
H             5.638280     -1.909399      9.234591
H             5.855716     -0.904647     10.657516
C             6.156882      0.119117      8.844832
O             7.291930      0.411889      9.129191
N             5.502078      0.621285      7.799030
C             6.098617      1.759274      7.146315
C             5.514930      3.076794      7.649682
H             5.013174      3.037228      8.615080
O             5.636959      4.094891      7.041084
C             6.002880      1.675018      5.620789
C             6.755249      0.506917      5.017467
C             8.120897      0.322917      5.238163
C             6.094905     -0.401623      4.195757
C             8.807270     -0.722520      4.635365
C             6.779038     -1.444694      3.591485
C             8.139125     -1.605637      3.801366
H             4.575199      0.334750      7.572736
H             7.134082      1.774541      7.451288
H             4.960665      1.610784      5.334377
H             6.381157      2.609453      5.220641
H             8.648975      0.979017      5.908743
H             5.034721     -0.305680      4.057633
H             9.857989     -0.849857      4.831180
H             6.246852     -2.136576      2.963236
H             8.659502     -2.418213      3.328337
N            12.613550      9.596234     -1.248802
H            11.873251      9.914036     -1.846107
C            12.304852      8.264740     -0.719487
C            11.274139      8.450687      0.386897
O            10.265877      7.779033      0.416424
C            13.562740      7.581006     -0.165338
C            14.645071      7.276096     -1.206565
S            14.017112      6.202108     -2.529901
C            15.499794      6.053606     -3.555345
H            13.433413      9.573022     -1.822259
H            11.834450      7.616518     -1.447338
H            14.001304      8.205423      0.607192
H            13.255687      6.653825      0.304791
```

```
H       15.045468    8.188561   -1.633889
H       15.470243    6.783690   -0.704559
H       15.248468    5.398939   -4.375892
H       15.798798    7.017676   -3.946031
H       16.318169    5.616741   -2.998068
N       11.504319    9.412913    1.296974
C       10.594348    9.607808    2.408562
C        9.277067   10.274789    2.010382
O        8.281020   10.053301    2.652714
C       11.276314   10.369119    3.552080
C       12.448034    9.564515    4.084489
O       13.582868    9.826309    3.744685
N       12.144421    8.562209    4.918899
H       12.328109    9.967780    1.206258
H       10.295558    8.637760    2.771295
H       11.657342   11.326350    3.218957
H       10.546139   10.537733    4.333500
H       12.888265    7.966850    5.213200
H       11.222925    8.175946    4.954709
N        9.263831   11.067722    0.922310
H       10.110797   11.229597    0.427324
C        8.003502   11.517470    0.368859
H        8.201222   12.260245   -0.393341
H        7.389103   11.966665    1.131940
C        7.186384   10.384439   -0.254324
O        5.986296   10.470082   -0.291509
N        7.862661    9.333318   -0.763260
C        7.170486    8.145603   -1.233909
C        6.634992    7.300679   -0.067270
O        5.534682    6.808152   -0.148207
C        8.048994    7.344301   -2.224229
C        8.343645    8.197264   -3.472928
C        7.379617    6.019178   -2.608125
C        9.428312    7.625246   -4.388996
H        8.823868    9.230810   -0.536735
H        6.279734    8.471644   -1.748344
H        8.988067    7.117937   -1.727344
H        7.421718    8.323212   -4.036260
H        8.645669    9.192928   -3.165447
H        7.997037    5.478886   -3.314349
H        7.228554    5.377877   -1.748308
H        6.409910    6.182774   -3.066625
H        9.622269    8.305829   -5.211634
H       10.363646    7.479783   -3.855321
H        9.148522    6.671165   -4.820127
N        7.387690    7.175308    1.043947
C        6.879166    6.458329    2.201997
C        5.528064    7.039473    2.639720
O        4.575678    6.332911    2.867084
C        7.833311    6.546165    3.403537
C        9.254785    5.997325    3.244210
O       10.146034    6.631439    3.802226
O        9.428387    4.935707    2.595214
H        8.347749    7.445033    1.022646
```

```
H         6.682584    5.429517    1.943351
H         7.921361    7.570767    3.733117
H         7.375197    5.992739    4.220726
N         5.489887    8.377412    2.786093
H         6.320922    8.912627    2.648912
C         4.299247    9.045888    3.249974
H         4.538272   10.090246    3.407640
H         3.957307    8.638803    4.188427
C         3.136661    8.987517    2.268580
O         1.995196    9.109753    2.649482
N         3.470198    8.824750    0.985184
H         4.431331    8.829481    0.735262
C         2.517191    8.905260   -0.078305
H         3.059845    9.091730   -0.996314
H         1.843947    9.738043    0.071942
C         1.609133    7.687020   -0.310194
O         0.561873    7.815080   -0.879456
N         2.094450    6.538077    0.201341
C         1.289915    5.364391    0.431052
C         0.472366    5.479579    1.721851
O        -0.722723    5.301849    1.692290
C         2.166640    4.112382    0.445787
H         3.001458    6.575677    0.616027
H         0.559459    5.290403   -0.357453
H         1.566371    3.234011    0.657559
H         2.636342    3.991439   -0.523469
H         2.946287    4.184370    1.196906
N         1.102526    5.751848    2.879338
C         0.364319    5.585295    4.123115
C        -0.756395    6.611745    4.274095
O        -1.792311    6.302365    4.823511
C         1.276044    5.511889    5.372777
C         1.867890    6.857325    5.819756
C         2.362792    4.445281    5.169577
C         2.699032    6.761047    7.101617
H         2.090115    5.889556    2.900976
H        -0.159805    4.642232    4.064891
H         0.615224    5.173609    6.165237
H         2.478240    7.266383    5.021419
H         1.064534    7.567480    5.986889
H         2.732497    4.092365    6.121113
H         1.969567    3.588565    4.631492
H         3.205860    4.834223    4.608992
H         3.004371    7.751979    7.424500
H         2.121225    6.317754    7.907248
H         3.593851    6.165168    6.971790
N        -0.542944    7.847430    3.801020
H         0.341450    8.089943    3.404838
C        -1.513861    8.887808    4.018652
H        -1.103685    9.816518    3.642483
H        -1.724196    9.016557    5.070574
C        -2.866972    8.618762    3.365891
O        -3.871054    8.837162    4.003288
N        -2.902689    8.108966    2.120757
```

```
C         -4.239994    8.013726    1.535324
C         -4.984708    6.749465    1.927653
O         -6.195848    6.756948    1.919618
C         -4.022033    8.221434    0.030819
C         -2.766445    9.099865    0.020044
C         -1.907276    8.456243    1.096073
H         -4.854277    8.814601    1.917373
H         -3.842473    7.278944   -0.469789
H         -4.886472    8.684732   -0.427141
H         -2.255835    9.142614   -0.923514
H         -3.027341   10.114526    0.311498
H         -1.420659    7.566150    0.740384
H         -1.151189    9.119065    1.484439
N         -4.269045    5.673722    2.291129
C         -4.938229    4.512552    2.850562
C         -5.638496    4.853563    4.169019
O         -6.741410    4.413657    4.408508
C         -3.973914    3.331409    3.086460
C         -3.635868    2.505617    1.861093
C         -4.333057    1.334853    1.576479
C         -2.597692    2.864521    1.013868
C         -4.007367    0.548080    0.481876
C         -2.270704    2.094030   -0.087806
C         -2.970649    0.931914   -0.361421
O         -2.617869    0.205930   -1.447938
H         -3.414679   -0.117521   -1.907982
H         -3.275174    5.750810    2.345605
H         -5.725643    4.209378    2.178910
H         -3.068234    3.710965    3.548120
H         -4.443647    2.682173    3.816457
H         -5.141779    1.028511    2.216703
H         -2.032745    3.752393    1.204716
H         -4.536277   -0.366724    0.289378
H         -1.479270    2.397027   -0.750216
N         -4.970839    5.618277    5.048350
H         -4.034974    5.911393    4.856556
C         -5.582770    6.005349    6.299222
H         -4.805964    6.385465    6.951082
H         -6.052320    5.158954    6.774277
C         -6.651496    7.084455    6.135958
O         -7.706406    7.028002    6.726261
N         -6.342147    8.091247    5.313906
H         -5.445016    8.139354    4.880110
C         -7.257687    9.220267    5.085304
H         -6.781278    9.871869    4.364214
H         -7.419712    9.775739    5.997421
C         -8.632477    8.840767    4.569520
O         -9.611434    9.429008    4.959464
N         -8.690359    7.854349    3.658146
C         -9.953936    7.451763    3.088772
C        -10.766323    6.542229    3.995444
O        -11.807588    6.068902    3.559164
C         -9.761142    6.816141    1.689857
O        -10.954305    6.903463    0.953096
```

```
C        -9.239228     5.382323     1.729283
H        -7.846118     7.427536     3.334427
H       -10.571638     8.331037     2.962658
H        -9.046018     7.439524     1.169869
H       -11.641554     6.500755     1.473063
H        -8.968246     5.061445     0.731314
H        -8.362371     5.288787     2.354506
H       -10.000981     4.705865     2.098583
N       -10.353337     6.327943     5.240900
C       -11.259093     5.732447     6.212924
C       -12.445055     6.656022     6.502021
O       -13.529726     6.188271     6.758159
C       -10.538025     5.397307     7.526339
C        -9.534251     4.235052     7.447051
C        -8.716965     4.184331     8.741485
C       -10.219852     2.887533     7.196333
H        -9.463052     6.657570     5.550986
H       -11.697039     4.842319     5.794053
H       -10.028513     6.287162     7.878980
H       -11.302479     5.156613     8.259987
H        -8.847326     4.423758     6.628936
H        -7.988825     3.378599     8.709769
H        -8.182037     5.114034     8.898260
H        -9.358174     4.014320     9.603606
H        -9.479647     2.092158     7.178431
H       -10.931103     2.661970     7.988749
H       -10.749907     2.858178     6.251424
N       -12.200391     7.978986     6.519086
C       -13.272865     8.918063     6.674440
C       -14.047138     9.177161     5.399140
H       -14.886714     9.870642     5.514692
O       -13.792339     8.699104     4.343338
H       -11.339485     8.323326     6.149711
H       -13.966260     8.566385     7.428079
H       -12.873616     9.866157     7.021585
N       -11.400139    10.603601    -3.013218
H       -11.151156    10.793474    -2.062887
C       -11.816457     9.206991    -3.172692
C       -11.420798     8.859979    -4.600829
O       -10.381146     8.297520    -4.846936
C       -13.274697     8.886451    -2.781041
C       -13.514119     9.244426    -1.302467
C       -13.600730     7.414103    -3.063630
C       -14.953833     9.045731    -0.824926
H       -12.140937    11.235286    -3.252954
H       -11.162892     8.607041    -2.555403
H       -13.942062     9.511037    -3.379532
H       -12.841512     8.656501    -0.682289
H       -13.250226    10.285287    -1.143016
H       -14.652033     7.209116    -2.907368
H       -13.370171     7.129386    -4.086046
H       -13.034226     6.765433    -2.402795
H       -15.063132     9.412642     0.189358
H       -15.657291     9.588307    -1.451064
```

```
H          -15.244825     8.001845    -0.821495
N          -12.217767     9.294713    -5.606266
H          -13.139261     9.595970    -5.384728
C          -11.972872     8.843880    -6.959258
H          -12.033858     7.767145    -7.039985
H          -12.729057     9.278149    -7.601893
C          -10.601635     9.233244    -7.517693
O           -9.996941     8.496732    -8.241527
N          -10.162845    10.477780    -7.182625
H          -10.711033    11.007782    -6.544746
C           -8.894341    11.023967    -7.637100
H           -8.789698    10.891248    -8.702753
H           -8.878586    12.081461    -7.407589
C           -7.672564    10.359017    -6.988311
O           -6.753256     9.935092    -7.630831
N           -7.659106    10.352139    -5.619638
H           -8.482447    10.628022    -5.133559
C           -6.495253     9.961264    -4.832348
H           -5.599197    10.377765    -5.264338
H           -6.616278    10.359356    -3.832341
C           -6.268292     8.463912    -4.714277
O           -5.133740     8.047342    -4.711620
N           -7.321026     7.642803    -4.555739
C           -7.043152     6.284788    -4.116082
C           -6.395851     5.449242    -5.213188
O           -5.518651     4.674607    -4.918154
C           -8.232787     5.569469    -3.436228
C           -9.368455     5.183222    -4.399094
C           -8.696989     6.392754    -2.228906
C          -10.549754     4.494052    -3.715373
H           -8.255737     7.997406    -4.569992
H           -6.261691     6.342851    -3.373705
H           -7.807220     4.647354    -3.048787
H           -9.726255     6.052590    -4.935186
H           -8.960269     4.506274    -5.145546
H           -9.400871     5.838844    -1.623588
H           -7.854458     6.643370    -1.592293
H           -9.177962     7.313866    -2.533136
H          -11.255362     4.128107    -4.455475
H          -10.227063     3.646217    -3.122038
H          -11.086774     5.173295    -3.065412
N           -6.814645     5.564832    -6.493391
C           -6.106146     4.822474    -7.516656
C           -4.760276     5.444472    -7.899124
O           -3.890142     4.715858    -8.316310
C           -6.967005     4.499892    -8.746620
C           -8.037268     3.418486    -8.510343
C           -7.451480     2.077185    -8.040920
N           -8.389453     0.950985    -8.074694
C           -9.199263     0.573556    -7.081841
N           -9.462412     1.417955    -6.095954
N           -9.735558    -0.626717    -7.080755
H           -7.489451     6.258135    -6.735507
H           -5.795978     3.898611    -7.061617
```

```
H           -7.450710     5.400293    -9.113841
H           -6.296125     4.164688    -9.530075
H           -8.778088     3.780379    -7.803276
H           -8.564870     3.256187    -9.444433
H           -6.626428     1.798640    -8.683547
H           -7.051772     2.146842    -7.038095
H           -8.319794     0.321084    -8.843001
H           -9.210871     2.374884    -6.171143
H           -9.976329     1.152578    -5.272038
H           -9.387084    -1.392206    -7.629023
H          -10.455807    -0.867823    -6.418365
N           -4.571383     6.757521    -7.721942
H           -5.312316     7.350933    -7.425893
C           -3.257811     7.352384    -8.011367
H           -2.933574     7.093271    -9.006665
H           -3.359220     8.427155    -7.938262
C           -2.194217     6.878612    -7.032962
O           -1.067310     6.644126    -7.392501
N           -2.610531     6.772888    -5.765801
C           -1.723040     6.464253    -4.671160
C           -1.444918     4.978008    -4.544815
O           -0.316663     4.592108    -4.340345
C           -2.307585     7.057556    -3.382764
C           -1.724829     6.352659    -2.158853
C           -2.065567     8.573569    -3.416052
H           -3.535171     7.069255    -5.539484
H           -0.755152     6.905360    -4.861604
H           -3.376293     6.879800    -3.397955
H           -1.939342     6.893014    -1.256070
H           -2.140806     5.355748    -2.055884
H           -0.652641     6.275472    -2.228560
H           -2.803049     9.103886    -2.830544
H           -1.078020     8.814606    -3.037070
H           -2.144829     8.956383    -4.426977
N           -2.474273     4.124764    -4.663714
C           -2.164153     2.717524    -4.734152
C           -1.265510     2.444099    -5.932795
O           -0.395981     1.606821    -5.875409
C           -3.408743     1.806684    -4.774960
C           -4.360591     1.970565    -3.580069
C           -5.448858     0.884496    -3.485617
N           -5.007890    -0.292811    -2.757433
C           -5.827923    -1.254618    -2.564041
N           -7.132237    -1.284998    -3.041769
N           -5.529119    -2.307063    -1.768413
H           -3.401435     4.451781    -4.832822
H           -1.576351     2.442291    -3.871004
H           -3.956426     1.985242    -5.698780
H           -3.044991     0.786278    -4.822098
H           -3.794925     1.985472    -2.660604
H           -4.853529     2.927337    -3.654816
H           -6.318404     1.325198    -2.997289
H           -5.765468     0.631256    -4.501251
H           -7.270081    -0.782411    -3.890557
```

```
H        -7.547128   -2.198086   -3.082229
H        -4.605959   -2.417238   -1.395079
H        -6.033341   -3.156447   -1.906095
N        -1.459759    3.193337   -7.035104
H        -2.264571    3.769495   -7.144196
C        -0.595832    2.983550   -8.179089
H        -0.940093    3.626187   -8.979240
H        -0.614272    1.957428   -8.516706
C         0.858332    3.326033   -7.861294
O         1.786295    2.682497   -8.284242
N         1.007444    4.396007   -7.064155
H         0.213477    4.959530   -6.860689
C         2.282309    4.880648   -6.640293
H         2.918466    5.143147   -7.476195
H         2.121614    5.773614   -6.050368
C         3.094243    3.926494   -5.780092
O         4.300279    4.045687   -5.701465
N         2.428061    2.983367   -5.101940
C         3.101641    2.134963   -4.154867
C         4.282232    1.343822   -4.722385
O         5.201595    1.052678   -3.983728
C         2.031200    1.219459   -3.563912
C         2.428279    0.354765   -2.373247
C         1.133390   -0.209096   -1.810740
N         1.349169   -1.210787   -0.786501
C         0.367647   -2.053261   -0.419186
N        -0.791012   -2.021153   -1.034330
N         0.596603   -2.937678    0.523714
H         1.433627    2.938694   -5.174913
H         3.548704    2.735628   -3.371667
H         1.222666    1.874552   -3.257062
H         1.623769    0.598124   -4.352985
H         3.090659   -0.451492   -2.668342
H         2.949837    0.944516   -1.622870
H         0.514363    0.599529   -1.433668
H         0.590414   -0.682622   -2.616027
H         2.060887   -1.044140   -0.110411
H        -1.129920   -1.169038   -1.431095
H        -1.541386   -2.654010   -0.762410
H         1.513623   -3.109826    0.865267
H        -0.183083   -3.502307    0.855343
N         4.281487    0.944901   -6.004141
H         3.540371    1.224938   -6.613418
C         5.363433    0.116651   -6.504612
H         5.138944   -0.142419   -7.531832
H         5.441320   -0.797906   -5.938283
C         6.736191    0.756138   -6.475981
O         7.738526    0.089729   -6.364633
N         6.766107    2.085332   -6.608300
H         5.912156    2.604985   -6.573201
C         8.005486    2.823552   -6.551065
H         8.709854    2.439480   -7.275335
H         7.800291    3.854267   -6.807648
C         8.748798    2.826930   -5.213288
```

```
O         9.889951     3.236315    -5.161832
N         8.104016     2.338735    -4.145826
C         8.685994     2.174934    -2.824675
C         9.950570     1.305606    -2.832425
O        10.769365     1.388724    -1.931295
C         7.582130     1.508244    -1.973555
C         7.780996     1.384533    -0.463019
C         7.441450     2.654429     0.326731
C         7.331647     2.387428     1.824123
N         8.671949     2.354031     2.481140
O        11.043286     1.617612     0.892719
H        10.893052     1.828799    -0.029133
H        11.521828     0.790985     0.841668
H         7.158097     2.035734    -4.258562
H         8.985493     3.132409    -2.419326
H         6.666251     2.053993    -2.156744
H         7.412207     0.521513    -2.388659
H         7.101135     0.605589    -0.131601
H         8.776310     1.030766    -0.228981
H         8.155104     3.454515     0.151101
H         6.476503     3.027407    -0.001413
H         6.771589     3.159838     2.327782
H         6.850425     1.441442     2.018892
H         9.382571     1.890136     1.923742
H         8.991993     3.324207     2.635424
H         8.618503     1.876206     3.368828
N        10.062466     0.448215    -3.837530
C        11.109251    -0.537627    -4.069489
C        12.486201     0.023149    -4.427400
O        13.443843    -0.705691    -4.301267
C        10.700734    -1.443986    -5.252716
C         9.562929    -2.442696    -4.997444
C         9.130895    -3.067095    -6.327950
C         9.990812    -3.533456    -4.015394
H         9.324093     0.446344    -4.510536
H        11.264778    -1.125208    -3.176521
H        10.433734    -0.798323    -6.084015
H        11.583574    -1.996197    -5.551819
H         8.706548    -1.911906    -4.593160
H         8.329842    -3.783539    -6.173274
H         8.773221    -2.310164    -7.014248
H         9.957864    -3.594938    -6.796442
H         9.201223    -4.263707    -3.875504
H        10.860934    -4.059422    -4.394350
H        10.243969    -3.136808    -3.036935
N        12.574050     1.261829    -4.919244
C        13.814624     1.844749    -5.393663
C        14.958706     2.033567    -4.379164
O        16.013158     2.414783    -4.812030
C        13.497165     3.171410    -6.086959
O        12.917964     4.088335    -5.202640
H        11.756425     1.833577    -4.917248
H        14.248583     1.177768    -6.129163
H        14.420310     3.590389    -6.455646
```

```
H              12.846429     2.977487    -6.933720
H              11.970594     4.079224    -5.267509
N              14.792295     1.722326    -3.074279
C              15.928362     1.647498    -2.176930
C              16.804572     0.409919    -2.402499
O              17.934404     0.393867    -1.987966
C              15.495618     1.578402    -0.707341
C              14.639479     2.735871    -0.198089
C              14.286573     2.489893     1.269178
N              13.261352     3.384797     1.791148
C              13.438934     4.535615     2.419219
N              14.656613     5.010999     2.656427
O              15.029030     7.467122     4.210782
H              14.745418     8.353611     3.971764
H              15.747617     7.573218     4.821163
N              12.388927     5.224132     2.815771
H              13.901038     1.419628    -2.758344
H              16.562064     2.504331    -2.346854
H              14.954846     0.650683    -0.551461
H              16.407670     1.509060    -0.124688
H              15.157351     3.683476    -0.314932
H              13.720110     2.817115    -0.766163
H              13.890152     1.490595     1.376643
H              15.166171     2.533074     1.900653
H              12.321274     3.089093     1.619810
H              15.464549     4.518437     2.358435
H              14.791850     5.865912     3.168142
H              11.433471     4.944259     2.657765
H              12.481277     6.067154     3.336059
N              16.200238    -0.648980    -2.967241
C              16.870796    -1.911508    -3.168197
C              17.669503    -1.889228    -4.468054
H              18.063310    -0.909605    -4.748073
O              17.875929    -2.852970    -5.130598
C              15.868043    -3.070177    -3.149486
C              15.324788    -3.399916    -1.791982
N              14.403993    -2.627607    -1.112039
O              12.508231    -0.434478    -0.465182
H              13.146153    -1.037701    -0.850756
H              12.028078    -0.039074    -1.184664
C              15.619975    -4.490463    -1.049798
C              14.172980    -3.251840    -0.003576
N              14.883327    -4.381982     0.103080
H              15.285740    -0.550231    -3.352920
H              17.596095    -2.030812    -2.370971
H              15.050489    -2.842195    -3.824652
H              16.368753    -3.943989    -3.544107
H              16.275364    -5.314216    -1.235376
H              13.494777    -2.922413     0.757508
H              14.782798    -5.064870     0.821794
O             -14.930263    -2.440032     5.283656
H             -15.070599    -1.569671     4.919845
H             -14.427377    -2.308097     6.080381
O             -12.772909    -1.570133     7.335218
```

```
H         -12.084057    -2.061818     7.768504
H         -12.983745    -0.851100     7.918196
O         -11.882749    -0.958225     4.588006
H         -11.986359    -1.108925     5.524154
H         -12.698520    -0.532607     4.317364
O         -13.470540    -1.498157     1.152937
H         -12.611059    -1.070319     1.129708
H         -13.392767    -2.234664     1.762625
O          -7.422954     2.197725    -0.103554
H          -7.555241     2.263681     0.846107
H          -6.505638     1.973138    -0.203986
O          -7.633101     2.150031     2.721242
H          -8.410996     1.691130     3.030900
H          -7.483601     2.881792     3.315363
O          -9.179938     0.349107    -1.275921
H          -8.589391     1.014686    -0.904213
H          -8.610112    -0.340096    -1.612837
O         -13.000534    -1.922754    -1.654893
H         -13.314173    -1.992742    -0.752983
H         -13.713485    -2.169468    -2.235174
O         -13.813492     1.512328     0.356300
H         -14.241974     0.684947     0.545414
H         -12.889308     1.336384     0.542820
O         -14.679677     5.246409     4.124678
H         -14.013136     5.837092     3.789073
H         -14.614470     5.349385     5.069876
O         -14.702604     2.662671     2.986968
H         -14.903288     3.540962     3.323119
H         -14.603995     2.715563     2.041436
O         -13.336717     0.833135    -2.396289
H         -13.633173     1.331508    -1.635880
H         -13.077609    -0.017755    -2.039831
O         -14.474167    -1.111029    -4.142076
H         -14.414266    -0.261514    -3.708659
H         -15.298045    -1.130810    -4.611127
O         -10.572453    -2.940303    -3.030550
H         -10.892042    -2.579323    -3.853822
H         -11.254298    -2.731306    -2.397683
O         -11.266099     0.240199     0.789947
H         -10.601653     0.163090     0.104503
H         -10.806266     0.507340     1.591006
O         -10.208600     0.991664     3.285826
H         -10.576781     0.309085     3.844813
H         -10.727483     1.781199     3.462469
O         -11.956667     3.113274     3.704819
H         -11.911088     4.063934     3.653883
H         -12.872018     2.889008     3.552708
O         -10.746486     0.954037    -3.524517
H         -11.593341     1.223036    -3.167728
H         -10.181604     0.783524    -2.763460
O         -14.430372    -0.068004     3.680697
H         -14.660205     0.857483     3.580218
H         -14.386823    -0.433381     2.802221
O         -13.081569    -3.333714     3.367965
```

```
H           -13.830636   -3.189182    3.954855
H           -12.406101   -2.767136    3.737728
O           -11.697057   -1.220663   -5.029854
H           -12.648835   -1.299146   -4.990487
H           -11.458123   -0.552442   -4.387631

Case 4:
voltage = -70mv Y266 negative, R300 positive; E183 neutral
N            15.580944   -6.451469    5.391416
H            14.780173   -6.202168    5.939086
C            15.508993   -5.856291    4.054355
C            14.783016   -6.782962    3.060452
O            14.052852   -6.345866    2.186400
C            16.918802   -5.575378    3.520809
H            16.389097   -6.126718    5.884712
H            14.943644   -4.933781    4.050735
H            16.887619   -5.188754    2.509159
H            17.402401   -4.830320    4.142393
H            17.526676   -6.473359    3.529226
N            15.032378   -8.085605    3.186405
H            15.611823   -8.361771    3.949887
C            14.398660   -9.092385    2.333233
H            14.482014   -8.823398    1.292932
H            14.891843  -10.042103    2.492271
C            12.912965   -9.221837    2.671557
O            12.082056   -9.427871    1.826217
N            12.582804   -9.055884    3.966878
H            13.302600   -8.983985    4.648750
C            11.195108   -9.170814    4.421375
H            11.188284   -9.158108    5.503134
H            10.743835  -10.088273    4.076932
C            10.336943   -8.010596    3.900736
O             9.191046   -8.186327    3.583715
N            10.946934   -6.814391    3.855551
C            10.317888   -5.633853    3.318203
C            10.125898   -5.690235    1.799945
O             9.095707   -5.261140    1.338159
C            11.092037   -4.387866    3.777168
O            12.478184   -4.570170    3.768239
H            11.905264   -6.743798    4.107138
H             9.314131   -5.553344    3.702087
H            10.799505   -3.539122    3.169275
H            10.827762   -4.175240    4.802705
H            12.803731   -4.753415    2.895470
N            11.098972   -6.199760    1.024173
C            10.842818   -6.395674   -0.400144
C             9.692691   -7.385274   -0.601544
O             8.944478   -7.280481   -1.545340
C            12.131172   -6.783268   -1.187427
C            12.334410   -8.288589   -1.402440
C            12.155830   -6.059826   -2.535219
```

```
H      11.922742    -6.586169     1.431210
H      10.472399    -5.461424    -0.790919
H      12.963052    -6.411050    -0.600020
H      13.288349    -8.450898    -1.893511
H      12.333844    -8.853758    -0.481177
H      11.567700    -8.697764    -2.053167
H      13.050308    -6.323364    -3.088571
H      11.295100    -6.333226    -3.135653
H      12.154727    -4.984474    -2.403241
N       9.570430    -8.340342     0.328312
H      10.314649    -8.474146     0.974145
C       8.593019    -9.421182     0.241228
H       8.932305   -10.216770     0.892269
H       8.535637    -9.806255    -0.765234
C       7.181203    -9.042464     0.638009
O       6.257797    -9.497564     0.003641
N       7.002492    -8.221880     1.679685
C       5.669019    -7.732021     1.946137
C       5.252621    -6.721873     0.882411
O       4.076751    -6.627532     0.602378
C       5.413666    -7.224142     3.379357
C       5.551149    -8.367537     4.389728
C       6.257926    -6.016989     3.792493
H       7.779501    -7.956116     2.244709
H       4.985221    -8.550829     1.784424
H       4.373722    -6.912400     3.370605
H       5.260481    -8.028316     5.378435
H       4.911189    -9.203555     4.126944
H       6.572303    -8.727453     4.447510
H       5.916474    -5.642739     4.751679
H       7.299292    -6.290260     3.902023
H       6.189352    -5.203276     3.078847
N       6.173165    -5.995446     0.240637
C       5.775650    -5.174300    -0.891224
C       5.339778    -6.055846    -2.069924
O       4.316554    -5.802555    -2.661062
C       6.869646    -4.156740    -1.271726
C       7.070893    -3.141320    -0.132717
C       6.498936    -3.444256    -2.577032
C       8.367774    -2.339676    -0.245832
H       7.134029    -6.050096     0.501393
H       4.879069    -4.636632    -0.621597
H       7.798320    -4.700086    -1.420947
H       6.218484    -2.463887    -0.116855
H       7.078334    -3.655467     0.819086
H       7.205586    -2.655636    -2.796593
H       6.497377    -4.118719    -3.423668
H       5.511966    -2.998986    -2.509092
H       8.470914    -1.666203     0.598331
H       9.226825    -3.000371    -0.244817
H       8.406844    -1.740406    -1.148542
N       6.110937    -7.098823    -2.408166
H       7.004041    -7.224504    -1.980349
C       5.766224    -7.936500    -3.536910
```

```
H         6.596769    -8.609256    -3.711161
H         5.607735    -7.348312    -4.428476
C         4.498689    -8.766736    -3.333203
O         3.658210    -8.824187    -4.199480
N         4.392527    -9.424027    -2.171602
H         5.130655    -9.378505    -1.501821
C         3.209146   -10.183380    -1.842411
H         2.968932   -10.893506    -2.619442
H         3.405813   -10.729233    -0.928135
C         1.974189    -9.306856    -1.644805
O         0.886786    -9.700155    -2.000101
N         2.187869    -8.107001    -1.107183
C         1.052634    -7.196951    -1.048103
C         0.601552    -6.713526    -2.428651
O        -0.582310    -6.579102    -2.644204
C         1.251457    -5.970574    -0.162525
O         1.412772    -6.294255     1.178745
H         3.086709    -7.815605    -0.796908
H         0.208799    -7.739346    -0.654199
H         2.078030    -5.371688    -0.529123
H         0.344443    -5.394923    -0.244733
H         2.319777    -6.527902     1.321835
N         1.523733    -6.420954    -3.355538
C         1.118171    -6.006582    -4.686094
C         0.366591    -7.131210    -5.393853
O        -0.616579    -6.878354    -6.058668
C         2.319655    -5.472227    -5.499986
C         2.696594    -4.078811    -4.969418
C         2.025603    -5.435939    -7.004976
C         4.006520    -3.527492    -5.529793
H         2.495346    -6.472438    -3.135669
H         0.385122    -5.217788    -4.597028
H         3.153949    -6.147638    -5.335701
H         1.887508    -3.387959    -5.200378
H         2.775597    -4.119379    -3.890688
H         2.876429    -5.040229    -7.545503
H         1.828531    -6.423019    -7.405382
H         1.167262    -4.809177    -7.225534
H         4.277310    -2.614058    -5.010545
H         4.817030    -4.236127    -5.392936
H         3.939404    -3.295428    -6.586923
N         0.812661    -8.380776    -5.239064
H         1.658723    -8.555543    -4.737799
C         0.043838    -9.489011    -5.752955
H        -0.063094    -9.437045    -6.826236
H         0.563431   -10.404330    -5.499466
C        -1.365267    -9.540096    -5.166931
O        -2.317049    -9.778914    -5.874508
N        -1.472644    -9.318876    -3.858022
C        -2.773861    -9.358770    -3.234318
C        -3.688320    -8.224570    -3.694876
O        -4.883588    -8.440681    -3.757278
C        -2.678486    -9.337286    -1.701918
O        -3.924134    -9.612073    -1.121424
```

```
H         -0.656065    -9.230217    -3.289993
H         -3.288400   -10.263844    -3.524706
H         -1.994466   -10.109348    -1.378045
H         -2.294276    -8.379929    -1.366622
H         -4.614315    -9.215993    -1.638945
N         -3.140832    -7.038300    -3.955706
C         -3.899323    -5.915086    -4.482726
C         -4.451814    -6.249649    -5.867267
O         -5.624687    -6.088408    -6.120172
C         -2.987397    -4.668975    -4.499857
C         -3.418161    -3.519203    -5.398608
C         -4.746416    -3.185602    -5.638959
C         -2.430755    -2.759753    -6.021257
C         -5.071658    -2.147370    -6.500352
C         -2.750066    -1.710393    -6.864747
C         -4.078759    -1.406114    -7.116393
H         -2.168236    -6.911483    -3.766018
H         -4.761149    -5.732637    -3.860641
H         -2.882581    -4.329553    -3.473769
H         -2.005629    -4.982126    -4.824976
H         -5.540598    -3.751901    -5.194369
H         -1.395513    -2.998702    -5.851834
H         -6.110294    -1.946391    -6.699659
H         -1.964466    -1.141973    -7.329659
H         -4.330924    -0.608516    -7.793667
N         -3.582681    -6.715106    -6.770418
C         -4.048127    -7.046157    -8.097915
C         -5.186090    -8.063944    -8.050111
O         -6.114137    -7.965980    -8.809499
C         -2.916689    -7.590810    -8.972520
S         -1.688348    -6.353753    -9.500059
H         -2.632397    -6.879591    -6.510637
H         -4.472223    -6.170046    -8.565992
H         -2.417255    -8.416480    -8.483860
H         -3.349076    -7.962399    -9.891961
H         -1.067496    -6.199102    -8.340924
N         -5.078560    -9.066910    -7.155609
H         -4.238851    -9.204214    -6.634586
C         -6.086025   -10.114377    -7.182687
H         -6.261083   -10.452751    -8.192667
H         -5.714273   -10.946867    -6.596749
C         -7.459139    -9.722627    -6.613157
O         -8.480633   -10.092015    -7.125244
N         -7.417940    -9.016150    -5.478415
C         -8.636115    -8.653971    -4.781504
C         -9.495657    -7.633536    -5.521422
O        -10.688492    -7.607651    -5.361495
C         -8.345059    -8.144470    -3.358873
C         -7.426847    -6.915813    -3.231955
C         -8.012652    -5.535819    -3.104155
O         -9.305154    -5.428966    -3.235643
H         -9.566081    -4.511826    -3.102773
O         -7.307092    -4.582490    -2.878392
H         -6.532786    -8.774357    -5.086263
```

```
H          -9.265048    -9.529213    -4.698554
H          -9.298086    -7.962025    -2.882575
H          -7.870967    -8.961678    -2.826041
H          -6.797836    -7.023546    -2.358218
H          -6.742023    -6.849089    -4.060794
N          -8.837281    -6.733946    -6.273830
C          -9.555208    -5.702634    -7.000846
C         -10.241246    -6.303328    -8.225725
H          -9.886965    -7.288064    -8.536094
O         -11.123214    -5.743558    -8.795675
C          -8.611510    -4.505222    -7.297455
O          -8.597017    -3.622202    -6.192752
C          -8.932695    -3.726903    -8.570353
H          -7.859762    -6.848467    -6.436592
H         -10.361686    -5.346750    -6.373685
H          -7.600579    -4.876156    -7.370234
H          -9.438003    -3.183621    -6.139031
H          -8.252828    -2.882799    -8.630093
H          -8.801458    -4.331156    -9.458851
H          -9.956879    -3.366625    -8.569319
N         -12.952099    -9.022565     3.174142
H         -13.213067    -9.553349     2.368879
C         -11.773563    -8.180068     2.898826
C         -10.790797    -8.116813     4.051837
O          -9.663991    -7.701334     3.858866
C         -12.146833    -6.729545     2.544429
O         -12.732347    -6.198257     3.704599
C         -13.100838    -6.633808     1.358713
H         -13.732180    -8.453398     3.440187
H         -11.218895    -8.607381     2.074186
H         -11.231022    -6.202436     2.314759
H         -12.851647    -5.257655     3.590089
H         -13.263333    -5.590693     1.110282
H         -12.705959    -7.131727     0.480806
H         -14.064686    -7.067140     1.597799
N         -11.197245    -8.533612     5.241880
H         -12.116436    -8.918186     5.272036
C         -10.412669    -8.535205     6.452135
H         -10.331223    -7.540879     6.861664
H         -10.939610    -9.140480     7.180255
C          -8.977467    -9.059643     6.382018
O          -8.146753    -8.509831     7.062118
N          -8.659576   -10.125588     5.624873
H          -9.368442   -10.596078     5.109017
C          -7.270783   -10.559509     5.524535
H          -7.254061   -11.535705     5.056973
H          -6.833680   -10.645053     6.506363
C          -6.387435    -9.608346     4.692696
O          -5.221533    -9.463486     4.952964
N          -7.045410    -9.018041     3.701147
C          -6.310791    -7.982307     2.986439
C          -6.265892    -6.696323     3.808100
O          -5.273688    -6.006844     3.776808
C          -6.639029    -7.804216     1.490455
```

```
C          -7.802289   -6.937044    1.054349
C          -7.783366   -5.559917    1.256384
C          -8.849303   -7.481404    0.321962
C          -8.793900   -4.754281    0.759851
C          -9.849025   -6.675063   -0.197929
C          -9.829848   -5.309104    0.027269
H          -8.039932   -9.013245    3.684374
H          -5.287260   -8.313220    2.998176
H          -5.738656   -7.393325    1.047002
H          -6.744883   -8.800772    1.078490
H          -6.960450   -5.105017    1.776080
H          -8.866993   -8.539436    0.124331
H          -8.754232   -3.691639    0.921090
H         -10.626298   -7.108815   -0.799445
H         -10.600457   -4.682964   -0.384771
N          -7.294316   -6.408986    4.609639
C          -7.227561   -5.302045    5.534434
C          -6.096903   -5.516946    6.539673
O          -5.444831   -4.565812    6.913838
C          -8.600103   -5.148622    6.222600
C          -8.841720   -3.835506    6.935796
C          -9.097852   -2.677055    6.207847
C          -8.880551   -3.771106    8.322263
C          -9.376988   -1.483019    6.849914
C          -9.169184   -2.579369    8.971176
C          -9.417835   -1.431224    8.237381
H          -8.162423   -6.882997    4.485443
H          -6.971403   -4.390630    5.012884
H          -9.352249   -5.262344    5.448756
H          -8.730460   -5.968286    6.915933
H          -9.085555   -2.706831    5.131767
H          -8.689596   -4.656063    8.903829
H          -9.561263   -0.597156    6.269776
H          -9.190687   -2.548983   10.045647
H          -9.622741   -0.501325    8.737904
N          -5.855191   -6.753756    6.970338
H          -6.508096   -7.477079    6.765111
C          -4.773232   -7.044576    7.884499
H          -4.809152   -6.412003    8.758724
H          -4.881918   -8.073724    8.202392
C          -3.380280   -6.867559    7.282620
O          -2.505144   -6.368755    7.949534
N          -3.183362   -7.298853    6.031826
C          -1.904254   -7.129798    5.359139
C          -1.649722   -5.662443    4.986657
O          -0.548371   -5.181647    5.149540
C          -1.826576   -8.075697    4.140312
C          -0.660442   -7.730824    3.212267
C          -1.730441   -9.532267    4.608212
H          -3.919237   -7.777663    5.558602
H          -1.112616   -7.373855    6.052609
H          -2.747525   -7.954226    3.574625
H          -0.600446   -8.463360    2.414463
H          -0.765634   -6.759153    2.750581
```

```
H         0.284622   -7.743232    3.746384
H        -1.747066  -10.200284    3.753147
H        -0.796988   -9.698143    5.140118
H        -2.547872   -9.813172    5.259657
N        -2.681891   -4.974529    4.488449
C        -2.544729   -3.598248    4.052412
C        -2.409400   -2.667629    5.255735
O        -1.517012   -1.850248    5.300323
C        -3.719613   -3.148952    3.171265
C        -3.946461   -3.930458    1.866772
C        -2.815028   -3.839906    0.855659
O        -1.692692   -4.192297    1.246563
O        -3.087204   -3.435732   -0.288369
H        -3.550456   -5.439000    4.341003
H        -1.630408   -3.497602    3.498921
H        -4.635599   -3.195680    3.752390
H        -3.554492   -2.100768    2.937811
H        -4.095971   -4.976696    2.089065
H        -4.855834   -3.559717    1.407102
N        -3.309499   -2.786699    6.250263
C        -3.093962   -2.024874    7.456968
C        -1.781427   -2.422854    8.139373
O        -1.112920   -1.584727    8.704382
C        -4.230130   -1.984010    8.489110
O        -4.436969   -3.210485    9.125309
C        -5.512456   -1.384185    7.924365
H        -3.985337   -3.517943    6.232720
H        -2.930707   -0.999966    7.162568
H        -3.859592   -1.324027    9.263496
H        -4.880343   -3.802171    8.529640
H        -6.246415   -1.303123    8.716555
H        -5.327951   -0.391537    7.525320
H        -5.931295   -1.996477    7.136884
N        -1.380404   -3.700488    8.043333
H        -2.031405   -4.418664    7.815121
C        -0.187103   -4.107119    8.740481
H        -0.237745   -3.891376    9.798432
H        -0.075398   -5.174668    8.604493
C         1.064398   -3.419527    8.218984
O         1.940509   -3.075711    8.979614
N         1.154868   -3.258584    6.883072
C         2.333323   -2.661209    6.317654
C         2.291863   -1.139270    6.162009
O         3.361073   -0.575733    6.019687
C         2.833850   -3.351996    5.048951
S         1.917574   -3.010657    3.521828
H         0.424128   -3.613200    6.298452
H         3.116994   -2.796536    7.047750
H         3.839399   -3.004126    4.860731
H         2.869582   -4.419143    5.218859
H         0.871393   -3.797789    3.723145
N         1.141585   -0.471381    6.273705
C         1.204073    0.968482    6.491382
C         1.754778    1.271816    7.870572
```

```
O          2.367382     2.289418     8.086251
C         -0.097125     1.738345     6.177633
C         -1.199708     1.569150     7.233116
C         -0.554722     1.428947     4.747046
C         -2.468607     2.370110     6.932475
H          0.273305    -0.966275     6.289291
H          1.951674     1.351075     5.815093
H          0.205235     2.779403     6.212995
H         -1.448695     0.524663     7.354476
H         -0.812453     1.897582     8.193255
H         -1.310148     2.133330     4.421954
H          0.278296     1.505066     4.053675
H         -0.965854     0.432503     4.657950
H         -3.141183     2.341901     7.784491
H         -2.243043     3.412937     6.724540
H         -3.006175     1.971272     6.079117
N          1.549774     0.323287     8.791241
H          0.889446    -0.401834     8.613645
C          1.977963     0.512992    10.159942
H          1.469328    -0.221878    10.771109
H          1.737815     1.502921    10.517661
C          3.481890     0.337003    10.305708
O          4.163989     1.127275    10.898753
N          3.999613    -0.753731     9.695240
H          3.390273    -1.464888     9.349179
C          5.424647    -0.938945     9.667146
H          5.637679    -1.909414     9.233983
H          5.855301    -0.905126    10.657241
C          6.156885     0.119116     8.844830
O          7.291893     0.411921     9.129599
N          5.502087     0.621274     7.799020
C          6.098611     1.759282     7.146322
C          5.516989     3.077428     7.650625
H          5.008623     3.037109     8.612516
O          5.647525     4.096724     7.045855
C          6.000490     1.675605     5.620851
C          6.752868     0.508291     5.015840
C          8.118857     0.324860     5.234851
C          6.092155    -0.399758     4.193756
C          8.805238    -0.719444     4.630036
C          6.776271    -1.441664     3.587522
C          8.136741    -1.601990     3.795788
H          4.575437     0.334273     7.572148
H          7.134546     1.773642     7.449604
H          4.957923     1.610670     5.335968
H          6.377511     2.610542     5.220655
H          8.647345     0.980755     5.905308
H          5.031812    -0.304257     4.056810
H          9.856335    -0.846128     4.824330
H          6.243756    -2.132916     2.958885
H          8.657125    -2.413641     3.321104
N         12.613550     9.596234    -1.248801
H         11.871931     9.914842    -1.844065
C         12.304852     8.264740    -0.719487
```

```
C        11.274074    8.450488    0.386932
O        10.266460    7.778079    0.416885
C        13.562635    7.581254   -0.164773
C        14.645809    7.276762   -1.205210
S        14.019107    6.202959   -2.529293
C        15.503560    6.052522   -3.551813
H        13.431761    9.572127   -1.824612
H        11.834595    7.616402   -1.447338
H        14.000549    8.205693    0.608096
H        13.255556    6.653940    0.305079
H        15.046417    8.189438   -1.631929
H        15.470590    6.784241   -0.702674
H        15.253538    5.397039   -4.372129
H        15.803961    7.015967   -3.943018
H        16.320662    5.615815   -2.992518
N        11.503854    9.413387    1.296494
C        10.594376    9.608173    2.408605
C         9.277505   10.276976    2.011975
O         8.282490   10.058380    2.656594
C        11.277548   10.367649    3.552607
C        12.448646    9.561571    4.084067
O        13.583823    9.823178    3.744914
N        12.144241    8.558465    4.917144
H        12.327517    9.968356    1.205447
H        10.294804    8.638030    2.770465
H        11.659335   11.324850    3.220226
H        10.547717   10.536155    4.334353
H        12.887664    7.961865    5.209970
H        11.222682    8.171961    4.951082
N         9.263954   11.068441    0.922631
H        10.110243   11.226365    0.425284
C         8.003503   11.517472    0.368854
H         8.201018   12.260056   -0.393596
H         7.389140   11.966802    1.131876
C         7.186384   10.384437   -0.254318
O         5.986775   10.472231   -0.295113
N         7.862978    9.332464   -0.761465
C         7.170138    8.144936   -1.231517
C         6.633636    7.301781   -0.064246
O         5.533035    6.810543   -0.144947
C         8.047996    7.341936   -2.220961
C         8.344874    8.194047   -3.469685
C         7.376586    6.017832   -2.604920
C         9.428444    7.619508   -4.385483
H         8.823031    9.228200   -0.530946
H         6.279797    8.471393   -1.746362
H         8.986449    7.114217   -1.723515
H         7.423341    8.321927   -4.033247
H         8.649049    9.189073   -3.162211
H         7.994109    5.475714   -3.309702
H         7.222745    5.377564   -1.744888
H         6.407909    6.182798   -3.065074
H         9.624223    8.299583   -5.208170
H        10.363320    7.471946   -3.851578
```

```
H         9.146586    6.666005   -4.816531
N         7.386090    7.177064    1.047533
C         6.878058    6.458044    2.204615
C         5.525417    7.035242    2.642676
O         4.575698    6.325710    2.869639
C         7.831823    6.546133    3.406336
C         9.253242    5.997872    3.245998
O        10.145552    6.633626    3.800850
O         9.426538    4.934993    2.599048
H         8.346713    7.444503    1.025500
H         6.683575    5.429172    1.944750
H         7.919361    7.570733    3.736175
H         7.374017    5.992117    4.223285
N         5.483175    8.373301    2.790396
H         6.312418    8.911141    2.653236
C         4.290495    9.037052    3.256214
H         4.528240   10.079997    3.424687
H         3.946128    8.620644    4.189680
C         3.130157    8.987765    2.271344
O         1.988696    9.117162    2.649945
N         3.466368    8.825158    0.988734
H         4.428136    8.815095    0.742262
C         2.517188    8.905263   -0.078309
H         3.062191    9.089692   -0.995140
H         1.843547    9.738082    0.069355
C         1.609135    7.687017   -0.310189
O         0.576528    7.817136   -0.907598
N         2.067321    6.540452    0.227893
C         1.247478    5.370795    0.435183
C         0.429320    5.471011    1.727210
O        -0.764889    5.294461    1.701385
C         2.105543    4.106074    0.428589
H         2.967725    6.569127    0.657195
H         0.519337    5.320351   -0.356785
H         1.488821    3.234769    0.619684
H         2.577659    3.996865   -0.540870
H         2.882861    4.150965    1.184548
N         1.068144    5.733430    2.884289
C         0.338240    5.567665    4.132151
C        -0.773547    6.602325    4.288484
O        -1.809091    6.306242    4.844108
C         1.259390    5.493095    5.375049
C         1.849954    6.839293    5.821976
C         2.348874    4.431293    5.161096
C         2.694150    6.741593    7.095152
H         2.056666    5.861681    2.900722
H        -0.191263    4.627355    4.077572
H         0.606028    5.150646    6.171767
H         2.450286    7.255113    5.019340
H         1.045016    7.544810    6.000464
H         2.725099    4.075955    6.109256
H         1.956046    3.575736    4.621135
H         3.187566    4.825930    4.597762
H         2.997228    7.732711    7.419839
```

```
H         2.126660     6.291255     7.904203
H         3.590948     6.151261     6.954100
N        -0.550448     7.835331     3.810099
H         0.334165     8.069694     3.410291
C        -1.514985     8.881093     4.023532
H        -1.098684     9.807618     3.648530
H        -1.729449     9.011353     5.074384
C        -2.866971     8.618756     3.365893
O        -3.871298     8.845797     4.000033
N        -2.902686     8.108974     2.120752
C        -4.238908     8.033144     1.526312
C        -4.983016     6.752713     1.867113
O        -6.195326     6.753512     1.841256
C        -4.018977     8.294981     0.029374
C        -2.743956     9.145315     0.042208
C        -1.900516     8.455929     1.102605
H        -4.854125     8.818720     1.936371
H        -3.864510     7.368756    -0.508147
H        -4.874696     8.795099    -0.406124
H        -2.231365     9.201413    -0.900536
H        -2.982282    10.157674     0.359873
H        -1.435726     7.562413     0.726469
H        -1.127588     9.090184     1.504427
N        -4.262083     5.675638     2.205270
C        -4.913756     4.487561     2.728202
C        -5.629669     4.783442     4.049083
O        -6.704184     4.276882     4.284658
C        -3.920509     3.321935     2.930975
C        -3.593074     2.501948     1.699298
C        -4.103263     1.217462     1.542430
C        -2.725310     2.957968     0.709553
C        -3.773125     0.420881     0.456677
C        -2.385976     2.179111    -0.376871
C        -2.891591     0.873949    -0.548713
O        -2.575077     0.154040    -1.578985
H        -3.268415     5.756907     2.260692
H        -5.692121     4.185194     2.045196
H        -3.015997     3.722001     3.380431
H        -4.359600     2.660499     3.669923
H        -4.769433     0.819062     2.291192
H        -2.286245     3.931681     0.795370
H        -4.149608    -0.585843     0.393077
H        -1.708846     2.562527    -1.121988
N        -5.003831     5.588445     4.927061
H        -4.072496     5.898401     4.741623
C        -5.608129     5.941694     6.191342
H        -4.823434     6.281171     6.856408
H        -6.091579     5.085964     6.633940
C        -6.660989     7.045619     6.082282
O        -7.715594     6.968719     6.671221
N        -6.342145     8.091259     5.313948
H        -5.447584     8.140444     4.874543
C        -7.257648     9.220183     5.085240
H        -6.772996     9.882451     4.379837
```

```
H          -7.440908     9.763387     6.000489
C          -8.621434     8.828988     4.542963
O          -9.606154     9.421356     4.911824
N          -8.673304     7.827129     3.645035
C          -9.943657     7.435616     3.076809
C         -10.836112     6.654154     4.038280
O         -11.951366     6.350768     3.656646
C          -9.775355     6.671601     1.743354
O         -10.966221     6.755487     0.999702
C          -9.335359     5.222831     1.911865
H          -7.828428     7.411577     3.308851
H         -10.508949     8.330803     2.856257
H          -9.027794     7.206342     1.172417
H         -11.678807     6.479207     1.568244
H          -9.141346     4.784207     0.940478
H          -8.432174     5.142045     2.500323
H         -10.115390     4.638921     2.389369
N         -10.364858     6.344445     5.246952
C         -11.209900     5.754972     6.261796
C         -12.250233     6.739905     6.800707
O         -13.299761     6.324066     7.234959
C         -10.393766     5.201063     7.438215
C          -9.459862     4.028309     7.101350
C          -8.608341     3.695033     8.330318
C         -10.218984     2.785840     6.625799
H          -9.431732     6.593809     5.498784
H         -11.788480     4.964027     5.813640
H          -9.808421     6.007047     7.865124
H         -11.106374     4.885791     8.194963
H          -8.787056     4.333148     6.307967
H          -7.910733     2.890830     8.114203
H          -8.035275     4.558331     8.650941
H          -9.230083     3.379019     9.165326
H          -9.522406     1.974178     6.436950
H         -10.930118     2.450629     7.378911
H         -10.765335     2.958213     5.705507
N         -11.913914     8.040303     6.829253
C         -12.864986     9.029249     7.245767
C         -13.878678     9.396399     6.182266
H         -14.628944    10.123921     6.511973
O         -13.888666     8.967770     5.077159
H         -11.120801     8.352755     6.310483
H         -13.401643     8.681093     8.119589
H         -12.337839     9.934594     7.531090
N         -11.495258    10.629858    -3.034777
H         -11.248077    10.848333    -2.090127
C         -11.816457     9.206991    -3.172691
C         -11.414026     8.858194    -4.598774
O         -10.375345     8.293058    -4.845137
C         -13.245719     8.788261    -2.767124
C         -13.495134     9.129862    -1.286499
C         -13.470822     7.296728    -3.047345
C         -14.921044     8.858600    -0.803354
H         -12.278577    11.208827    -3.272327
```

```
H            -11.118046     8.661113    -2.553749
H            -13.958980     9.364155    -3.361464
H            -12.791593     8.573797    -0.671323
H            -13.282903    10.182143    -1.125066
H            -14.499942     7.015742    -2.863719
H            -13.248956     7.033394    -4.077557
H            -12.840088     6.689705    -2.405270
H            -15.044255     9.216650     0.212390
H            -15.653383     9.367357    -1.424742
H            -15.159411     7.801297    -0.802281
N            -12.214010     9.288635    -5.604063
H            -13.131104     9.601756    -5.380554
C            -11.972873     8.843878    -6.959262
H            -12.035595     7.767552    -7.044834
H            -12.729370     9.282577    -7.598478
C            -10.601634     9.233249    -7.517685
O             -9.997336     8.495992    -8.242649
N            -10.162498    10.476538    -7.182377
H            -10.709239    11.005111    -6.541746
C             -8.894338    11.023969    -7.637102
H             -8.790001    10.891226    -8.702742
H             -8.880302    12.081585    -7.408178
C             -7.672571    10.358998    -6.988330
O             -6.759069     9.923372    -7.632194
N             -7.657768    10.354083    -5.620722
H             -8.475389    10.646641    -5.135096
C             -6.495249     9.961279    -4.832307
H             -5.598013    10.374782    -5.264527
H             -6.616236    10.361980    -3.833380
C             -6.268288     8.463922    -4.714299
O             -5.135168     8.046485    -4.751247
N             -7.314744     7.639629    -4.521267
C             -7.020319     6.271198    -4.124299
C             -6.465820     5.440474    -5.275172
O             -5.668775     4.563531    -5.043674
C             -8.156724     5.541156    -3.368995
C             -9.345704     5.122690    -4.249903
C             -8.567333     6.358545    -2.138727
C            -10.386214     4.279190    -3.509429
H             -8.251834     7.988603    -4.514215
H             -6.188423     6.314123    -3.437886
H             -7.685830     4.629992    -3.010188
H             -9.825096     5.993879    -4.678492
H             -8.968657     4.536094    -5.084857
H             -9.222673     5.791693    -1.491862
H             -7.697748     6.634611    -1.551106
H             -9.089360     7.266113    -2.418740
H            -11.138813     3.914238    -4.202010
H             -9.930522     3.416713    -3.033396
H            -10.898388     4.848340    -2.743523
N             -6.882017     5.679794    -6.537956
C             -6.242877     4.969527    -7.628380
C             -4.821659     5.460691    -7.917387
O             -3.990473     4.676202    -8.307804
```

```
C         -7.108198    4.971753   -8.897586
C         -8.277947    3.978321   -8.838637
C         -7.800139    2.523672   -8.902302
N         -8.881613    1.534874   -8.891460
C         -9.434749    0.982245   -7.816715
N         -9.166716    1.438150   -6.605629
N        -10.274741   -0.031477   -7.966914
H         -7.493953    6.446051   -6.719521
H         -6.080804    3.958127   -7.296394
H         -7.494577    5.969805   -9.079059
H         -6.473446    4.727161   -9.743172
H         -8.868341    4.149107   -7.943185
H         -8.939450    4.160827   -9.678931
H         -7.254451    2.358428   -9.821294
H         -7.115529    2.281763   -8.100881
H         -9.153769    1.158705   -9.772140
H         -8.640528    2.272789   -6.487212
H         -9.639822    1.093020   -5.784688
H        -10.333485   -0.492681   -8.847730
H        -10.589868   -0.568969   -7.173564
N         -4.571419    6.757537   -7.721986
H         -5.302822    7.370440   -7.441701
C         -3.257792    7.352338   -8.011302
H         -2.929120    7.090538   -9.004951
H         -3.360599    8.426642   -7.939669
C         -2.202314    6.874116   -7.024551
O         -1.073683    6.639153   -7.377141
N         -2.634097    6.758718   -5.763275
C         -1.761616    6.448145   -4.656398
C         -1.491280    4.960888   -4.524187
O         -0.363855    4.577069   -4.319321
C         -2.350380    7.048033   -3.373471
C         -1.763409    6.354104   -2.145479
C         -2.113584    8.563997   -3.410440
H         -3.557786    7.064290   -5.547223
H         -0.788704    6.881447   -4.838418
H         -3.418831    6.866648   -3.387816
H         -1.980538    6.901562   -1.247451
H         -2.173217    5.356047   -2.033377
H         -0.690739    6.283819   -2.215385
H         -2.838959    9.089537   -2.805586
H         -1.118867    8.805764   -3.051243
H         -2.215567    8.949902   -4.418163
N         -2.517791    4.099989   -4.645961
C         -2.183943    2.693865   -4.750294
C         -1.220401    2.507869   -5.924342
O         -0.293656    1.732576   -5.842545
C         -3.407935    1.776689   -4.871585
C         -4.365142    1.887237   -3.676898
C         -5.406639    0.763185   -3.611684
N         -4.878991   -0.427510   -2.949865
C         -5.583335   -1.519506   -2.748470
N         -6.812937   -1.655781   -3.294816
N         -5.154037   -2.493880   -1.972978
```

```
H    -3.439509    4.421220   -4.850424
H    -1.629168    2.394292   -3.874460
H    -3.944310    1.990555   -5.794151
H    -3.028749    0.764647   -4.951019
H    -3.802460    1.900235   -2.756953
H    -4.904655    2.820111   -3.734796
H    -6.276650    1.112905   -3.065381
H    -5.727669    0.511016   -4.619263
H    -3.984166   -0.324710   -2.456363
H    -6.974347   -1.177487   -4.151327
H    -7.208358   -2.578475   -3.266250
H    -4.300366   -2.472857   -1.440328
H    -5.664429   -3.351690   -1.949454
N    -1.459790    3.193298   -7.035093
H    -2.285219    3.739318   -7.148812
C    -0.596550    2.997474   -8.183376
H    -0.949630    3.640655   -8.979121
H    -0.611029    1.972250   -8.524591
C     0.855931    3.344672   -7.871776
O     1.781598    2.711140   -8.313085
N     1.005371    4.405090   -7.063925
H     0.212146    4.964338   -6.846605
C     2.282350    4.880661   -6.640316
H     2.919001    5.137295   -7.477590
H     2.128458    5.777142   -6.053742
C     3.092840    3.928367   -5.775178
O     4.297669    4.059457   -5.688187
N     2.428015    2.983364   -5.101894
C     3.096572    2.136985   -4.148066
C     4.270175    1.327688   -4.711094
O     5.184059    1.025872   -3.970246
C     2.008341    1.259942   -3.526779
C     2.372020    0.441550   -2.293976
C     1.055758    0.031691   -1.642286
N     1.235378   -1.004799   -0.647430
C     0.298798   -1.926790   -0.373999
N    -0.855083   -1.920612   -0.991963
N     0.588351   -2.879746    0.489881
H     1.438713    2.900979   -5.211350
H     3.557525    2.746655   -3.380324
H     1.210708    1.943596   -3.257011
H     1.592591    0.612085   -4.289465
H     2.946770   -0.439481   -2.558613
H     2.977062    1.024247   -1.603012
H     0.561443    0.897278   -1.214220
H     0.398150   -0.361469   -2.402202
H     1.939830   -0.874312    0.043445
H    -1.298717   -1.062577   -1.309484
H    -1.527138   -2.646256   -0.776918
H     1.506800   -2.990418    0.850720
H    -0.161269   -3.479286    0.821989
N     4.280002    0.941732   -5.998662
H     3.540833    1.225199   -6.608051
C     5.363424    0.116638   -6.504622
```

```
H         5.137151   -0.137697   -7.532686
H         5.442866   -0.800473   -5.942810
C         6.736121    0.755923   -6.476823
O         7.738867    0.089514   -6.368823
N         6.766124    2.085339   -6.608293
H         5.911903    2.604351   -6.566390
C         8.004993    2.823920   -6.549089
H         8.710109    2.440732   -7.273087
H         7.799709    3.854876   -6.804647
C         8.746900    2.826001   -5.210612
O         9.887685    3.236397   -5.157394
N         8.101358    2.335228   -4.144928
C         8.680903    2.169883   -2.822859
C         9.947179    1.302961   -2.830367
O        10.764503    1.385616   -1.927871
C         7.576267    1.499495   -1.975518
C         7.769338    1.380008   -0.464046
C         7.422858    2.651184    0.320627
C         7.309505    2.388446    1.818384
N         8.648575    2.358116    2.478467
O        11.029331    1.615187    0.898018
H        10.881146    1.825071   -0.024448
H        11.506317    0.787662    0.849543
H         7.155638    2.032457   -4.259047
H         8.977804    3.127170   -2.415175
H         6.658632    2.040632   -2.163162
H         7.411904    0.511411   -2.389546
H         7.090270    0.600060   -0.133568
H         8.764776    1.029803   -0.225093
H         8.134561    3.453008    0.144581
H         6.457656    3.019727   -0.011584
H         6.747001    3.161349    2.318600
H         6.829051    1.442376    2.014474
H         9.360209    1.892163    1.924320
H         8.969585    3.328263    2.631120
H         8.593559    1.882790    3.367436
N        10.062468    0.448216   -3.837538
C        11.109355   -0.537517   -4.069387
C        12.486379    0.023157   -4.427133
O        13.444198   -0.705546   -4.301401
C        10.701431   -1.443076   -5.253560
C         9.563249   -2.441793   -5.000147
C         9.132350   -3.064984   -6.331588
C         9.990084   -3.533487   -4.018744
H         9.324190    0.445767   -4.510612
H        11.264635   -1.125579   -3.176687
H        10.435176   -0.796633   -6.084514
H        11.584405   -1.995209   -5.552449
H         8.706564   -1.911333   -4.596274
H         8.331207   -3.781573   -6.178100
H         8.775123   -2.307430   -7.017426
H         9.959751   -3.592416   -6.799865
H         9.199967   -4.263282   -3.879665
H        10.860108   -4.059658   -4.397726
```

```
H       10.242959      -3.137666      -3.039915
N       12.574050       1.261829      -4.919242
C       13.814406       1.845262      -5.393366
C       14.958878       2.033284      -4.379087
O       16.013534       2.413871      -4.812175
C       13.496569       3.172498      -6.085440
O       12.918096       4.089114      -5.200443
H       11.756268       1.833402      -4.916258
H       14.248272       1.179028      -6.129607
H       14.419539       3.591691      -6.454352
H       12.845354       2.979136      -6.931993
H       11.970528       4.078373      -5.262637
N       14.792294       1.722326      -3.074279
C       15.927713       1.647124      -2.176233
C       16.804430       0.410056      -2.402579
O       17.934558       0.394323      -1.988747
C       15.493359       1.575964      -0.707144
C       14.634954       2.731677      -0.197567
C       14.278275       2.482164       1.268234
N       13.252919       3.376635       1.790435
C       13.430800       4.528269       2.417003
N       14.648910       5.003840       2.652460
O       15.028274       7.461559       4.203992
H       14.743853       8.348419       3.967260
H       15.746592       7.566788       4.814797
N       12.381272       5.217168       2.813488
H       13.900680       1.420521      -2.758456
H       16.561246       2.504414      -2.344561
H       14.953644       0.647366      -0.552984
H       16.404775       1.507146      -0.123410
H       15.152401       3.679935      -0.311134
H       13.716902       2.813455      -0.767682
H       13.880352       1.483080       1.371699
H       15.156521       2.522513       1.901889
H       12.312653       3.080244       1.620849
H       15.456406       4.510167       2.355170
H       14.784908       5.859087       3.163319
H       11.425389       4.937390       2.656475
H       12.473703       6.061002       3.332453
N       16.200238      -0.648980      -2.967242
C       16.870752      -1.911587      -3.167518
C       17.670006      -1.889781      -4.467008
H       18.062386      -0.909889      -4.748118
O       17.878402      -2.854052      -5.128230
C       15.867932      -3.070228      -3.148719
C       15.323739      -3.398988      -1.791343
N       14.402798      -2.625854      -1.112620
O       12.505039      -0.433470      -0.461488
H       13.142221      -1.035718      -0.849492
H       12.022768      -0.037399      -1.179250
C       15.618082      -4.489097      -1.048180
C       14.170786      -3.249195      -0.003865
N       14.880727      -4.379499       0.104143
H       15.285196      -0.550771      -3.351807
```

```
H         17.595790    -2.030750    -2.370018
H         15.050801    -2.842595    -3.824522
H         16.368880    -3.944319    -3.542447
H         16.273237    -5.313268    -1.232733
H         13.492091    -2.918983     0.756432
H         14.779374    -5.062039     0.823108
O        -15.022332    -2.472904     4.971587
H        -15.215300    -1.663969     4.504968
H        -14.530546    -2.214795     5.744220
O        -12.909024    -1.269273     6.930563
H        -12.222363    -1.721546     7.408263
H        -13.143100    -0.515308     7.458159
O        -12.057046    -0.850324     4.159021
H        -12.178973    -0.901868     5.104213
H        -12.869358    -0.460195     3.833426
O        -13.394460    -1.861170     0.708884
H        -12.619910    -1.296729     0.765568
H        -13.316722    -2.508005     1.414714
O         -7.286814     1.298494    -0.241241
H         -7.499537     1.634533     0.633103
H         -6.357823     1.097608    -0.204946
O         -7.887148     2.075832     2.453266
H         -8.798915     1.921491     2.688293
H         -7.560872     2.752423     3.040734
O         -9.343424    -0.025500    -1.609512
H         -8.581061     0.392263    -1.188839
H         -9.094469    -0.923898    -1.806345
O        -12.844866    -2.342602    -2.020416
H        -13.121070    -2.378383    -1.101268
H        -13.457236    -2.853965    -2.536527
O        -14.033873     0.923434    -0.583660
H        -14.378102     0.061001    -0.376872
H        -13.123337     0.882180    -0.275039
O        -14.251532     4.927624     4.735036
H        -13.682965     5.569178     4.315612
H        -14.290144     5.212331     5.643516
O        -15.231526     2.152017     1.699077
H        -15.858627     2.863324     1.685254
H        -14.946794     1.987269     0.801499
O        -13.156822     0.240589    -3.245781
H        -13.606222     0.770980    -2.590586
H        -12.938664    -0.563428    -2.772435
O        -13.770736    -1.932656    -4.984021
H        -13.939482    -1.069812    -4.612480
H        -14.497007    -2.146080    -5.555548
O        -10.158816    -2.763951    -2.954684
H        -10.300592    -2.610371    -3.882439
H        -11.021513    -2.687230    -2.548996
O        -11.486467     0.211533     0.369975
H        -10.759889     0.130538    -0.249436
H        -11.146307     0.652745     1.155701
O        -10.714271     1.340138     2.829264
H        -10.880652     0.580670     3.384821
H        -11.395603     1.966258     3.090879
```

```
O          -13.057024    2.613282    3.631853
H          -13.358808    3.352220    4.162904
H          -13.713452    2.533207    2.947629
O          -10.472382    0.624522   -4.111174
H          -11.377476    0.788837   -3.840270
H           -9.985111    0.491478   -3.290428
O          -14.689701   -0.343782    3.041473
H          -15.011369    0.525584    2.812373
H          -14.564724   -0.802618    2.217301
O          -13.086346   -3.361956    3.148089
H          -13.858824   -3.221180    3.705238
H          -12.460173   -2.723946    3.489388
O          -11.006560   -1.654552   -5.649329
H          -11.936215   -1.878970   -5.566085
H          -10.849057   -0.947846   -5.019580

Case 5:
voltage -35 mV; Y266 negative' R300 positive; E183 neutral
N           15.580944   -6.451469    5.391416
H           14.781091   -6.200103    5.939456
C           15.509751   -5.856564    4.054164
C           14.783016   -6.782962    3.060452
O           14.052231   -6.345849    2.186818
C           16.919877   -5.577004    3.520677
H           16.390237   -6.128756    5.884185
H           14.945257   -4.933528    4.050277
H           16.889205   -5.190740    2.508870
H           17.403974   -4.832107    4.142045
H           17.527065   -6.475436    3.529506
N           15.032378   -8.085605    3.186405
H           15.612411   -8.361775    3.949450
C           14.398660   -9.092385    2.333233
H           14.482160   -8.823545    1.292899
H           14.891694  -10.042153    2.492416
C           12.912930   -9.221560    2.671507
O           12.081797   -9.427103    1.826194
N           12.582804   -9.055884    3.966878
H           13.302624   -8.984073    4.648759
C           11.195108   -9.170814    4.421375
H           11.188291   -9.158045    5.503134
H           10.743837  -10.088296    4.076981
C           10.336943   -8.010596    3.900736
O            9.190873   -8.186196    3.583978
N           10.946985   -6.814479    3.855537
C           10.318198   -5.633870    3.318106
C           10.125791   -5.690670    1.799932
O            9.095030   -5.262426    1.338387
C           11.092794   -4.387934    3.776743
O           12.478717   -4.571005    3.769569
H           11.905387   -6.744053    4.106970
H            9.314538   -5.552986    3.702167
```

```
H          10.801350    -3.539534     3.167806
H          10.827492    -4.174283     4.801806
H          12.805484    -4.754836     2.897355
N          11.098814    -6.199773     1.024010
C          10.842605    -6.395616    -0.400299
C           9.692264    -7.385027    -0.601336
O           8.943209    -7.279826    -1.544493
C          12.130941    -6.783382    -1.187602
C          12.333878    -8.288693    -1.402909
C          12.155886    -6.059627    -2.535221
H          11.923312    -6.585041     1.430728
H          10.472291    -5.461328    -0.791112
H          12.962862    -6.411440    -0.600078
H          13.287694    -8.451074    -1.894160
H          12.333414    -8.854024    -0.481751
H          11.567004    -8.697623    -2.053617
H          13.050400    -6.323168    -3.088487
H          11.295217    -6.332776    -3.135875
H          12.154970    -4.984307    -2.402985
N           9.570430    -8.340342     0.328312
H          10.315146    -8.474590     0.973538
C           8.593019    -9.421182     0.241228
H           8.932376   -10.216809     0.892183
H           8.535552    -9.806223    -0.765248
C           7.181275    -9.042549     0.638193
O           6.257642    -9.497245     0.003736
N           7.002656    -8.222457     1.680250
C           5.669209    -7.732900     1.947061
C           5.252360    -6.722864     0.883408
O           4.076325    -6.628877     0.603483
C           5.414160    -7.225036     3.380337
C           5.552431    -8.368373     4.390677
C           6.258144    -6.017592     3.793165
H           7.779778    -7.956827     2.245245
H           4.985483    -8.551816     1.785555
H           4.374079    -6.913683     3.371905
H           5.261938    -8.029263     5.379481
H           4.912756    -9.204679     4.128075
H           6.573742    -8.727837     4.448151
H           5.917098    -5.643641     4.752620
H           7.299669    -6.290423     3.902014
H           6.188808    -5.203807     3.079677
N           6.172629    -5.996260     0.241633
C           5.775138    -5.175126    -0.890224
C           5.339353    -6.056734    -2.068895
O           4.315679    -5.803894    -2.659636
C           6.869138    -4.157520    -1.270669
C           7.070399    -3.142261    -0.131528
C           6.498412    -3.444914    -2.575899
C           8.367244    -2.340569    -0.244584
H           7.133553    -6.050773     0.502348
H           4.878541    -4.637464    -0.620622
H           7.797785    -4.700839    -1.419964
H           6.217942    -2.464864    -0.115509
```

```
H         7.077954    -3.656580     0.820178
H         7.205179    -2.656424    -2.795501
H         6.496710    -4.119325    -3.422578
H         5.511513    -2.999455    -2.507823
H         8.470448    -1.667319     0.599745
H         9.226289    -3.001250    -0.243796
H         8.406234    -1.741067    -1.147137
N         6.110770    -7.099307    -2.407461
H         7.004114    -7.224662    -1.979936
C         5.766130    -7.937018    -3.536188
H         6.596660    -8.609801    -3.710364
H         5.607740    -7.348885    -4.427809
C         4.498577    -8.767159    -3.332400
O         3.657706    -8.824310    -4.198409
N         4.392510    -9.424584    -2.170883
H         5.130817    -9.379417    -1.501210
C         3.209017   -10.183686    -1.841694
H         2.968869   -10.894126    -2.618465
H         3.405377   -10.729151    -0.927119
C         1.974189    -9.306856    -1.644805
O         0.886706    -9.699804    -2.000519
N         2.187869    -8.107001    -1.107183
C         1.053184    -7.196290    -1.049133
C         0.602296    -6.713916    -2.430128
O        -0.581645    -6.579853    -2.645997
C         1.252922    -5.968692    -0.165429
O         1.412852    -6.289994     1.176558
H         3.086662    -7.815834    -0.796443
H         0.208948    -7.737541    -0.654474
H         2.080444    -5.371463    -0.532552
H         0.346632    -5.392140    -0.249368
H         2.319532    -6.524414     1.320781
N         1.524545    -6.421357    -3.356796
C         1.119340    -6.007312    -4.687560
C         0.366941    -7.131822    -5.394547
O        -0.616594    -6.878978    -6.058973
C         2.321339    -5.474700    -5.501861
C         2.700149    -4.081492    -4.972089
C         2.027002    -5.438763    -7.006801
C         4.010930    -3.532444    -5.532708
H         2.496147    -6.472601    -3.136624
H         0.386994    -5.217779    -4.599022
H         3.154808    -6.151073    -5.337426
H         1.892034    -3.389647    -5.203569
H         2.779037    -4.121497    -3.893337
H         2.878141    -5.044252    -7.547683
H         1.828756    -6.425797    -7.406757
H         1.169328    -4.811086    -7.227436
H         4.283091    -2.619153    -5.013969
H         4.820371    -4.242192    -5.395364
H         3.944286    -3.300907    -6.589979
N         0.812833    -8.381381    -5.239583
H         1.659106    -8.556154    -4.738586
C         0.043777    -9.489598    -5.753025
```

```
H           -0.063177    -9.438059    -6.826330
H            0.563129   -10.404949    -5.499179
C           -1.365267    -9.540096    -5.166931
O           -2.317300    -9.778481    -5.874449
N           -1.472644    -9.318876    -3.858022
C           -2.773824    -9.358721    -3.234357
C           -3.688263    -8.224473    -3.694805
O           -4.883636    -8.440525    -3.757268
C           -2.678523    -9.337272    -1.701948
O           -3.924501    -9.611371    -1.121610
H           -0.656034    -9.230247    -3.289963
H           -3.288434   -10.263746    -3.524794
H           -1.994932   -10.109663    -1.377982
H           -2.293917    -8.380107    -1.366616
H           -4.614336    -9.215571    -1.639796
N           -3.140792    -7.038199    -3.955302
C           -3.899230    -5.914910    -4.482136
C           -4.452023    -6.249501    -5.866547
O           -5.625075    -6.088448    -6.119169
C           -2.987087    -4.668947    -4.499399
C           -3.417506    -3.519152    -5.398291
C           -4.745639    -3.185576    -5.639365
C           -2.429784    -2.759580    -6.020311
C           -5.070431    -2.147296    -6.500877
C           -2.748658    -1.710165    -6.863894
C           -4.077226    -1.405944    -7.116296
H           -2.168081    -6.911521    -3.765855
H           -4.760930    -5.732366    -3.859883
H           -2.882251    -4.329419    -3.473346
H           -2.005354    -4.982331    -4.824377
H           -5.540022    -3.751971    -5.195256
H           -1.394633    -2.998469    -5.850315
H           -6.108971    -1.946338    -6.700821
H           -1.962807    -1.141650    -7.328248
H           -4.329005    -0.608301    -7.793668
N           -3.583096    -6.714759    -6.769885
C           -4.048698    -7.045628    -8.097358
C           -5.186661    -8.063368    -8.049386
O           -6.115432    -7.964920    -8.807918
C           -2.917288    -7.590043    -8.972181
S           -1.689356    -6.352640    -9.499737
H           -2.632651    -6.879001    -6.510381
H           -4.472881    -6.169452    -8.565247
H           -2.417606    -8.415632    -8.483638
H           -3.349733    -7.961716    -9.891567
H           -1.067005    -6.199593    -8.341190
N           -5.078560    -9.066910    -7.155609
H           -4.238567    -9.204541    -6.635057
C           -6.086052   -10.114294    -7.182808
H           -6.261190   -10.452571    -8.192815
H           -5.714335   -10.946880    -6.596984
C           -7.459139    -9.722627    -6.613157
O           -8.480744   -10.092011    -7.125161
N           -7.417940    -9.016150    -5.478415
```

```
C            -8.636054     -8.653965     -4.781521
C            -9.495623     -7.633624     -5.521555
O           -10.688582     -7.608002     -5.362047
C            -8.344878     -8.144478     -3.358911
C            -7.426713     -6.915793     -3.232024
C            -8.012745     -5.535991     -3.103580
O            -9.305367     -5.429296     -3.235178
H            -9.566295     -4.512283     -3.101628
O            -7.307557     -4.582583     -2.877283
H            -6.532739     -8.774317     -5.086305
H            -9.265015     -9.529190     -4.698509
H            -9.297852     -7.962122     -2.882458
H            -7.870667     -8.961722     -2.826246
H            -6.797312     -7.023712     -2.358611
H            -6.742167     -6.848728     -4.061081
N            -8.837281     -6.733946     -6.273830
C            -9.555208     -5.702634     -7.000846
C           -10.241444     -6.303436     -8.225512
H            -9.886122     -7.287502     -8.536798
O           -11.124482     -5.744364     -8.794575
C            -8.611262     -4.505520     -7.297856
O            -8.596367     -3.622330     -6.193337
C            -8.932608     -3.727460     -8.570875
H            -7.859620     -6.848089     -6.436168
H           -10.361529     -5.346458     -6.373644
H            -7.600419     -4.876651     -7.370771
H            -9.437618     -3.184416     -6.138697
H            -8.252636     -2.883477     -8.630984
H            -8.801590     -4.331958     -9.459248
H            -9.956788     -3.367087     -8.569729
N           -12.952099     -9.022565      3.174142
H           -13.212983     -9.553275      2.368796
C           -11.773563     -8.180068      2.898826
C           -10.790963     -8.116278      4.051923
O            -9.664563     -7.699910      3.859086
C           -12.147103     -6.729778      2.543795
O           -12.733777     -6.198346      3.703356
C           -13.100358     -6.634674      1.357417
H           -13.732123     -8.453221      3.439978
H           -11.218696     -8.607536      2.074398
H           -11.231354     -6.202371      2.314642
H           -12.852378     -5.257660      3.588813
H           -13.263234     -5.591661      1.108782
H           -12.704662     -7.132439      0.479807
H           -14.064139     -7.068482      1.595973
N           -11.197245     -8.533612      5.241880
H           -12.116114     -8.918914      5.271876
C           -10.412641     -8.534803      6.452134
H           -10.330954     -7.540281      6.861132
H           -10.939748     -9.139572      7.180551
C            -8.977528     -9.059639      6.382431
O            -8.147112     -8.510269      7.063166
N            -8.659647    -10.125342      5.624950
H            -9.368430    -10.595275      5.108491
```

```
C         -7.270846    -10.559425     5.524585
H         -7.254236    -11.535653     5.057095
H         -6.833782    -10.644954     6.506426
C         -6.387435     -9.608346     4.692696
O         -5.221622     -9.463504     4.953059
N         -7.045410     -9.018041     3.701147
C         -6.310644     -7.982221     2.986667
C         -6.265755     -6.696391     3.808545
O         -5.273914     -6.006573     3.776925
C         -6.638572     -7.803893     1.490680
C         -7.801944     -6.936852     1.054646
C         -7.783273     -5.559758     1.256976
C         -8.848770     -7.481221     0.321991
C         -8.793848     -4.754182     0.760441
C         -9.848520     -6.674923    -0.197906
C         -9.829590     -5.309009     0.027572
H         -8.039897     -9.013221     3.684354
H         -5.287134     -8.313207     2.998513
H         -5.738222     -7.392705     1.047483
H         -6.744121     -8.800366     1.078440
H         -6.960446     -5.104826     1.776764
H         -8.866169     -8.539197     0.124031
H         -8.754263     -3.691552     0.921788
H        -10.625514     -7.108648    -0.799807
H        -10.600112     -4.682866    -0.384637
N         -7.293899     -6.409573     4.610793
C         -7.226774     -5.302748     5.535658
C         -6.096069     -5.517837     6.540834
O         -5.444773     -4.566561     6.915796
C         -8.599152     -5.148978     6.224076
C         -8.840646     -3.835326     6.936330
C         -9.095369     -2.677111     6.207499
C         -8.880576     -3.770076     8.322719
C         -9.374103     -1.482488     6.848635
C         -9.168904     -2.577761     8.970707
C         -9.416084     -1.429845     8.236054
H         -8.162020     -6.883468     4.486674
H         -6.970431     -4.391381     5.014129
H         -9.351528     -5.263245     5.450526
H         -8.729316     -5.968136     6.918061
H         -9.081971     -2.707479     5.131451
H         -8.690468     -4.654767     8.904957
H         -9.556950     -0.596768     6.267818
H         -9.191023     -2.546691    10.045147
H         -9.620419     -0.499437     8.735873
N         -5.853788     -6.754869     6.970525
H         -6.506301     -7.478288     6.764593
C         -4.771469     -7.046114     7.884182
H         -4.807238     -6.414045     8.758773
H         -4.879903     -8.075481     8.201436
C         -3.378763     -6.868401     7.281895
O         -2.503741     -6.369400     7.948712
N         -3.182169     -7.299127     6.030851
C         -1.903339     -7.129367     5.357699
```

```
C         -1.649485    -5.661756     4.985820
O         -0.548494    -5.180483     5.149104
C         -1.825810    -8.074680     4.138419
C         -0.659949    -7.729133     3.210260
C         -1.729287    -9.531460     4.605591
H         -3.917955    -7.778217     5.557820
H         -1.111378    -7.373486     6.050772
H         -2.746930    -7.953082     3.572991
H         -0.599900    -8.461363     2.412187
H         -0.765505    -6.757335     2.748919
H          0.285210    -7.741500     3.744188
H         -1.745998   -10.199062     3.750208
H         -0.795673    -9.697428     5.137152
H         -2.546467    -9.812858     5.257167
N         -2.682005    -4.974019     4.487882
C         -2.545164    -3.597631     4.052069
C         -2.409867    -2.667118     5.255465
O         -1.518125    -1.849152     5.299780
C         -3.720352    -3.148186     3.171366
C         -3.947925    -3.929451     1.866852
C         -2.816703    -3.839267     0.855428
O         -1.694637    -4.192671     1.245875
O         -3.088981    -3.434340    -0.288347
H         -3.550187    -5.438910     4.339790
H         -1.630955    -3.496690     3.498427
H         -4.636093    -3.194818     3.752918
H         -3.555192    -2.100004     2.937947
H         -4.097726    -4.975661     2.089032
H         -4.857302    -3.558258     1.407512
N         -3.309689    -2.786764     6.250329
C         -3.093962    -2.024874     7.456968
C         -1.781672    -2.422988     8.139766
O         -1.113697    -1.584913     8.705368
C         -4.230241    -1.983044     8.488898
O         -4.437644    -3.208907     9.126153
C         -5.512339    -1.383379     7.923457
H         -3.984636    -3.518795     6.233181
H         -2.930236    -1.000161     7.162253
H         -3.859682    -1.322574     9.262834
H         -4.879909    -3.801439     8.530532
H         -6.246381    -1.301433     8.715494
H         -5.327528    -0.391141     7.523527
H         -5.931168    -1.996298     7.136449
N         -1.380404    -3.700488     8.043333
H         -2.031003    -4.418665     7.814160
C         -0.186948    -4.107261     8.740209
H         -0.237554    -3.891848     9.798220
H         -0.075182    -5.174752     8.603858
C          1.064490    -3.419348     8.218883
O          1.940102    -3.075052     8.979783
N          1.155313    -3.258516     6.882966
C          2.333919    -2.661015     6.317786
C          2.292315    -1.139101     6.162096
O          3.361308    -0.575434     6.019565
```

```
C         2.834703    -3.351539     5.049051
S         1.918436    -3.009871     3.521955
H         0.424945    -3.613498     6.298139
H         3.117469    -2.796350     7.048011
H         3.840189    -3.003438     4.861002
H         2.870646    -4.418722     5.218663
H         0.872619    -3.797578     3.722815
N         1.141841    -0.471357     6.273787
C         1.204073     0.968482     6.491382
C         1.753690     1.272376     7.870910
O         2.364358     2.290991     8.087017
C        -0.097046     1.738100     6.176832
C        -1.200206     1.568193     7.231597
C        -0.553630     1.428961     4.745869
C        -2.469273     2.368698     6.930463
H         0.273681    -0.966378     6.289418
H         1.952058     1.351106     5.815539
H         0.204996     2.779216     6.212705
H        -1.448849     0.523567     7.352494
H        -0.813653     1.896478     8.192054
H        -1.309057     2.133214     4.420446
H         0.279830     1.505532     4.053097
H        -0.964394     0.432418     4.656221
H        -3.142326     2.340022     7.782109
H        -2.243999     3.411664     6.722902
H        -3.006174     1.969830     6.076689
N         1.549774     0.323287     8.791241
H         0.890691    -0.402779     8.613169
C         1.977802     0.512702    10.160033
H         1.469272    -0.222454    10.770953
H         1.737385     1.502491    10.517938
C         3.481795     0.337028    10.305837
O         4.163620     1.127585    10.898746
N         3.999649    -0.753675     9.695463
H         3.390372    -1.464979     9.349663
C         5.424736    -0.938883     9.667283
H         5.637718    -1.909422     9.234270
H         5.855436    -0.904929    10.657348
C         6.156885     0.119116     8.844830
O         7.291754     0.412091     9.129721
N         5.502087     0.621274     7.799020
C         6.098611     1.759282     7.146322
C         5.515755     3.077093     7.649950
H         5.009274     3.036938     8.612868
O         5.642887     4.095919     7.043695
C         6.001799     1.675241     5.620806
C         6.754289     0.507592     5.016621
C         8.119984     0.323554     5.236933
C         6.093947    -0.400178     4.193981
C         8.806411    -0.721137     4.632864
C         6.778104    -1.442485     3.588462
C         8.138271    -1.603448     3.798047
H         4.575710     0.333815     7.571699
H         7.134274     1.774212     7.450535
```

```
H            4.959444     1.610593     5.335055
H            6.379367     2.609976     5.220685
H            8.648202     0.979222     5.907817
H            5.033800    -0.304076     4.055849
H            9.857250    -0.848320     4.828159
H            6.245884    -2.133503     2.959315
H            8.658738    -2.415364     3.323940
N           12.613550     9.596234    -1.248801
H           11.873304     9.913777    -1.846311
C           12.304852     8.264740    -0.719487
C           11.273689     8.450546     0.386591
O           10.265530     7.778767     0.415804
C           13.562549     7.581221    -0.164596
C           14.645707     7.276548    -1.205017
S           14.018810     6.203150    -2.529344
C           15.503354     6.052419    -3.551773
H           13.433529     9.572929    -1.822100
H           11.834820     7.616386    -1.447461
H           14.000496     8.205656     0.608263
H           13.255359     6.653971     0.305291
H           15.046679     8.189118    -1.631593
H           15.470297     6.783724    -0.702487
H           15.253298     5.397008    -4.372129
H           15.803968     7.015827    -3.942864
H           16.320271     5.615506    -2.992399
N           11.503479     9.412865     1.296722
C           10.593734     9.607380     2.408635
C            9.276928    10.276086     2.011659
O            8.281457    10.056921     2.655491
C           11.276507    10.366957     3.552816
C           12.448227     9.561470     4.083903
O           13.582992     9.823427     3.743914
N           12.144690     8.558477     4.917500
H           12.327486     9.967466     1.206418
H           10.294206     8.637194     2.770427
H           11.657728    11.324458     3.220662
H           10.546636    10.534855     4.334660
H           12.888577     7.962509     5.210493
H           11.223226     8.171947     4.952502
N            9.263775    11.068227     0.922800
H           10.110275    11.226553     0.425903
C            8.003503    11.517472     0.368854
H            8.201268    12.259938    -0.393645
H            7.389040    11.966937     1.131722
C            7.186384    10.384437    -0.254318
O            5.986674    10.472166    -0.294965
N            7.862838     9.332497    -0.761515
C            7.169921     8.145082    -1.231698
C            6.632700     7.302320    -0.064480
O            5.531496     6.812191    -0.145111
C            8.048023     7.341579    -2.220518
C            8.345826     8.193291    -3.469295
C            7.376281     6.017628    -2.604436
C            9.429443     7.618030    -4.384588
```

```
H            8.822974     9.228191    -0.531268
H            6.279891     8.471639    -1.747020
H            8.986130     7.113644    -1.722574
H            7.424555     8.321605    -4.033217
H            8.650434     9.188198    -3.161869
H            7.993928     5.475068    -3.308756
H            7.221789     5.377635    -1.744299
H            6.407889     6.182861    -3.065124
H            9.625971     8.297919    -5.207242
H           10.364005     7.469971    -3.850308
H            9.147214     6.664653    -4.815676
N            7.385225     7.176512     1.047012
C            6.877042     6.457524     2.204057
C            5.524559     7.035060     2.642101
O            4.574308     6.325967     2.868652
C            7.830848     6.545496     3.405779
C            9.252392     5.997242     3.245545
O           10.144252     6.632337     3.801686
O            9.425757     4.935103     2.597434
H            8.346014     7.443499     1.024955
H            6.682390     5.428675     1.944175
H            7.918404     7.570066     3.735686
H            7.373052     5.991421     4.222696
N            5.482753     8.373085     2.790280
H            6.312206     8.910754     2.653481
C            4.290219     9.036913     3.256193
H            4.527987    10.079896     3.424426
H            3.945919     8.620697     4.189781
C            3.129854     8.987455     2.271387
O            1.988204     9.116366     2.649901
N            3.466171     8.825313     0.988812
H            4.427981     8.815249     0.742385
C            2.517188     8.905263    -0.078309
H            3.062362     9.089581    -0.995062
H            1.843510     9.738111     0.069104
C            1.609135     7.687017    -0.310189
O            0.575988     7.817118    -0.906827
N            2.067776     6.540439     0.227359
C            1.248377     5.370549     0.434794
C            0.429813     5.470884     1.726571
O           -0.764475     5.294262     1.700373
C            2.107097     4.106256     0.428929
H            2.968602     6.569054     0.655859
H            0.520431     5.319457    -0.357322
H            1.490830     3.234670     0.620250
H            2.579505     3.996917    -0.540369
H            2.884237     4.151893     1.184999
N            1.068191     5.733375     2.883787
C            0.338029     5.567457     4.131461
C           -0.773870     6.602006     4.287546
O           -1.809965     6.305745     4.842241
C            1.258920     5.492817     5.374590
C            1.849360     6.838951     5.821881
C            2.348493     4.431102     5.160687
```

```
C         2.693295     6.740957     7.095211
H         2.056692     5.862019     2.900468
H        -0.191362     4.627095     4.076667
H         0.605403     5.150203     6.171125
H         2.449867     7.254926     5.019470
H         1.044378     7.544444     6.000363
H         2.724610     4.075758     6.108871
H         1.955797     3.575567     4.620583
H         3.187257     4.825824     4.597549
H         2.996390     7.731996     7.420105
H         2.125610     6.290510     7.904070
H         3.590054     6.150574     6.954232
N        -0.550527     7.835200     3.809811
H         0.334405     8.069848     3.410757
C        -1.515000     8.880907     4.023560
H        -1.098733     9.807511     3.648703
H        -1.729412     9.010996     5.074454
C        -2.866971     8.618756     3.365893
O        -3.871476     8.845921     3.999867
N        -2.902686     8.108974     2.120752
C        -4.238787     8.033194     1.526231
C        -4.983202     6.753155     1.867818
O        -6.195626     6.754318     1.842582
C        -4.018520     8.294005     0.029162
C        -2.743446     9.144276     0.041792
C        -1.900223     8.455342     1.102623
H        -4.853889     8.819190     1.935695
H        -3.863914     7.367408    -0.507685
H        -4.874122     8.793873    -0.406893
H        -2.230657     9.199871    -0.900853
H        -2.981774    10.156806     0.358919
H        -1.435443     7.561610     0.727030
H        -1.127335     9.089719     1.504288
N        -4.262497     5.675999     2.205931
C        -4.914197     4.488141     2.729283
C        -5.629842     4.784577     4.050116
O        -6.704975     4.279094     4.285713
C        -3.921052     3.322434     2.932266
C        -3.594390     2.501999     1.700733
C        -4.106842     1.218509     1.543310
C        -2.725103     2.956670     0.711682
C        -3.777512     0.421568     0.457558
C        -2.386559     2.177468    -0.374700
C        -2.894361     0.873216    -0.547092
O        -2.578338     0.152933    -1.577154
H        -3.268757     5.756994     2.260898
H        -5.692673     4.185650     2.046458
H        -3.016284     3.722511     3.381139
H        -4.359935     2.661397     3.671706
H        -4.774113     0.821132     2.291646
H        -2.284183     3.929499     0.798091
H        -4.155513    -0.584569     0.393644
H        -1.708100     2.559722    -1.119194
N        -5.003331     5.588851     4.928170
```

```
H           -4.071671    5.898081    4.742882
C           -5.607624    5.942419    6.192324
H           -4.823026    6.282676    6.857091
H           -6.090651    5.086742    6.635504
C           -6.660906    7.045825    6.082512
O           -7.715840    6.968813    6.671022
N           -6.342145    8.091259    5.313948
H           -5.447376    8.140725    4.874874
C           -7.257648    9.220183    5.085240
H           -6.773142    9.882285    4.379589
H           -7.440761    9.763554    6.000424
C           -8.621452    8.828637    4.543285
O           -9.606440    9.420459    4.912576
N           -8.673257    7.827086    3.645025
C           -9.943614    7.435666    3.076848
C          -10.836112    6.654154    4.038280
O          -11.951735    6.351439    3.656934
C           -9.775573    6.671908    1.743215
O          -10.966834    6.755852    1.000025
C           -9.335384    5.223187    1.911330
H           -7.828303    7.411929    3.308412
H          -10.508955    8.330879    2.856496
H           -9.028293    7.206830    1.172109
H          -11.679080    6.479466    1.568919
H           -9.141858    4.784639    0.939817
H           -8.431870    5.142469    2.499257
H          -10.115082    4.639142    2.389249
N          -10.364715    6.343808    5.246677
C          -11.209900    5.754972    6.261796
C          -12.249540    6.740696    6.800523
O          -13.299314    6.325844    7.235299
C          -10.393773    5.201421    7.438426
C           -9.459963    4.028445    7.102085
C           -8.608442    3.695712    8.331194
C          -10.219219    2.785819    6.627183
H           -9.431308    6.592549    5.498266
H          -11.788811    4.964004    5.814091
H           -9.808306    6.007527    7.864915
H          -11.106388    4.886558    8.195353
H           -8.787143    4.332846    6.308556
H           -7.910898    2.891370    8.115457
H           -8.035281    4.559110    8.651339
H           -9.230183    3.380162    9.166390
H           -9.522693    1.974053    6.438632
H          -10.930266    2.450990    7.380572
H          -10.765692    2.957793    5.706887
N          -11.912361    8.040824    6.828433
C          -12.861882    9.030441    7.246619
C          -13.878980    9.396814    6.186155
H          -14.627930   10.124808    6.517977
O          -13.893172    8.967294    5.081459
H          -11.118791    8.352760    6.310016
H          -13.395984    8.683690    8.122608
H          -12.333526    9.935970    7.529092
```

```
N          -11.496190    10.630062    -3.035009
H          -11.247733    10.848697    -2.090743
C          -11.816457     9.206991    -3.172691
C          -11.414235     8.858125    -4.598813
O          -10.375790     8.292762    -4.845254
C          -13.245339     8.787459    -2.766642
C          -13.494301     9.128697    -1.285863
C          -13.469848     7.295863    -3.047016
C          -14.919980     8.857049    -0.802254
H          -12.280122    11.208570    -3.271591
H          -11.117532     8.661615    -2.553899
H          -13.959090     9.363117    -3.360659
H          -12.790450     8.572654    -0.671044
H          -13.282206    10.180979    -1.124282
H          -14.498742     7.014321    -2.862893
H          -13.248375     7.032851    -4.077384
H          -12.838464     6.689036    -2.405414
H          -15.042877     9.214780     0.213647
H          -15.652653     9.365844    -1.423251
H          -15.158148     7.799696    -0.801399
N          -12.214200     9.288719    -5.604122
H          -13.131169     9.602190    -5.380670
C          -11.972873     8.843878    -6.959262
H          -12.035531     7.767542    -7.044753
H          -12.729290     9.282532    -7.598613
C          -10.601634     9.233249    -7.517685
O           -9.997808     8.496125    -8.243133
N          -10.162516    10.476492    -7.182350
H          -10.708599    11.004447    -6.540654
C           -8.894338    11.023969    -7.637102
H           -8.790084    10.891100    -8.702725
H           -8.880399    12.081618    -7.408308
C           -7.672571    10.358998    -6.988330
O           -6.759378     9.922967    -7.632279
N           -7.657747    10.354147    -5.620749
H           -8.474939    10.647805    -5.135105
C           -6.495249     9.961279    -4.832307
H           -5.597967    10.374767    -5.264456
H           -6.616235    10.362002    -3.833391
C           -6.268288     8.463922    -4.714299
O           -5.135244     8.046505    -4.751044
N           -7.314867     7.639575    -4.521584
C           -7.020622     6.271133    -4.124525
C           -6.467326     5.439528    -5.275336
O           -5.672345     4.560888    -5.043417
C           -8.156626     5.541630    -3.368127
C           -9.346222     5.122976    -4.248120
C           -8.566288     6.359529    -2.137897
C          -10.386105     4.279503    -3.506728
H           -8.251879     7.988654    -4.514505
H           -6.188186     6.314002    -3.438769
H           -7.685647     4.630542    -3.009289
H           -9.825976     5.994079    -4.676481
H           -8.969838     4.536264    -5.083273
```

```
H         -9.221411    5.793059   -1.490471
H         -7.696283    6.635527   -1.550876
H         -9.088194    7.267188   -2.417891
H        -11.139265    3.914479   -4.198687
H         -9.929940    3.417091   -3.031033
H        -10.897643    4.848657   -2.740383
N         -6.882379    5.679822   -6.538388
C         -6.242751    4.969450   -7.628480
C         -4.821713    5.460812   -7.917777
O         -3.990853    4.676350   -8.308830
C         -7.107766    4.970248   -8.897881
C         -8.275025    3.973956   -8.839301
C         -7.793513    2.520527   -8.903265
N         -8.872639    1.529185   -8.891127
C         -9.426215    0.978437   -7.815749
N         -9.158400    1.435841   -6.605222
N        -10.266984   -0.034923   -7.964820
H         -7.492448    6.447501   -6.720172
H         -6.080259    3.958401   -7.295770
H         -7.496482    5.967443   -9.079069
H         -6.472335    4.727412   -9.743447
H         -8.865782    4.143041   -7.943760
H         -8.937018    4.154881   -9.679564
H         -7.248442    2.356428   -9.822809
H         -7.107343    2.280401   -8.102644
H         -9.143815    1.151075   -9.771265
H         -8.629247    2.268681   -6.487321
H         -9.631602    1.091880   -5.783876
H        -10.325757   -0.497121   -8.845103
H        -10.582363   -0.571456   -7.170949
N         -4.571419    6.757537   -7.721986
H         -5.302695    7.370349   -7.441242
C         -3.257792    7.352338   -8.011302
H         -2.929240    7.090628   -9.005007
H         -3.360549    8.426645   -7.939570
C         -2.202231    6.873958   -7.024718
O         -1.073852    6.638604   -7.377578
N         -2.633783    6.758811   -5.763283
C         -1.761007    6.448293   -4.656574
C         -1.491000    4.961024   -4.524131
O         -0.363792    4.576977   -4.319024
C         -2.349255    7.048560   -3.373588
C         -1.762508    6.354340   -2.145654
C         -2.111635    8.564410   -3.410714
H         -3.557144    7.065119   -5.546989
H         -0.788051    6.881331   -4.838967
H         -3.417824    6.867712   -3.387781
H         -1.979380    6.901791   -1.247581
H         -2.172663    5.356412   -2.033629
H         -0.689873    6.283703   -2.215540
H         -2.836841    9.090480   -2.806099
H         -1.116882    8.805684   -3.051328
H         -2.213132    8.950231   -4.418533
N         -2.517723    4.100220   -4.645904
```

```
C         -2.183874    2.694098   -4.750258
C         -1.220401    2.507869   -5.924342
O         -0.294163    1.732064   -5.842626
C         -3.407847    1.776927   -4.871499
C         -4.364888    1.887265   -3.676696
C         -5.406675    0.763417   -3.611904
N         -4.879475   -0.427261   -2.949793
C         -5.583840   -1.519150   -2.748491
N         -6.813258   -1.655527   -3.295771
N         -5.155159   -2.493137   -1.972250
H         -3.439230    4.421566   -4.850950
H         -1.629052    2.394610   -3.874428
H         -3.944330    1.990896   -5.793998
H         -3.028660    0.764904   -4.951115
H         -3.802055    1.899610   -2.756841
H         -4.904106    2.820331   -3.734069
H         -6.276908    1.113315   -3.066024
H         -5.727354    0.511292   -4.619620
H         -3.984955   -0.324272   -2.455368
H         -6.972757   -1.178773   -4.153531
H         -7.208521   -2.578273   -3.267051
H         -4.301483   -2.472079   -1.439477
H         -5.665078   -3.351236   -1.949303
N         -1.459790    3.193298   -7.035093
H         -2.284850    3.739868   -7.148495
C         -0.596621    2.997529   -8.183469
H         -0.949748    3.640778   -8.979139
H         -0.611196    1.972323   -8.524722
C          0.855924    3.344619   -7.871933
O          1.781357    2.710999   -8.313423
N          1.005385    4.404934   -7.063953
H          0.212185    4.964248   -6.846845
C          2.282350    4.880661   -6.640316
H          2.918995    5.137222   -7.477606
H          2.128365    5.777235   -6.053918
C          3.092850    3.928537   -5.775022
O          4.297527    4.059879   -5.687843
N          2.428015    2.983364   -5.101894
C          3.096637    2.136984   -4.148030
C          4.270120    1.327443   -4.710931
O          5.183441    1.025017   -3.969758
C          2.008539    1.259742   -3.526825
C          2.372558    0.440796   -2.294456
C          1.056393    0.029098   -1.643842
N          1.236414   -1.007170   -0.648781
C          0.299013   -1.927750   -0.373780
N         -0.855343   -1.920458   -0.991016
N          0.587706   -2.880258    0.490821
H          1.438768    2.900947   -5.211376
H          3.557572    2.746662   -3.380300
H          1.210866    1.943186   -3.256602
H          1.592799    0.612104   -4.289713
H          2.948408   -0.439377   -2.559505
H          2.976597    1.023650   -1.602782
```

```
H            0.560578     0.894051    -1.216187
H            0.400052    -0.365038    -2.404392
H            1.942194    -0.877371     0.040877
H           -1.297768    -1.062276    -1.309198
H           -1.528394    -2.645016    -0.775507
H            1.506491    -2.992819     0.850273
H           -0.162067    -3.479780     0.822424
N            4.280031     0.941778    -5.998578
H            3.541252     1.225888    -6.608101
C            5.363424     0.116638    -6.504622
H            5.137067    -0.137711    -7.532661
H            5.442848    -0.800432    -5.942754
C            6.736144     0.755908    -6.476882
O            7.738718     0.089366    -6.368969
N            6.766124     2.085339    -6.608293
H            5.911948     2.604339    -6.566385
C            8.005021     2.823956    -6.549017
H            8.710228     2.440517    -7.272781
H            7.799820     3.854828    -6.804955
C            8.746724     2.826456    -5.210435
O            9.887175     3.237466    -5.157199
N            8.101429     2.335406    -4.144636
C            8.681535     2.170358    -2.822752
C            9.947628     1.303165    -2.830509
O           10.765101     1.385821    -1.928230
C            7.577232     1.500558    -1.974588
C            7.771989     1.380062    -0.463379
C            7.427640     2.651039     0.322537
C            7.315792     2.387413     1.820281
N            8.655413     2.356332     2.479120
O           11.033153     1.615196     0.896933
H           10.884685     1.825393    -0.025412
H           11.511050     0.788172     0.847886
H            7.155893     2.032135    -4.258577
H            8.978832     3.127729    -2.415574
H            6.659843     2.042620    -2.160986
H            7.411669     0.512780    -2.388883
H            7.092628     0.600517    -0.132528
H            8.767415     1.028857    -0.225864
H            8.139802     3.452371     0.146214
H            6.462347     3.020645    -0.008273
H            6.754126     3.160288     2.321462
H            6.835166     1.441414     2.016282
H            9.366719     1.891048     1.923897
H            8.976059     3.326528     2.632108
H            8.601182     1.880366     3.367772
N           10.062468     0.448216    -3.837538
C           11.109366    -0.537544    -4.069419
C           12.486372     0.023143    -4.427254
O           13.444004    -0.705739    -4.301638
C           10.701363    -1.443263    -5.253397
C            9.563117    -2.441846    -4.999769
C            9.132182    -3.065331    -6.331061
C            9.989883    -3.533367    -4.018140
```

```
H        9.324246    0.445928   -4.510632
H       11.264674   -1.125458   -3.176633
H       10.435185   -0.796967   -6.084494
H       11.584298   -1.995497   -5.552179
H        8.706457   -1.911239   -4.595973
H        8.330940   -3.781801   -6.177422
H        8.775111   -2.307920   -7.017149
H        9.959527   -3.592977   -6.799164
H        9.199757   -4.263157   -3.878961
H       10.859908   -4.059608   -4.396973
H       10.242747   -3.137364   -3.039386
N       12.574050    1.261829   -4.919242
C       13.814453    1.845370   -5.393357
C       14.958485    2.034739   -4.378827
O       16.012412    2.417529   -4.811576
C       13.496419    3.171929   -6.086597
O       12.917141    4.088890   -5.202404
H       11.756570    1.833767   -4.915474
H       14.248723    1.178604   -6.128872
H       14.419368    3.591232   -6.455377
H       12.845628    2.977723   -6.933290
H       11.969675    4.078408   -5.265672
N       14.792294    1.722326   -3.074279
C       15.927992    1.647230   -2.176435
C       16.804495    0.409874   -2.402229
O       17.934274    0.393859   -1.987681
C       15.494230    1.576952   -0.707148
C       14.636405    2.733139   -0.197767
C       14.280846    2.484613    1.268455
N       13.255547    3.379242    1.790687
C       13.433452    4.530654    2.417606
N       14.651388    5.006220    2.653408
O       15.028043    7.462722    4.207124
H       14.745070    8.349635    3.968965
H       15.748001    7.567698    4.816085
N       12.383758    5.219596    2.814043
H       13.901416    1.417906   -2.758955
H       16.561534    2.504336   -2.345514
H       14.954330    0.648557   -0.552268
H       16.405896    1.508159   -0.123829
H       15.153949    3.681221   -0.312268
H       13.717995    2.814690   -0.767353
H       13.883321    1.485481    1.373081
H       15.159484    2.525723    1.901468
H       12.315325    3.083056    1.620748
H       15.459068    4.512343    2.356887
H       14.787188    5.861152    3.164897
H       11.428069    4.939741    2.656779
H       12.476311    6.063013    3.333653
N       16.200238   -0.648980   -2.967242
C       16.870485   -1.911748   -3.168205
C       17.667865   -1.889273   -4.468826
H       18.067003   -0.910804   -4.745289
O       17.867599   -2.851332   -5.135873
```

```
C         15.867590    -3.070231    -3.148676
C         15.324123    -3.399076    -1.791036
N         14.402671    -2.626651    -1.112116
O         12.506042    -0.433892    -0.462206
H         13.143122    -1.036871    -0.849396
H         12.024862    -0.037603    -1.180542
C         15.619497    -4.488899    -1.047858
C         14.171505    -3.250081    -0.003229
N         14.882335    -4.379818     0.104639
H         15.285715    -0.550257    -3.352830
H         17.596346    -2.030782    -2.371442
H         15.050132    -2.842573    -3.824055
H         16.368086    -3.944387    -3.542813
H         16.275389    -5.312474    -1.232455
H         13.492857    -2.920320     0.757312
H         14.781793    -5.062261     0.823783
O        -15.020213    -2.471018     4.973947
H        -15.212269    -1.661971     4.507127
H        -14.526681    -2.213501     5.745680
O        -12.902805    -1.270687     6.931187
H        -12.213112    -1.722239     7.405267
H        -13.132674    -0.515337     7.458650
O        -12.055753    -0.850867     4.158774
H        -12.176337    -0.902832     5.104108
H        -12.868116    -0.459842     3.834508
O        -13.393747    -1.856946     0.710307
H        -12.618284    -1.293680     0.766814
H        -13.315966    -2.505070     1.414951
O         -7.286578     1.302460    -0.239832
H         -7.498670     1.637792     0.634927
H         -6.357470     1.101754    -0.204429
O         -7.886340     2.078825     2.455260
H         -8.797907     1.924358     2.690781
H         -7.559723     2.755410     3.042588
O         -9.340583    -0.024714    -1.607137
H         -8.578916     0.394601    -1.186585
H         -9.089673    -0.922547    -1.804081
O        -12.844468    -2.336472    -2.019732
H        -13.120890    -2.372994    -1.100710
H        -13.457233    -2.846557    -2.536626
O        -14.029220     0.929631    -0.581729
H        -14.374084     0.067218    -0.375890
H        -13.118598     0.887120    -0.273370
O        -14.252289     4.929717     4.735637
H        -13.682720     5.570161     4.315809
H        -14.289917     5.214909     5.644026
O        -15.228230     2.155337     1.701339
H        -15.850248     2.871084     1.687644
H        -14.941090     1.991837     0.804275
O        -13.151682     0.247400    -3.244027
H        -13.600362     0.778017    -2.588539
H        -12.933673    -0.556693    -2.770726
O        -13.770241    -1.924623    -4.981136
H        -13.935128    -1.061137    -4.609276
```

```
H            -14.496296   -2.133600   -5.554536
O            -10.158517   -2.764133   -2.953315
H            -10.299828   -2.610238   -3.881079
H            -11.020981   -2.685373   -2.547665
O            -11.483664    0.213499    0.371431
H            -10.756842    0.131628   -0.247651
H            -11.143523    0.654620    1.157231
O            -10.713541    1.341352    2.831460
H            -10.879054    0.581125    3.386200
H            -11.394559    1.967053    3.094746
O            -13.056431    2.612688    3.638284
H            -13.358288    3.352888    4.167517
H            -13.711662    2.532419    2.952964
O            -10.465899    0.624989   -4.110407
H            -11.370355    0.792546   -3.839565
H             -9.979019    0.491443   -3.289473
O            -14.689453   -0.342247    3.042901
H            -15.010244    0.527555    2.814387
H            -14.562674   -0.799930    2.218332
O            -13.086321   -3.361499    3.147701
H            -13.858109   -3.220234    3.705614
H            -12.459215   -2.724240    3.488715
O            -11.004346   -1.654020   -5.646469
H            -11.934456   -1.876332   -5.563288
H            -10.845416   -0.947014   -5.017375

Case 6
voltage 0; Y266 neutral; R300 positive; E183 negative, S4 free
N             15.958147   -6.023564    5.657444
H             15.147782   -5.795492    6.200200
C             15.873943   -5.442505    4.316654
C             15.177275   -6.393150    3.326435
O             14.461900   -5.970346    2.434395
C             17.277751   -5.130217    3.785588
H             16.753341   -5.671351    6.152955
H             15.284259   -4.535343    4.302529
H             17.239319   -4.760150    2.767752
H             17.739718   -4.363804    4.397720
H             17.909513   -6.011284    3.807316
N             15.450096   -7.689618    3.466657
H             16.022065   -7.948242    4.241686
C             14.839060   -8.716481    2.620908
H             14.929710   -8.460293    1.578332
H             15.342753   -9.657453    2.798311
C             13.352509   -8.858691    2.950720
O             12.526965   -9.068953    2.100620
N             13.015302   -8.697267    4.246017
H             13.732041   -8.621581    4.930668
```

```
C        11.627875    -8.833339     4.695514
H        11.615416    -8.808202     5.776975
H        11.194416    -9.762705     4.359798
C        10.750839    -7.694514     4.159327
O         9.604591    -7.888926     3.852514
N        11.344477    -6.490028     4.088719
C        10.680391    -5.320826     3.569254
C        10.460803    -5.368059     2.055070
O         9.406882    -4.966261     1.621740
C        11.424799    -4.056358     4.023165
O        12.815532    -4.188005     3.969016
H        12.304795    -6.401851     4.327159
H         9.681844    -5.268118     3.970412
H        11.076757    -3.209266     3.441143
H        11.186619    -3.869733     5.060088
H        13.124233    -4.373550     3.090359
N        11.430559    -5.834027     1.245109
C        11.113355    -6.022213    -0.167301
C        10.052968    -7.110384    -0.339755
O         9.277668    -7.062781    -1.267658
C        12.362345    -6.238582    -1.058016
C        12.888073    -7.677544    -1.079824
C        12.081986    -5.753705    -2.484963
H        12.271630    -6.215373     1.622931
H        10.629992    -5.118414    -0.499269
H        13.131033    -5.600771    -0.638354
H        13.829427    -7.709830    -1.618151
H        13.053523    -8.081724    -0.090166
H        12.200434    -8.340562    -1.596120
H        12.963161    -5.886783    -3.102211
H        11.266103    -6.305796    -2.936461
H        11.827785    -4.699200    -2.497124
N        10.007054    -8.074201     0.587087
H        10.768456    -8.185455     1.218068
C         9.050240    -9.173820     0.506580
H         9.381793    -9.943398     1.191771
H         9.019509    -9.589902    -0.488455
C         7.629189    -8.786352     0.860301
O         6.718182    -9.259726     0.221821
N         7.430677    -7.930038     1.870761
C         6.091796    -7.436153     2.098284
C         5.689004    -6.449166     1.006143
O         4.515930    -6.368929     0.706811
C         5.801531    -6.904580     3.516630
C         5.918416    -8.030707     4.548849
C         6.631916    -5.688508     3.931023
H         8.195060    -7.654657     2.449040
H         5.414202    -8.260424     1.937687
H         4.760715    -6.597298     3.479208
H         5.603638    -7.675791     5.524478
H         5.287372    -8.873237     4.285369
H         6.939058    -8.385854     4.636706
H         6.260155    -5.295887     4.871581
H         7.668630    -5.959960     4.079110
```

```
H         6.590363   -4.889206    3.199097
N         6.610423   -5.729660    0.359685
C         6.223281   -4.948504   -0.804880
C         5.768865   -5.866735   -1.947967
O         4.750064   -5.616476   -2.548742
C         7.340697   -3.972538   -1.230551
C         7.523064   -2.877075   -0.164837
C         7.027791   -3.357031   -2.598930
C         8.813898   -2.073321   -0.321679
H         7.566442   -5.763662    0.640642
H         5.337465   -4.381898   -0.558208
H         8.263633   -4.537811   -1.313973
H         6.664706   -2.207593   -0.208284
H         7.519006   -3.323309    0.821392
H         7.772106   -2.617430   -2.864410
H         7.022545   -4.099269   -3.386326
H         6.056385   -2.873330   -2.596036
H         8.893700   -1.322000    0.456566
H         9.683738   -2.714607   -0.242022
H         8.871532   -1.554852   -1.271249
N         6.529688   -6.926511   -2.249936
H         7.410214   -7.054304   -1.797864
C         6.191738   -7.793133   -3.357696
H         7.030382   -8.458362   -3.521746
H         6.021080   -7.225859   -4.260341
C         4.937662   -8.634322   -3.127445
O         4.084771   -8.710184   -3.980201
N         4.854869   -9.277047   -1.956386
H         5.599569   -9.212148   -1.295379
C         3.689396  -10.054251   -1.607263
H         3.456245  -10.778962   -2.372806
H         3.903525  -10.583983   -0.687502
C         2.439211   -9.201118   -1.409874
O         1.357526   -9.622547   -1.748531
N         2.628174   -7.992185   -0.883447
C         1.476015   -7.102191   -0.805182
C         1.015973   -6.585646   -2.172836
O        -0.169145   -6.430953   -2.370525
C         1.653169   -5.915607    0.145765
O         1.881315   -6.327966    1.453301
H         3.528290   -7.674487   -0.603201
H         0.643619   -7.679076   -0.437123
H         2.443076   -5.261700   -0.205323
H         0.723618   -5.366367    0.137845
H         2.805409   -6.520079    1.536885
N         1.932208   -6.298387   -3.109202
C         1.519965   -5.907323   -4.445322
C         0.790140   -7.053261   -5.145852
O        -0.200055   -6.816799   -5.805346
C         2.707842   -5.349656   -5.268125
C         3.020080   -3.921467   -4.786725
C         2.430783   -5.381036   -6.776709
C         4.290258   -3.314696   -5.384475
H         2.905559   -6.366998   -2.902916
```

```
H         0.767523    -5.136663    -4.362346
H         3.569110    -5.979988    -5.071712
H         2.170966    -3.284218    -5.031361
H         3.113155    -3.925726    -3.707708
H         3.276880    -4.984148    -7.323445
H         2.267184    -6.388445    -7.139109
H         1.558131    -4.788914    -7.033364
H         4.510227    -2.357656    -4.919253
H         5.143701    -3.964057    -5.219970
H         4.200985    -3.141231    -6.450906
N         1.260957    -8.293549    -4.992664
H         2.117118    -8.451082    -4.503095
C         0.529286    -9.419980    -5.525678
H         0.423105    -9.354586    -6.598332
H         1.081367   -10.320078    -5.286750
C        -0.879037    -9.531543    -4.944238
O        -1.816757    -9.810078    -5.653177
N        -0.996537    -9.299243    -3.638120
C        -2.296512    -9.382777    -3.015902
C        -3.245251    -8.289587    -3.501193
O        -4.426836    -8.549301    -3.580477
C        -2.205061    -9.325902    -1.484924
O        -3.443331    -9.611058    -0.896617
H        -0.182302    -9.205607    -3.067331
H        -2.776835   -10.311110    -3.291161
H        -1.504058   -10.074618    -1.142350
H        -1.840417    -8.352437    -1.173055
H        -4.147707    -9.264460    -1.431352
N        -2.735855    -7.082942    -3.762997
C        -3.533630    -5.998610    -4.310990
C        -4.041716    -6.335976    -5.711675
O        -5.202659    -6.168608    -6.000396
C        -2.690470    -4.703286    -4.318033
C        -3.174240    -3.633075    -5.281992
C        -4.511109    -3.249594    -5.353114
C        -2.272869    -3.044432    -6.160355
C        -4.928944    -2.320485    -6.289982
C        -2.686241    -2.106303    -7.094344
C        -4.020163    -1.747470    -7.167118
H        -1.766584    -6.925522    -3.585380
H        -4.415756    -5.855558    -3.708465
H        -2.664959    -4.328661    -3.299007
H        -1.675278    -4.960205    -4.585001
H        -5.238786    -3.712647    -4.714787
H        -1.239259    -3.343370    -6.139687
H        -5.974183    -2.072247    -6.348687
H        -1.968508    -1.672764    -7.765854
H        -4.348940    -1.037918    -7.906234
N        -3.141073    -6.794341    -6.590201
C        -3.571178    -7.138775    -7.926036
C        -4.680642    -8.188027    -7.897217
O        -5.565625    -8.147234    -8.709553
C        -2.413327    -7.657575    -8.780390
S        -1.194773    -6.398787    -9.280227
```

```
H         -2.199749    -6.957332    -6.302061
H         -4.009697    -6.274769    -8.402909
H         -1.908877    -8.477399    -8.286809
H         -2.822455    -8.031872    -9.709061
H         -0.615498    -6.221976    -8.103070
N         -4.591211    -9.147236    -6.954078
H         -3.767718    -9.247201    -6.401883
C         -5.577115   -10.215868    -6.969857
H         -5.735691   -10.573744    -7.975548
H         -5.192826   -11.030342    -6.366805
C         -6.961689    -9.841575    -6.415510
O         -7.970189   -10.236427    -6.934406
N         -6.938965    -9.122778    -5.287844
C         -8.171367    -8.778832    -4.603032
C         -9.043492    -7.810764    -5.383239
O        -10.240446    -7.844467    -5.280821
C         -7.941568    -8.220781    -3.185095
C         -7.031739    -6.990699    -3.020888
C         -7.629816    -5.591440    -2.954692
O         -8.808734    -5.381417    -3.279371
O         -6.836061    -4.690580    -2.605539
H         -6.063697    -8.858927    -4.890249
H         -8.774882    -9.670808    -4.509607
H         -8.920530    -8.014493    -2.777531
H         -7.515278    -9.030348    -2.600654
H         -6.447062    -7.101738    -2.116289
H         -6.307592    -6.944459    -3.817779
N         -8.396504    -6.875797    -6.112705
C         -9.167326    -5.878473    -6.830337
C         -9.714577    -6.388704    -8.166400
H         -9.379435    -7.383445    -8.467288
O        -10.489392    -5.764674    -8.819039
C         -8.369870    -4.555337    -6.875228
O         -8.582119    -3.858969    -5.659674
C         -8.733056    -3.590242    -7.992215
H         -7.414638    -6.952496    -6.270548
H        -10.059348    -5.691674    -6.247589
H         -7.316380    -4.793718    -6.960041
H         -8.536462    -4.465532    -4.916707
H         -8.180596    -2.668381    -7.837296
H         -8.469399    -3.988547    -8.962891
H         -9.793684    -3.374881    -7.995944
N        -12.512024    -9.145072     3.339860
H        -12.743508    -9.686953     2.533271
C        -11.364979    -8.260438     3.074139
C        -10.308904    -8.227279     4.212189
O         -9.174013    -7.858306     4.014874
C        -11.773140    -6.796364     2.826856
O        -12.306118    -6.344900     4.043943
C        -12.779243    -6.636243     1.692123
H        -13.317259    -8.611195     3.602047
H        -10.833162    -8.619947     2.205561
H        -10.871190    -6.246534     2.583327
H        -12.417115    -5.396407     4.020045
```

```
H            -12.958692     -5.582034      1.510798
H            -12.420088     -7.076101      0.769674
H            -13.730113     -7.089371      1.946109
N            -10.776104     -8.603192      5.410250
H            -11.712157     -8.943330      5.418663
C            -10.020203     -8.630141      6.639226
H             -9.998137     -7.657185      7.106958
H            -10.522269     -9.306481      7.321775
C             -8.557704     -9.068715      6.551804
O             -7.736247     -8.441615      7.175793
N             -8.204525    -10.156126      5.843150
H             -8.900017    -10.681538      5.363159
C             -6.803167    -10.542583      5.750152
H             -6.753501    -11.531071      5.311458
H             -6.360473    -10.585661      6.732591
C             -5.944663     -9.594019      4.893164
O             -4.777297     -9.432639      5.147296
N             -6.608774     -9.025916      3.892815
C             -5.877247     -8.052606      3.091064
C             -5.877539     -6.670768      3.745729
O             -4.918843     -5.943397      3.600119
C             -6.173960     -8.057379      1.580907
C             -7.441551     -7.415518      1.061536
C             -7.567515     -6.029619      1.019311
C             -8.450837     -8.188482      0.505650
C             -8.671225     -5.434327      0.435879
C             -9.554333     -7.594975     -0.088451
C             -9.666387     -6.216445     -0.129742
H             -7.601365     -9.064707      3.845787
H             -4.846158     -8.352152      3.163415
H             -5.327156     -7.562394      1.118552
H             -6.125113     -9.091767      1.262079
H             -6.778856     -5.410277      1.409248
H             -8.358381     -9.261239      0.498915
H             -8.735800     -4.362811      0.381404
H            -10.306336     -8.208618     -0.551773
H            -10.498708     -5.756243     -0.627940
N             -6.908609     -6.346994      4.527295
C             -6.891083     -5.201229      5.403478
C             -5.790904     -5.358706      6.451814
O             -5.127250     -4.395983      6.775011
C             -8.289970     -5.060680      6.038743
C             -8.509525     -3.824445      6.883736
C             -8.862424     -2.617225      6.290494
C             -8.410433     -3.881973      8.268995
C             -9.103737     -1.492604      7.062348
C             -8.653268     -2.760442      9.046475
C             -9.002393     -1.561638      8.444902
H             -7.731696     -6.907198      4.498264
H             -6.640586     -4.304275      4.855546
H             -9.006395     -5.069114      5.224608
H             -8.480272     -5.942464      6.635966
H             -8.959844     -2.552936      5.220539
H             -8.148189     -4.809807      8.747379
```

```
H         -9.368956   -0.567412    6.584536
H         -8.564130   -2.822743   10.115880
H         -9.176367   -0.685423    9.044545
N         -5.586508   -6.566740    6.972241
H         -6.242830   -7.295052    6.795543
C         -4.540693   -6.819257    7.937545
H         -4.586468   -6.128257    8.765966
H         -4.682553   -7.822828    8.317691
C         -3.132967   -6.703698    7.361209
O         -2.268097   -6.159126    8.004994
N         -2.906568   -7.218792    6.146318
C         -1.608235   -7.068968    5.511934
C         -1.351778   -5.605148    5.136261
O         -0.268934   -5.105830    5.352332
C         -1.482184   -8.029494    4.309991
C         -0.279633   -7.690129    3.427470
C         -1.388678   -9.479024    4.800390
H         -3.608707   -7.771965    5.701982
H         -0.835382   -7.298144    6.231616
H         -2.381192   -7.922508    3.707794
H         -0.193376   -8.419120    2.629359
H         -0.355442   -6.714615    2.967363
H          0.644433   -7.709999    3.996427
H         -1.372285  -10.157710    3.953856
H         -0.470313   -9.628153    5.362501
H         -2.223086   -9.759511    5.428773
N         -2.357821   -4.918329    4.577710
C         -2.166263   -3.549133    4.144384
C         -2.037480   -2.584600    5.328350
O         -1.174991   -1.727383    5.299012
C         -3.298757   -3.082419    3.218478
C         -3.366759   -3.767443    1.840212
C         -2.249292   -3.362291    0.892494
O         -1.093647   -3.605170    1.229331
O         -2.567137   -2.801072   -0.196230
H         -3.224639   -5.373393    4.382412
H         -1.230306   -3.483537    3.618729
H         -4.248264   -3.236526    3.718082
H         -3.185272   -2.010934    3.084557
H         -3.326923   -4.841870    1.961167
H         -4.320431   -3.529781    1.384960
N         -2.903835   -2.706316    6.340707
C         -2.761842   -1.906055    7.541265
C         -1.604336   -2.365800    8.424855
O         -1.103170   -1.558713    9.177038
C         -4.032330   -1.712603    8.381697
O         -4.421939   -2.885163    9.035177
C         -5.169022   -1.073709    7.593118
H         -3.571171   -3.446869    6.322949
H         -2.456649   -0.923510    7.221554
H         -3.732231   -1.035112    9.169502
H         -4.826134   -3.475931    8.411379
H         -5.999065   -0.875451    8.259225
H         -4.852138   -0.133842    7.151228
```

```
H         -5.521082    -1.722678     6.801045
N         -1.063248    -3.560996     8.206098
H         -1.607882    -4.303860     7.828742
C          0.182982    -3.908727     8.837784
H          0.194988    -3.651582     9.885713
H          0.323959    -4.976655     8.732484
C          1.363363    -3.196026     8.186203
O          2.276232    -2.756837     8.847849
N          1.339117    -3.097516     6.843245
C          2.455573    -2.505702     6.150384
C          2.510245    -0.985065     6.287512
O          3.597089    -0.448536     6.305682
C          2.551069    -2.965415     4.678397
S          1.864142    -1.861709     3.400043
H          0.618357    -3.561666     6.329564
H          3.360544    -2.833878     6.641101
H          3.597496    -3.036023     4.416304
H          2.113716    -3.947404     4.572754
H          0.584914    -1.901332     3.741132
N          1.369104    -0.295193     6.402722
C          1.452563     1.141057     6.608340
C          2.010264     1.454174     7.999103
O          2.618503     2.476803     8.202664
C          0.139907     1.907439     6.339855
C         -0.922355     1.723326     7.436366
C         -0.370907     1.613689     4.924177
C         -2.232250     2.464557     7.162640
H          0.491571    -0.759312     6.284952
H          2.191320     1.525247     5.921930
H          0.437003     2.949816     6.373122
H         -1.127343     0.672827     7.599043
H         -0.516723     2.091914     8.372743
H         -1.116161     2.340922     4.628259
H          0.440545     1.667770     4.203901
H         -0.817099     0.631231     4.845532
H         -2.874594     2.425898     8.036443
H         -2.056124     3.511160     6.928052
H         -2.780349     2.027122     6.334988
N          1.788530     0.523639     8.926312
H          1.193095    -0.246792     8.725719
C          2.204820     0.716780    10.298358
H          1.689594    -0.014883    10.907591
H          1.954682     1.707246    10.646781
C          3.707364     0.550971    10.474153
O          4.357422     1.322929    11.125203
N          4.256112    -0.509665     9.842554
H          3.673528    -1.173142     9.377940
C          5.685088    -0.664444     9.824901
H          5.921805    -1.632159     9.398327
H          6.100326    -0.618039    10.820419
C          6.398318     0.405485     9.002386
O          7.521478     0.739391     9.291916
N          5.739184     0.884907     7.948658
C          6.400050     2.007737     7.321672
```

```
C         6.319292      3.248703     8.210167
H         5.946336      3.092196     9.220941
O         6.674364      4.318401     7.824448
C         6.002622      2.221907     5.860673
C         6.664779      1.186471     4.966420
C         8.058431      1.105221     4.896341
C         5.920646      0.264981     4.240162
C         8.684115      0.119411     4.148280
C         6.545376     -0.717644     3.481909
C         7.926678     -0.800487     3.435099
H         4.804820      0.622883     7.727341
H         7.455039      1.778756     7.340054
H         4.926688      2.181782     5.748590
H         6.324700      3.218218     5.579199
H         8.661068      1.796811     5.459679
H         4.850506      0.281850     4.308002
H         9.759189      0.060920     4.138362
H         5.946771     -1.433031     2.945595
H         8.404974     -1.580872     2.871125
N        12.725229      9.898977    -1.157802
H        12.008805     10.176727    -1.801998
C        12.438766      8.567360    -0.616433
C        11.361226      8.730958     0.444861
O        10.345250      8.070100     0.406122
C        13.700434      7.935548    -0.009412
C        14.808471      7.607888    -1.017077
S        14.233969      6.440947    -2.285262
C        15.742632      6.255629    -3.268258
H        13.578129      9.898477    -1.681060
H        12.017782      7.893401    -1.351314
H        14.112220      8.602870     0.741344
H        13.407722      7.023775     0.498863
H        15.186630      8.506140    -1.491958
H        15.640452      7.170676    -0.476928
H        15.546734      5.504050    -4.019848
H        16.006344      7.186042    -3.753979
H        16.567824      5.922324    -2.651646
N        11.554270      9.662415     1.392550
C        10.614098      9.798338     2.485803
C         9.311847     10.494662     2.083329
O         8.301847     10.259897     2.699990
C        11.276242     10.465467     3.698776
C        12.412618      9.585521     4.190770
O        13.546572      9.762769     3.796339
N        12.077931      8.613147     5.048939
H        12.402262     10.186634     1.386024
H        10.296510      8.809603     2.777289
H        11.689687     11.432990     3.442125
H        10.528918     10.598938     4.470164
H        12.781776      7.941323     5.266743
H        11.140071      8.270779     5.082521
N         9.322333     11.309474     1.011300
H        10.180989     11.472506     0.536730
C         8.073006     11.750550     0.419861
```

```
H         8.290520    12.476374    -0.352941
H         7.439479    12.212534     1.159178
C         7.279881    10.596360    -0.195437
O         6.076055    10.642438    -0.207484
N         7.973684     9.575087    -0.736619
C         7.296347     8.394984    -1.245279
C         6.734791     7.527988    -0.107153
O         5.643993     7.021124    -0.223160
C         8.201941     7.609359    -2.225578
C         8.604856     8.507241    -3.411035
C         7.503446     6.335549    -2.716148
C         9.641893     7.888704    -4.352358
H         8.939144     9.483210    -0.523593
H         6.417843     8.727881    -1.776770
H         9.101116     7.319525    -1.690357
H         7.711435     8.766904    -3.975082
H         9.001394     9.443958    -3.034235
H         8.162713     5.774105    -3.365934
H         7.217268     5.687943    -1.897688
H         6.600793     6.573236    -3.271095
H         9.946046     8.612670    -5.101286
H        10.533678     7.576628    -3.815876
H         9.259858     7.022805    -4.880107
N         7.460659     7.392363     1.020852
C         6.934323     6.639113     2.146001
C         5.601478     7.225387     2.620229
O         4.637501     6.523202     2.823823
C         7.890200     6.648572     3.344584
C         9.294670     6.088277     3.123870
O        10.210017     6.691603     3.675303
O         9.430398     5.050589     2.429431
H         8.416074     7.680039     1.028361
H         6.714484     5.627970     1.841188
H         7.997636     7.652778     3.727419
H         7.432622     6.058261     4.135036
N         5.590422     8.552067     2.830457
H         6.424749     9.083392     2.694713
C         4.404766     9.224852     3.301130
H         4.643552    10.271885     3.439372
H         4.072512     8.831197     4.248810
C         3.241866     9.136952     2.323580
O         2.095784     9.210264     2.701964
N         3.586261     8.991037     1.041106
H         4.545093     9.062803     0.790978
C         2.638057     9.032729    -0.025191
H         3.181888     9.210572    -0.943941
H         1.948151     9.856721     0.101482
C         1.753792     7.795583    -0.248807
O         0.723764     7.873335    -0.861861
N         2.240962     6.667458     0.308875
C         1.462344     5.468734     0.483817
C         0.644217     5.487292     1.779540
O        -0.517030     5.149451     1.757137
C         2.360198     4.231368     0.439480
```

```
H         3.141310      6.719092      0.736465
H         0.732737      5.422462     -0.308623
H         1.773518      3.333948      0.603746
H         2.839074      4.167906     -0.531231
H         3.131934      4.281623      1.200502
N         1.247303      5.850769      2.925075
C         0.527814      5.689575      4.179175
C        -0.610371      6.698864      4.328443
O        -1.638476      6.373091      4.882955
C         1.451662      5.647907      5.422056
C         2.021861      7.008902      5.849784
C         2.554984      4.598465      5.221631
C         2.885727      6.939185      7.111685
H         2.206881      6.123010      2.927984
H         0.021854      4.737188      4.135917
H         0.804487      5.307374      6.224539
H         2.601906      7.431370      5.036191
H         1.205171      7.699150      6.034391
H         2.939841      4.265776      6.175740
H         2.171567      3.726465      4.701944
H         3.380252      4.994900      4.639813
H         3.163414      7.938516      7.432086
H         2.347647      6.469662      7.929286
H         3.800182      6.379055      6.956474
N        -0.423764      7.932814      3.841266
H         0.456491      8.188871      3.444613
C        -1.408004      8.964677      4.052625
H        -1.006873      9.893975      3.667885
H        -1.619205      9.099260      5.103652
C        -2.755765      8.681014      3.397791
O        -3.770697      8.901759      4.014471
N        -2.776237      8.158183      2.157718
C        -4.084438      7.968654      1.534728
C        -4.828300      6.744904      2.040737
O        -6.036756      6.745538      2.064013
C        -3.780312      7.911067      0.029446
C        -2.532717      8.786649     -0.089814
C        -1.731342      8.375123      1.138652
H        -4.726369      8.808632      1.757543
H        -3.555079      6.892925     -0.270110
H        -4.623137      8.256579     -0.555137
H        -1.970592      8.636642     -0.999639
H        -2.807793      9.835784     -0.026184
H        -1.197843      7.458934      0.955651
H        -1.024515      9.127775      1.450587
N        -4.096453      5.681875      2.418650
C        -4.741956      4.543411      3.044323
C        -5.402920      4.920587      4.370954
O        -6.482401      4.458780      4.659828
C        -3.753685      3.389724      3.296779
C        -3.488176      2.473044      2.120903
C        -4.079604      1.216875      2.070151
C        -2.605132      2.819606      1.104250
C        -3.777511      0.317237      1.058352
```

```
C         -2.293016    1.930342    0.091609
C         -2.859322    0.668186    0.080143
O         -2.503069   -0.197944   -0.906003
H         -3.105562    5.780237    2.481380
H         -5.547049    4.206640    2.410011
H         -2.822154    3.805190    3.665968
H         -4.160858    2.790500    4.102379
H         -4.782151    0.926014    2.831494
H         -2.117463    3.772985    1.118565
H         -4.225136   -0.660370    1.044737
H         -1.585144    2.195618   -0.673025
H         -2.480148   -1.100747   -0.563364
N         -4.708945    5.731559    5.185491
H         -3.786112    6.021868    4.938510
C         -5.251004    6.174885    6.447961
H         -4.445666    6.617019    7.021270
H         -5.658073    5.346675    7.006498
C         -6.370948    7.201319    6.306147
O         -7.336305    7.167163    7.031720
N         -6.229617    8.108638    5.335552
H         -5.366231    8.153918    4.840004
C         -7.164678    9.218186    5.091495
H         -6.697903    9.866058    4.361134
H         -7.340888    9.782167    5.995393
C         -8.524879    8.796086    4.574067
O         -9.521176    9.365299    4.948768
N         -8.558424    7.790013    3.682322
C         -9.821475    7.388977    3.107488
C        -10.723208    6.654595    4.091745
O        -11.863429    6.409351    3.754754
C         -9.633711    6.550065    1.823090
O        -10.819275    6.564713    1.072933
C         -9.167162    5.121838    2.078295
H         -7.705241    7.383759    3.356617
H        -10.376226    8.274292    2.829209
H         -8.894241    7.065250    1.223762
H        -11.537139    6.315413    1.647092
H         -8.963371    4.632144    1.133345
H         -8.265805    5.093585    2.675650
H         -9.938621    4.550627    2.584343
N        -10.207083    6.270767    5.262614
C        -11.048455    5.695123    6.295287
C        -12.087778    6.678692    6.838840
O        -13.123105    6.254883    7.298271
C        -10.209025    5.185544    7.476839
C         -9.231980    4.042672    7.162267
C         -8.389176    3.746524    8.406740
C         -9.941727    2.774275    6.680810
H         -9.265780    6.503617    5.488770
H        -11.623043    4.889816    5.867476
H         -9.650734    6.020510    7.887193
H        -10.906903    4.860562    8.242788
H         -8.553154    4.364471    6.380665
H         -7.660603    2.966316    8.204933
```

```
H          -7.852198     4.631578     8.730151
H          -9.012472     3.412378     9.233476
H          -9.215567     1.987795     6.496192
H         -10.645486     2.413069     7.428624
H         -10.488419     2.927771     5.757716
N         -11.762118     7.981094     6.859894
C         -12.706036     8.959644     7.316662
C         -13.777108     9.309588     6.304784
H         -14.519049    10.027715     6.672200
O         -13.838326     8.878628     5.202741
H         -10.980290     8.305104     6.331469
H         -13.194630     8.610258     8.217988
H         -12.176319     9.872600     7.570958
N         -14.325465     9.683300    -4.682042
H         -14.241103    10.360580    -5.414952
C         -13.041566     9.396581    -4.053960
C         -12.331550     8.307411    -4.853461
O         -11.120536     8.224784    -4.894561
C         -13.257671     8.969682    -2.573442
C         -13.593625    10.212349    -1.722346
C         -12.063760     8.215637    -1.977994
C         -14.253553     9.891816    -0.379821
H         -14.987439    10.041853    -4.024694
H         -12.362783    10.245351    -4.065796
H         -14.119798     8.304536    -2.570574
H         -12.676625    10.772685    -1.554941
H         -14.252916    10.880387    -2.270366
H         -12.216356     8.012581    -0.927728
H         -11.901061     7.260110    -2.462075
H         -11.150973     8.794096    -2.074306
H         -14.490520    10.807374     0.151013
H         -15.180827     9.344197    -0.521128
H         -13.613567     9.301619     0.264577
N         -13.119236     7.442742    -5.519904
H         -14.102709     7.580976    -5.429824
C         -12.603869     6.195689    -6.034144
H         -12.226142     5.560593    -5.245481
H         -13.415720     5.672096    -6.524903
C         -11.473303     6.351904    -7.038530
O         -10.555880     5.568188    -7.062392
N         -11.556099     7.368736    -7.914824
H         -12.330214     7.990919    -7.871022
C         -10.509491     7.569343    -8.889377
H         -10.379547     6.693432    -9.506586
H         -10.788509     8.397524    -9.528546
C          -9.153433     7.886025    -8.267823
O          -8.135563     7.485103    -8.776033
N          -9.165162     8.668757    -7.174558
H         -10.015005     8.787889    -6.668293
C          -7.930996     9.022917    -6.517095
H          -7.245709     9.471744    -7.218633
H          -8.157589     9.744986    -5.742581
C          -7.203842     7.843048    -5.887301
O          -6.010549     7.713098    -6.031561
```

```
N           -7.940153    6.978615   -5.166919
C           -7.282439    5.870085   -4.504788
C           -6.837253    4.792619   -5.495093
O           -5.867265    4.115873   -5.231040
C           -8.052661    5.309609   -3.288939
C           -9.399398    4.656316   -3.634432
C           -8.184436    6.399786   -2.218453
C          -10.048800    3.936008   -2.448659
H           -8.917073    7.141000   -5.044263
H           -6.343528    6.242177   -4.124035
H           -7.404596    4.532587   -2.892931
H          -10.084698    5.406150   -4.019826
H           -9.254291    3.939002   -4.437940
H           -8.569050    5.990809   -1.294362
H           -7.220115    6.848597   -2.000741
H           -8.858979    7.189550   -2.531892
H          -10.895674    3.347472   -2.780805
H           -9.349344    3.250813   -1.978183
H          -10.395930    4.629302   -1.691205
N           -7.497146    4.646499   -6.656328
C           -6.955434    3.790353   -7.695435
C           -5.584994    4.270936   -8.181064
O           -4.695212    3.475990   -8.368905
C           -7.925115    3.678649   -8.882947
C           -9.147100    2.789986   -8.618425
C           -8.792888    1.301407   -8.647476
N           -9.924610    0.423658   -8.363139
C          -10.316798    0.031936   -7.151338
N           -9.767963    0.539598   -6.063491
N          -11.279255   -0.869045   -7.031647
H           -8.308613    5.197585   -6.841150
H           -6.751015    2.816150   -7.276571
H           -8.250694    4.673499   -9.163147
H           -7.372961    3.284977   -9.731602
H           -9.614228    3.060637   -7.678203
H           -9.891623    2.975994   -9.385661
H           -8.437725    1.028967   -9.632239
H           -7.989179    1.061069   -7.962809
H          -10.339702   -0.044900   -9.136731
H           -9.184043    1.339617   -6.136101
H          -10.175598    0.375658   -5.144654
H          -11.572614   -1.370822   -7.840785
H          -11.349185   -1.384239   -6.160572
N           -5.437808    5.584980   -8.404204
H           -6.200077    6.213622   -8.269221
C           -4.172638    6.109182   -8.857730
H           -3.863276    5.645046   -9.782324
H           -4.290505    7.172230   -9.024452
C           -3.036846    5.894906   -7.863905
O           -1.941083    5.560613   -8.253380
N           -3.308176    6.109592   -6.569768
C           -2.280558    5.949053   -5.558695
C           -1.840851    4.491663   -5.423531
O           -0.669248    4.226723   -5.264845
```

```
C        -2.752176     6.526203    -4.207560
C        -1.798644     6.152641    -3.067592
C        -2.897825     8.049753    -4.300640
H        -4.211907     6.446897    -6.309741
H        -1.388033     6.472249    -5.873403
H        -3.725936     6.092646    -3.990897
H        -2.128473     6.615701    -2.148325
H        -1.761368     5.083217    -2.897924
H        -0.789188     6.496503    -3.262295
H        -3.286333     8.445424    -3.368263
H        -1.931507     8.513425    -4.477386
H        -3.572579     8.354850    -5.089810
N        -2.785068     3.539733    -5.451110
C        -2.405779     2.151642    -5.286810
C        -1.536974     1.641604    -6.446497
O        -0.719857     0.779645    -6.231839
C        -3.616174     1.248279    -5.017044
C        -4.319607     1.586178    -3.694304
C        -5.440222     0.610232    -3.319311
N        -4.903611    -0.600437    -2.703796
C        -5.580005    -1.728323    -2.497206
N        -6.794759    -1.914194    -2.996513
N        -5.068662    -2.694382    -1.756815
H        -3.743963     3.782514    -5.591284
H        -1.741079     2.083389    -4.436740
H        -4.321121     1.331199    -5.839288
H        -3.262140     0.225117    -5.005791
H        -3.595653     1.622364    -2.885395
H        -4.760864     2.570196    -3.758504
H        -6.132076     1.089314    -2.634872
H        -5.989367     0.343471    -4.214599
H        -4.037748    -0.502299    -2.212065
H        -7.151760    -1.318637    -3.705178
H        -7.158958    -2.857872    -2.946628
H        -4.185638    -2.630124    -1.285420
H        -5.538408    -3.584281    -1.773792
N        -1.693919     2.203548    -7.659635
H        -2.434861     2.853041    -7.813075
C        -0.782876     1.910839    -8.744047
H        -1.252760     2.212990    -9.671955
H        -0.572839     0.854467    -8.790203
C         0.571331     2.620905    -8.637269
O         1.602375     2.035839    -8.880235
N         0.547745     3.911968    -8.268518
H        -0.324513     4.376513    -8.123722
C         1.758142     4.679816    -8.115958
H         2.345083     4.684350    -9.024485
H         1.484546     5.702086    -7.890156
C         2.689962     4.174159    -7.018870
O         3.863530     4.473991    -7.019218
N         2.134455     3.354312    -6.119817
C         2.855751     2.609176    -5.118344
C         4.035116     1.827352    -5.683420
O         4.945548     1.525206    -4.933221
```

```
C         1.837821    1.651598   -4.480832
C         2.323421    0.787624   -3.319311
C         1.232773   -0.225269   -2.990598
N         1.582972   -0.995771   -1.802453
C         0.755062   -1.867047   -1.219299
N        -0.297809   -2.327048   -1.873180
N         1.000743   -2.307672   -0.006293
H         1.137645    3.326301   -6.079790
H         3.288729    3.265686   -4.372344
H         1.006824    2.259064   -4.140803
H         1.444320    1.019042   -5.265512
H         3.239620    0.265704   -3.565154
H         2.527431    1.409678   -2.449728
H         0.287975    0.282224   -2.826809
H         1.109953   -0.895381   -3.832312
H         2.338149   -0.662992   -1.247429
H        -0.545998   -1.941104   -2.754746
H        -1.043861   -2.759586   -1.344959
H         1.738290   -1.933299    0.545758
H         0.305395   -2.876294    0.475894
N         4.048648    1.484991   -6.971734
H         3.270969    1.698345   -7.566193
C         5.210550    0.841254   -7.539628
H         5.056502    0.750865   -8.607213
H         5.356141   -0.150021   -7.134802
C         6.521892    1.587079   -7.305181
O         7.563326    0.972601   -7.244040
N         6.475481    2.921189   -7.165026
H         5.600693    3.407455   -7.204822
C         7.672401    3.672050   -6.884269
H         8.426673    3.507615   -7.641214
H         7.423623    4.725239   -6.888416
C         8.343285    3.361582   -5.549619
O         9.482150    3.734701   -5.345074
N         7.643106    2.670340   -4.646852
C         8.198808    2.237806   -3.383690
C         9.371507    1.266376   -3.547780
O        10.148407    1.109809   -2.619457
C         7.066373    1.577140   -2.566069
C         7.274969    1.470372   -1.052355
C         6.986616    2.774004   -0.295060
C         6.971171    2.589923    1.220527
N         8.341553    2.635672    1.811074
O        10.640659    1.764232    0.190819
H        10.482183    1.912765   -0.740219
H        11.081160    0.914255    0.183597
H         6.712507    2.376193   -4.867102
H         8.608445    3.090669   -2.858951
H         6.161077    2.138965   -2.754400
H         6.885252    0.590048   -2.977564
H         6.580835    0.717813   -0.687660
H         8.266233    1.100320   -0.828018
H         7.692214    3.557024   -0.558735
H         6.007531    3.142483   -0.583548
```

```
H         6.402896     3.366813     1.708619
H         6.537572     1.640006     1.493862
H         9.044163     2.181251     1.233629
H         8.660844     3.603813     1.979296
H         8.348317     2.172638     2.709109
N         9.489581     0.614283    -4.703485
C        10.575573    -0.305852    -5.001144
C        11.947333     0.360147    -5.140079
O        12.926228    -0.351046    -5.178643
C        10.276219    -1.064592    -6.308562
C         9.186773    -2.145235    -6.206514
C         8.709761    -2.529387    -7.610329
C         9.690347    -3.382865    -5.456501
H         8.814610     0.776825    -5.424335
H        10.685352    -1.003420    -4.182984
H         9.996480    -0.334449    -7.061605
H        11.198753    -1.525590    -6.639049
H         8.333510    -1.736251    -5.669837
H         7.949025    -3.302565    -7.562386
H         8.287941    -1.672847    -8.123139
H         9.531514    -2.913513    -8.209076
H         8.909461    -4.132875    -5.384843
H        10.531569    -3.831058    -5.977560
H        10.013711    -3.154396    -4.445491
N        12.016310     1.693814    -5.229239
C        13.216647     2.392553    -5.654992
C        14.384790     2.423025    -4.665310
O        15.418090     2.917302    -5.033118
C        12.851735     3.818288    -6.074752
O        12.314695     4.545227    -5.009278
H        11.170882     2.220565    -5.199224
H        13.629609     1.881541    -6.517432
H        13.752812     4.310043    -6.407255
H        12.158019     3.772875    -6.908360
H        11.366436     4.532657    -5.049725
N        14.244709     1.858581    -3.447960
C        15.394891     1.632051    -2.595867
C        16.202206     0.374371    -2.942027
O        17.285726     0.220136    -2.441351
C        14.993654     1.481006    -1.125052
C        14.242080     2.667883    -0.532839
C        13.879537     2.370651     0.919843
N        12.968760     3.344932     1.504487
C        13.293825     4.435668     2.181902
N        14.563230     4.743270     2.435426
O        14.839592     7.243417     4.013729
H        14.590200     8.153281     3.833300
H        15.510059     7.288100     4.683994
N        12.343962     5.226082     2.626306
H        13.365364     1.468527    -3.197915
H        16.068700     2.465945    -2.715960
H        14.385083     0.588955    -1.020991
H        15.907592     1.302013    -0.571027
H        14.833958     3.575511    -0.601688
```

```
H              13.328791       2.858447      -1.083875
H              13.377631       1.415570       0.978580
H              14.761954       2.284960       1.542685
H              12.004971       3.218153       1.272666
H              15.308476       4.242582       2.013505
H              14.778564       5.588773       2.932063
H              11.359458       5.037358       2.509350
H              12.552390       6.013580       3.198162
N              15.599779      -0.536970      -3.727008
C              16.154350      -1.845509      -3.977565
C              16.568461      -1.958209      -5.439343
H              16.949390      -1.030135      -5.873859
O              16.514203      -2.961905      -6.070665
C              15.192999      -2.975821      -3.580079
C              15.023147      -3.169855      -2.100869
N              14.146059      -2.434677      -1.328658
O              11.931793      -0.489578      -0.947327
H              12.620548      -1.094460      -1.228306
H              11.424866      -0.260513      -1.718903
C              15.644438      -4.079300      -1.315252
C              14.257069      -2.898181      -0.126209
N              15.154892      -3.889389      -0.045603
H              14.710250      -0.331583      -4.130490
H              17.067694      -1.912344      -3.398192
H              14.225531      -2.788229      -4.033444
H              15.569185      -3.893524      -4.012757
H              16.378496      -4.823357      -1.538426
H              13.704377      -2.550874       0.723010
H              15.313768      -4.465716       0.751079
O             -14.664764      -2.587899       5.657014
H             -14.904863      -1.842393       5.112328
H             -14.188158      -2.224294       6.395793
O             -12.603144      -1.192596       7.488951
H             -11.903456      -1.668192       7.922436
H             -12.843089      -0.482425       8.071472
O             -11.807406      -0.826721       4.676906
H             -11.935936      -0.808371       5.621841
H             -12.651533      -0.553201       4.312568
O             -13.191779      -2.261510       1.389614
H             -12.456015      -1.646903       1.346915
H             -13.035699      -2.830317       2.145493
O              -7.290826       1.110690       0.127370
H              -7.527309       1.516899       0.963566
H              -6.369373       0.906843       0.219818
O              -7.753731       2.024902       2.842836
H              -8.674401       1.853347       3.038210
H              -7.491955       2.757352       3.392147
O              -9.432838      -0.372862      -1.129667
H              -8.629005      -0.000608      -0.764131
H              -9.280560      -1.292785      -1.344290
O             -12.622759      -3.172299      -1.344950
H             -12.824222      -3.080160      -0.416535
H             -11.671571      -3.241756      -1.433822
O             -14.129668       0.183480      -0.135912
```

```
H           -14.340806   -0.698985    0.152076
H           -13.206006    0.284451    0.109294
O           -14.160008    4.925944    4.813007
H           -13.632326    5.597040    4.388263
H           -14.162246    5.182311    5.730412
O           -15.190343    1.695506    2.005406
H           -15.950755    2.253139    1.907646
H           -14.973166    1.339612    1.142679
O           -13.298191   -0.794540   -2.785924
H           -13.785032   -0.262934   -2.162341
H           -13.013219   -1.555435   -2.277457
O           -13.562051   -3.472586   -4.134422
H           -13.890486   -2.590885   -3.992917
H           -13.469955   -3.829543   -3.256252
O            -9.902635   -3.023078   -2.197938
H           -10.250834   -2.693289   -3.025739
H            -9.477798   -3.856843   -2.442568
O           -11.481243   -0.096122    0.826425
H           -10.785773   -0.184352    0.171059
H           -11.124980    0.434269    1.543397
O           -10.568024    1.296591    3.141989
H           -10.744541    0.589398    3.760649
H           -11.277531    1.926170    3.303685
O           -12.945219    2.630598    3.680639
H           -13.281303    3.378339    4.177846
H           -13.636640    2.416924    3.062051
O           -10.851536    0.529805   -3.493900
H           -11.701653    0.114468   -3.338528
H           -10.328999    0.335606   -2.716641
O           -14.448952   -0.593604    3.570986
H           -14.817569    0.233551    3.265283
H           -14.307354   -1.119869    2.789173
O           -12.672399   -3.504015    3.913369
H           -13.438216   -3.375425    4.482630
H           -12.077761   -2.807095    4.186447
O           -11.010902   -2.604005   -4.847480
H           -11.812510   -3.112560   -4.692592
H           -10.330446   -3.182956   -5.195606
```

Case 7:
Voltage -70 mV,Y266 neutral,R300 positive; E183 negative
```
N            15.580944   -6.451469    5.391415
H            14.785021   -6.202855    5.946150
C            15.502157   -5.853385    4.056239
C            14.783017   -6.782962    3.060453
O            14.049936   -6.347324    2.187893
C            16.908605   -5.556335    3.522538
H            16.393783   -6.132236    5.880519
H            14.926056   -4.937087    4.055154
H            16.873357   -5.166701    2.512237
H            17.385293   -4.808865    4.146467
```

```
H         17.525125     -6.448468      3.527564
N         15.032376     -8.085605      3.186404
H         15.613995     -8.361941      3.948242
C         14.398661     -9.092388      2.333235
H         14.483994     -8.825420      1.292572
H         14.889302    -10.042967      2.494876
C         12.912184     -9.216795      2.670679
O         12.078850     -9.413014      1.824879
N         12.582803     -9.055884      3.966872
H         13.303111     -8.992448      4.649211
C         11.195101     -9.170813      4.421394
H         11.187863     -9.157795      5.503118
H         10.742827    -10.087813      4.077044
C         10.336957     -8.010601      3.900715
O          9.187625     -8.187424      3.594460
N         10.943091     -6.812644      3.842472
C         10.298641     -5.641820      3.300751
C         10.119909     -5.700436      1.781122
O          9.089280     -5.281513      1.309377
C         11.040627     -4.380353      3.764011
O         12.431303     -4.516879      3.720508
H         11.904883     -6.736779      4.079992
H          9.290882     -5.581491      3.676837
H         10.700048     -3.531586      3.179855
H         10.794878     -4.196404      4.799781
H         12.741828     -4.692398      2.841067
N         11.103321     -6.202518      1.014352
C         10.856282     -6.407444     -0.410126
C          9.704113     -7.394663     -0.609331
O          8.960894     -7.293116     -1.557963
C         12.147946     -6.803021     -1.188940
C         12.337678     -8.309066     -1.411493
C         12.190103     -6.071683     -2.532224
H         11.926343     -6.584013      1.428135
H         10.489955     -5.475303     -0.809727
H         12.978610     -6.442645     -0.592488
H         13.292334     -8.477860     -1.898700
H         12.327687     -8.878784     -0.493118
H         11.570545     -8.707137     -2.068617
H         13.087958     -6.337744     -3.078614
H         11.333645     -6.336699     -3.142595
H         12.194573     -4.996937     -2.394499
N          9.570414     -8.340327      0.328325
H         10.315278     -8.476771      0.973469
C          8.593031     -9.421194      0.241221
H          8.937059    -10.220512      0.885312
H          8.531140     -9.801463     -0.766745
C          7.183165     -9.048833      0.647817
O          6.257091     -9.499143      0.014086
N          7.005605     -8.234324      1.694904
C          5.672926     -7.742221      1.956147
C          5.263457     -6.735497      0.886388
O          4.085265     -6.614576      0.623565
C          5.411119     -7.228569      3.386154
```

```
C        5.548284   -8.366370    4.402828
C        6.249468   -6.016562    3.797878
H        7.781669   -7.973699    2.264124
H        4.987402   -8.559854    1.794714
H        4.369752   -6.921526    3.372145
H        5.252547   -8.023076    5.388471
H        4.912310   -9.205814    4.141654
H        6.570309   -8.722482    4.466210
H        5.894886   -5.632793    4.748561
H        7.288815   -6.289625    3.924349
H        6.193254   -5.209355    3.074876
N        6.192691   -6.040018    0.222666
C        5.810873   -5.239013   -0.928245
C        5.380097   -6.134264   -2.097802
O        4.350547   -5.889434   -2.683393
C        6.909958   -4.228953   -1.313853
C        7.075336   -3.183320   -0.198361
C        6.563762   -3.552205   -2.644273
C        8.365563   -2.368697   -0.286706
H        7.154063   -6.112755    0.477530
H        4.911910   -4.696554   -0.677276
H        7.844152   -4.769390   -1.428829
H        6.218415   -2.510961   -0.234752
H        7.052411   -3.670517    0.767701
H        7.263087   -2.757578   -2.863956
H        6.592715   -4.245163   -3.475003
H        5.567351   -3.122420   -2.612945
H        8.431204   -1.674955    0.545559
H        9.230862   -3.018589   -0.240625
H        8.427894   -1.788110   -1.200274
N        6.153730   -7.173520   -2.432180
H        7.049867   -7.292155   -2.007908
C        5.802095   -8.026518   -3.547887
H        6.621603   -8.716873   -3.703661
H        5.659531   -7.453507   -4.451995
C        4.518399   -8.827429   -3.332630
O        3.678831   -8.884478   -4.199072
N        4.393731   -9.454836   -2.156214
H        5.136308   -9.422220   -1.490543
C        3.200819  -10.197518   -1.824418
H        2.956735  -10.915276   -2.593078
H        3.385397  -10.732958   -0.901532
C        1.974192   -9.306850   -1.644816
O        0.885650   -9.692198   -2.000856
N        2.187868   -8.107003   -1.107172
C        1.067348   -7.177714   -1.097647
C        0.633953   -6.722406   -2.495885
O       -0.542764   -6.542164   -2.709934
C        1.316595   -5.900137   -0.311136
O        1.460230   -6.079143    1.063953
H        3.095652   -7.806832   -0.833743
H        0.209361   -7.678030   -0.679415
H        2.174087   -5.389733   -0.727744
H        0.439265   -5.287200   -0.447472
```

```
H         2.308353    -6.473521     1.225962
N         1.571161    -6.469596    -3.419189
C         1.189951    -6.071326    -4.763294
C         0.416401    -7.191702    -5.450522
O        -0.560360    -6.930888    -6.117929
C         2.411320    -5.590763    -5.580531
C         2.832630    -4.198619    -5.081292
C         2.123151    -5.578342    -7.087336
C         4.164071    -3.708300    -5.648993
H         2.538332    -6.567554    -3.197631
H         0.476822    -5.262097    -4.697121
H         3.223558    -6.287978    -5.398270
H         2.048647    -3.486567    -5.334939
H         2.905562    -4.217662    -4.001012
H         2.986952    -5.220853    -7.633781
H         1.897392    -6.567256    -7.466480
H         1.284298    -4.931899    -7.324786
H         4.469599    -2.792706    -5.152732
H         4.947487    -4.442678    -5.491950
H         4.109250    -3.500187    -6.711344
N         0.840355    -8.445451    -5.268678
H         1.682955    -8.626020    -4.764947
C         0.044204    -9.547645    -5.751679
H        -0.060686    -9.523773    -6.826027
H         0.538316   -10.468835    -5.470118
C        -1.365265    -9.540099    -5.166932
O        -2.321156    -9.747003    -5.876935
N        -1.472641    -9.318885    -3.858021
C        -2.779476    -9.325927    -3.246195
C        -3.677519    -8.205982    -3.765931
O        -4.865607    -8.426703    -3.853082
C        -2.697449    -9.238741    -1.716406
O        -3.948230    -9.470513    -1.131684
H        -0.657342    -9.264450    -3.284872
H        -3.299116   -10.238362    -3.503205
H        -2.025269   -10.004008    -1.352226
H        -2.302522    -8.271923    -1.421794
H        -4.635949    -9.102095    -1.673654
N        -3.119584    -7.024876    -4.043112
C        -3.864304    -5.902895    -4.595444
C        -4.405272    -6.232215    -5.985557
O        -5.559050    -6.013108    -6.267223
C        -2.934930    -4.671684    -4.626800
C        -3.395754    -3.498328    -5.477006
C        -4.729209    -3.119502    -5.590920
C        -2.442384    -2.764455    -6.176855
C        -5.093276    -2.050557    -6.395674
C        -2.800120    -1.687385    -6.970177
C        -4.133888    -1.330501    -7.087546
H        -2.146532    -6.906278    -3.857472
H        -4.727902    -5.701015    -3.982308
H        -2.783759    -4.353900    -3.599099
H        -1.971184    -4.990296    -4.999597
H        -5.497120    -3.669808    -5.084019
```

```
H        -1.405943    -3.048335    -6.114881
H        -6.136300    -1.804398    -6.490910
H        -2.042152    -1.142175    -7.504099
H        -4.419641    -0.511092    -7.724224
N        -3.537051    -6.747892    -6.865978
C        -4.002526    -7.110558    -8.184570
C        -5.143093    -8.125283    -8.119673
O        -6.031712    -8.076085    -8.927474
C        -2.873229    -7.678453    -9.045859
S        -1.627290    -6.462431    -9.582638
H        -2.609753    -6.976093    -6.577206
H        -4.423349    -6.244067    -8.672205
H        -2.385047    -8.503317    -8.544312
H        -3.305644    -8.056527    -9.962340
H        -1.036681    -6.271335    -8.413212
N        -5.078560    -9.066906    -7.155629
H        -4.254254    -9.180362    -6.607446
C        -6.081132   -10.120388    -7.166259
H        -6.245462   -10.479933    -8.170409
H        -5.709428   -10.937807    -6.559223
C        -7.459089    -9.722699    -6.613094
O        -8.475534   -10.093070    -7.133958
N        -7.418039    -9.015884    -5.478394
C        -8.618010    -8.617445    -4.761832
C        -9.496235    -7.642413    -5.528484
O       -10.689800    -7.642058    -5.391606
C        -8.313976    -8.009342    -3.373434
C        -7.340433    -6.814995    -3.275733
C        -7.860948    -5.385747    -3.159234
O        -9.030795    -5.113416    -3.476233
O        -7.019484    -4.535418    -2.797074
H        -6.529986    -8.768045    -5.098646
H        -9.240663    -9.488335    -4.609484
H        -9.269885    -7.742213    -2.947889
H        -7.910226    -8.814158    -2.767180
H        -6.691550    -6.961260    -2.422453
H        -6.679726    -6.791988    -4.127072
N        -8.837318    -6.734108    -6.273932
C        -9.553892    -5.702551    -7.001622
C       -10.155236    -6.201982    -8.316521
H        -9.841903    -7.199129    -8.634189
O       -10.938392    -5.563328    -8.942857
C        -8.639719    -4.464230    -7.138145
O        -8.676689    -3.734662    -5.925805
C        -9.007020    -3.482919    -8.241079
H        -7.863977    -6.858754    -6.450332
H       -10.409136    -5.415107    -6.404328
H        -7.624985    -4.803333    -7.306792
H        -8.710269    -4.327140    -5.171401
H        -8.364201    -2.613268    -8.144071
H        -8.855454    -3.910880    -9.222713
H       -10.042635    -3.173524    -8.167677
N       -12.952208    -9.022351     3.174334
H       -13.208292    -9.557741     2.370724
```

```
C            -11.773679       -8.184330        2.897671
C            -10.728602       -8.154325        4.045625
O             -9.589253       -7.797737        3.850980
C            -12.122649       -6.717443        2.588063
O            -12.631020       -6.187409        3.784351
C            -13.125896       -6.567985        1.449441
H            -13.734650       -8.453874        3.433861
H            -11.243676       -8.593789        2.049417
H            -11.200748       -6.216201        2.318547
H            -12.747705       -5.243810        3.686863
H            -13.265311       -5.516210        1.226028
H            -12.786279       -7.059629        0.545752
H            -14.092406       -6.974064        1.723604
N            -11.197369       -8.533629        5.242004
H            -12.133712       -8.874113        5.246101
C            -10.436278       -8.584864        6.466908
H            -10.391098       -7.615919        6.940815
H            -10.949903       -9.254617        7.147284
C             -8.983893       -9.054293        6.370312
O             -8.148158       -8.454576        7.001844
N             -8.654475      -10.138343        5.644662
H             -9.360988      -10.642602        5.158190
C             -7.261102      -10.550098        5.542456
H             -7.231408      -11.534033        5.091831
H             -6.816416      -10.613470        6.522880
C             -6.387469       -9.608355        4.692670
O             -5.215440       -9.474882        4.943167
N             -7.045367       -9.017982        3.701150
C             -6.295611       -8.049283        2.909681
C             -6.268440       -6.678279        3.585879
O             -5.290769       -5.973025        3.460720
C             -6.586302       -8.033855        1.398610
C             -7.829852       -7.351651        0.873235
C             -7.904178       -5.962565        0.817026
C             -8.863335       -8.092598        0.318204
C             -8.979188       -5.332961        0.215926
C             -9.938583       -7.464998       -0.291951
C             -9.997166       -6.083720       -0.351314
H             -8.038901       -9.034088        3.660295
H             -5.270993       -8.370031        2.980552
H             -5.722545       -7.562522        0.943304
H             -6.566025       -9.067200        1.073297
H             -7.095810       -5.369096        1.207068
H             -8.809738       -9.168107        0.321673
H             -9.001882       -4.260585        0.147700
H            -10.707846       -8.055336       -0.757078
H            -10.804259       -5.598564       -0.867215
N             -7.295258       -6.344331        4.368310
C             -7.255302       -5.218373        5.268686
C             -6.160092       -5.418497        6.315382
O             -5.505436       -4.466729        6.685733
C             -8.653123       -5.065058        5.904903
C             -8.878776       -3.813700        6.725582
C             -9.209007       -2.613978        6.103976
```

```
C         -8.820184    -3.850026     8.113523
C         -9.468806    -1.476774     6.850851
C         -9.083883    -2.716071     8.866253
C         -9.410576    -1.525593     8.236772
H         -8.132201    -6.882311     4.323665
H         -6.986849    -4.314748     4.740473
H         -9.370158    -5.087911     5.091647
H         -8.842989    -5.936480     6.517269
H         -9.276369    -2.566192     5.030810
H         -8.575377    -4.771061     8.613800
H         -9.716726    -0.558363     6.351001
H         -9.029841    -2.763147     9.938903
H         -9.603171    -0.640860     8.817699
N         -5.947434    -6.647872     6.780459
H         -6.610812    -7.364419     6.585308
C         -4.902083    -6.940179     7.735327
H         -4.936409    -6.272209     8.582792
H         -5.056042    -7.952124     8.087416
C         -3.494272    -6.827420     7.156640
O         -2.624497    -6.290174     7.800602
N         -3.272166    -7.337433     5.938674
C         -1.978588    -7.176100     5.296956
C         -1.715475    -5.703150     4.968912
O         -0.620994    -5.222846     5.169977
C         -1.874258    -8.087931     4.054793
C         -0.672959    -7.729276     3.175019
C         -1.799752    -9.558159     4.482833
H         -3.983642    -7.872330     5.486589
H         -1.199565    -7.439587     5.998357
H         -2.775099    -7.942041     3.463878
H         -0.600517    -8.436466     2.355416
H         -0.745282    -6.739607     2.744123
H          0.253407    -7.777635     3.739933
H         -1.796135   -10.201001     3.608575
H         -0.882464    -9.744714     5.035566
H         -2.636424    -9.852500     5.102068
N         -2.731953    -4.993853     4.462229
C         -2.541382    -3.621981     4.042019
C         -2.424609    -2.666595     5.232647
O         -1.569310    -1.800340     5.211966
C         -3.658688    -3.157644     3.097996
C         -3.679884    -3.843315     1.717460
C         -2.491046    -3.465364     0.847494
O         -1.384349    -3.891138     1.193166
O         -2.692889    -2.734612    -0.153788
H         -3.599497    -5.444420     4.260096
H         -1.599553    -3.554918     3.528196
H         -4.618960    -3.318470     3.573759
H         -3.546124    -2.085800     2.970132
H         -3.687068    -4.916686     1.843023
H         -4.595860    -3.562472     1.211405
N         -3.278891    -2.802110     6.250271
C         -3.097635    -2.026500     7.460765
C         -1.898256    -2.497718     8.276556
```

```
O         -1.355397    -1.698391     9.008897
C         -4.330795    -1.877677     8.363760
O         -4.684190    -3.080192     8.982507
C         -5.503720    -1.212566     7.653717
H         -3.947091    -3.541791     6.234934
H         -2.821376    -1.030817     7.155813
H         -3.999779    -1.231502     9.165584
H         -5.121301    -3.642000     8.354089
H         -6.307936    -1.056354     8.361573
H         -5.212199    -0.248201     7.248679
H         -5.882127    -1.823444     6.844004
N         -1.380394    -3.700443     8.043325
H         -1.953663    -4.436417     7.693766
C         -0.146512    -4.077234     8.683129
H         -0.137056    -3.828186     9.733250
H         -0.025560    -5.146881     8.569973
C          1.045451    -3.381567     8.040949
O          1.971271    -2.968398     8.700895
N          1.011155    -3.267375     6.700333
C          2.126201    -2.678433     6.006829
C          2.215179    -1.168954     6.199352
O          3.309994    -0.657503     6.261151
C          2.148696    -3.049853     4.514550
S          1.274055    -1.936526     3.367962
H          0.285226    -3.725281     6.187851
H          3.039372    -3.054186     6.446601
H          3.177791    -3.010201     4.182407
H          1.787538    -4.061167     4.390514
H          0.053658    -2.006304     3.875460
N          1.083014    -0.458322     6.294713
C          1.196019     0.973827     6.493589
C          1.784615     1.258868     7.872419
O          2.423969     2.262552     8.080728
C         -0.105029     1.769082     6.255199
C         -1.161950     1.578002     7.355466
C         -0.635179     1.516324     4.839009
C         -2.456649     2.356458     7.113937
H          0.197361    -0.899873     6.159278
H          1.928591     1.344993     5.793107
H          0.211712     2.805461     6.308627
H         -1.390097     0.528140     7.487236
H         -0.741280     1.909262     8.299158
H         -1.367223     2.265429     4.566648
H          0.169127     1.566552     4.110701
H         -1.104982     0.546307     4.745217
H         -3.092266     2.308032     7.992198
H         -2.258378     3.405070     6.907223
H         -3.021801     1.955001     6.279607
N          1.549786     0.323264     8.791266
H          0.923997    -0.423306     8.592383
C          1.980720     0.495222    10.161894
H          1.457930    -0.234171    10.767396
H          1.750673     1.485977    10.523432
C          3.481337     0.301983    10.326891
```

```
O         4.150152    1.061208   10.974214
N         4.007440   -0.769496    9.694199
H         3.411841   -1.412808    9.217692
C         5.433404   -0.945622    9.663451
H         5.651842   -1.914471    9.229648
H         5.857917   -0.911709   10.655684
C         6.156883    0.119119    8.844832
O         7.288307    0.441807    9.117267
N         5.502080    0.621278    7.799024
C         6.099985    1.759183    7.147111
C         6.127065    2.971177    8.081538
H         5.699676    2.831840    9.073113
O         6.593560    4.008457    7.728905
C         5.411190    2.089556    5.825967
C         5.559185    1.045604    4.730895
C         6.583201    0.102871    4.703112
C         4.626396    1.023760    3.696286
C         6.663728   -0.833907    3.681128
C         4.701023    0.092490    2.676893
C         5.723333   -0.846360    2.665139
H         4.573808    0.337101    7.574383
H         7.145705    1.543895    6.970444
H         4.357330    2.251125    6.011128
H         5.813751    3.036218    5.479655
H         7.296684    0.054811    5.504468
H         3.814871    1.728808    3.705942
H         7.449036   -1.570135    3.696076
H         3.950745    0.090883    1.905710
H         5.782016   -1.584268    1.886263
N        12.613548    9.596235   -1.248798
H        11.868651    9.918628   -1.837894
C        12.304854    8.264737   -0.719490
C        11.280789    8.458704    0.391593
O        10.270261    7.790383    0.428870
C        13.564016    7.577453   -0.173437
C        14.641679    7.276546   -1.220645
S        14.006590    6.210540   -2.547094
C        15.484327    6.065378   -3.580289
H        13.427425    9.571615   -1.830747
H        11.827855    7.618321   -1.444695
H        14.006146    8.197780    0.600300
H        13.257428    6.648440    0.293134
H        15.041780    8.190638   -1.644588
H        15.468209    6.780218   -0.724789
H        15.226916    5.417808   -4.404524
H        15.784331    7.031546   -3.964871
H        16.303799    5.621712   -3.030137
N        11.517460    9.426975    1.292757
C        10.614326    9.633799    2.407651
C         9.291141   10.286996    2.007164
O         8.297462   10.062563    2.652547
C        11.299956   10.416970    3.534385
C        12.477580    9.624330    4.071835
O        13.606987    9.869599    3.702802
```

```
N        12.184782     8.650819     4.943054
H        12.343656     9.977115     1.194948
H        10.323772     8.668552     2.788639
H        11.675834    11.369738     3.183039
H        10.573839    10.596402     4.317158
H        12.929542     8.058766     5.241636
H        11.261880     8.273061     5.008011
N         9.268406    11.071441     0.913418
H        10.113490    11.244350     0.418868
C         8.003498    11.517477     0.368851
H         8.193586    12.264042    -0.391673
H         7.391681    11.961877     1.136819
C         7.186389    10.384432    -0.254316
O         5.985324    10.464096    -0.282421
N         7.863792     9.337546    -0.769733
C         7.174532     8.144908    -1.231856
C         6.652245     7.301342    -0.057631
O         5.557984     6.793570    -0.130865
C         8.051433     7.346052    -2.225527
C         8.338173     8.198846    -3.476267
C         7.384831     6.018647    -2.605837
C         9.421267     7.629690    -4.395982
H         8.827013     9.240595    -0.548811
H         6.278713     8.464401    -1.741707
H         8.993349     7.123212    -1.732143
H         7.413535     8.321021    -4.036015
H         8.637949     9.195811    -3.171212
H         7.999932     5.480917    -3.315986
H         7.240986     5.376895    -1.745277
H         6.412048     6.179241    -3.058944
H         9.609786     8.310348    -5.219735
H        10.359110     7.487865    -3.865800
H         9.143039     6.674403    -4.825556
N         7.410517     7.196682     1.051299
C         6.916572     6.489997     2.220833
C         5.564635     7.066595     2.662530
O         4.622983     6.354032     2.917495
C         7.881556     6.597475     3.412689
C         9.306951     6.057577     3.251097
O        10.192679     6.689290     3.817094
O         9.488211     5.002532     2.590802
H         8.368121     7.474148     1.021987
H         6.723419     5.457838     1.974910
H         7.964161     7.625890     3.731277
H         7.437088     6.049896     4.241178
N         5.512890     8.406459     2.779547
H         6.338987     8.946176     2.629960
C         4.319354     9.072780     3.238174
H         4.550718    10.121496     3.377376
H         3.985043     8.679063     4.185048
C         3.152293     8.990537     2.263559
O         2.011433     9.100236     2.650183
N         3.478598     8.822190     0.978486
H         4.438076     8.845404     0.721746
```

```
C         2.517177    8.905268   -0.078293
H         3.052846    9.096728   -0.999396
H         1.843617    9.736394    0.079918
C         1.609145    7.687011   -0.310212
O         0.561942    7.813198   -0.879966
N         2.094042    6.538075    0.203123
C         1.292851    5.360703    0.425345
C         0.475793    5.458325    1.718345
O        -0.712641    5.241430    1.694715
C         2.172954    4.110693    0.430361
H         3.004897    6.574244    0.609571
H         0.562221    5.289936   -0.363374
H         1.574252    3.229150    0.633171
H         2.645561    3.999958   -0.538721
H         2.951285    4.178212    1.183482
N         1.105452    5.755333    2.871338
C         0.372742    5.605585    4.119553
C        -0.747903    6.630862    4.262533
O        -1.782920    6.320555    4.813389
C         1.280783    5.556440    5.372406
C         1.878648    6.908093    5.792560
C         2.360276    4.480055    5.197937
C         2.731543    6.830467    7.061331
H         2.086253    5.934921    2.884772
H        -0.148298    4.660478    4.076136
H         0.615922    5.237943    6.169710
H         2.474422    7.309420    4.979882
H         1.076781    7.618141    5.965150
H         2.721708    4.149461    6.161223
H         1.959769    3.611736    4.684666
H         3.204151    4.849092    4.624452
H         3.027881    7.826297    7.375661
H         2.178258    6.379093    7.879130
H         3.635497    6.250871    6.915940
N        -0.537763    7.864521    3.786565
H         0.346750    8.110623    3.392596
C        -1.512684    8.900779    4.009617
H        -1.111345    9.831115    3.628268
H        -1.716024    9.030426    5.062729
C        -2.866994    8.618767    3.365898
O        -3.868012    8.832389    4.006555
N        -2.902658    8.108970    2.120752
C        -4.243490    8.013012    1.538274
C        -4.976678    6.736957    1.916706
O        -6.185644    6.716773    1.877592
C        -4.036055    8.245951    0.036082
C        -2.778403    9.121517    0.028296
C        -1.913925    8.465923    1.093496
H        -4.860639    8.804068    1.935643
H        -3.864775    7.311542   -0.482143
H        -4.902851    8.719520   -0.406377
H        -2.273239    9.175016   -0.918152
H        -3.035554   10.133163    0.332679
H        -1.429002    7.578783    0.726766
```

| | | | |
|---|---:|---:|---:|
| H | -1.154662 | 9.124341 | 1.482676 |
| N | -4.240641 | 5.682820 | 2.306944 |
| C | -4.877750 | 4.536077 | 2.928105 |
| C | -5.553856 | 4.910085 | 4.252940 |
| O | -6.640714 | 4.455316 | 4.527284 |
| C | -3.873965 | 3.399687 | 3.202938 |
| C | -3.535268 | 2.481545 | 2.044886 |
| C | -4.034743 | 1.185773 | 2.017406 |
| C | -2.653292 | 2.854544 | 1.035870 |
| C | -3.648752 | 0.279262 | 1.040384 |
| C | -2.249269 | 1.957575 | 0.062960 |
| C | -2.726521 | 0.657321 | 0.075149 |
| O | -2.287264 | -0.209041 | -0.872352 |
| H | -3.254112 | 5.802674 | 2.398420 |
| H | -5.672060 | 4.185700 | 2.288026 |
| H | -2.966729 | 3.835505 | 3.607895 |
| H | -4.296245 | 2.790436 | 3.993271 |
| H | -4.733018 | 0.867562 | 2.772097 |
| H | -2.237756 | 3.840576 | 1.024896 |
| H | -4.031671 | -0.723621 | 1.042365 |
| H | -1.543458 | 2.252732 | -0.692996 |
| H | -2.400283 | -1.120315 | -0.564317 |
| N | -4.866793 | 5.708217 | 5.087484 |
| H | -3.936110 | 5.989207 | 4.858906 |
| C | -5.405489 | 6.110702 | 6.367626 |
| H | -4.590691 | 6.501419 | 6.964520 |
| H | -5.837497 | 5.268565 | 6.885022 |
| C | -6.496044 | 7.176848 | 6.275957 |
| O | -7.446820 | 7.159377 | 7.019818 |
| N | -6.342226 | 8.091166 | 5.313866 |
| H | -5.481295 | 8.120700 | 4.813031 |
| C | -7.257630 | 9.220227 | 5.085274 |
| H | -6.780273 | 9.871399 | 4.364593 |
| H | -7.424646 | 9.774668 | 5.996719 |
| C | -8.624765 | 8.829168 | 4.562223 |
| O | -9.608105 | 9.419760 | 4.936827 |
| N | -8.681124 | 7.827664 | 3.665323 |
| C | -9.956017 | 7.458637 | 3.093943 |
| C | -10.867476 | 6.733123 | 4.076335 |
| O | -12.014543 | 6.519225 | 3.742877 |
| C | -9.797800 | 6.635812 | 1.794679 |
| O | -10.978548 | 6.723174 | 1.038552 |
| C | -9.399938 | 5.182174 | 2.023280 |
| H | -7.840203 | 7.396162 | 3.341301 |
| H | -10.492314 | 8.360185 | 2.833006 |
| H | -9.032098 | 7.126925 | 1.208617 |
| H | -11.707823 | 6.480508 | 1.601147 |
| H | -9.209213 | 4.704257 | 1.069745 |
| H | -8.504307 | 5.100018 | 2.625000 |
| H | -10.200931 | 4.637030 | 2.511729 |
| N | -10.353364 | 6.327996 | 5.240897 |
| C | -11.196946 | 5.750629 | 6.270642 |
| C | -12.228329 | 6.737734 | 6.822496 |
| O | -13.270561 | 6.319563 | 7.270373 |

```
C         -10.357739    5.228097    7.446817
C          -9.397498    4.072497    7.125756
C          -8.547053    3.769330    8.363411
C         -10.127896    2.811781    6.655047
H          -9.405292    6.535775    5.461811
H         -11.777019    4.951791    5.838724
H          -9.787258    6.055285    7.856140
H         -11.055259    4.909951    8.215965
H          -8.722430    4.383274    6.335838
H          -7.831434    2.978264    8.157693
H          -7.995415    4.648088    8.679156
H          -9.167390    3.445948    9.196485
H          -9.414585    2.013383    6.472463
H         -10.833060    2.466455    7.409002
H         -10.675756    2.968205    5.733158
N         -11.886359    8.036076    6.865730
C         -12.819293    9.019425    7.334859
C         -13.868566    9.416879    6.317735
H         -14.605881   10.134213    6.695777
O         -13.917282    9.022284    5.201510
H         -11.100564    8.357347    6.341875
H         -13.326742    8.653955    8.219030
H         -12.277727    9.915614    7.622226
N         -11.959310   10.656095   -3.211148
H         -11.756992   11.055526   -2.315912
C         -11.842239    9.204609   -3.171081
C         -11.353632    8.843153   -4.564439
O         -10.285459    8.321559   -4.783665
C         -13.115643    8.453855   -2.719544
C         -13.472123    8.844088   -1.273409
C         -12.951342    6.938683   -2.867598
C         -14.811419    8.291739   -0.781789
H         -12.889812   10.949319   -3.441557
H         -11.030282    8.950125   -2.503213
H         -13.937057    8.774653   -3.364112
H         -12.678449    8.506119   -0.611936
H         -13.506876    9.926409   -1.192162
H         -13.883261    6.425814   -2.666921
H         -12.641581    6.654631   -3.868565
H         -12.210957    6.568236   -2.167296
H         -15.036424    8.681907    0.204382
H         -15.624633    8.576848   -1.444127
H         -14.804351    7.210719   -0.706177
N         -12.160019    9.244791   -5.566153
H         -13.062722    9.584116   -5.326516
C         -11.972870    8.843873   -6.959244
H         -12.064848    7.773938   -7.082933
H         -12.737726    9.331845   -7.549424
C         -10.600034    9.232099   -7.514369
O          -9.960378    8.473155   -8.185333
N         -10.204761   10.496241   -7.241758
H         -10.748603   11.023513   -6.596436
C          -8.894340   11.023987   -7.637097
H          -8.768161   10.915375   -8.703124
```

```
H            -8.877075    12.077107    -7.388559
C            -7.678102    10.347205    -6.988124
O            -6.770577     9.898375    -7.634004
N            -7.659043    10.373885    -5.639634
H            -8.479869    10.660652    -5.156814
C            -6.495260     9.961251    -4.832307
H            -5.596539    10.377017    -5.257671
H            -6.629590    10.360625    -3.834911
C            -6.267882     8.464578    -4.716021
O            -5.129594     8.057800    -4.765368
N            -7.299170     7.633426    -4.488287
C            -6.983907     6.276427    -4.061634
C            -6.419131     5.426128    -5.193055
O            -5.624176     4.548484    -4.941790
C            -8.115535     5.546142    -3.299342
C            -9.312244     5.133604    -4.172253
C            -8.517748     6.362006    -2.065141
C           -10.316682     4.236735    -3.446274
H            -8.241628     7.969734    -4.474371
H            -6.154285     6.346052    -3.374014
H            -7.642581     4.633597    -2.946546
H            -9.814187     6.012434    -4.557037
H            -8.943506     4.586466    -5.036374
H            -9.167203     5.792016    -1.415264
H            -7.645116     6.641669    -1.482815
H            -9.046230     7.267454    -2.340944
H           -11.109453     3.936963    -4.124745
H            -9.848588     3.331529    -3.082046
H           -10.780967     4.740294    -2.608749
N            -6.827616     5.657315    -6.455922
C            -6.208680     4.940208    -7.552196
C            -4.798435     5.448228    -7.863727
O            -3.944929     4.667227    -8.213997
C            -7.108565     4.931841    -8.796768
C            -8.307980     3.981851    -8.657693
C            -7.903914     2.509061    -8.787792
N            -9.002467     1.573390    -8.561784
C            -9.329610     1.001168    -7.400905
N            -8.739234     1.363814    -6.273743
N           -10.252947     0.060887    -7.377157
H            -7.465298     6.402677    -6.635976
H            -6.034997     3.930005    -7.220158
H            -7.464699     5.937519    -8.997743
H            -6.507992     4.639596    -9.652635
H            -8.809997     4.150908    -7.710222
H            -9.034546     4.209706    -9.430336
H            -7.538130     2.317143    -9.787908
H            -7.093257     2.248777    -8.120629
H            -9.493808     1.253074    -9.365930
H            -8.192197     2.192476    -6.245792
H            -9.089911     1.038813    -5.377476
H           -10.585011    -0.326780    -8.232199
H           -10.411486    -0.508383    -6.548477
N            -4.571387     6.757464    -7.722062
```

```
H           -5.311766    7.371691   -7.468095
C           -3.257810    7.352404   -8.011229
H           -2.928258    7.094733   -9.005831
H           -3.359949    8.426478   -7.935407
C           -2.203542    6.870096   -7.026102
O           -1.075343    6.627374   -7.376442
N           -2.633714    6.760030   -5.763094
C           -1.755545    6.458266   -4.660052
C           -1.480548    4.973327   -4.515783
O           -0.353320    4.592677   -4.302718
C           -2.338870    7.066930   -3.378339
C           -1.754906    6.374353   -2.148134
C           -2.092119    8.581512   -3.423935
H           -3.554411    7.075725   -5.545705
H           -0.783733    6.891118   -4.849327
H           -3.408216    6.892599   -3.391010
H           -1.968159    6.924681   -1.251524
H           -2.171235    5.378484   -2.034269
H           -0.682910    6.295941   -2.218010
H           -2.820594    9.117321   -2.832177
H           -1.099555    8.820905   -3.057051
H           -2.181661    8.959197   -4.435759
N           -2.505543    4.106271   -4.624831
C           -2.157526    2.705716   -4.726733
C           -1.220454    2.507829   -5.924345
O           -0.293613    1.731288   -5.835409
C           -3.374048    1.767899   -4.755303
C           -4.196355    1.830013   -3.456962
C           -5.290848    0.758531   -3.332481
N           -4.770827   -0.497260   -2.789696
C           -5.483963   -1.613916   -2.643376
N           -6.717661   -1.713638   -3.117771
N           -4.993152   -2.653610   -1.994614
H           -3.425901    4.419148   -4.852286
H           -1.550740    2.440579   -3.873736
H           -4.011915    2.009861   -5.601170
H           -3.000640    0.763113   -4.918914
H           -3.539142    1.765467   -2.596831
H           -4.685089    2.792495   -3.395439
H           -6.077124    1.125425   -2.682093
H           -5.722947    0.570248   -4.309346
H           -3.875818   -0.470581   -2.345702
H           -7.082513   -1.044863   -3.753975
H           -7.149462   -2.628955   -3.088040
H           -4.146702   -2.626324   -1.454316
H           -5.526218   -3.506121   -2.017281
N           -1.459732    3.193320   -7.035078
H           -2.288144    3.735775   -7.153967
C           -0.593844    2.994125   -8.182060
H           -0.942484    3.639758   -8.977545
H           -0.611850    1.969389   -8.524037
C            0.857576    3.336479   -7.860396
O            1.785652    2.691426   -8.279303
N            1.004409    4.405631   -7.062311
```

```
H           0.215540    4.983132   -6.875072
C           2.282305    4.880660   -6.640296
H           2.919052    5.138696   -7.477241
H           2.129192    5.776726   -6.052632
C           3.088947    3.930390   -5.778979
O           4.293688    4.057118   -5.690944
N           2.428055    2.983359   -5.101943
C           3.131916    2.162407   -4.156040
C           4.314666    1.385358   -4.738432
O           5.257362    1.122981   -4.016908
C           2.127156    1.213068   -3.498181
C           2.726061    0.381306   -2.372324
C           1.673769   -0.513011   -1.744588
N           2.317598   -1.384729   -0.771740
C           1.666864   -2.363598   -0.134879
N           0.383078   -2.503231   -0.248759
N           2.406357   -3.222512    0.582203
H           1.437524    2.892464   -5.197662
H           3.587980    2.787554   -3.397807
H           1.318980    1.825287   -3.111716
H           1.689444    0.572928   -4.256113
H           3.532712   -0.235892   -2.752119
H           3.155761    1.026759   -1.611543
H           0.912862    0.075407   -1.247147
H           1.189465   -1.106532   -2.514826
H           3.294848   -1.532304   -0.891640
H          -0.144230   -1.908328   -0.848226
H          -0.191055   -3.121392    0.347140
H           3.205221   -2.840320    1.037655
H           1.918559   -3.906501    1.129204
N           4.284443    0.949941   -6.005865
H           3.532797    1.215919   -6.608759
C           5.363432    0.116659   -6.504616
H           5.139588   -0.143264   -7.531622
H           5.438985   -0.797145   -5.936371
C           6.736062    0.756835   -6.471872
O           7.736889    0.089822   -6.353202
N           6.766108    2.085324   -6.608293
H           5.913396    2.606419   -6.582055
C           8.006705    2.822900   -6.554611
H           8.708415    2.438047   -7.280999
H           7.801084    3.853324   -6.812009
C           8.756623    2.829179   -5.220282
O           9.899668    3.234032   -5.179248
N           8.118371    2.351042   -4.144121
C           8.715977    2.201407   -2.827968
C           9.967145    1.313387   -2.836904
O          10.791003    1.391978   -1.940070
C           7.617280    1.570293   -1.944362
C           7.848339    1.453450   -0.436866
C           7.590620    2.744454    0.351187
C           7.454573    2.488042    1.849393
N           8.781455    2.430675    2.535898
O          11.102317    1.646320    0.899340
```

```
H          10.933841        1.858903       -0.018781
H          11.590861        0.825826        0.835482
H           7.170254        2.049471       -4.245647
H           9.036730        3.161382       -2.445667
H           6.710975        2.137682       -2.112045
H           7.415636        0.583368       -2.345605
H           7.139058        0.712535       -0.080846
H           8.830611        1.053029       -0.223560
H           8.350565        3.499901        0.173439
H           6.650441        3.175208        0.021831
H           6.906501        3.275682        2.339839
H           6.948374        1.557817        2.040308
H           9.491932        1.941222        2.000079
H           9.114638        3.400608        2.678933
H           8.690062        1.982566        3.433751
N          10.062467        0.448216       -3.837536
C          11.108312       -0.537707       -4.075206
C          12.485979        0.024385       -4.425696
O          13.443052       -0.704232       -4.294743
C          10.704073       -1.434480       -5.267216
C           9.584533       -2.454017       -5.016053
C           9.149702       -3.067036       -6.350929
C          10.039397       -3.551113       -4.052800
H           9.322330        0.451608       -4.508457
H          11.261810       -1.131945       -3.186321
H          10.421513       -0.782728       -6.088337
H          11.592253       -1.970239       -5.579974
H           8.724421       -1.940778       -4.596639
H           8.361293       -3.798151       -6.199408
H           8.774594       -2.306420       -7.023736
H           9.980268       -3.575507       -6.833967
H           9.265392       -4.299065       -3.919134
H          10.917261       -4.054740       -4.444251
H          10.291653       -3.164171       -3.070216
N          12.574049        1.261829       -4.919242
C          13.815659        1.843027       -5.394229
C          14.958224        2.035776       -4.378901
O          16.011408        2.420545       -4.811142
C          13.499706        3.166561       -6.093935
O          12.918193        4.086714       -5.214361
H          11.756632        1.833723       -4.922010
H          14.250487        1.172645       -6.126152
H          14.423664        3.583886       -6.462316
H          12.850845        2.969146       -6.941311
H          11.971324        4.080215       -5.284933
N          14.792295        1.722324       -3.074280
C          15.929125        1.648971       -2.177483
C          16.802176        0.407362       -2.395476
O          17.928268        0.388209       -1.971236
C          15.499122        1.594726       -0.706586
C          14.660726        2.767803       -0.203098
C          14.325320        2.547234        1.272313
N          13.311493        3.456514        1.792793
C          13.501604        4.599755        2.432088
```

```
N     14.723179     5.061159     2.673451
O     15.058100     7.524418     4.217067
H     14.774274     8.405614     3.959302
H     15.772401     7.643399     4.830251
N     12.458836     5.296190     2.835108
H     13.901830     1.417200    -2.758504
H     16.565549     2.502212    -2.355004
H     14.946670     0.675091    -0.542466
H     16.411848     1.517347    -0.126171
H     15.186666     3.708250    -0.339409
H     13.735460     2.850395    -0.761459
H     13.926537     1.551446     1.402751
H     15.213258     2.596467     1.890977
H     12.368489     3.170257     1.624730
H     15.527430     4.566134     2.369660
H     14.863123     5.916272     3.184023
H     11.500696     5.028470     2.674497
H     12.563153     6.130238     3.367723
N     16.200238    -0.648979    -2.967243
C     16.867568    -1.914108    -3.163580
C     17.662808    -1.897649    -4.465560
H     18.075065    -0.923570    -4.738310
O     17.847284    -2.859275    -5.137433
C     15.862260    -3.070218    -3.138248
C     15.323269    -3.395831    -1.778048
N     14.410401    -2.618241    -1.093234
O     12.500740    -0.439539    -0.454995
H     13.151583    -1.033962    -0.833114
H     12.029171    -0.045277    -1.180788
C     15.615888    -4.488080    -1.037289
C     14.181988    -3.241116     0.016632
N     14.885758    -4.375551     0.119250
H     15.287960    -0.547880    -3.357395
H     17.594139    -2.030705    -2.367154
H     15.043080    -2.842040    -3.811358
H     16.359103    -3.946349    -3.532607
H     16.266053    -5.315254    -1.225990
H     13.510960    -2.906992     0.782088
H     14.785423    -5.057914     0.838617
O    -15.054769    -2.361876     4.927268
H    -15.193367    -1.575748     4.404421
H    -14.623511    -2.070956     5.723451
O    -13.042214    -1.185100     7.012070
H    -12.395288    -1.695796     7.485433
H    -13.285341    -0.468232     7.584997
O    -12.031412    -0.770520     4.287519
H    -12.244477    -0.796093     5.217014
H    -12.822199    -0.427703     3.865776
O    -13.267630    -1.869830     0.841902
H    -12.491972    -1.307941     0.897093
H    -13.174602    -2.536491     1.523026
O     -7.067206     1.067228     0.191068
H     -7.337333     1.488555     1.009640
H     -6.141525     0.895753     0.301811
```

```
O      -7.724652   2.065606   2.820518
H      -8.648661   1.883296   2.986219
H      -7.492634   2.809691   3.368613
O      -9.180305  -0.244913  -1.159426
H      -8.364668   0.008446  -0.724088
H      -9.129554  -1.158558  -1.439114
O     -12.649354  -2.305534  -1.995623
H     -12.926314  -2.381746  -1.085309
H     -11.733958  -2.586808  -2.040269
O     -13.772649   0.924952  -0.639808
H     -14.146396   0.051321  -0.599522
H     -12.900373   0.823449  -0.246528
O     -14.274450   4.976797   4.775093
H     -13.764325   5.672539   4.368948
H     -14.295458   5.217078   5.696486
O     -14.976844   2.186427   1.577128
H     -15.694965   2.795848   1.469483
H     -14.676937   1.925794   0.704679
O     -12.567888   0.255336  -3.263652
H     -13.033603   0.835854  -2.669433
H     -12.519639  -0.572927  -2.783424
O     -13.355938  -2.129459  -4.867875
H     -13.457680  -1.211708  -4.637703
H     -13.420490  -2.584514  -4.033466
O      -9.888840  -2.634441  -2.578677
H     -10.051630  -2.213666  -3.421417
H      -9.571201  -3.523120  -2.802835
O     -11.358573   0.189545   0.592200
H     -10.615449   0.057440  -0.001460
H     -11.024201   0.649258   1.366154
O     -10.551911   1.375201   3.048805
H     -10.814489   0.632199   3.590074
H     -11.228579   2.036981   3.217688
O     -12.863847   2.858686   3.503576
H     -13.260439   3.520242   4.071636
H     -13.506008   2.714747   2.815050
O      -9.759239   0.835312  -3.714540
H     -10.705317   0.685418  -3.684265
H      -9.448006   0.562595  -2.850608
O     -14.542270  -0.283496   2.985156
H     -14.828810   0.584763   2.707491
H     -14.370293  -0.771654   2.183031
O     -12.973723  -3.362862   3.324795
H     -13.784663  -3.195151   3.814735
H     -12.374885  -2.699054   3.663044
O     -10.614893  -1.831985  -5.344368
H     -11.532030  -2.113709  -5.282517
H     -10.077412  -2.578219  -5.623219
```

```
Case 8:
voltage= +70; Y266 neutral; R300 positive; E183 negative
N             85.286997     87.466003    -74.794008
H             84.521457     87.517653    -75.438004
C             86.104233     88.680342    -74.850977
C             87.214040     88.568017    -75.912971
O             87.557921     89.523290    -76.582050
C             86.751467     88.944044    -73.486993
H             84.895364     87.339730    -73.881505
H             85.525210     89.548516    -75.137569
H             87.404320     89.807970    -73.522856
H             85.983492     89.144719    -72.748430
H             87.326335     88.086360    -73.154453
N             87.797954     87.375922    -76.027013
H             87.430089     86.636320    -75.468272
C             88.848022     87.107098    -77.010987
H             89.612486     87.866170    -76.977707
H             89.293737     86.146193    -76.790458
C             88.265913     87.089304    -78.425289
O             88.876716     87.511237    -79.372153
N             87.015032     86.602955    -78.545073
H             86.577400     86.182072    -77.758087
C             86.350843     86.518984    -79.847938
H             85.432669     85.959634    -79.727835
H             86.974357     86.022080    -80.574689
C             86.020153     87.912039    -80.398996
O             86.101021     88.140855    -81.576853
N             85.611470     88.816963    -79.494278
C             85.348615     90.192422    -79.840303
C             86.617212     90.959767    -80.218849
O             86.577644     91.706220    -81.168060
C             84.545440     90.863383    -78.716077
O             84.985755     90.504466    -77.439128
H             85.618705     88.585167    -78.528001
H             84.747445     90.228642    -80.733671
H             84.556028     91.938758    -78.860067
H             83.521152     90.525788    -78.780440
H             85.862553     90.823684    -77.263203
N             87.736133     90.783969    -79.496567
C             88.975486     91.417379    -79.934019
C             89.368671     90.914789    -81.326723
O             89.963413     91.634764    -82.095945
C             90.110916     91.274661    -78.874060
C             91.107736     90.140366    -79.143995
C             90.857077     92.602014    -78.724808
H             87.754157     90.151680    -78.725392
H             88.771931     92.466988    -80.074912
H             89.615981     91.069086    -77.930629
H             91.812405     90.082463    -78.320702
H             90.631322     89.174950    -79.243856
H             91.683824     90.331975    -80.043972
H             91.641199     92.512628    -77.980890
H             91.314488     92.894262    -79.663681
H             90.189557     93.394168    -78.405652
```

```
N         88.997817      89.665836      -81.628976
H         88.671356      89.080206      -80.894279
C         89.356146      88.993124      -82.873967
H         89.295926      87.927243      -82.693580
H         90.370279      89.230124      -83.156134
C         88.490025      89.325783      -84.070062
O         89.017599      89.439167      -85.152748
N         87.173235      89.481165      -83.894957
C         86.397621      90.000023      -84.997405
C         86.733920      91.469501      -85.228023
O         86.617067      91.923733      -86.347289
C         84.879726      89.737318      -84.937831
C         84.589150      88.234465      -85.000480
C         84.160428      90.389979      -83.755695
H         86.758468      89.281704      -83.010544
H         86.755349      89.524501      -85.897607
H         84.495436      90.186805      -85.848477
H         83.520218      88.064316      -85.072855
H         85.054925      87.779592      -85.868586
H         84.949842      87.722493      -84.115477
H         83.087924      90.276565      -83.872666
H         84.439083      89.916589      -82.823555
H         84.373000      91.450970      -83.676342
N         87.198678      92.218137      -84.223723
C         87.706152      93.554019      -84.487730
C         89.014558      93.490282      -85.283540
O         89.146873      94.181327      -86.267310
C         87.841572      94.378918      -83.194521
C         86.459325      94.620332      -82.567614
C         88.543175      95.709817      -83.485941
C         86.520937      95.016997      -81.094911
H         87.231568      91.851592      -83.296708
H         87.018421      94.052920      -85.153856
H         88.447311      93.812280      -82.494817
H         85.954906      95.400702      -83.136151
H         85.851909      93.729054      -82.650393
H         88.523145      96.350863      -82.615410
H         89.580550      95.573895      -83.763233
H         88.053151      96.236099      -84.299253
H         85.521215      95.142806      -80.695616
H         87.010942      94.246948      -80.513491
H         87.055941      95.946653      -80.941795
N         89.980259      92.660763      -84.870289
H         89.898891      92.188387      -83.993965
C         91.240856      92.592482      -85.578217
H         91.897540      91.938615      -85.018098
H         91.705828      93.564430      -85.652241
C         91.114823      92.058803      -87.004520
O         91.675050      92.612929      -87.920247
N         90.371246      90.957119      -87.165861
H         89.970493      90.502968      -86.372678
C         90.140476      90.393304      -88.474269
H         91.067919      90.227873      -89.001593
H         89.643885      89.439769      -88.345006
```

```
C        89.278971    91.284930   -89.363997
O        89.511303    91.367192   -90.547844
N        88.299013    91.957093   -88.762057
C        87.581807    92.941779   -89.557042
C        88.448299    94.129718   -89.989551
O        88.258583    94.618232   -91.080366
C        86.366799    93.544611   -88.866729
O        85.328710    92.645876   -88.626480
H        88.135188    91.884207   -87.783618
H        87.258209    92.474973   -90.473078
H        86.681344    94.028437   -87.951627
H        85.975738    94.287279   -89.543669
H        85.589322    92.072807   -87.915881
N        89.342961    94.635384   -89.131085
C        90.202230    95.738141   -89.527025
C        91.117158    95.312584   -90.670197
O        91.317534    96.063191   -91.600524
C        90.979664    96.316897   -88.321704
C        90.017102    97.136146   -87.445867
C        92.180155    97.161458   -88.768867
C        90.614441    97.579503   -86.110544
H        89.432555    94.255318   -88.213453
H        89.591055    96.520571   -89.953421
H        91.354169    95.479839   -87.740415
H        89.697599    98.013560   -88.006284
H        89.130401    96.547117   -87.248081
H        92.695715    97.569064   -87.908418
H        92.904795    96.577782   -89.322931
H        91.869672    97.989861   -89.397739
H        89.846630    98.018453   -85.481935
H        91.033723    96.735116   -85.572784
H        91.397257    98.319199   -86.232790
N        91.659796    94.093523   -90.611760
H        91.524939    93.517030   -89.807846
C        92.391039    93.582830   -91.745143
H        93.262987    94.182030   -91.960846
H        92.708908    92.574008   -91.514571
C        91.543021    93.558969   -93.013005
O        92.001689    93.949426   -94.061698
N        90.308984    93.070034   -92.904018
C        89.479707    92.990832   -94.083106
C        89.137493    94.365021   -94.655902
O        89.000973    94.468411   -95.856345
C        88.182605    92.213567   -93.822053
O        87.517032    91.934998   -95.021572
H        89.999780    92.656895   -92.049035
H        90.017425    92.486620   -94.873762
H        88.422732    91.265540   -93.360340
H        87.546149    92.770329   -93.141489
H        87.601897    92.668364   -95.619161
N        88.951550    95.370725   -93.798717
C        88.690894    96.740665   -94.210441
C        89.862674    97.301116   -95.013098
O        89.677058    97.846915   -96.075032
```

```
C         88.410274    97.570406   -92.940271
C         88.460428    99.084590   -93.064753
C         88.080239    99.779559   -94.206107
C         88.897817    99.819524   -91.964918
C         88.153864   101.163907   -94.246388
C         88.960565   101.201019   -91.996735
C         88.593864   101.880660   -93.148164
H         88.985849    95.170439   -92.821794
H         87.842181    96.767681   -94.876393
H         87.440506    97.260979   -92.561798
H         89.138139    97.282728   -92.195810
H         87.757630    99.257766   -95.084506
H         89.206253    99.301681   -91.073157
H         87.872047   101.668791   -95.152857
H         89.300118   101.741907   -91.131705
H         88.663113   102.954073   -93.185326
N         91.081278    97.170406   -94.474554
C         92.231374    97.659286   -95.199652
C         92.351641    96.996173   -96.571124
O         92.701396    97.642251   -97.523642
C         93.525842    97.437288   -94.414609
S         93.723781    98.494671   -92.944725
H         91.198925    96.660132   -93.624858
H         92.111474    98.712721   -95.403178
H         93.631293    96.398145   -94.133108
H         94.359026    97.692964   -95.055528
H         92.821610    97.910714   -92.171143
N         92.084981    95.676016   -96.645012
H         91.919907    95.139128   -95.821071
C         92.362040    94.996052   -97.899839
H         93.336599    95.268145   -98.276107
H         92.347954    93.929456   -97.708421
C         91.367020    95.276992   -99.035985
O         91.733868    95.425584  -100.169393
N         90.080985    95.255993   -98.669014
C         88.967142    95.434141   -99.589917
C         88.867950    96.813554  -100.231547
O         88.334949    96.958673  -101.304337
C         87.602634    95.138752   -98.922929
C         87.161899    95.972224   -97.693783
C         86.177645    97.125732   -97.860895
O         86.207459    97.826246   -98.889694
O         85.420515    97.359913   -96.891030
H         89.865973    95.112023   -97.705026
H         89.082437    94.747189  -100.417065
H         86.866791    95.217308   -99.707688
H         87.622609    94.097865   -98.621253
H         86.726089    95.307053   -96.963708
H         88.017466    96.415543   -97.204396
N         89.303025    97.826014   -99.471063
C         89.276718    99.220339   -99.890910
C         90.444029    99.585991  -100.809996
H         91.311950    98.923132  -100.770066
O         90.406502   100.526499  -101.534715
```

```
C         89.208485   100.123339   -98.632598
O         87.876667   100.181993   -98.161425
C         89.643332   101.566643   -98.842286
H         89.821485    97.604764   -98.648505
H         88.381314    99.385445  -100.473457
H         89.833300    99.683441   -97.865164
H         87.449985    99.337270   -98.283166
H         89.465560   102.108424   -97.916437
H         90.698000   101.646150   -99.068016
H         89.088607   102.036522   -99.645726
N         81.460930    91.680026  -102.958969
H         82.352965    91.683058  -103.407941
C         81.522267    92.339632  -101.643921
C         80.849083    91.535949  -100.497016
O         81.132563    91.714314   -99.336487
C         80.867823    93.734772  -101.676607
O         79.539455    93.514618  -102.073344
C         81.577644    94.695604  -102.624096
H         80.797301    92.137019  -103.551613
H         82.551277    92.448860  -101.339888
H         80.900879    94.144763  -100.670877
H         79.077268    94.340047  -102.203654
H         81.081638    95.658098  -102.602617
H         82.610231    94.846043  -102.337648
H         81.550355    94.333156  -103.645368
N         79.912065    90.673974  -100.914032
H         79.808572    90.587768  -101.900682
C         79.061514    89.893903  -100.046531
H         78.228148    90.478449   -99.684288
H         78.657613    89.071660  -100.626046
C         79.713671    89.304078   -98.797313
O         79.111876    89.360340   -97.753268
N         80.907880    88.688229   -98.876016
H         81.379625    88.632698   -99.750204
C         81.525099    88.145172   -97.676595
H         82.357142    87.520225   -97.975939
H         80.820419    87.531478   -97.138251
C         82.061883    89.201077   -96.694917
O         82.127845    88.947538   -95.517744
N         82.451025    90.333986   -97.269086
C         82.979287    91.379854   -96.401293
C         81.883191    92.335187   -95.922327
O         81.915218    92.736841   -94.780359
C         84.290129    92.034831   -96.871178
C         84.378987    92.621159   -98.261842
C         83.914489    93.904394   -98.536967
C         85.031589    91.921340   -99.269333
C         84.119431    94.478888   -99.777747
C         85.243666    92.495168  -100.515384
C         84.798126    93.780379  -100.767776
H         82.253148    90.523881   -98.225107
H         83.248053    90.880279   -95.487783
H         84.523669    92.800263   -96.138626
H         85.059157    91.279938   -96.764652
```

```
H            83.425977      94.474090     -97.765534
H            85.416538      90.936540     -99.067457
H            83.787678      95.481463     -99.967692
H            85.783702      91.951033    -101.270109
H            85.002102      94.255813    -101.709404
N            80.876373      92.614112     -96.751979
C            79.611344      93.138375     -96.293183
C            78.996100      92.219126     -95.237032
O            78.386360      92.702796     -94.308183
C            78.695210      93.303343     -97.523111
C            77.372362      93.993023     -97.276544
C            77.306611      95.380237     -97.198161
C            76.195968      93.262552     -97.179418
C            76.093059      96.022693     -97.026832
C            74.977382      93.901941     -97.010990
C            74.922961      95.282938     -96.936954
H            80.930669      92.302028     -97.694973
H            79.746100      94.091102     -95.802726
H            79.259896      93.868533     -98.256287
H            78.520112      92.324756     -97.954008
H            78.206550      95.964704     -97.287223
H            76.225738      92.188354     -97.246807
H            76.061796      97.095696     -96.971622
H            74.074904      93.321464     -96.943525
H            73.978315      95.780015     -96.813856
N            79.167843      90.903031     -95.373540
H            79.480984      90.542625     -96.246355
C            78.641179      89.941069     -94.429339
H            77.597000      90.113589     -94.216738
H            78.742356      88.962677     -94.880055
C            79.358127      89.946846     -93.082061
O            78.711873      89.901946     -92.062644
N            80.696051      90.008690     -93.074259
C            81.415126      90.153864     -91.820776
C            81.069297      91.489753     -91.158425
O            80.900039      91.546377     -89.959385
C            82.933435      89.972822     -92.041626
C            83.757379      90.462647     -90.846554
C            83.250377      88.502077     -92.336727
H            81.204947      89.909953     -93.926789
H            81.067402      89.408318     -91.119199
H            83.212229      90.571989     -92.904697
H            84.806986      90.248050     -91.019379
H            83.667396      91.527970     -90.680405
H            83.459306      89.957260     -89.932574
H            84.303340      88.387195     -92.573608
H            83.038118      87.888686     -91.464637
H            82.682342      88.115661     -93.171911
N            80.957856      92.561953     -91.951928
C            80.685927      93.867317     -91.390159
C            79.249529      93.983580     -90.867743
O            79.052225      94.554058     -89.809983
C            80.972847      94.984920     -92.400987
C            82.454509      95.164955     -92.788638
```

```
C         83.323950    95.660907   -91.643840
O         83.499828    94.890154   -90.697196
O         83.805234    96.822043   -91.716631
H         81.172817    92.484218   -92.923786
H         81.314474    94.003475   -90.528650
H         80.402986    94.799872   -93.304580
H         80.593008    95.907324   -91.974546
H         82.857675    94.225556   -93.137891
H         82.506830    95.869625   -93.610329
N         78.259207    93.472333   -91.599747
C         76.900154    93.441155   -91.100612
C         76.702956    92.388916   -90.010991
O         75.813781    92.555745   -89.204534
C         75.821083    93.327511   -92.187672
O         75.944722    92.143921   -92.922693
C         75.763140    94.559820   -93.083284
H         78.476980    93.015779   -92.458285
H         76.726660    94.370784   -90.584319
H         74.886281    93.243204   -91.649603
H         76.627983    92.251291   -93.573481
H         74.954388    94.448654   -93.794745
H         75.581582    95.454957   -92.496536
H         76.682670    94.695022   -93.639854
N         77.605976    91.420996   -89.881922
H         78.139141    91.126668   -90.670751
C         77.555091    90.545769   -88.740733
H         76.554670    90.194584   -88.539009
H         78.195721    89.695882   -88.939892
C         78.052664    91.242198   -87.480747
O         77.524008    91.062237   -86.407680
N         79.116349    92.052477   -87.631844
C         79.684260    92.704164   -86.480102
C         78.811378    93.833474   -85.944657
O         78.769226    94.012875   -84.749804
C         81.125059    93.184154   -86.726513
S         81.342086    94.885430   -87.341399
H         79.583919    92.089008   -88.514494
H         79.716731    91.990995   -85.668694
H         81.635937    93.186466   -85.772647
H         81.628719    92.488335   -87.383143
H         80.658425    94.800952   -88.471516
N         78.117231    94.585225   -86.810901
C         77.290079    95.647505   -86.271475
C         76.123006    95.050010   -85.492025
O         75.604272    95.660284   -84.587826
C         76.803619    96.683993   -87.305893
C         75.720103    96.149084   -88.256555
C         77.995876    97.325368   -88.025312
C         75.240074    97.170835   -89.288850
H         78.243726    94.468687   -87.794798
H         77.876189    96.178966   -85.536705
H         76.331926    97.451058   -86.700611
H         76.065276    95.257186   -88.763775
H         74.864278    95.840221   -87.665525
```

```
H            77.692717     98.232127    -88.532959
H            78.774944     97.592400    -87.317342
H            78.431784     96.664657    -88.762863
H            74.373522     96.789343    -89.819375
H            74.953017     98.107618    -88.817964
H            76.004050     97.390578    -90.026758
N            75.745013     93.830999    -85.873966
H            76.130779     93.425943    -86.695701
C            74.609361     93.174298    -85.264983
H            74.321991     92.348387    -85.903316
H            73.774202     93.850613    -85.164010
C            74.919857     92.645284    -83.871676
O            74.157662     92.793605    -82.954961
N            76.101522     92.004740    -83.738108
H            76.681118     91.853588    -84.535830
C            76.558542     91.616573    -82.432086
H            77.440042     90.996299    -82.546091
H            75.806962     91.047856    -81.905457
C            76.917957     92.797999    -81.535040
O            76.812745     92.707064    -80.334745
N            77.396069     93.885989    -82.136990
C            77.549444     95.060852    -81.305519
C            76.174972     95.572000    -80.848988
H            75.310636     95.023174    -81.221567
O            76.065215     96.513766    -80.129484
C            78.371521     96.157167    -81.977972
C            79.831907     95.825738    -82.243459
C            80.465428     94.677503    -81.781683
C            80.583194     96.724435    -82.998445
C            81.797800     94.428448    -82.079900
C            81.910412     96.480471    -83.299331
C            82.526864     95.323614    -82.842867
H            77.498665     93.943187    -83.126341
H            78.042035     94.764698    -80.388296
H            77.902679     96.424996    -82.917370
H            78.329094     97.032477    -81.339410
H            79.915772     93.948214    -81.218442
H            80.115060     97.621908    -83.362494
H            82.259235     93.521410    -81.727403
H            82.457016     97.188121    -83.899078
H            83.557396     95.121889    -83.075907
N            82.399005    104.821991    -73.868985
H            82.477568    105.532486    -74.573225
C            82.521978    103.494014    -74.477021
C            81.197907    103.224902    -75.190397
O            81.152739    102.715220    -76.285496
C            82.779623    102.416484    -73.414555
C            84.077556    102.587594    -72.617963
S            85.528160    102.628273    -73.708830
C            86.851186    102.781545    -72.485241
H            83.155446    104.993253    -73.235415
H            83.288192    103.441648    -75.239637
H            81.949611    102.398796    -72.714641
H            82.799242    101.457713    -73.918089
```

```
H          84.046699   103.483980   -72.008530
H          84.167820   101.752944   -71.931905
H          87.776787   102.820729   -73.038589
H          86.740622   103.689307   -71.906261
H          86.873238   101.925640   -71.823638
N          80.083765   103.634029   -74.550245
C          78.771555   103.441228   -75.128868
C          78.561630   104.292140   -76.386830
O          77.946446   103.858464   -77.324563
C          77.667375   103.754973   -74.118643
C          77.592291   102.746600   -72.983570
O          78.073448   101.644012   -73.078800
N          76.943654   103.164661   -71.888493
H          80.208702   104.123901   -73.691624
H          78.671415   102.417979   -75.452144
H          77.775024   104.759479   -73.720034
H          76.712041   103.723378   -74.634726
H          76.778738   102.518728   -71.147991
H          76.482929   104.044837   -71.844437
N          79.074260   105.540604   -76.372258
H          79.590508   105.844725   -75.579416
C          78.990020   106.395985   -77.537987
H          79.300562   107.394587   -77.259358
H          77.976116   106.445049   -77.902478
C          79.862988   105.934024   -78.706020
O          79.570732   106.254751   -79.826596
N          80.965504   105.215161   -78.404526
C          81.840741   104.676097   -79.433876
C          81.215044   103.460443   -80.130631
O          81.230646   103.391480   -81.335759
C          83.250574   104.395136   -78.865456
C          83.913610   105.715064   -78.428982
C          84.114228   103.647924   -79.888686
C          85.204632   105.545398   -77.624714
H          81.045871   104.832919   -77.491847
H          81.918397   105.416966   -80.214666
H          83.135585   103.755411   -77.992311
H          84.115271   106.312521   -79.314998
H          83.210837   106.290275   -77.834092
H          85.103977   103.470662   -79.488214
H          83.690505   102.684392   -80.146063
H          84.219326   104.215599   -80.807395
H          85.572333   106.512341   -77.297186
H          85.047359   104.936233   -76.738721
H          85.994612   105.080877   -78.202778
N          80.629171   102.518806   -79.369805
C          79.890602   101.416856   -79.972448
C          78.790892   101.935459   -80.906716
O          78.639304   101.460962   -82.007260
C          79.334058   100.491064   -78.866466
C          80.290862    99.327059   -78.593326
O          80.471949    98.556818   -79.523630
O          80.841072    99.213229   -77.461558
H          80.684121   102.569754   -78.375489
```

```
H        80.533468   100.843710   -80.621220
H        79.112727   101.056425   -77.970449
H        78.413979   100.038402   -79.217270
N        78.020241   102.944119   -80.457768
H        78.099557   103.252250   -79.512564
C        76.976303   103.493900   -81.291080
H        76.450834   104.247762   -80.718069
H        76.262540   102.739536   -81.584215
C        77.484345   104.162432   -82.566093
O        76.774429   104.250459   -83.540844
N        78.723682   104.660792   -82.509803
H        79.216397   104.622653   -81.648825
C        79.295016   105.461005   -83.550997
H        80.097988   106.043885   -83.118233
H        78.564815   106.158046   -83.938452
C        79.857980   104.737991   -84.785004
O        80.007633   105.335160   -85.814544
N        80.110869   103.430154   -84.586501
C        80.289808   102.478644   -85.654382
C        78.949187   102.031675   -86.245688
O        78.772438   102.084853   -87.440183
C        81.101291   101.278151   -85.167118
H        79.977669   103.079089   -83.661883
H        80.805972   102.965737   -86.465125
H        81.201058   100.543831   -85.959322
H        82.088859   101.609739   -84.867940
H        80.630246   100.801232   -84.314308
N        77.993727   101.561322   -85.422110
C        76.835100   100.907076   -86.010789
C        75.949765   101.881582   -86.782346
O        75.386422   101.516743   -87.792520
C        76.028204   100.049972   -85.004949
C        75.179441   100.852939   -84.006598
C        76.957073    99.052182   -84.299136
C        74.359933    99.973802   -83.058516
H        78.150674   101.510067   -84.438272
H        77.196137   100.232634   -86.773420
H        75.341825    99.484692   -85.628568
H        75.820568   101.510209   -83.428787
H        74.494627   101.494864   -84.551016
H        76.393260    98.206361   -83.931943
H        77.707120    98.669107   -84.983261
H        77.472419    99.507680   -83.460070
H        73.695944   100.587751   -82.457642
H        73.745714    99.268435   -83.609824
H        74.984619    99.405264   -82.379774
N        75.805103   103.123655   -86.299885
H        76.225755   103.377620   -85.429866
C        74.879433   104.034871   -86.923486
H        74.843717   104.940328   -86.330951
H        73.882764   103.619478   -86.959670
C        75.228003   104.378990   -88.370006
O        74.342209   104.394822   -89.190084
N        76.512996   104.620999   -88.689010
```

```
C        76.724141    105.088919    -90.061174
C        76.838141    103.964435    -91.076983
O        76.588818    104.192195    -92.237567
C        77.928475    106.034741    -89.971028
C        77.815104    106.543352    -88.530040
C        77.443195    105.283673    -87.764573
H        75.859588    105.649777    -90.381718
H        78.855996    105.498673    -90.124396
H        77.865321    106.815161    -90.718235
H        78.707218    106.996984    -88.140765
H        77.008225    107.269032    -88.461505
H        78.296541    104.653853    -87.587573
H        76.979442    105.492110    -86.814305
N        77.189553    102.744498    -90.632101
C        77.047597    101.589317    -91.499829
C        75.581460    101.343518    -91.872520
O        75.277093    101.059351    -93.009114
C        77.593822    100.303460    -90.850449
C        79.093952    100.091463    -90.890482
C        79.645419     99.151354    -91.752039
C        79.951895    100.745226    -90.011665
C        81.000616     98.851730    -91.726347
C        81.299058    100.439833    -89.960497
C        81.825263     99.473111    -90.801369
O        83.147719     99.181439    -90.708960
H        77.288691    102.609701    -89.648158
H        77.556300    101.785838    -92.430488
H        77.243627    100.262460    -89.824422
H        77.127515     99.470288    -91.363191
H        79.013285     98.632904    -92.452657
H        79.566973    101.470237    -89.326030
H        81.410901     98.121667    -92.397860
H        81.944238    100.930546    -89.253766
H        83.320573     98.285460    -91.036158
N        74.680188    101.408711    -90.880124
H        74.973801    101.579815    -89.941012
C        73.274850    101.177305    -91.127253
H        72.781778    101.071477    -90.168966
H        73.124815    100.271066    -91.692909
C        72.595145    102.302994    -91.902169
O        71.754222    102.068760    -92.736577
N        72.999773    103.542039    -91.607700
H        73.614168    103.676451    -90.835031
C        72.424111    104.764980    -92.189098
H        72.854750    105.596562    -91.645812
H        71.353062    104.785769    -92.048335
C        72.663921    104.985381    -93.662291
O        71.809876    105.472464    -94.356801
N        73.880342    104.653118    -94.123777
C        74.223797    104.908462    -95.500403
C        73.598046    103.889943    -96.443017
O        73.782940    103.997371    -97.641996
C        75.769994    104.992471    -95.647821
O        76.148588    105.875548    -96.660521
```

```
C         76.448563    103.634312    -95.777362
H         74.568569    104.290973    -93.497280
H         73.822728    105.868740    -95.791258
H         76.109462    105.457122    -94.731916
H         75.969037    105.527328    -97.531923
H         77.524218    103.752816    -95.755437
H         76.167142    102.974145    -94.965963
H         76.191888    103.159131    -96.715705
N         72.937934    102.846044    -95.934006
C         72.440819    101.797976    -96.805842
C         71.513753    102.321278    -97.900387
O         71.550962    101.824840    -99.002218
C         71.700667    100.707814    -96.014164
C         72.587100     99.785401    -95.161826
C         71.691823     98.851603    -94.341455
C         73.573712     98.973923    -96.006270
H         72.815489    102.753920    -94.950121
H         73.269877    101.361069    -97.336496
H         70.960299    101.181782    -95.378125
H         71.159378    100.100088    -96.734230
H         73.158484    100.387468    -94.466321
H         72.289246     98.194301    -93.716021
H         71.029882     99.418011    -93.696442
H         71.082550     98.223937    -94.987963
H         74.133892     98.290992    -95.375611
H         73.054448     98.386094    -96.760002
H         74.297422     99.601594    -96.515276
N         70.606278    103.247654    -97.543990
C         69.681554    103.785829    -98.500219
C         70.228969    104.908189    -99.356950
H         69.530068    105.256578   -100.125999
O         71.297792    105.403918    -99.233680
H         70.742055    103.738361    -96.687233
H         69.342744    102.994530    -99.156886
H         68.806014    104.164017    -97.980057
N         77.567008    111.194025    -97.210978
H         76.663822    111.005632    -96.822628
C         78.252493    109.952415    -97.547820
C         79.719300    110.284505    -97.317438
O         80.406144    109.757926    -96.473230
C         77.938491    109.387727    -98.952229
C         76.433504    109.095189    -99.092735
C         78.783856    108.144897    -99.247675
C         75.994316    108.747583   -100.515983
H         77.407161    111.762565    -98.020904
H         77.993202    109.210221    -96.805072
H         78.194239    110.156796    -99.684689
H         76.164127    108.282518    -98.424786
H         75.869770    109.966382    -98.772489
H         78.649340    107.823927   -100.273178
H         79.845324    108.327455    -99.106573
H         78.490106    107.325481    -98.600867
H         74.914545    108.663962   -100.560676
H         76.298534    109.516688   -101.222342
```

```
H      76.395152    107.797615   -100.842866
N      80.181005    111.304920    -98.067075
H      79.599476    111.632229    -98.803117
C      81.591984    111.675074    -98.159941
H      82.191061    110.869818    -98.561493
H      81.665711    112.529070    -98.820758
C      82.208091    112.045644    -96.808798
O      83.298187    111.664004    -96.494354
N      81.461006    112.873968    -96.043002
H      80.526234    113.054316    -96.332784
C      81.856988    113.283023    -94.690984
H      82.839006    113.728680    -94.724287
H      81.147817    114.028472    -94.354826
C      81.927718    112.171939    -93.634943
O      82.894493    112.004538    -92.943718
N      80.791023    111.463969    -93.469038
H      80.063590    111.565720    -94.139871
C      80.594017    110.483993    -92.384011
H      80.975349    110.892393    -91.462569
H      79.528660    110.319167    -92.280109
C      81.256633    109.135958    -92.571268
O      81.776387    108.617075    -91.609168
N      81.184411    108.500368    -93.752233
C      81.450589    107.069551    -93.744727
C      82.922696    106.749837    -93.523891
O      83.218283    105.720065    -92.959381
C      80.848817    106.285897    -94.932670
C      81.523354    106.557326    -96.286306
C      79.329811    106.486225    -94.966534
C      81.095913    105.569651    -97.373553
H      80.777305    108.941075    -94.553168
H      80.986652    106.664588    -92.857277
H      81.035771    105.246048    -94.679050
H      81.326161    107.571333    -96.610002
H      82.601997    106.484247    -96.159289
H      78.862630    105.842852    -95.697157
H      78.887829    106.251545    -94.003141
H      79.066986    107.505926    -95.222317
H      81.656920    105.738080    -98.288376
H      81.262929    104.543718    -97.067038
H      80.046513    105.660517    -97.619031
N      83.863610    107.607546    -93.968915
C      85.256201    107.369758    -93.655050
C      85.708421    107.976673    -92.317858
O      86.635170    107.446261    -91.747667
C      86.208692    107.683519    -94.818568
C      86.066955    106.738956    -96.030505
C      86.198316    105.251680    -95.666307
N      86.459625    104.354598    -96.798671
C      85.563495    103.563639    -97.408036
N      84.275452    103.869272    -97.363763
N      85.962083    102.492035    -98.052325
H      83.578955    108.445338    -94.429308
H      85.330501    106.319280    -93.439431
```

```
H                   86.070599   108.707243   -95.153662
H                   87.219699   107.614268   -94.431989
H                   85.122513   106.916952   -96.535712
H                   86.845547   106.993322   -96.741972
H                   87.028263   105.118951   -94.982772
H                   85.314789   104.895834   -95.153954
H                   87.413165   104.118034   -96.963833
H                   83.994005   104.783626   -97.096532
H                   83.581526   103.313065   -97.846120
H                   86.860096   102.070977   -97.907728
H                   85.307555   101.860956   -98.490893
N                   85.040015   109.006039   -91.779960
H                   84.280621   109.433224   -92.259298
C                   85.314994   109.414025   -90.393986
H                   86.357380   109.659055   -90.268811
H                   84.715996   110.290099   -90.183788
C                   84.968480   108.309362   -89.406558
O                   85.653819   108.087571   -88.439709
N                   83.849734   107.632140   -89.697835
C                   83.282434   106.642830   -88.814869
C                   83.959535   105.290968   -88.939445
O                   84.267870   104.682454   -87.941456
C                   81.771331   106.563279   -89.063410
C                   81.221550   105.231084   -88.556235
C                   81.126931   107.799241   -88.419205
H                   83.280152   107.953594   -90.450431
H                   83.460161   106.934373   -87.789608
H                   81.612485   106.610540   -90.134120
H                   80.149462   105.242609   -88.505480
H                   81.513855   104.419675   -89.215194
H                   81.577384   105.016420   -87.562362
H                   80.199343   108.060518   -88.908039
H                   80.931283   107.626106   -87.366220
H                   81.773576   108.664538   -88.507158
N                   84.210125   104.797750   -90.166508
C                   85.046048   103.620779   -90.234030
C                   86.395005   103.924027   -89.576066
O                   86.936898   103.059496   -88.919712
C                   85.188635   103.035671   -91.646175
C                   83.855951   102.507224   -92.198423
C                   83.961745   101.754577   -93.533224
N                   84.246844   100.329297   -93.347265
C                   84.619191    99.505836   -94.317382
N                   84.820893    99.976484   -95.554901
N                   84.844186    98.227197   -94.081375
H                   83.934228   105.284098   -90.993307
H                   84.618711   102.856566   -89.601541
H                   85.595757   103.783705   -92.321645
H                   85.916709   102.234551   -91.588727
H                   83.384389   101.865204   -91.463600
H                   83.182035   103.337637   -92.354971
H                   83.030068   101.860726   -94.074705
H                   84.750000   102.179427   -94.142247
H                   84.039658    99.931718   -92.453102
```

```
H         84.172315   100.665232   -95.874727
H         85.056511    99.282436   -96.242500
H         84.609834    97.785978   -93.207134
H         85.052402    97.639199   -94.867727
N         86.940010   105.123008   -89.737966
H         86.560450   105.799672   -90.364008
C         88.212505   105.406861   -89.101626
H         88.484850   106.425371   -89.345680
H         88.989495   104.740169   -89.446333
C         88.133118   105.259447   -87.584340
O         89.024886   104.778409   -86.931027
N         86.983288   105.699622   -87.051089
H         86.348706   106.211807   -87.621253
C         86.702010   105.635016   -85.652968
H         87.430140   106.179492   -85.064910
H         85.734072   106.089771   -85.486241
C         86.636557   104.245593   -85.054972
O         86.800080   104.086513   -83.861807
N         86.359970   103.221986   -85.872013
C         86.128823   101.916787   -85.315282
C         87.283881   101.366462   -84.475027
O         87.035882   100.603293   -83.563231
C         85.777631   100.957738   -86.455148
C         85.378247    99.569338   -85.974438
C         85.014693    98.676594   -87.145573
N         84.788958    97.325417   -86.653833
C         84.589353    96.280882   -87.456942
N         84.473045    96.435037   -88.740427
N         84.575798    95.062562   -86.890881
H         86.232549   103.376652   -86.850691
H         85.304491   101.959800   -84.613852
H         84.956013   101.405509   -87.004720
H         86.611942   100.899299   -87.145597
H         86.195926    99.114501   -85.426824
H         84.538044    99.633667   -85.289026
H         84.121361    99.032048   -87.644325
H         85.825151    98.675771   -87.868658
H         85.126440    97.115351   -85.741585
H         84.557890    97.335024   -89.157514
H         84.134883    95.708871   -89.387702
H         84.157077    95.009896   -85.987816
H         84.341781    94.290325   -87.485623
N         88.553907   101.676187   -84.777850
H         88.749436   102.349648   -85.489956
C         89.637998   101.053997   -84.040000
H         90.573337   101.425791   -84.439201
H         89.622808    99.982081   -84.162464
C         89.630603   101.316709   -82.548031
O         90.098165   100.520882   -81.767576
N         89.115010   102.487008   -82.162003
H         88.632430   103.061036   -82.823094
C         89.058657   102.864311   -80.768358
H         90.039708   102.802252   -80.319354
H         88.735586   103.895341   -80.709448
```

```
C         88.137565   102.046750   -79.859018
O         88.254721   102.139073   -78.652880
N         87.247805   101.238655   -80.443251
C         86.370009   100.315857   -79.737068
C         87.124733    99.349981   -78.815259
O         86.581631    98.868850   -77.840607
C         85.588847    99.566950   -80.841655
C         84.858680    98.279909   -80.477313
C         83.687065    98.395822   -79.510349
C         82.945878    97.063082   -79.468608
N         82.255077    96.880851   -78.157549
O         85.103643    96.462317   -77.143221
H         85.545181    97.267622   -77.404227
H         85.829183    95.895357   -76.881379
H         87.206535   101.225808   -81.442249
H         85.699817   100.858096   -79.084582
H         84.895730   100.277360   -81.280485
H         86.296117    99.320832   -81.622140
H         84.493068    97.861931   -81.411831
H         85.562602    97.558534   -80.082009
H         84.051131    98.662901   -78.525077
H         82.996729    99.168668   -79.821149
H         82.194874    97.010157   -80.234768
H         83.626733    96.231196   -79.559210
H         82.947153    96.642834   -77.460071
H         81.732906    97.731477   -77.888236
H         81.577692    96.138210   -78.218545
N         88.376128    99.066973   -79.165214
C         89.297966    98.188569   -78.474283
C         89.612173    98.582140   -77.037298
O         90.014538    97.707355   -76.294191
C         90.655418    98.172081   -79.213928
C         90.709104    97.389292   -80.533787
C         92.024743    97.699566   -81.254392
C         90.578579    95.884607   -80.293416
H         88.738055    99.521611   -79.977696
H         88.881440    97.194065   -78.413439
H         90.945657    99.202661   -79.392851
H         91.388944    97.749007   -78.536887
H         89.898718    97.721798   -81.175545
H         92.089993    97.149372   -82.188203
H         92.103297    98.754853   -81.482599
H         92.878444    97.413581   -80.644908
H         90.656152    95.333530   -81.224510
H         91.372831    95.536774   -79.640011
H         89.632694    95.616581   -79.834178
N         89.503149    99.858069   -76.666431
C         89.937503   100.344889   -75.365312
C         89.286734    99.737837   -74.102480
O         89.769289   100.032441   -73.041249
C         89.822106   101.870863   -75.349981
O         88.503397   102.298125   -75.535425
H         89.098744   100.507938   -77.307446
H         90.984010   100.093492   -75.237345
```

```
H            90.161046    102.224245    -74.388467
H            90.476573    102.278944    -76.113622
H            88.324841    102.468489    -76.452909
N            88.267900     98.862353    -74.213061
C            87.758936     98.093031    -73.101704
C            88.717325     96.980845    -72.649441
O            88.565271     96.490486    -71.561590
C            86.405544     97.456429    -73.453568
C            85.268125     98.449371    -73.731768
C            84.020672     97.688119    -74.186990
N            82.910071     98.545180    -74.626217
C            81.746668     98.729288    -74.010306
N            81.579476     98.362296    -72.743736
O            78.771497     99.031979    -72.329792
H            78.427100     99.923109    -72.372259
H            78.023845     98.448995    -72.364344
N            80.744059     99.295751    -74.650698
H            87.884602     98.701713    -75.116403
H            87.656584     98.739070    -72.241402
H            86.544051     96.812744    -74.317870
H            86.136906     96.812754    -72.623337
H            85.053580     99.036407    -72.842315
H            85.551921     99.154194    -74.504952
H            84.276222     97.062488    -75.028090
H            83.657913     97.030938    -73.407489
H            82.963096     98.918438    -75.550621
H            82.352948     98.120128    -72.170830
H            80.683869     98.498904    -72.314582
H            80.739753     99.396401    -75.660413
H            79.881797     99.411835    -74.160077
N            89.665394     96.584170    -73.525510
C            90.764916     95.739975    -73.102060
C            91.862268     96.607504    -72.483626
H            91.504252     97.514196    -71.990094
O            93.016979     96.332961    -72.508076
C            91.316038     94.869565    -74.241230
C            90.607398     93.563479    -74.434254
N            89.366522     93.441683    -75.022737
O            87.755351     95.491002    -76.398737
H            88.175505     94.777130    -75.913469
H            88.415958     96.174857    -76.398662
C            91.058441     92.336822    -74.090016
C            89.094145     92.178753    -75.026018
N            90.078245     91.453456    -74.471719
H            89.780820     97.094121    -74.375007
H            90.396214     95.112382    -72.300485
H            91.302759     95.437289    -75.165804
H            92.352817     94.661234    -74.015360
H            91.964225     92.019349    -73.620245
H            88.211223     91.729295    -75.427424
H            90.102594     90.464895    -74.380407
O            74.507540    101.730003    -99.520699
H            73.589744    101.470441    -99.592652
H            74.451621    102.574723    -99.080409
```

```
O      76.042036   97.186134  -101.466936
H      75.801427   98.065754  -101.745329
H      76.425376   97.319459  -100.605090
O      79.916150   98.022617  -101.655532
H      79.368247   98.247733  -100.899658
H      79.410089   97.376599  -102.145825
O      79.996857  101.141808   -94.579665
H      79.087427  100.893071   -94.766842
H      80.179334  100.773721   -93.724907
O      77.368167  100.182968   -94.988023
H      77.122977  100.309920   -95.902527
H      76.626213  100.464885   -94.459065
O      82.022836  100.672670   -96.457386
H      81.276149  100.777168   -95.860302
H      82.045694   99.764909   -96.757115
O      82.575844   98.052207  -100.620397
H      83.351700   97.773272  -101.111045
H      81.788372   97.799739  -101.100442
O      80.243882  100.619729  -102.889961
H      80.127842   99.719558  -102.588273
H      81.022498  100.933591  -102.437316
O      77.693695  103.927556  -100.940946
H      77.705328  103.297192  -101.663449
H      78.007827  104.746876  -101.299784
O      77.870373  101.976211  -102.994171
H      77.723884  102.145155  -103.915970
H      78.728441  101.531560  -102.936823
O      75.619680  105.398776   -99.446315
H      74.726834  105.125255   -99.263336
H      76.041278  104.698614   -99.934046
O      82.507087  100.877071  -101.017734
H      82.465882   99.938170  -100.820607
H      83.442355  101.043577  -101.056380
O      85.195660   97.256261  -101.473089
H      85.912758   97.218678  -102.095585
H      85.635296   97.387065  -100.631952
O      84.614806  100.200120   -99.055257
H      85.249067   99.479963   -99.106980
H      83.854666   99.790726   -98.652409
O      82.814343   98.252510   -97.759345
H      82.637231   97.980084   -98.658428
H      83.628384   97.820426   -97.491469
O      77.876608   98.736582   -99.816693
H      77.798909   99.168148   -98.968305
H      77.380454   99.308511  -100.403393
O      77.039320  100.675910   -97.862634
H      76.233166  100.923593   -98.307431
H      77.674849  101.373910   -98.043770
O      78.911034  102.710480   -98.576782
H      78.608221  103.185963   -99.346478
H      79.861775  102.682283   -98.622197
O      82.054875  102.422503   -98.610613
H      82.145826  101.930311   -99.427723
H      81.953386  101.771986   -97.910244
```

```
O           76.107880   100.215502 -101.496631
H           75.553534   100.773736 -100.953831
H           76.578895   100.797375 -102.090013
O           78.173697    96.008124 -102.796895
H           77.972370    95.945942 -103.721819
H           77.368646    96.328253 -102.370430

Case 9:
Voltage = +70 mV;Y266 neutral; R300 neutral; E183 neutral
N           15.580942    -6.451471    5.391415
H           14.780349    -6.201766    5.939146
C           15.509367    -5.856426    4.054303
C           14.783016    -6.782958    3.060460
O           14.052521    -6.345772    2.186629
C           16.919361    -5.576248    3.520821
H           16.389372    -6.127384    5.884687
H           14.944498    -4.933617    4.050508
H           16.888494    -5.189753    2.509108
H           17.403232    -4.831330    4.142353
H           17.526831    -6.474492    3.529436
N           15.032384    -8.085605    3.186393
H           15.612054    -8.361765    3.949721
C           14.398655    -9.092385    2.333235
H           14.482396    -8.823759    1.292869
H           14.891390   -10.042255    2.492729
C           12.912798    -9.220863    2.671372
O           12.081474    -9.424771    1.825885
N           12.582805    -9.055889    3.966871
H           13.302698    -8.986571    4.648940
C           11.195100    -9.170804    4.421405
H           11.188267    -9.157896    5.503158
H           10.743722   -10.088319    4.077196
C           10.336963    -8.010603    3.900697
O            9.190772    -8.186042    3.584573
N           10.947108    -6.814557    3.855284
C           10.318434    -5.634083    3.317568
C           10.126894    -5.691133    1.799331
O            9.096452    -5.262764    1.337093
C           11.092693    -4.388012    3.776312
O           12.478661    -4.570737    3.769134
H           11.905473    -6.744232    4.106887
H            9.314542    -5.553322    3.701121
H           10.801112    -3.539672    3.167344
H           10.827326    -4.174458    4.801380
H           12.805421    -4.755040    2.897012
N           11.100275    -6.200613    1.024177
C           10.844930    -6.396984   -0.400192
C            9.694191    -7.385718   -0.601590
O            8.946504    -7.280192   -1.545833
C           12.133418    -6.785894   -1.186814
C           12.333972    -8.291280   -1.403984
```

```
C           12.160705   -6.060462   -2.533508
H           11.924268   -6.586240    1.431556
H           10.475301   -5.462697   -0.791660
H           12.965434   -6.416169   -0.598032
H           13.287800   -8.454577   -1.894898
H           12.332078   -8.857756   -0.483538
H           11.566859   -8.698071   -2.055726
H           13.055508   -6.324363   -3.086116
H           11.300428   -6.332005   -3.135436
H           12.160806   -4.985273   -2.400041
N            9.570408   -8.340330    0.328361
H           10.314480   -8.474432    0.974388
C            8.593032   -9.421193    0.241195
H            8.935861  -10.219590    0.886968
H            8.532017   -9.802139   -0.766597
C            7.181944   -9.047549    0.645357
O            6.257521   -9.505331    0.014451
N            7.004056   -8.228491    1.688362
C            5.670081   -7.742066    1.958390
C            5.250599   -6.730681    0.897072
O            4.073692   -6.634174    0.621488
C            5.416214   -7.237024    3.392842
C            5.559791   -8.381152    4.401516
C            6.257028   -6.027261    3.805597
H            7.781723   -7.961726    2.252011
H            4.987432   -8.561748    1.796393
H            4.375059   -6.929201    3.386405
H            5.269882   -8.044251    5.391231
H            4.922390   -9.219159    4.138918
H            6.582310   -8.737422    4.456785
H            5.916438   -5.655352    4.766019
H            7.299625   -6.296792    3.912474
H            6.184055   -5.212892    3.092997
N            6.171151   -6.006054    0.252849
C            5.775815   -5.188296   -0.881848
C            5.344935   -6.072896   -2.060131
O            4.321599   -5.821913   -2.652478
C            6.867606   -4.168045   -1.261631
C            7.064145   -3.150823   -0.123538
C            6.496595   -3.458456   -2.568499
C            8.356860   -2.342647   -0.237676
H            7.132427   -6.063805    0.511463
H            4.877360   -4.652106   -0.615760
H            7.798107   -4.708570   -1.408911
H            6.208311   -2.477482   -0.107810
H            7.074430   -3.664394    0.828510
H            7.201410   -2.668340   -2.788559
H            6.497445   -4.133933   -3.414165
H            5.508088   -3.015876   -2.502185
H            8.457266   -1.668840    0.606528
H            9.219071   -2.999136   -0.237414
H            8.392266   -1.743091   -1.140341
N            6.117711   -7.114918   -2.395362
H            7.012137   -7.235987   -1.968883
```

```
C     5.776849   -7.956143   -3.522959
H     6.607482   -8.630335   -3.690883
H     5.623490   -7.371145   -4.417519
C     4.507232   -8.783923   -3.322025
O     3.670649   -8.842852   -4.191822
N     4.393877   -9.435505   -2.157905
H     5.130851   -9.392153   -1.486720
C     3.206567  -10.189471   -1.830312
H     2.967419  -10.903250   -2.604295
H     3.396800  -10.730478   -0.911826
C     1.974201   -9.306854   -1.644807
O     0.887592   -9.694727   -2.007218
N     2.187865   -8.107004   -1.107166
C     1.065875   -7.180082   -1.078547
C     0.623278   -6.721166   -2.470731
O    -0.558585   -6.582566   -2.692751
C     1.286703   -5.920711   -0.247405
O     1.414449   -6.167423    1.112892
H     3.087781   -7.818076   -0.797810
H     0.211730   -7.692961   -0.667269
H     2.134735   -5.363828   -0.631909
H     0.397434   -5.326895   -0.377869
H     2.304978   -6.441587    1.286480
N     1.552246   -6.437224   -3.393002
C     1.156398   -6.032021   -4.729693
C     0.385423   -7.153760   -5.418744
O    -0.601390   -6.897794   -6.075867
C     2.369640   -5.538668   -5.551767
C     2.784659   -4.147221   -5.045031
C     2.073813   -5.517231   -7.056783
C     4.114937   -3.649070   -5.608139
H     2.522156   -6.493743   -3.167532
H     0.439066   -5.227110   -4.653854
H     3.186486   -6.232980   -5.378815
H     1.998346   -3.437083   -5.295685
H     2.856144   -4.168576   -3.965206
H     2.933347   -5.151736   -7.604695
H     1.849975   -6.504239   -7.442685
H     1.231972   -4.871160   -7.285116
H     4.414406   -2.735690   -5.104875
H     4.901076   -4.381227   -5.453355
H     4.061563   -3.435277   -6.669765
N     0.823559   -8.405150   -5.255441
H     1.669666   -8.581384   -4.754991
C     0.043390   -9.512153   -5.753610
H    -0.063568   -9.473792   -6.827452
H     0.553346  -10.429292   -5.487533
C    -1.365316   -9.540114   -5.166944
O    -2.319938   -9.767288   -5.874474
N    -1.472594   -9.318871   -3.858013
C    -2.776806   -9.346623   -3.236465
C    -3.690591   -8.211640   -3.700453
O    -4.888648   -8.419516   -3.743742
C    -2.680572   -9.315384   -1.703977
```

```
O         -3.925575    -9.579870    -1.118921
H         -0.656820    -9.233313    -3.288687
H         -3.294817   -10.251748    -3.520567
H         -1.999587   -10.088365    -1.375700
H         -2.290650    -8.357371    -1.376577
H         -4.614992    -9.178316    -1.633687
N         -3.140671    -7.031631    -3.981802
C         -3.902800    -5.911611    -4.509848
C         -4.444401    -6.236116    -5.902315
O         -5.608320    -6.045095    -6.172322
C         -3.003179    -4.655368    -4.508644
C         -3.446430    -3.519778    -5.415770
C         -4.768369    -3.093147    -5.495387
C         -2.499107    -2.874292    -6.202386
C         -5.125967    -2.067161    -6.354301
C         -2.850203    -1.834810    -7.048372
C         -4.171554    -1.432517    -7.132535
H         -2.163960    -6.909757    -3.812366
H         -4.770939    -5.740614    -3.893718
H         -2.925428    -4.312966    -3.481048
H         -2.010683    -4.954438    -4.814888
H         -5.527531    -3.578735    -4.912488
H         -1.472615    -3.193968    -6.163074
H         -6.159685    -1.773371    -6.422763
H         -2.096161    -1.352588    -7.644053
H         -4.452444    -0.639105    -7.803271
N         -3.570568    -6.729386    -6.788376
C         -4.012389    -7.071542    -8.120729
C         -5.136283    -8.106206    -8.097705
O         -6.014865    -8.052897    -8.918488
C         -2.861979    -7.603003    -8.979078
S         -1.645138    -6.351001    -9.496656
H         -2.629879    -6.913493    -6.509016
H         -4.438183    -6.201578    -8.597974
H         -2.357333    -8.419729    -8.480725
H         -3.278715    -7.983739    -9.901750
H         -1.045060    -6.180606    -8.329095
N         -5.078532    -9.066905    -7.155631
H         -4.257250    -9.188153    -6.602517
C         -6.085178   -10.116164    -7.180937
H         -6.256484   -10.456631    -8.190330
H         -5.714002   -10.946528    -6.591481
C         -7.459147    -9.722831    -6.613014
O         -8.481305   -10.087403    -7.126876
N         -7.418017    -9.015655    -5.478475
C         -8.641664    -8.655583    -4.788433
C         -9.503025    -7.657707    -5.553989
O        -10.700950    -7.657359    -5.447580
C         -8.362789    -8.115649    -3.374852
C         -7.517819    -6.836019    -3.258320
C         -8.186702    -5.483367    -3.283259
O         -9.487484    -5.484213    -3.352610
H         -9.837195    -4.582227    -3.306516
O         -7.537218    -4.468877    -3.215821
```

```
H          -6.535760    -8.773291    -5.079991
H          -9.261399    -9.536011    -4.691970
H          -9.318829    -7.979455    -2.890166
H          -7.839592    -8.900272    -2.838741
H          -6.984949    -6.853250    -2.314144
H          -6.752209    -6.796859    -4.014506
N          -8.837317    -6.734064    -6.273863
C          -9.562353    -5.706192    -6.995284
C         -10.429282    -6.285649    -8.104962
H         -10.447834    -7.368339    -8.211261
O         -11.058382    -5.575959    -8.833188
C          -8.634934    -4.592890    -7.533973
O          -9.429234    -3.468192    -7.883582
C          -7.760018    -4.998713    -8.712559
H          -7.855346    -6.834553    -6.418399
H         -10.250263    -5.232842    -6.304175
H          -8.009124    -4.274700    -6.711601
H         -10.132138    -3.782364    -8.446313
H          -7.080829    -4.186024    -8.947006
H          -7.170316    -5.879379    -8.504966
H          -8.365083    -5.194266    -9.591490
N         -12.952321    -9.022071     3.174379
H         -13.222242    -9.545279     2.367114
C         -11.777508    -8.184389     2.899330
C         -10.788044    -8.138268     4.046003
O          -9.654030    -7.745285     3.846063
C         -12.130734    -6.717775     2.577797
O         -12.657555    -6.188825     3.767839
C         -13.123326    -6.574357     1.430132
H         -13.731255    -8.457326     3.453056
H         -11.232605    -8.601484     2.063655
H         -11.209285    -6.212204     2.322373
H         -12.680798    -5.236074     3.711916
H         -13.268773    -5.522576     1.207991
H         -12.771579    -7.061413     0.528011
H         -14.088830    -6.989115     1.694065
N         -11.197267    -8.533701     5.241859
H         -12.124328    -8.896923     5.283816
C         -10.409167    -8.524250     6.449423
H         -10.331035    -7.527086     6.853346
H         -10.928864    -9.129539     7.182800
C          -8.972188    -9.040182     6.371330
O          -8.135436    -8.471324     7.028298
N          -8.657852   -10.121190     5.634760
H          -9.368763   -10.601945     5.131375
C          -7.269059   -10.552673     5.533184
H          -7.251934   -11.532386     5.073001
H          -6.828276   -10.630382     6.514139
C          -6.387526    -9.608322     4.692701
O          -5.219497    -9.466997     4.948475
N          -7.045373    -9.017976     3.701159
C          -6.304931    -8.012558     2.952689
C          -6.268090    -6.697759     3.725473
O          -5.276112    -6.007922     3.677171
```

```
C         -6.623074    -7.894316     1.449060
C         -7.795882    -7.059029     0.980280
C         -7.766000    -5.670957     1.082999
C         -8.866773    -7.649269     0.322235
C         -8.787869    -4.896672     0.560020
C         -9.880657    -6.876416    -0.220950
C         -9.848945    -5.497693    -0.097189
H         -8.039868    -9.019905     3.675706
H         -5.281581    -8.342691     2.985193
H         -5.724588    -7.489368     0.996675
H         -6.714187    -8.906752     1.073838
H         -6.925939    -5.186139     1.546078
H         -8.892693    -8.718667     0.200367
H         -8.738272    -3.825285     0.637789
H        -10.679672    -7.347393    -0.763725
H        -10.627261    -4.898628    -0.532836
N         -7.302440    -6.389884     4.510376
C         -7.241904    -5.272589     5.420553
C         -6.127059    -5.479045     6.445228
O         -5.497086    -4.519332     6.836380
C         -8.622454    -5.098367     6.088776
C         -8.872837    -3.753989     6.737360
C         -9.064940    -2.621771     5.950646
C         -8.980564    -3.631491     8.115883
C         -9.345869    -1.395886     6.527256
C         -9.270126    -2.406751     8.699869
C         -9.450419    -1.284332     7.908326
H         -8.162416    -6.883321     4.408297
H         -6.972272    -4.370777     4.889804
H         -9.367819    -5.249171     5.314884
H         -8.757533    -5.887072     6.816818
H         -8.996843    -2.697121     4.878790
H         -8.840289    -4.495153     8.742009
H         -9.481307    -0.533112     5.901194
H         -9.342837    -2.330540     9.769933
H         -9.649214    -0.328205     8.359915
N         -5.873327    -6.714919     6.870140
H         -6.522006    -7.441169     6.663147
C         -4.799226    -7.005427     7.793460
H         -4.840126    -6.373275     8.667686
H         -4.911987    -8.034134     8.111036
C         -3.402321    -6.831884     7.200144
O         -2.527365    -6.342320     7.874602
N         -3.202406    -7.252132     5.946462
C         -1.926492    -7.066544     5.273132
C         -1.685041    -5.592029     4.920417
O         -0.587000    -5.103790     5.087924
C         -1.845186    -7.996026     4.042074
C         -0.683925    -7.630368     3.116008
C         -1.736960    -9.457739     4.490962
H         -3.935891    -7.731633     5.470520
H         -1.130946    -7.313462     5.961147
H         -2.767813    -7.872914     3.479766
H         -0.619047    -8.353756     2.310238
```

```
H        -0.800258   -6.654658    2.665480
H         0.262332   -7.640071    3.648170
H        -1.751106  -10.114767    3.627348
H        -0.800602   -9.623412    5.017899
H        -2.550214   -9.753041    5.141056
N        -2.723898   -4.903800    4.437579
C        -2.597873   -3.516586    4.035586
C        -2.446008   -2.619124    5.261663
O        -1.559369   -1.795104    5.313952
C        -3.790373   -3.045871    3.189272
C        -4.037835   -3.782688    1.862411
C        -2.904281   -3.687319    0.855894
O        -1.795219   -4.103278    1.227963
O        -3.154229   -3.214010   -0.265332
H        -3.585472   -5.375546    4.270945
H        -1.693274   -3.397091    3.469942
H        -4.695979   -3.115782    3.784263
H        -3.635079   -1.989268    2.988289
H        -4.209364   -4.831126    2.054315
H        -4.937775   -3.375838    1.415311
N        -3.328770   -2.774724    6.264389
C        -3.116859   -2.037315    7.484959
C        -1.791620   -2.427091    8.149505
O        -1.125457   -1.586158    8.713508
C        -4.245902   -2.040988    8.526077
O        -4.451830   -3.295548    9.104538
C        -5.530771   -1.412628    7.998546
H        -4.008877   -3.501147    6.228544
H        -2.972408   -1.002401    7.215625
H        -3.869398   -1.417516    9.327519
H        -4.908463   -3.852384    8.485526
H        -6.261226   -1.373828    8.797319
H        -5.347703   -0.399615    7.653558
H        -5.953429   -1.981335    7.181070
N        -1.380408   -3.700466    8.043355
H        -2.027778   -4.421552    7.812090
C        -0.184104   -4.103740    8.736422
H        -0.230722   -3.886129    9.794177
H        -0.071111   -5.171594    8.602335
C         1.063837   -3.415977    8.208148
O         1.943392   -3.068667    8.963047
N         1.148099   -3.258383    6.871248
C         2.325512   -2.662498    6.301899
C         2.289608   -1.139382    6.162378
O         3.359845   -0.576666    6.026494
C         2.805039   -3.337807    5.016948
S         1.888257   -2.938296    3.504645
H         0.413717   -3.611876    6.290069
H         3.116173   -2.811897    7.021600
H         3.817145   -3.008490    4.830336
H         2.816852   -4.409073    5.162126
H         0.819535   -3.695637    3.703000
N         1.139670   -0.470506    6.273406
C         1.204805    0.968264    6.492091
```

```
C         1.756128     1.271223     7.870957
O         2.369161     2.288717     8.085908
C        -0.095209     1.740211     6.181737
C        -1.202670     1.553130     7.228888
C        -0.544879     1.447272     4.745363
C        -2.470298     2.359126     6.937464
H         0.270099    -0.963389     6.287514
H         1.952635     1.350962     5.815916
H         0.205266     2.781508     6.231591
H        -1.452777     0.506845     7.328849
H        -0.819884     1.863788     8.196479
H        -1.301750     2.153662     4.427816
H         0.291009     1.536211     4.056776
H        -0.952151     0.450631     4.642613
H        -3.144614     2.318997     7.787465
H        -2.244102     3.404899     6.745042
H        -3.006451     1.971509     6.078076
N         1.549780     0.323275     8.791255
H         0.890444    -0.402546     8.613047
C         1.977874     0.512251    10.160280
H         1.469252    -0.222981    10.770964
H         1.737371     1.501937    10.518378
C         3.481809     0.336581    10.305781
O         4.163662     1.126803    10.899107
N         3.999635    -0.753621     9.694705
H         3.390680    -1.463356     9.345149
C         5.424640    -0.938719     9.667230
H         5.638010    -1.909229     9.234347
H         5.854673    -0.904705    10.657557
C         6.156890     0.119123     8.844832
O         7.292104     0.411098     9.129501
N         5.502074     0.621268     7.799018
C         6.105378     1.756770     7.150115
C         5.525754     3.075976     7.654228
H         5.023096     3.036937     8.619263
O         5.650481     4.094321     7.046575
C         6.011748     1.676271     5.624080
C         6.760831     0.506655     5.019708
C         8.126684     0.320974     5.237773
C         6.097513    -0.401591     4.200082
C         8.810437    -0.725805     4.634383
C         6.779052    -1.446022     3.595243
C         8.139346    -1.608653     3.802465
H         4.574642     0.337206     7.571870
H         7.140280     1.766917     7.456520
H         4.969721     1.616607     5.335781
H         6.394175     2.610107     5.226508
H         8.656905     0.977099     5.906681
H         5.037177    -0.304328     4.064082
H         9.861415    -0.854335     4.828031
H         6.244709    -2.137611     2.968497
H         8.657635    -2.422252     3.328888
N        12.613548     9.596235    -1.248797
H        11.872888     9.914331    -1.845504
```

```
C        12.304855    8.264733   -0.719494
C        11.274118    8.450599    0.386898
O        10.265933    7.778869    0.416405
C        13.562743    7.581057   -0.165268
C        14.645171    7.276174   -1.206399
S        14.017413    6.202010   -2.529687
C        15.500160    6.053615   -3.555024
H        13.432955    9.572824   -1.822901
H        11.834482    7.616492   -1.447350
H        14.001230    8.205515    0.607268
H        13.255717    6.653873    0.304879
H        15.045479    8.188658   -1.633772
H        15.470364    6.783888   -0.704308
H        15.248997    5.398790   -4.375499
H        15.799031    7.017675   -3.945849
H        16.318596    5.616979   -2.997646
N        11.504255    9.412828    1.296996
C        10.594230    9.607677    2.408576
C         9.277006   10.274769    2.010397
O         8.280961   10.053319    2.652723
C        11.276203   10.368772    3.552226
C        12.447520    9.563777    4.084885
O        13.582550    9.825193    3.745343
N        12.143330    8.561481    4.919061
H        12.328022    9.967725    1.206308
H        10.295334    8.637602    2.771156
H        11.657595   11.325890    3.219193
H        10.545941   10.537611    4.333516
H        12.886830    7.965718    5.213395
H        11.221666    8.175583    4.954638
N         9.263820   11.067708    0.922336
H        10.110806   11.229613    0.427402
C         8.003507   11.517472    0.368847
H         8.201268   12.260205   -0.393381
H         7.389112   11.966735    1.131893
C         7.186380   10.384432   -0.254307
O         5.986325   10.470163   -0.291674
N         7.862646    9.333243   -0.763184
C         7.170360    8.145621   -1.233907
C         6.634551    7.300767   -0.067358
O         5.534115    6.808599   -0.148417
C         8.048803    7.344196   -2.224169
C         8.343623    8.197115   -3.472857
C         7.379229    6.019179   -2.608087
C         9.428272    7.624966   -4.388863
H         8.823800    9.230617   -0.536510
H         6.279715    8.471812   -1.748434
H         8.987821    7.117690   -1.727244
H         7.421746    8.323158   -4.036246
H         8.645736    9.192750   -3.165369
H         7.996585    5.478783   -3.314292
H         7.228035    5.377911   -1.748270
H         6.409560    6.182911   -3.066614
H         9.622315    8.305493   -5.211529
```

```
H         10.363575      7.479448     -3.855143
H          9.148417      6.670885     -4.819956
N          7.387121      7.175059      1.043940
C          6.878148      6.458328      2.201978
C          5.527294      7.040202      2.639577
O          4.574441      6.334199      2.866783
C          7.832292      6.545691      3.403562
C          9.253664      5.996720      3.243963
O         10.145200      6.631053      3.801319
O          9.427037      4.934897      2.595238
H          8.347205      7.444682      1.022838
H          6.681042      5.429620      1.943308
H          7.920530      7.570200      3.733380
H          7.374144      5.992143      4.220647
N          5.489861      8.378154      2.785912
H          6.321206      8.912914      2.648845
C          4.299565      9.047333      3.249675
H          4.538969     10.091718      3.406566
H          3.957722      8.641020      4.188490
C          3.136817      8.988654      2.268517
O          1.995394      9.111339      2.649451
N          3.470151      8.825006      0.985194
H          4.431261      8.828978      0.735190
C          2.517174      8.905264     -0.078282
H          3.059635      9.091925     -0.996343
H          1.843680      9.737834      0.072013
C          1.609145      7.687007     -0.310220
O          0.563858      7.814945     -0.883301
N          2.090528      6.538674      0.206384
C          1.283381      5.366494      0.435195
C          0.467875      5.480594      1.727562
O         -0.727393      5.301042      1.700501
C          2.156685      4.111974      0.446513
H          2.996864      6.575239      0.622653
H          0.551923      5.295989     -0.352655
H          1.554370      3.234745      0.657176
H          2.624951      3.991708     -0.523505
H          2.937305      4.180443      1.196957
N          1.100166      5.752938      2.883496
C          0.363908      5.588936      4.128803
C         -0.756441      6.616054      4.277688
O         -1.793301      6.307749      4.825540
C          1.277017      5.517294      5.377563
C          1.869171      6.863332      5.822217
C          2.363575      4.450435      5.174566
C          2.702782      6.768521      7.102590
H          2.087549      5.892476      2.903068
H         -0.160392      4.645887      4.073071
H          0.617019      5.179838      6.171072
H          2.477859      7.271919      5.022362
H          1.065842      7.573320      5.990239
H          2.734478      4.098798      6.126137
H          1.969683      3.592879      4.638409
H          3.205836      4.838596      4.612193
```

```
H      3.008334    7.759834    7.424055
H      2.126734    6.325663    7.909692
H      3.597632    6.172900    6.971536
N     -0.542353    7.851148    3.802925
H      0.342661    8.093820    3.408237
C     -1.514248    8.891052    4.018116
H     -1.105100    9.819454    3.640120
H     -1.725168    9.021780    5.069665
C     -2.867021    8.618723    3.365915
O     -3.871177    8.835136    4.003743
N     -2.902653    8.109029    2.120712
C     -4.238324    8.007130    1.534100
C     -4.968810    6.730607    1.912165
O     -6.180377    6.723992    1.899404
C     -4.023439    8.231576    0.030898
C     -2.767606    9.109700    0.024624
C     -1.907994    8.459364    1.096316
H     -4.860642    8.797264    1.925061
H     -3.845864    7.294605   -0.480577
H     -4.888944    8.700104   -0.419709
H     -2.258156    9.157674   -0.919537
H     -3.027601   10.122702    0.322427
H     -1.423342    7.570099    0.735637
H     -1.150039    9.118658    1.486793
N     -4.242059    5.660668    2.268948
C     -4.898311    4.489696    2.822075
C     -5.601942    4.809620    4.143697
O     -6.682156    4.319140    4.387286
C     -3.922513    3.315617    3.044840
C     -3.602555    2.490632    1.814460
C     -4.319968    1.333086    1.526420
C     -2.563037    2.837884    0.964040
C     -4.010876    0.545635    0.427664
C     -2.253031    2.067610   -0.142613
C     -2.971130    0.916942   -0.418016
O     -2.631708    0.190538   -1.507631
H     -3.433330   -0.132565   -1.959934
H     -3.248637    5.746501    2.321824
H     -5.684982    4.182788    2.150843
H     -3.011435    3.701906    3.489434
H     -4.374530    2.664717    3.784611
H     -5.129536    1.032893    2.169558
H     -1.985421    3.717452    1.156964
H     -4.552745   -0.361458    0.235267
H     -1.460382    2.361604   -0.807624
N     -4.963985    5.610804    5.012229
H     -4.035727    5.925115    4.817424
C     -5.577479    5.988549    6.265746
H     -4.801211    6.365451    6.920165
H     -6.043964    5.136989    6.734166
C     -6.651474    7.065805    6.114413
O     -7.709535    6.987358    6.695136
N     -6.341993    8.091462    5.314137
H     -5.441195    8.148173    4.889085
```

```
C         -7.257699    9.220084    5.085090
H         -6.773806    9.881174    4.377832
H         -7.436862    9.765711    5.999948
C         -8.620673    8.834837    4.545706
O         -9.605806    9.430594    4.903538
N         -8.666130    7.828883    3.652697
C         -9.928735    7.427001    3.085873
C        -10.814144    6.631211    4.032578
O        -11.919862    6.304410    3.637505
C         -9.727464    6.646934    1.766796
O        -10.949256    6.584046    1.057330
C         -9.151248    5.252108    1.967152
H         -7.818197    7.408776    3.331769
H        -10.503099    8.315235    2.856328
H         -9.047528    7.228489    1.158208
H        -11.656023    6.539435    1.697411
H         -8.976723    4.785046    1.004296
H         -8.208905    5.278259    2.490920
H         -9.838835    4.627571    2.524251
N        -10.353503    6.328342    5.241144
C        -11.230850    5.723473    6.228817
C        -12.252751    6.705790    6.808551
O        -13.256428    6.277675    7.316005
C        -10.435281    5.093264    7.379626
C         -9.488829    3.950679    6.982356
C         -8.704761    3.492370    8.215910
C        -10.221569    2.767193    6.343598
H         -9.434347    6.600099    5.520645
H        -11.827490    4.972438    5.737201
H         -9.859024    5.870540    7.870016
H        -11.161263    4.730326    8.100330
H         -8.772341    4.325287    6.260793
H         -7.994812    2.711349    7.958815
H         -8.150205    4.316678    8.652154
H         -9.369959    3.098927    8.981775
H         -9.518635    1.971544    6.116205
H        -10.980974    2.365598    7.012467
H        -10.707399    3.037330    5.412318
N        -11.935445    8.016698    6.787516
C        -12.884695    9.000950    7.215647
C        -13.851947    9.432812    6.133866
H        -14.612175   10.146788    6.470712
O        -13.816147    9.067787    5.006124
H        -11.174724    8.331014    6.225290
H        -13.454872    8.617012    8.051621
H        -12.360078    9.886922    7.562692
N        -11.450004   10.622759   -3.030048
H        -11.209347   10.836129   -2.082610
C        -11.805542    9.208255   -3.171992
C        -11.410598    8.856561   -4.599609
O        -10.372854    8.290278   -4.846281
C        -13.244706    8.824652   -2.765620
C        -13.482122    9.163105   -1.282303
C        -13.513810    7.342058   -3.054218
```

```
C         -14.912145    8.922492   -0.794921
H         -12.216211   11.220192   -3.277637
H         -11.119634    8.644120   -2.555652
H         -13.943154    9.424616   -3.353506
H         -12.790744    8.584378   -0.673583
H         -13.242391   10.208439   -1.114120
H         -14.550839    7.090861   -2.872760
H         -13.298105    7.076750   -4.084999
H         -12.903473    6.712301   -2.414402
H         -15.024839    9.283592    0.220942
H         -15.634322    9.447362   -1.414760
H         -15.172680    7.870776   -0.791579
N         -12.211997    9.288083   -5.602657
H         -13.130153    9.596772   -5.377247
C         -11.972873    8.843875   -6.959238
H         -12.037064    7.767786   -7.045600
H         -12.729661    9.284178   -7.596912
C         -10.601375    9.232961   -7.517198
O          -9.996255    8.496520   -8.240744
N         -10.162429   10.477356   -7.181602
H         -10.711675   11.008704   -6.545682
C          -8.894348   11.024005   -7.637086
H          -8.790089   10.891089   -8.702746
H          -8.878869   12.081540   -7.407775
C          -7.672362   10.359334   -6.988510
O          -6.752644    9.936648   -7.631237
N          -7.659719   10.350048   -5.619869
H          -8.482692   10.627738   -5.134437
C          -6.495283    9.961236   -4.832291
H          -5.599773   10.376146   -5.266681
H          -6.615414   10.362680   -3.833476
C          -6.266714    8.464815   -4.710673
O          -5.130741    8.052021   -4.722029
N          -7.313250    7.638968   -4.532827
C          -7.014200    6.281823   -4.104247
C          -6.385416    5.453733   -5.218218
O          -5.500749    4.680846   -4.941532
C          -8.169386    5.547942   -3.385495
C          -9.334503    5.136864   -4.300510
C          -8.608947    6.363530   -2.163665
C         -10.433979    4.357572   -3.576425
H          -8.251741    7.983907   -4.545552
H          -6.212499    6.347950   -3.384840
H          -7.712696    4.634775   -3.012828
H          -9.766409    6.006148   -4.780323
H          -8.936982    4.509370   -5.094599
H          -9.268349    5.788334   -1.528201
H          -7.751178    6.646670   -1.562057
H          -9.133613    7.266301   -2.453200
H         -11.164557    3.981301   -4.285647
H         -10.030402    3.505615   -3.040225
H         -10.965010    4.979204   -2.865860
N          -6.821822    5.574666   -6.491817
C          -6.122655    4.837517   -7.526349
```

```
C         -4.766885    5.445593   -7.897275
O         -3.900145    4.709900   -8.308774
C         -6.985394    4.553869   -8.764595
C         -8.063608    3.474780   -8.557474
C         -7.486040    2.123528   -8.106951
N         -8.413891    0.992753   -8.200354
C         -9.247023    0.580826   -7.240055
N         -9.565917    1.405081   -6.254299
N         -9.750419   -0.632651   -7.274070
H         -7.500531    6.267906   -6.723047
H         -5.828205    3.900592   -7.087760
H         -7.460499    5.467402   -9.109894
H         -6.316779    4.232109   -9.555651
H         -8.808220    3.828074   -7.849781
H         -8.583797    3.332195   -9.498826
H         -6.639544    1.867317   -8.730819
H         -7.118616    2.168991   -7.090414
H         -8.306663    0.377692   -8.976215
H         -9.344245    2.370532   -6.314187
H        -10.097070    1.113256   -5.449550
H         -9.353985   -1.381590   -7.810395
H        -10.478295   -0.911510   -6.634346
N         -4.571397    6.757500   -7.722194
H         -5.311895    7.353674   -7.430382
C         -3.257798    7.352383   -8.011132
H         -2.933305    7.094249   -9.006697
H         -3.359275    8.427002   -7.936831
C         -2.193936    6.877304   -7.033115
O         -1.066806    6.644459   -7.393122
N         -2.610473    6.768838   -5.766290
C         -1.723951    6.459952   -4.670781
C         -1.446916    4.973453   -4.542400
O         -0.318988    4.587468   -4.336016
C         -2.308861    7.054869   -3.383162
C         -1.727790    6.351443   -2.157648
C         -2.066160    8.570611   -3.416451
H         -3.535867    7.063198   -5.540745
H         -0.755605    6.900148   -4.860973
H         -3.377683    6.877671   -3.399219
H         -1.942950    6.894074   -1.256222
H         -2.144316    5.354939   -2.053116
H         -0.655569    6.273740   -2.226170
H         -2.801922    9.100706   -2.828503
H         -1.077657    8.810853   -3.039486
H         -2.147842    8.954143   -4.426881
N         -2.476509    4.120594   -4.661332
C         -2.166304    2.713538   -4.735604
C         -1.263021    2.445154   -5.932153
O         -0.390765    1.610691   -5.874936
C         -3.409253    1.801827   -4.796543
C         -4.384750    1.957396   -3.620250
C         -5.461605    0.858243   -3.552073
N         -5.024352   -0.309686   -2.806819
C         -5.834088   -1.285026   -2.637352
```

```
N         -7.120142    -1.337912    -3.162328
N         -5.545208    -2.328950    -1.829388
H         -3.402241     4.448062    -4.837198
H         -1.584983     2.432899    -3.869933
H         -3.941363     1.981098    -5.729249
H         -3.042255     0.782421    -4.840102
H         -3.838795     1.977506    -2.689048
H         -4.885565     2.909052    -3.704949
H         -6.351102     1.287798    -3.090722
H         -5.745348     0.597643    -4.575397
H         -7.226829    -0.855583    -4.027469
H         -7.521277    -2.257450    -3.202406
H         -4.636224    -2.422398    -1.418379
H         -6.040804    -3.182966    -1.968389
N         -1.459772     3.193335    -7.035054
H         -2.267192     3.765881    -7.144744
C         -0.596016     2.984490    -8.179296
H         -0.940318     3.627846    -8.978864
H         -0.614398     1.958679    -8.517973
C          0.858094     3.326702    -7.861330
O          1.786035     2.683375    -8.284624
N          1.007247     4.396249    -7.063579
H          0.213446     4.960297    -6.860737
C          2.282316     4.880650    -6.640295
H          2.918456     5.142312    -7.476474
H          2.122221     5.774091    -6.050920
C          3.094582     3.926515    -5.779533
O          4.300635     4.045656    -5.701005
N          2.428034     2.983354    -5.101941
C          3.098460     2.130521    -4.156539
C          4.281984     1.344235    -4.723594
O          5.201684     1.054362    -3.984935
C          2.026255     1.207884    -3.579586
C          2.417310     0.331931    -2.395276
C          1.121230    -0.251754    -1.856780
N          1.333350    -1.247206    -0.825494
C          0.341807    -2.072958    -0.445718
N         -0.814787    -2.039801    -1.064745
N          0.558512    -2.942064     0.513940
H          1.433643     2.939341    -5.176232
H          3.540110     2.726300    -3.366546
H          1.214694     1.858336    -3.271280
H          1.624632     0.593830    -4.377160
H          3.092263    -0.463522    -2.691370
H          2.922286     0.917915    -1.630625
H          0.483946     0.548459    -1.492586
H          0.599758    -0.734748    -2.670949
H          2.042725    -1.074796    -0.148314
H         -1.141342    -1.192989    -1.482243
H         -1.575377    -2.653242    -0.776559
H          1.472900    -3.119781     0.859926
H         -0.227485    -3.495225     0.850629
N          4.281642     0.945295    -6.005331
H          3.540514     1.224776    -6.614846
```

```
C         5.363426    0.116658   -6.504603
H         5.139423   -0.143515   -7.531654
H         5.441111   -0.797366   -5.937306
C         6.736237    0.756120   -6.476105
O         7.738575    0.089675   -6.364842
N         6.766121    2.085332   -6.608323
H         5.912179    2.604995   -6.573019
C         8.005509    2.823538   -6.551101
H         8.709834    2.439534   -7.275449
H         7.800300    3.854283   -6.807558
C         8.748871    2.826732   -5.213347
O         9.890154    3.235828   -5.161891
N         8.103920    2.338653   -4.145971
C         8.685569    2.174685   -2.824698
C         9.950210    1.305456   -2.832305
O        10.768807    1.388487   -1.930967
C         7.581462    1.507752   -1.974036
C         7.779545    1.384045   -0.463395
C         7.439220    2.653824    0.326219
C         7.328841    2.386785    1.823564
N         8.668931    2.353589    2.481038
O        11.041280    1.616744    0.893292
H        10.891336    1.827886   -0.028619
H        11.519641    0.790022    0.842503
H         7.157843    2.036144   -4.258798
H         8.984887    3.132101   -2.419072
H         6.665568    2.053317   -2.157635
H         7.411935    0.520992   -2.389253
H         7.099674    0.604947   -0.132388
H         8.774806    1.030482   -0.228813
H         8.152705    3.454140    0.150925
H         6.474304    3.026469   -0.002372
H         6.768461    3.159097    2.327029
H         6.847680    1.440721    2.018165
H         9.379709    1.889596    1.923959
H         8.989016    3.323756    2.635285
H         8.615271    1.875867    3.368782
N        10.062478    0.448235   -3.837531
C        11.109288   -0.537620   -4.069164
C        12.486185    0.023066   -4.427535
O        13.443856   -0.705807   -4.301670
C        10.700469   -1.444643   -5.251751
C         9.562495   -2.442934   -4.995550
C         9.129826   -3.067907   -6.325579
C         9.990453   -3.533264   -4.013046
H         9.324245    0.446354   -4.510697
H        11.265001   -1.124722   -3.175914
H        10.433405   -0.799467   -6.083409
H        11.583144   -1.997215   -5.550670
H         8.706388   -1.911705   -4.591231
H         8.328652   -3.784069   -6.170262
H         8.772094   -2.311226   -7.012127
H         9.956502   -3.596183   -6.794105
H         9.200793   -4.263319   -3.872571
```

```
H         10.860385    -4.059559    -4.391976
H         10.243938    -3.136158    -3.034849
N         12.574038     1.261815    -4.919239
C         13.814598     1.844700    -5.393665
C         14.958833     2.033070    -4.379249
O         16.013518     2.413551    -4.812238
C         13.497199     3.171601    -6.086541
O         12.918242     4.088351    -5.201903
H         11.756392     1.833549    -4.917202
H         14.248406     1.177890    -6.129408
H         14.420350     3.590566    -6.455251
H         12.846329     2.977977    -6.933265
H         11.970851     4.079335    -5.266599
N         14.792294     1.722322    -3.074271
C         15.928267     1.647463    -2.176843
C         16.804560     0.409991    -2.402623
O         17.934510     0.394028    -1.988350
C         15.495334     1.578022    -0.707310
C         14.638736     2.735142    -0.197993
C         14.285265     2.488615     1.269059
N         13.259778     3.383220     1.790963
C         13.437031     4.534134     2.418972
N         14.654643     5.009739     2.656236
O         15.027936     7.465602     4.210742
H         14.744204     8.352164     3.972083
H         15.745893     7.571535     4.821879
N         12.386879     5.222483     2.815285
H         13.900806     1.420456    -2.758176
H         16.561947     2.504372    -2.346507
H         14.954826     0.650121    -0.551663
H         16.407320     1.508875    -0.124521
H         15.156423     3.682909    -0.314349
H         13.719550     2.816364    -0.766363
H         13.888865     1.489253     1.375952
H         15.164630     2.531614     1.900891
H         12.319759     3.087225     1.619716
H         15.462651     4.517082     2.358613
H         14.789755     5.864579     3.168103
H         11.431450     4.942520     2.657106
H         12.478890     6.065589     3.335508
N         16.200243    -0.648980    -2.967249
C         16.870911    -1.911444    -3.168063
C         17.670040    -1.889279    -4.467661
H         18.062080    -0.909278    -4.748851
O         17.878681    -2.853544    -5.128762
C         15.868239    -3.070209    -3.149497
C         15.324776    -3.399867    -1.792053
N         14.403859    -2.627488    -1.112382
O         12.508213    -0.434098    -0.465247
H         13.145673    -1.037493    -0.851266
H         12.027814    -0.038330    -1.184376
C         15.619918    -4.490272    -1.049640
C         14.172683    -3.251551    -0.003857
N         14.883081    -4.381637     0.103104
```

```
H       15.285615   -0.550381   -3.352678
H       17.596037   -2.030736   -2.370677
H       15.050771   -2.842318   -3.824804
H       16.369122   -3.943997   -3.543957
H       16.275343   -5.314047   -1.234992
H       13.494340   -2.922022    0.757056
H       14.782430   -5.064459    0.821874
O      -14.798003   -2.305642    5.077112
H      -15.090888   -1.616867    4.485156
H      -14.377476   -1.862811    5.806964
O      -12.886313   -0.760382    6.902943
H      -12.159802   -1.250025    7.273334
H      -13.153007   -0.137559    7.567914
O      -12.140772   -0.228791    4.090461
H      -12.285224   -0.223791    5.032520
H      -13.017444   -0.218849    3.702739
O      -13.359306   -2.024100    0.907285
H      -12.590290   -1.456457    0.826721
H      -13.169281   -2.625977    1.626936
O       -7.462894    2.078515   -0.250981
H       -7.582969    2.166579    0.697926
H       -6.548672    1.845155   -0.359334
O       -7.781878    2.147925    2.587834
H       -8.712664    1.982022    2.720968
H       -7.530710    2.835293    3.200006
O       -9.219556    0.268046   -1.543300
H       -8.654417    0.920861   -1.116260
H       -8.625594   -0.395291   -1.892853
O      -13.104983   -2.225893   -1.908178
H      -13.351355   -2.375510   -0.993819
H      -13.802447   -2.556578   -2.463598
O      -13.484588    1.744872   -0.191105
H      -12.857556    1.055739    0.036926
H      -12.999598    2.554874   -0.018958
O      -14.181231    4.914670    2.449534
H      -13.801872    5.401488    3.175814
H      -14.711660    4.211922    2.806152
O      -15.446011    2.145884    1.861697
H      -16.298797    2.345407    1.497799
H      -14.850135    2.051818    1.115481
O      -13.497749    0.498958   -2.728436
H      -13.706908    1.081122   -1.997612
H      -13.279855   -0.342565   -2.324130
O      -14.513559   -1.463913   -4.511132
H      -14.515837   -0.619962   -4.063929
H      -15.310446   -1.513964   -5.022410
O      -10.574390   -2.969205   -3.202248
H      -10.866490   -2.657676   -4.055481
H      -11.318892   -2.817667   -2.624688
O      -11.340505   -0.095921    0.389263
H      -10.663811   -0.135502   -0.288446
H      -10.954446    0.397240    1.116547
O      -10.645926    1.635178    2.510120
H      -11.151041    1.180674    3.185435
```

| | | | |
|---|---|---|---|
| H | -11.153101 | 2.384321 | 2.199776 |
| O | -12.086338 | 3.782159 | 1.051177 |
| H | -12.806634 | 4.185138 | 1.549342 |
| H | -11.540705 | 4.511703 | 0.772887 |
| O | -10.888322 | 0.832962 | -3.747451 |
| H | -11.766127 | 0.977606 | -3.390586 |
| H | -10.308683 | 0.683062 | -2.994652 |
| O | -14.735511 | -0.392918 | 2.923478 |
| H | -15.121325 | 0.439349 | 2.649763 |
| H | -14.531491 | -0.883649 | 2.129065 |
| O | -12.750242 | -3.277893 | 3.454330 |
| H | -13.513525 | -3.074626 | 4.006952 |
| H | -12.113276 | -2.610803 | 3.692382 |
| O | -11.716553 | -1.375614 | -5.293525 |
| H | -12.662337 | -1.516011 | -5.271893 |
| H | -11.531061 | -0.699335 | -4.643453 |

Case 10:
Voltage -70mV; R297 neutral; R300 positive; Y266 neutral; E183 neutral

| | | | |
|---|---|---|---|
| N | 15.580944 | -6.451469 | 5.391416 |
| H | 14.782832 | -6.196337 | 5.940199 |
| C | 15.511605 | -5.857197 | 4.053821 |
| C | 14.783016 | -6.782962 | 3.060452 |
| O | 14.050303 | -6.345718 | 2.188087 |
| C | 16.922531 | -5.580949 | 3.520537 |
| H | 16.392404 | -6.133008 | 5.883385 |
| H | 14.949323 | -4.932793 | 4.049274 |
| H | 16.893283 | -5.195456 | 2.508377 |
| H | 17.407782 | -4.836530 | 4.141514 |
| H | 17.527991 | -6.480509 | 3.530406 |
| N | 15.032378 | -8.085605 | 3.186405 |
| H | 15.613916 | -8.361722 | 3.948379 |
| C | 14.398660 | -9.092385 | 2.333233 |
| H | 14.483649 | -8.824955 | 1.292643 |
| H | 14.889838 | -10.042789 | 2.494249 |
| C | 12.912291 | -9.217675 | 2.670855 |
| O | 12.079485 | -9.415597 | 1.825046 |
| N | 12.582804 | -9.055884 | 3.966877 |
| H | 13.302950 | -8.991613 | 4.649305 |
| C | 11.195108 | -9.170813 | 4.421378 |
| H | 11.188158 | -9.157638 | 5.503122 |
| H | 10.743550 | -10.088393 | 4.077472 |
| C | 10.336944 | -8.010597 | 3.900733 |
| O | 9.190090 | -8.185982 | 3.585436 |
| N | 10.946503 | -6.814815 | 3.855488 |
| C | 10.318587 | -5.633861 | 3.318193 |
| C | 10.125972 | -5.692625 | 1.800329 |
| O | 9.093403 | -5.267864 | 1.337702 |
| C | 11.095880 | -4.388400 | 3.775379 |
| O | 12.480642 | -4.576241 | 3.775153 |

```
H         11.904440    -6.744133     4.109092
H          9.315080    -5.551484     3.702653
H         10.810141    -3.541382     3.161721
H         10.826622    -4.169772     4.798328
H         12.812492    -4.760227     2.904811
N         11.099831    -6.200494     1.025787
C         10.845908    -6.397088    -0.398610
C          9.692193    -7.382489    -0.598046
O          8.938680    -7.270512    -1.537477
C         12.134340    -6.790106    -1.184095
C         12.326945    -8.295537    -1.408224
C         12.168942    -6.058386    -2.527233
H         11.926436    -6.581206     1.432858
H         10.478921    -5.462351    -0.791613
H         12.967055    -6.427953    -0.591651
H         13.279521    -8.461207    -1.900545
H         12.323387    -8.866061    -0.490335
H         11.557591    -8.695594    -2.061577
H         13.064520    -6.323141    -3.077966
H         11.309865    -6.324251    -3.133521
H         12.172770    -4.983785    -2.388962
N          9.570429    -8.340341     0.328313
H         10.317411    -8.477743     0.970708
C          8.593020    -9.421183     0.241228
H          8.938957   -10.221455     0.883024
H          8.529024    -9.799398    -0.767446
C          7.183046    -9.051326     0.651615
O          6.256027    -9.500598     0.018322
N          7.006982    -8.242711     1.703279
C          5.672757    -7.764598     1.982476
C          5.245834    -6.744947     0.932315
O          4.064671    -6.629984     0.679929
C          5.421989    -7.271619     3.421906
C          5.578107    -8.421083     4.422546
C          6.255897    -6.058412     3.839292
H          7.784801    -7.983695     2.270846
H          4.991836    -8.585061     1.816006
H          4.378224    -6.972364     3.422568
H          5.289724    -8.092428     5.415343
H          4.945933    -9.262091     4.156977
H          6.603363    -8.769904     4.471337
H          5.911493    -5.689342     4.799396
H          7.299545    -6.322650     3.947399
H          6.182573    -5.243022     3.127170
N          6.164890    -6.029080     0.275759
C          5.770072    -5.213307    -0.860195
C          5.347411    -6.096054    -2.042917
O          4.317254    -5.850966    -2.627325
C          6.855696    -4.183736    -1.232577
C          7.038496    -3.167175    -0.091958
C          6.487832    -3.476931    -2.541721
C          8.319733    -2.340575    -0.200328
H          7.129221    -6.104574     0.518552
H          4.865271    -4.686414    -0.597427
```

```
H         7.791690   -4.715255   -1.374712
H         6.174029   -2.502179   -0.076232
H         7.052934   -3.683172    0.858675
H         7.186715   -2.680881   -2.758537
H         6.497850   -4.151953   -3.387350
H         5.493954   -3.043581   -2.483179
H         8.408427   -1.668565    0.646299
H         9.189581   -2.986423   -0.199342
H         8.349605   -1.738399   -1.101241
N         6.128270   -7.126727   -2.387644
H         7.024927   -7.244000   -1.964099
C         5.789163   -7.967920   -3.516368
H         6.618179   -8.644925   -3.680129
H         5.641448   -7.383546   -4.412301
C         4.515381   -8.789090   -3.317512
O         3.678682   -8.841814   -4.187138
N         4.394747   -9.437883   -2.152257
H         5.135079   -9.408404   -1.483824
C         3.204838  -10.190943   -1.832486
H         2.965798  -10.900276   -2.610543
H         3.390063  -10.737169   -0.916088
C         1.974189   -9.306854   -1.644809
O         0.885733   -9.694433   -1.998734
N         2.187870   -8.107002   -1.107178
C         1.062559   -7.183939   -1.093339
C         0.625644   -6.728287   -2.490052
O        -0.553371   -6.564369   -2.705302
C         1.306396   -5.909002   -0.303522
O         1.425876   -6.090887    1.073695
H         3.093245   -7.807585   -0.825694
H         0.207337   -7.689602   -0.675997
H         2.171890   -5.402376   -0.705835
H         0.434524   -5.290644   -0.453725
H         2.275644   -6.474839    1.251619
N         1.560075   -6.459190   -3.410895
C         1.175198   -6.057120   -4.752834
C         0.403905   -7.178331   -5.441239
O        -0.578272   -6.921050   -6.102342
C         2.394188   -5.571585   -5.570489
C         2.818567   -4.183024   -5.063528
C         2.099905   -5.548292   -7.075991
C         4.155520   -3.697045   -5.621981
H         2.528111   -6.544668   -3.187946
H         0.460238   -5.249683   -4.683541
H         3.206173   -6.271214   -5.395823
H         2.039236   -3.466766   -5.319259
H         2.885717   -4.205344   -3.982975
H         2.960283   -5.183822   -7.623167
H         1.875615   -6.534966   -7.462053
H         1.258316   -4.902315   -7.304658
H         4.460782   -2.782825   -5.123111
H         4.935350   -4.434388   -5.459971
H         4.108871   -3.488465   -6.684605
N         0.834495   -8.430959   -5.266447
```

| | | | |
|---|---:|---:|---:|
| H | 1.680689 | -8.609175 | -4.767666 |
| C | 0.043002 | -9.534250 | -5.754547 |
| H | -0.064765 | -9.504044 | -6.828438 |
| H | 0.542575 | -10.454744 | -5.480359 |
| C | -1.365270 | -9.540099 | -5.166932 |
| O | -2.320705 | -9.759469 | -5.873830 |
| N | -1.472643 | -9.318873 | -3.858023 |
| C | -2.777018 | -9.339307 | -3.239862 |
| C | -3.683781 | -8.203538 | -3.711059 |
| O | -4.881860 | -8.404585 | -3.729419 |
| C | -2.682961 | -9.300733 | -1.708000 |
| O | -3.931466 | -9.538744 | -1.120043 |
| H | -0.656879 | -9.250793 | -3.287101 |
| H | -3.298758 | -10.243281 | -3.521532 |
| H | -2.011731 | -10.080442 | -1.374186 |
| H | -2.279247 | -8.345514 | -1.388190 |
| H | -4.619728 | -9.184632 | -1.670852 |
| N | -3.124698 | -7.034042 | -4.020132 |
| C | -3.871663 | -5.911455 | -4.567907 |
| C | -4.422117 | -6.255810 | -5.951122 |
| O | -5.591691 | -6.097103 | -6.204226 |
| C | -2.948540 | -4.677092 | -4.594832 |
| C | -3.400815 | -3.497340 | -5.442305 |
| C | -4.734137 | -3.157859 | -5.641914 |
| C | -2.424107 | -2.712674 | -6.051406 |
| C | -5.076868 | -2.078151 | -6.444326 |
| C | -2.760707 | -1.626439 | -6.839481 |
| C | -4.095292 | -1.309104 | -7.044239 |
| H | -2.142627 | -6.928952 | -3.879626 |
| H | -4.734488 | -5.716874 | -3.951359 |
| H | -2.799767 | -4.365221 | -3.564509 |
| H | -1.982346 | -4.993242 | -4.963472 |
| H | -5.524754 | -3.743878 | -5.216772 |
| H | -1.385234 | -2.964190 | -5.920539 |
| H | -6.117700 | -1.865842 | -6.607831 |
| H | -1.985094 | -1.040440 | -7.300264 |
| H | -4.364405 | -0.481094 | -7.676299 |
| N | -3.550551 | -6.732514 | -6.849316 |
| C | -4.020122 | -7.089742 | -8.168719 |
| C | -5.157548 | -8.107697 | -8.103663 |
| O | -6.058204 | -8.049159 | -8.896311 |
| C | -2.891848 | -7.652717 | -9.035265 |
| S | -1.652032 | -6.433448 | -9.578139 |
| H | -2.605573 | -6.910890 | -6.583753 |
| H | -4.444975 | -6.222682 | -8.652448 |
| H | -2.399541 | -8.476884 | -8.536502 |
| H | -3.326831 | -8.030908 | -9.950558 |
| H | -1.055179 | -6.239888 | -8.412230 |
| N | -5.078560 | -9.066911 | -7.155611 |
| H | -4.240183 | -9.198161 | -6.632391 |
| C | -6.083457 | -10.117608 | -7.174241 |
| H | -6.252950 | -10.468510 | -8.180675 |
| H | -5.713328 | -10.942051 | -6.575574 |
| C | -7.459136 | -9.722605 | -6.613168 |

```
O        -8.477419   -10.100642   -7.123748
N        -7.417946   -9.016179    -5.478396
C        -8.627703   -8.635962    -4.777114
C        -9.494348   -7.638588    -5.533638
O        -10.689137  -7.633172    -5.384573
C        -8.325557   -8.065997    -3.379858
C        -7.408255   -6.828502    -3.308384
C        -7.986523   -5.451668    -3.112663
O        -9.270101   -5.314718    -3.264470
H        -9.518242   -4.398148    -3.089607
O        -7.269642   -4.521453    -2.820302
H        -6.531614   -8.755964    -5.102782
H        -9.252280   -9.509632    -4.653406
H        -9.277621   -7.860721    -2.913135
H        -7.853021   -8.857275    -2.808065
H        -6.708915   -6.944926    -2.492731
H        -6.795363   -6.738640    -4.191553
N        -8.837269   -6.733933    -6.273830
C        -9.555216   -5.702643    -7.000855
C        -10.256318  -6.316855    -8.208730
H        -9.862462   -7.279427    -8.543296
O        -11.178256  -5.796996    -8.750786
C        -8.586540   -4.546834    -7.345439
O        -8.268224   -3.847213    -6.167853
C        -9.112808   -3.587440    -8.409027
H        -7.853022   -6.817040    -6.406801
H        -10.340047  -5.310415    -6.367506
H        -7.652720   -4.970832    -7.689479
H        -9.030598   -3.343613    -5.896190
H        -8.422071   -2.755542    -8.489487
H        -9.196396   -4.062468    -9.377814
H        -10.090267  -3.200176    -8.143366
N        -12.952094  -9.022576     3.174130
H        -13.216901  -9.547583     2.366247
C        -11.773583  -8.180047     2.898863
C        -10.786093  -8.135624     4.047223
O        -9.649525   -7.747387     3.850698
C        -12.147817  -6.720270     2.577984
O        -12.697467  -6.209299     3.764260
C        -13.131863  -6.595668     1.419815
H        -13.730880  -8.454001     3.444854
H        -11.227247  -8.594061     2.063427
H        -11.234208  -6.194510     2.333023
H        -12.845903  -5.269987     3.664258
H        -13.292646  -5.547357     1.194535
H        -12.764938  -7.079791     0.522697
H        -14.092855  -7.025286     1.676157
N        -11.197211  -8.533634     5.241858
H        -12.126273  -8.892747     5.279768
C        -10.412578  -8.537930     6.450859
H        -10.339123  -7.546921     6.870680
H        -10.931211  -9.156445     7.174028
C        -8.974225   -9.047392     6.366825
O        -8.135712   -8.474959     7.019450
```

| | | | |
|---|---:|---:|---:|
| N | -8.657398 | -10.127885 | 5.631027 |
| H | -9.367470 | -10.613651 | 5.131026 |
| C | -7.267629 | -10.555077 | 5.531398 |
| H | -7.246342 | -11.534208 | 5.070133 |
| H | -6.827791 | -10.632406 | 6.512835 |
| C | -6.387485 | -9.608345 | 4.692714 |
| O | -5.218169 | -9.468574 | 4.948101 |
| N | -7.045385 | -9.018031 | 3.701138 |
| C | -6.306850 | -8.018586 | 2.940025 |
| C | -6.288934 | -6.674501 | 3.665630 |
| O | -5.321224 | -5.954560 | 3.552431 |
| C | -6.605722 | -7.941937 | 1.430798 |
| C | -7.822154 | -7.186177 | 0.941639 |
| C | -7.869949 | -5.796289 | 1.008111 |
| C | -8.862543 | -7.853766 | 0.310483 |
| C | -8.936268 | -5.095519 | 0.472303 |
| C | -9.923842 | -7.154338 | -0.242090 |
| C | -9.966682 | -5.773452 | -0.158825 |
| H | -8.039921 | -9.028712 | 3.668541 |
| H | -5.279958 | -8.335221 | 2.994033 |
| H | -5.725940 | -7.490773 | 0.985684 |
| H | -6.631674 | -8.962844 | 1.069055 |
| H | -7.055392 | -5.254366 | 1.454467 |
| H | -8.828901 | -8.925759 | 0.217073 |
| H | -8.947769 | -4.021331 | 0.518170 |
| H | -10.701020 | -7.685418 | -0.761221 |
| H | -10.779089 | -5.230818 | -0.605379 |
| N | -7.308710 | -6.373959 | 4.470858 |
| C | -7.262144 | -5.248549 | 5.373303 |
| C | -6.158832 | -5.450615 | 6.411412 |
| O | -5.518936 | -4.494565 | 6.796814 |
| C | -8.651681 | -5.083363 | 6.024518 |
| C | -8.893457 | -3.778673 | 6.753046 |
| C | -9.151201 | -2.612193 | 6.038946 |
| C | -8.930198 | -3.730745 | 8.140414 |
| C | -9.430403 | -1.425690 | 6.694870 |
| C | -9.216609 | -2.546231 | 8.803180 |
| C | -9.466407 | -1.389760 | 8.082576 |
| H | -8.156142 | -6.894665 | 4.400566 |
| H | -6.993456 | -4.347303 | 4.841354 |
| H | -9.382832 | -5.178463 | 5.228976 |
| H | -8.811105 | -5.911384 | 6.702096 |
| H | -9.146187 | -2.630416 | 4.962667 |
| H | -8.741306 | -4.623164 | 8.711498 |
| H | -9.621564 | -0.535090 | 6.124612 |
| H | -9.237943 | -2.528376 | 9.878076 |
| H | -9.672962 | -0.466163 | 8.593917 |
| N | -5.924518 | -6.684716 | 6.852051 |
| H | -6.579405 | -7.405856 | 6.646032 |
| C | -4.866585 | -6.981788 | 7.791277 |
| H | -4.901033 | -6.332816 | 8.653278 |
| H | -5.004584 | -8.002242 | 8.125133 |
| C | -3.465861 | -6.843414 | 7.200179 |
| O | -2.593785 | -6.310949 | 7.844689 |
| N | -3.256297 | -7.332595 | 5.971490 |
| C | -1.973017 | -7.163148 | 5.310820 |
| C | -1.726686 | -5.692940 | 4.956473 |
| O | -0.628667 | -5.208263 | 5.121851 |
| C | -1.875484 | -8.093466 | 4.081531 |
| C | -0.687726 | -7.740646 | 3.181573 |
| C | -1.786344 | -9.556404 | 4.531098 |
| H | -3.972980 | -7.862306 | 5.521919 |
| H | -1.183317 | -7.408606 | 6.006493 |
| H | -2.784403 | -7.963488 | 3.499032 |
| H | -0.621505 | -8.459108 | 2.371330 |
| H | -0.771242 | -6.757659 | 2.737987 |
| H | 0.245909 | -7.775128 | 3.735147 |
| H | -1.786505 | -10.212517 | 3.666813 |
| H | -0.862713 | -9.728195 | 5.077893 |
| H | -2.615053 | -9.847094 | 5.162825 |
| N | -2.758495 | -4.996092 | 4.464391 |
| C | -2.595515 | -3.622938 | 4.035444 |
| C | -2.473797 | -2.669150 | 5.225948 |
| O | -1.614235 | -1.808868 | 5.210449 |
| C | -3.733998 | -3.171539 | 3.109922 |
| C | -3.782268 | -3.862314 | 1.732549 |
| C | -2.628467 | -3.465903 | 0.823727 |
| O | -1.503384 | -3.862060 | 1.146447 |
| O | -2.878794 | -2.754466 | -0.178807 |
| H | -3.628493 | -5.453211 | 4.293832 |
| H | -1.662253 | -3.541483 | 3.507575 |
| H | -4.683425 | -3.333494 | 3.606390 |
| H | -3.629625 | -2.099678 | 2.974451 |
| H | -3.764594 | -4.935259 | 1.861759 |
| H | -4.717400 | -3.599612 | 1.252188 |
| N | -3.322598 | -2.798411 | 6.252692 |
| C | -3.093977 | -2.024905 | 7.456984 |
| C | -1.770250 | -2.419841 | 8.118195 |
| O | -1.080725 | -1.569856 | 8.638609 |
| C | -4.228084 | -2.006714 | 8.492193 |
| O | -4.451738 | -3.260511 | 9.065621 |
| C | -5.501765 | -1.364557 | 7.954337 |
| H | -3.991038 | -3.537687 | 6.255820 |
| H | -2.939552 | -0.999826 | 7.158257 |
| H | -3.848640 | -1.387571 | 9.295800 |
| H | -4.941623 | -3.800442 | 8.456879 |
| H | -6.238344 | -1.316874 | 8.746765 |
| H | -5.305382 | -0.354045 | 7.609717 |
| H | -5.924491 | -1.930325 | 7.134453 |
| N | -1.380395 | -3.700450 | 8.043326 |
| H | -2.037075 | -4.426279 | 7.854158 |
| C | -0.123636 | -4.078780 | 8.640191 |
| H | -0.076668 | -3.831240 | 9.691302 |
| H | -0.013673 | -5.149386 | 8.525951 |
| C | 1.060830 | -3.392291 | 7.973497 |
| O | 2.010897 | -3.010526 | 8.616609 |
| N | 0.994944 | -3.257262 | 6.636331 |
| C | 2.100088 | -2.676025 | 5.920370 |
| C | 2.210290 | −1.166170 | 6.105519 |
| O | 3.310629 | −0.655277 | 6.068740 |
| C | 2.065932 | −3.038324 | 4.424320 |
| S | 1.213057 | −1.870162 | 3.314856 |
| H | 0.260249 | −3.714683 | 6.135777 |
| H | 3.020693 | −3.064959 | 6.332515 |
| H | 3.085948 | −3.050682 | 4.062298 |
| H | 1.652644 | −4.029383 | 4.302568 |
| H | 0.009263 | −1.880556 | 3.866893 |
| N | 1.089820 | −0.460022 | 6.279344 |
| C | 1.204070 | 0.968487 | 6.491353 |
| C | 1.793252 | 1.251591 | 7.860497 |
| O | 2.449274 | 2.245324 | 8.064324 |
| C | −0.084412 | 1.782993 | 6.237232 |
| C | −1.150112 | 1.646674 | 7.336467 |
| C | −0.619632 | 1.506876 | 4.827843 |
| C | −2.404965 | 2.486042 | 7.085351 |
| H | 0.203274 | −0.918743 | 6.281692 |
| H | 1.941931 | 1.333103 | 5.793456 |
| H | 0.256136 | 2.812070 | 6.265931 |
| H | −1.427393 | 0.610107 | 7.471391 |
| H | −0.716338 | 1.964095 | 8.279535 |
| H | −1.335766 | 2.264397 | 4.536626 |
| H | 0.184783 | 1.520727 | 4.098589 |
| H | −1.110169 | 0.544821 | 4.761493 |
| H | −3.041027 | 2.482259 | 7.965003 |
| H | −2.157113 | 3.520431 | 6.861802 |
| H | −2.991507 | 2.100056 | 6.258571 |
| N | 1.549782 | 0.323269 | 8.791262 |
| H | 0.896480 | −0.405693 | 8.605850 |
| C | 1.981104 | 0.509769 | 10.158324 |
| H | 1.455787 | −0.209800 | 10.773727 |
| H | 1.758396 | 1.505735 | 10.510791 |
| C | 3.479650 | 0.307555 | 10.320961 |
| O | 4.154478 | 1.061788 | 10.967399 |
| N | 3.997964 | −0.764135 | 9.681464 |
| H | 3.397714 | −1.409685 | 9.213509 |
| C | 5.422868 | −0.948401 | 9.652385 |
| H | 5.636920 | −1.913285 | 9.207836 |
| H | 5.845795 | −0.926991 | 10.645481 |
| C | 6.156883 | 0.119117 | 8.844832 |
| O | 7.290225 | 0.414952 | 9.132108 |
| N | 5.502084 | 0.621277 | 7.799023 |
| C | 6.098613 | 1.759279 | 7.146320 |
| C | 5.521189 | 3.079027 | 7.649995 |
| H | 5.033582 | 3.044820 | 8.622585 |
| O | 5.635975 | 4.093661 | 7.034342 |
| C | 6.000411 | 1.671809 | 5.621389 |
| C | 6.758884 | 0.503759 | 5.027020 |
| C | 8.132640 | 0.352090 | 5.221768 |
| C | 6.097469 | −0.441019 | 4.249131 |
| C | 8.826854 | −0.695658 | 4.633359 |
| C | 6.790177 | −1.486997 | 3.659973 |
| C | 8.157457 | −1.614938 | 3.840168 |
| H | 4.568535 | 0.346971 | 7.585218 |
| H | 7.134963 | 1.771457 | 7.449061 |
| H | 4.957808 | 1.601779 | 5.336521 |
| H | 6.374162 | 2.605828 | 5.216566 |
| H | 8.660992 | 1.036948 | 5.863146 |
| H | 5.031944 | -0.371548 | 4.137594 |
| H | 9.883823 | -0.797813 | 4.809045 |
| H | 6.259361 | -2.211242 | 3.067736 |
| H | 8.683193 | -2.430190 | 3.378028 |
| N | 12.613550 | 9.596234 | -1.248801 |
| H | 11.872611 | 9.915204 | -1.844664 |
| C | 12.304852 | 8.264740 | -0.719487 |
| C | 11.276183 | 8.452035 | 0.388350 |
| O | 10.267876 | 7.780403 | 0.420039 |
| C | 13.563363 | 7.580163 | -0.167897 |
| C | 14.643963 | 7.276000 | -1.211139 |
| S | 14.014180 | 6.201837 | -2.533509 |
| C | 15.493465 | 6.058480 | -3.564712 |
| H | 13.432661 | 9.573379 | -1.823317 |
| H | 11.832776 | 7.617000 | -1.446708 |
| H | 14.003246 | 8.203770 | 0.604517 |
| H | 13.256668 | 6.652637 | 0.301777 |
| H | 15.043007 | 8.188649 | -1.639202 |
| H | 15.470479 | 6.784132 | -0.710827 |
| H | 15.239926 | 5.405902 | -4.386232 |
| H | 15.789386 | 7.024127 | -3.953754 |
| H | 16.314579 | 5.621394 | -3.011691 |
| N | 11.507903 | 9.415436 | 1.296489 |
| C | 10.598979 | 9.613245 | 2.408323 |
| C | 9.279899 | 10.275368 | 2.008260 |
| O | 8.283691 | 10.050110 | 2.649193 |
| C | 11.280769 | 10.381044 | 3.547652 |
| C | 12.453370 | 9.579695 | 4.082990 |
| O | 13.587352 | 9.837605 | 3.737775 |
| N | 12.151254 | 8.583846 | 4.925733 |
| H | 12.331269 | 9.970542 | 1.203123 |
| H | 10.302625 | 8.644307 | 2.775812 |
| H | 11.660860 | 11.336911 | 3.209616 |
| H | 10.550987 | 10.553048 | 4.328714 |
| H | 12.895109 | 7.990321 | 5.223649 |
| H | 11.229168 | 8.200230 | 4.968705 |
| N | 9.264766 | 11.067881 | 0.920135 |
| H | 10.111985 | 11.236630 | 0.427846 |
| C | 8.003502 | 11.517473 | 0.368853 |
| H | 8.199841 | 12.261065 | -0.392909 |
| H | 7.389704 | 11.965715 | 1.133008 |
| C | 7.186385 | 10.384437 | -0.254317 |
| O | 5.985862 | 10.466595 | -0.287055 |
| N | 7.863647 | 9.335915 | -0.767247 |
| C | 7.173994 | 8.147390 | -1.239255 |
| C | 6.642709 | 7.297574 | -0.074031 |
| O | 5.546047 | 6.797146 | -0.156916 |
| C | 8.054036 | 7.351661 | -2.232783 |
| C | 8.343918 | 8.208705 | -3.479841 |
| C | 7.389660 | 6.024776 | -2.619158 |
| C | 9.428868 | 7.642600 | -4.399230 |
| H | 8.826092 | 9.236574 | -0.544585 |
| H | 6.281544 | 8.471908 | -1.751791 |
| H | 8.994579 | 7.127692 | -1.737429 |
| H | 7.420681 | 8.333241 | -4.041301 |
| H | 8.643264 | 9.204484 | -3.170322 |
| H | 8.008556 | 5.488934 | -3.327523 |
| H | 7.242213 | 5.380879 | -1.760618 |
| H | 6.418764 | 6.185578 | -3.076124 |
| H | 9.620286 | 8.326698 | -5.219474 |
| H | 10.365162 | 7.497217 | -3.867186 |
| H | 9.150553 | 6.689604 | -4.833807 |
| N | 7.395512 | 7.176335 | 1.037396 |
| C | 6.888845 | 6.461942 | 2.197503 |
| C | 5.537648 | 7.043881 | 2.634270 |
| O | 4.584655 | 6.338019 | 2.861016 |
| C | 7.843569 | 6.553388 | 3.398435 |
| C | 9.266713 | 6.008056 | 3.241915 |
| O | 10.154083 | 6.639935 | 3.807509 |
| O | 9.445929 | 4.949936 | 2.587572 |
| H | 8.353495 | 7.453252 | 1.017465 |
| H | 6.692261 | 5.432436 | 1.941474 |
| H | 7.929379 | 7.578500 | 3.726787 |
| H | 7.386631 | 6.000629 | 4.216793 |
| N | 5.500383 | 8.381884 | 2.779173 |
| H | 6.331966 | 8.916498 | 2.642954 |
| C | 4.310302 | 9.051724 | 3.242305 |
| H | 4.547952 | 10.097967 | 3.389412 |
| H | 3.972367 | 8.652748 | 4.185685 |
| C | 3.145239 | 8.982071 | 2.264987 |
| O | 2.003692 | 9.093130 | 2.649079 |
| N | 3.475626 | 8.822516 | 0.980212 |
| H | 4.435722 | 8.843337 | 0.726370 |
| C | 2.517189 | 8.905262 | -0.078310 |
| H | 3.055177 | 9.094409 | -0.998554 |
| H | 1.843648 | 9.736855 | 0.077017 |
| C | 1.609134 | 7.687018 | -0.310189 |
| O | 0.556730 | 7.818082 | -0.869314 |
| N | 2.097339 | 6.532777 | 0.186432 |
| C | 1.293115 | 5.354489 | 0.393886 |
| C | 0.469958 | 5.442074 | 1.683315 |
| O | -0.716663 | 5.213445 | 1.656285 |
| C | 2.172201 | 4.103766 | 0.394725 |
| H | 3.008794 | 6.562818 | 0.592222 |
| H | 0.567591 | 5.290491 | -0.399998 |
| H | 1.571887 | 3.221107 | 0.587172 |
| H | 2.650486 | 4.000487 | -0.572497 |
| H | 2.945022 | 4.164852 | 1.153870 |
| N | 1.095606 | 5.744175 | 2.836284 |
| C | 0.370048 | 5.577868 | 4.085933 |
| C | -0.747050 | 6.605756 | 4.249910 |
| O | -1.779113 | 6.292778 | 4.806033 |
| C | 1.291025 | 5.500017 | 5.328748 |
| C | 1.896886 | 6.841132 | 5.769543 |
| C | 2.367473 | 4.424527 | 5.118375 |
| C | 2.725583 | 6.742232 | 7.052782 |
| H | 2.076909 | 5.921410 | 2.850031 |
| H | -0.155757 | 4.636139 | 4.029979 |
| H | 0.632959 | 5.168978 | 6.126570 |
| H | 2.512732 | 7.238755 | 4.970077 |
| H | 1.101440 | 7.561025 | 5.932534 |
| H | 2.746328 | 4.073368 | 6.067270 |
| H | 1.962202 | 3.568308 | 4.588320 |
| H | 3.206178 | 4.805542 | 4.546220 |
| H | 3.041282 | 7.731432 | 7.370854 |
| H | 2.141909 | 6.310053 | 7.860272 |
| H | 3.614010 | 6.136111 | 6.927325 |
| N | -0.540129 | 7.842723 | 3.780466 |
| H | 0.342488 | 8.089408 | 3.382471 |
| C | -1.510278 | 8.881664 | 4.014688 |
| H | -1.102103 | 9.814010 | 3.645563 |
| H | -1.713744 | 8.999729 | 5.069139 |
| C | -2.866972 | 8.618759 | 3.365892 |
| O | -3.867795 | 8.839803 | 4.004953 |
| N | -2.902685 | 8.108971 | 2.120754 |
| C | -4.242863 | 8.013553 | 1.537563 |
| C | -4.979283 | 6.742722 | 1.927160 |
| O | -6.188991 | 6.730051 | 1.905224 |
| C | -4.031733 | 8.233296 | 0.034025 |
| C | -2.775505 | 9.110844 | 0.023665 |
| C | -1.911792 | 8.462517 | 1.093799 |
| H | -4.858322 | 8.809717 | 1.927431 |
| H | -3.856474 | 7.294889 | -0.475512 |
| H | -4.897694 | 8.701300 | -0.415825 |
| H | -2.268925 | 9.159450 | -0.922029 |
| H | -3.034920 | 10.123815 | 0.321839 |
| H | -1.424615 | 7.574589 | 0.732325 |
| H | -1.154979 | 9.124576 | 1.481825 |
| N | -4.246478 | 5.682921 | 2.308284 |
| C | -4.884764 | 4.537854 | 2.931312 |
| C | -5.556245 | 4.913475 | 4.258304 |
| O | -6.645329 | 4.464496 | 4.535352 |
| C | -3.882699 | 3.397274 | 3.198839 |
| C | -3.568052 | 2.472268 | 2.039147 |
| C | -4.097059 | 1.188235 | 2.012444 |
| C | -2.684657 | 2.827278 | 1.024591 |
| C | -3.742249 | 0.276240 | 1.028472 |
| C | -2.311984 | 1.924378 | 0.044502 |
| C | -2.821629 | 0.636263 | 0.054760 |
| O | -2.415336 | -0.235216 | -0.902961 |
| H | -3.257638 | 5.793738 | 2.385916 |
| H | -5.683185 | 4.190018 | 2.294708 |
| H | -2.967446 | 3.829704 | 3.588631 |
| H | -4.296973 | 2.793750 | 3.997790 |
| H | -4.795192 | 0.883462 | 2.773001 |
| H | -2.244936 | 3.803103 | 1.013504 |
| H | -4.149887 | -0.716961 | 1.029997 |

```
H         -1.607096      2.205925     -0.717307
H         -2.543911     -1.145958     -0.599150
N         -4.864690      5.708448      5.091712
H         -3.932045      5.983803      4.863035
C         -5.402216      6.114321      6.371185
H         -4.587472      6.510357      6.964601
H         -5.830945      5.273358      6.893353
C         -6.496076      7.176796      6.275894
O         -7.449839      7.157415      7.016600
N         -6.342148      8.091257      5.313951
H         -5.480905      8.121945      4.813221
C         -7.257646      9.220182      5.085235
H         -6.781034      9.869618      4.362564
H         -7.422016      9.776332      5.996157
C         -8.628120      8.831988      4.565388
O         -9.610237      9.421708      4.945185
N         -8.684899      7.833489      3.665690
C         -9.953904      7.451896      3.088789
C        -10.836113      6.654160      4.038285
O        -11.951069      6.348522      3.664953
C         -9.772535      6.689610      1.754830
O        -10.957216      6.762936      1.004373
C         -9.322259      5.242596      1.928269
H         -7.841589      7.403173      3.345557
H        -10.519588      8.347563      2.872919
H         -9.025525      7.229021      1.187647
H        -11.673682      6.473150      1.560981
H         -9.117617      4.807962      0.957337
H         -8.422924      5.168382      2.524941
H        -10.102048      4.651927      2.397531
N        -10.353416      6.327970      5.240915
C        -11.209899      5.754970      6.261794
C        -12.269831      6.740729      6.760583
O        -13.321353      6.321678      7.185952
C        -10.389988      5.272841      7.468074
C         -9.409266      4.122282      7.193927
C         -8.575510      3.863618      8.452802
C        -10.115725      2.839753      6.745629
H         -9.422935      6.582664      5.487088
H        -11.765663      4.937261      5.833105
H         -9.837994      6.116888      7.868437
H        -11.099077      4.965000      8.230932
H         -8.725974      4.421859      6.406758
H         -7.849245      3.074090      8.280969
H         -8.036877      4.756509      8.750142
H         -9.205526      3.558001      9.285371
H         -9.389914      2.044449      6.602781
H        -10.833559      2.508931      7.494141
H        -10.645189      2.961927      5.807698
N        -11.944653      8.043886      6.777024
C        -12.908463      9.031276      7.168608
C        -13.893086      9.396663      6.077540
H        -14.654689     10.121410      6.386173
O        -13.869382      8.968959      4.972325
```

| | | | |
|---|---:|---:|---:|
| H | -11.143636 | 8.358925 | 6.272308 |
| H | -13.466990 | 8.681268 | 8.027662 |
| H | -12.390550 | 9.937375 | 7.468124 |
| N | -11.959308 | 10.656091 | -3.211148 |
| H | -11.755756 | 11.058973 | -2.317761 |
| C | -11.816457 | 9.206991 | -3.172691 |
| C | -11.341663 | 8.851087 | -4.571361 |
| O | -10.272480 | 8.336218 | -4.800009 |
| C | -13.067033 | 8.428786 | -2.704147 |
| C | -13.417284 | 8.815512 | -1.255785 |
| C | -12.866578 | 6.917168 | -2.851237 |
| C | -14.739368 | 8.236114 | -0.748921 |
| H | -12.896506 | 10.933787 | -3.433623 |
| H | -10.990230 | 8.968208 | -2.516626 |
| H | -13.902678 | 8.727074 | -3.341181 |
| H | -12.610390 | 8.497223 | -0.601005 |
| H | -13.475327 | 9.897152 | -1.178082 |
| H | -13.779878 | 6.381397 | -2.625804 |
| H | -12.575753 | 6.638382 | -3.859161 |
| H | -12.098827 | 6.567345 | -2.169367 |
| H | -14.962711 | 8.623965 | 0.238622 |
| H | -15.565385 | 8.501528 | -1.403633 |
| H | -14.708115 | 7.155597 | -0.670753 |
| N | -12.160013 | 9.244786 | -5.566149 |
| H | -13.062339 | 9.581757 | -5.322268 |
| C | -11.972873 | 8.843878 | -6.959263 |
| H | -12.063075 | 7.773719 | -7.082244 |
| H | -12.739200 | 9.329967 | -7.549200 |
| C | -10.601636 | 9.233245 | -7.517684 |
| O | -9.966047 | 8.477163 | -8.194487 |
| N | -10.204772 | 10.496260 | -7.241744 |
| H | -10.745741 | 11.021179 | -6.592200 |
| C | -8.894336 | 11.023975 | -7.637100 |
| H | -8.768163 | 10.911620 | -8.702720 |
| H | -8.880083 | 12.078516 | -7.393800 |
| C | -7.672570 | 10.358996 | -6.988335 |
| O | -6.754917 | 9.935251 | -7.636302 |
| N | -7.659042 | 10.373849 | -5.639645 |
| H | -8.490112 | 10.631744 | -5.158174 |
| C | -6.495251 | 9.961279 | -4.832301 |
| H | -5.597170 | 10.376228 | -5.259419 |
| H | -6.627732 | 10.362067 | -3.835155 |
| C | -6.268288 | 8.463918 | -4.714305 |
| O | -5.127482 | 8.062503 | -4.733447 |
| N | -7.303763 | 7.629682 | -4.518437 |
| C | -6.995293 | 6.262618 | -4.118861 |
| C | -6.397875 | 5.446125 | -5.259534 |
| O | -5.550376 | 4.616864 | -5.016950 |
| C | -8.138406 | 5.504852 | -3.400912 |
| C | -9.329115 | 5.122114 | -4.296478 |
| C | -8.556184 | 6.273530 | -2.141988 |
| C | -10.286891 | 4.121107 | -3.645544 |
| H | -8.246869 | 7.963362 | -4.526703 |
| H | -6.182269 | 6.318022 | -3.410787 |

```
H           -7.667346    4.579888   -3.078107
H           -9.870791    6.010801   -4.594097
H           -8.955119    4.671202   -5.210008
H           -9.230739    5.688348   -1.531861
H           -7.693427    6.514493   -1.528388
H           -9.062598    7.198710   -2.391998
H          -11.082575    3.862865   -4.337858
H           -9.776980    3.202278   -3.381239
H          -10.751996    4.519164   -2.752028
N           -6.833422    5.653023   -6.515054
C           -6.209522    4.945425   -7.618242
C           -4.788994    5.445304   -7.874125
O           -3.916767    4.674116   -8.198284
C           -7.073807    5.016307   -8.886477
C           -8.327259    4.134335   -8.799993
C           -8.023065    2.640288   -8.977824
N           -9.143085    1.783638   -8.625299
C           -9.257772    1.084606   -7.452695
N           -8.456742    1.524117   -6.428013
N          -10.027711    0.084965   -7.273331
H           -7.514663    6.359338   -6.691394
H           -6.075039    3.917907   -7.321122
H           -7.358332    6.047883   -9.073647
H           -6.464020    4.705915   -9.729707
H           -8.829298    4.295538   -7.851211
H           -9.034232    4.441087   -9.563850
H           -7.770202    2.449756  -10.014650
H           -7.161635    2.342352   -8.396422
H           -9.673651    1.417052   -9.382221
H           -8.316214    2.508125   -6.372549
H           -8.685693    1.125312   -5.538928
H          -10.470187   -0.207140   -8.124039
N           -4.571417    6.757546   -7.721981
H           -5.324518    7.371876   -7.509670
C           -3.257793    7.352338   -8.011311
H           -2.927685    7.095601   -9.006206
H           -3.358948    8.426670   -7.934769
C           -2.200462    6.873890   -7.028854
O           -1.071246    6.634407   -7.380783
N           -2.624841    6.770007   -5.764313
C           -1.743731    6.467238   -4.664736
C           -1.469179    4.982033   -4.520918
O           -0.343872    4.598273   -4.299683
C           -2.325901    7.074500   -3.381697
C           -1.744078    6.378310   -2.152690
C           -2.077058    8.588864   -3.425734
H           -3.552406    7.065772   -5.547400
H           -0.772855    6.901278   -4.856273
H           -3.395250    6.902016   -3.395401
H           -1.956834    6.926113   -1.254464
H           -2.162121    5.382760   -2.042325
H           -0.672325    6.297826   -2.222264
H           -2.807112    9.125895   -2.836912
H           -1.085283    8.827496   -3.055954
```

| | | | |
|---|---:|---:|---:|
| H | -2.163395 | 8.965961 | -4.437996 |
| N | -2.497025 | 4.121563 | -4.635814 |
| C | -2.164150 | 2.717547 | -4.734186 |
| C | -1.220412 | 2.507859 | -5.924341 |
| O | -0.294519 | 1.730163 | -5.831241 |
| C | -3.398800 | 1.806095 | -4.782331 |
| C | -4.255388 | 1.912116 | -3.510630 |
| C | -5.384345 | 0.877091 | -3.412802 |
| N | -4.911125 | -0.384469 | -2.837073 |
| C | -5.641349 | -1.487664 | -2.723430 |
| N | -6.868149 | -1.556099 | -3.215965 |
| N | -5.167912 | -2.546342 | -2.084122 |
| H | -3.410474 | 4.441689 | -4.882412 |
| H | -1.572280 | 2.442473 | -3.873490 |
| H | -4.008476 | 2.048084 | -5.648015 |
| H | -3.045266 | 0.790241 | -4.918391 |
| H | -3.627534 | 1.838156 | -2.628754 |
| H | -4.715886 | 2.888893 | -3.477228 |
| H | -6.178671 | 1.270717 | -2.788470 |
| H | -5.795289 | 0.688071 | -4.397354 |
| H | -4.016300 | -0.384548 | -2.391594 |
| H | -7.283987 | -0.820002 | -3.742352 |
| H | -7.345473 | -2.437609 | -3.193052 |
| H | -4.318191 | -2.542449 | -1.549922 |
| H | -5.707893 | -3.388640 | -2.099643 |
| N | -1.459768 | 3.193313 | -7.035090 |
| H | -2.291401 | 3.731055 | -7.157331 |
| C | -0.593747 | 2.993272 | -8.181725 |
| H | -0.941346 | 3.639611 | -8.977063 |
| H | -0.612453 | 1.968830 | -8.524740 |
| C | 0.857709 | 3.334431 | -7.858295 |
| O | 1.786996 | 2.686949 | -8.271991 |
| N | 1.004645 | 4.406026 | -7.063224 |
| H | 0.216422 | 4.986749 | -6.881609 |
| C | 2.282318 | 4.880664 | -6.640292 |
| H | 2.919615 | 5.140381 | -7.476401 |
| H | 2.128573 | 5.775463 | -6.050859 |
| C | 3.089706 | 3.928944 | -5.780075 |
| O | 4.295254 | 4.052723 | -5.693914 |
| N | 2.428054 | 2.983357 | -5.101936 |
| C | 3.125745 | 2.157358 | -4.158665 |
| C | 4.307275 | 1.377717 | -4.739097 |
| O | 5.245113 | 1.107231 | -4.013913 |
| C | 2.108535 | 1.213076 | -3.516019 |
| C | 2.684684 | 0.366309 | -2.391931 |
| C | 1.604206 | -0.501734 | -1.777021 |
| N | 2.215258 | -1.390979 | -0.801973 |
| C | 1.543007 | -2.356343 | -0.174298 |
| N | 0.258345 | -2.481552 | -0.311342 |
| N | 2.259203 | -3.221607 | 0.557411 |
| H | 1.436860 | 2.897295 | -5.197136 |
| H | 3.580579 | 2.775743 | -3.393985 |
| H | 1.303606 | 1.831249 | -3.132486 |
| H | 1.670052 | 0.584170 | -4.282571 |
| H | 3.478434 | -0.269402 | -2.769093 |
| H | 3.123712 | 0.998938 | -1.625682 |
| H | 0.853947 | 0.106288 | -1.286505 |
| H | 1.112971 | -1.081058 | -2.553359 |
| H | 3.198724 | -1.524183 | -0.874375 |
| H | -0.249761 | -1.869076 | -0.909957 |
| H | -0.330109 | -3.092244 | 0.278098 |
| H | 3.070523 | -2.854257 | 1.001802 |
| H | 1.759324 | -3.887447 | 1.114600 |
| N | 4.283097 | 0.949471 | -6.008939 |
| H | 3.535735 | 1.220842 | -6.615074 |
| C | 5.363433 | 0.116654 | -6.504616 |
| H | 5.140510 | -0.146352 | -7.531095 |
| H | 5.439318 | -0.795651 | -5.934055 |
| C | 6.736092 | 0.756728 | -6.473846 |
| O | 7.737272 | 0.089520 | -6.357457 |
| N | 6.766109 | 2.085327 | -6.608293 |
| H | 5.913119 | 2.606424 | -6.582392 |
| C | 8.006711 | 2.823080 | -6.556646 |
| H | 8.708195 | 2.437596 | -7.282942 |
| H | 7.800881 | 3.853376 | -6.814337 |
| C | 8.754961 | 2.828364 | -5.221498 |
| O | 9.898041 | 3.232398 | -5.175672 |
| N | 8.111610 | 2.347489 | -4.149769 |
| C | 8.697702 | 2.187912 | -2.830082 |
| C | 9.957675 | 1.311939 | -2.837098 |
| O | 10.778688 | 1.394874 | -1.938032 |
| C | 7.593837 | 1.530625 | -1.971682 |
| C | 7.799561 | 1.403320 | -0.461893 |
| C | 7.459568 | 2.669199 | 0.334045 |
| C | 7.360411 | 2.397025 | 1.831508 |
| N | 8.704557 | 2.372381 | 2.479585 |
| O | 11.066288 | 1.633041 | 0.883914 |
| H | 10.916396 | 1.845603 | -0.037572 |
| H | 11.543707 | 0.805677 | 0.831947 |
| H | 7.163621 | 2.048804 | -4.258477 |
| H | 9.003138 | 3.145746 | -2.430044 |
| H | 6.681596 | 2.084013 | -2.150212 |
| H | 7.415303 | 0.545346 | -2.387081 |
| H | 7.122948 | 0.621892 | -0.128865 |
| H | 8.796512 | 1.050899 | -0.232930 |
| H | 8.169180 | 3.472388 | 0.156872 |
| H | 6.491542 | 3.040054 | 0.012710 |
| H | 6.797490 | 3.164177 | 2.339924 |
| H | 6.886554 | 1.447117 | 2.027472 |
| H | 9.414912 | 1.912513 | 1.918340 |
| H | 9.019342 | 3.345795 | 2.631192 |
| H | 8.659266 | 1.895150 | 3.367755 |
| N | 10.062466 | 0.448216 | -3.837536 |
| C | 11.109101 | -0.537837 | -4.068290 |
| C | 12.485570 | 0.022300 | -4.430189 |
| O | 13.442152 | -0.708288 | -4.306832 |
| C | 10.697003 | -1.450049 | -5.245174 |
| C | 9.557196 | -2.444193 | -4.980866 |

```
C         9.121105   -3.076977   -6.306079
C         9.984271   -3.528786   -3.991472
H         9.325524    0.449453   -4.512115
H        11.266964   -1.121191   -3.172975
H        10.429691   -0.809293   -6.080123
H        11.578006   -2.006067   -5.542437
H         8.702892   -1.907963   -4.578702
H         8.318584   -3.790603   -6.145378
H         8.764202   -2.323936   -6.997047
H         9.945818   -3.609952   -6.772479
H         9.194832   -4.258665   -3.847595
H        10.854362   -4.057244   -4.366543
H        10.237401   -3.125943   -3.015392
N        12.574050    1.261828   -4.919242
C        13.815635    1.842778   -5.394482
C        14.959966    2.030505   -4.379949
O        16.015586    2.407437   -4.813149
C        13.500190    3.169358   -6.088651
O        12.920229    4.086018   -5.204398
H        11.757489    1.835028   -4.917551
H        14.248513    1.174471   -6.129335
H        14.424129    3.587469   -6.456383
H        12.850283    2.975915   -6.936087
H        11.973459    4.083109   -5.276725
N        14.792294    1.722326   -3.074279
C        15.927642    1.647623   -2.175962
C        16.802954    0.408621   -2.398334
O        17.930612    0.391106   -1.978325
C        15.493748    1.583077   -0.706587
C        14.646418    2.747776   -0.198835
C        14.299328    2.511246    1.271428
N        13.278030    3.411239    1.792438
C        13.459620    4.558886    2.425749
N        14.678357    5.029548    2.665225
O        15.035329    7.483892    4.221498
H        14.752239    8.368927    3.976386
H        15.752817    7.593707    4.832595
N        12.411528    5.248657    2.825510
H        13.898964    1.426892   -2.757310
H        16.562901    2.503112   -2.347108
H        14.946313    0.659234   -0.549637
H        16.405076    1.507813   -0.123636
H        15.169266    3.691709   -0.322412
H        13.725083    2.831995   -0.763390
H        13.902116    1.513079    1.387419
H        15.181785    2.556843    1.898491
H        12.336971    3.119919    1.620287
H        15.485258    4.538316    2.362231
H        14.814923    5.883744    3.178093
H        11.455456    4.972143    2.666048
H        12.507489    6.089644    3.348496
N        16.200238   -0.648980   -2.967242
C        16.870493   -1.911561   -3.169274
C        17.666752   -1.888601   -4.470651
```

```
H         18.068929   -0.910857   -4.745210
O         17.862293   -2.849893   -5.139899
C         15.867612   -3.069988   -3.149678
C         15.326236   -3.400607   -1.791630
N         14.404882   -2.630104   -1.110143
O         12.495269   -0.448167   -0.464717
H         13.143018   -1.043126   -0.846976
H         12.027721   -0.044212   -1.187658
C         15.624239   -4.490901   -1.050154
C         14.176585   -3.255071   -0.001535
N         14.888763   -4.384181    0.103649
H         15.287617   -0.548406   -3.356786
H         17.596849   -2.030625   -2.373010
H         15.049269   -2.841549   -3.823707
H         16.367458   -3.943710   -3.545564
H         16.281273   -5.313134   -1.236668
H         13.499281   -2.926844    0.760896
H         14.791125   -5.066805    0.822887
O        -15.173069   -2.387988    4.913834
H        -15.316558   -1.600474    4.394292
H        -14.733857   -2.099793    5.706586
O        -13.115423   -1.181806    6.976833
H        -12.458245   -1.649435    7.479508
H        -13.347691   -0.417639    7.490170
O        -12.158798   -0.851456    4.234040
H        -12.358012   -0.872234    5.167083
H        -12.941268   -0.474824    3.826903
O        -13.434395   -1.918101    0.799254
H        -12.673618   -1.339924    0.893840
H        -13.354192   -2.581669    1.485840
O         -7.169646    1.025612    0.338688
H         -7.409427    1.441455    1.169781
H         -6.225449    0.946658    0.363438
O         -7.794203    2.046415    2.949863
H         -8.722331    1.885942    3.111572
H         -7.551704    2.813777    3.460568
O         -9.253341   -0.313555   -0.958221
H         -8.443934   -0.031169   -0.525761
H         -9.135697   -1.186227   -1.325175
O        -12.669156   -2.237416   -1.918220
H        -13.092490   -2.357058   -1.067974
H        -13.240841   -2.558827   -2.608700
O        -13.694209    1.068297   -0.750922
H        -14.130352    0.225026   -0.790552
H        -12.879180    0.889268   -0.271106
O        -14.286039    4.932814    4.722182
H        -13.721328    5.572768    4.296790
H        -14.316405    5.219814    5.630028
O        -14.997355    2.180191    1.524645
H        -15.638613    2.873211    1.437735
H        -14.649494    1.991948    0.652190
O        -12.282264    0.283649   -3.288564
H        -12.724424    0.948164   -2.769320
H        -12.228402   -0.468436   -2.702124
```

```
O                                     -13.353785   -1.919087   -4.757537
H                                     -13.208447   -0.997828   -4.549714
H                                     -14.014970   -1.955861   -5.436175
O                                      -9.944794   -2.649887   -2.654314
H                                     -10.012987   -2.281468   -3.539767
H                                     -10.837544   -2.625270   -2.311629
O                                     -11.483407    0.132133    0.686980
H                                     -10.716858   -0.028786    0.130641
H                                     -11.164179    0.598343    1.463943
O                                     -10.642780    1.332890    3.131370
H                                     -10.900813    0.561499    3.633575
H                                     -11.338196    1.968426    3.324684
O                                     -13.037077    2.625322    3.676747
H                                     -13.395041    3.358236    4.180532
H                                     -13.613843    2.549006    2.922905
O                                      -9.287224    0.600990   -3.663131
H                                     -10.218343    0.502323   -3.846561
H                                      -9.229602    0.496194   -2.714565
O                                     -14.698542   -0.323312    2.944305
H                                     -14.959437    0.554388    2.673718
H                                     -14.551307   -0.811127    2.137885
O                                     -13.130657   -3.408778    3.275034
H                                     -13.933144   -3.240116    3.778510
H                                     -12.527699   -2.743326    3.604859
O                                     -10.448034   -2.040510   -5.342020
H                                     -11.390189   -2.184640   -5.374342
H                                     -10.265025   -1.281014   -5.908652

Case 11
Voltage = 0; R297 neutral; E183 neutral; Y266 neutral; R300 positive
N                                      15.766799   -6.199193    5.278994
H                                      14.964272   -5.968081    5.831968
C                                      15.681412   -5.592052    3.948131
C                                      14.971469   -6.522888    2.946057
O                                      14.227506   -6.092485    2.079840
C                                      17.084708   -5.278677    3.415251
H                                      16.571885   -5.867940    5.772982
H                                      15.098698   -4.680401    3.954756
H                                      17.044730   -4.882903    2.407435
H                                      17.554619   -4.530472    4.043437
H                                      17.710022   -6.164607    3.414114
N                                      15.249756   -7.821004    3.057322
H                                      15.838798   -8.092485    3.815185
C                                      14.636571   -8.832199    2.194429
H                                      14.713474   -8.551702    1.156655
H                                      15.148880   -9.773240    2.344197
C                                      13.154033   -8.993858    2.533425
```

```
O        12.324048   -9.200896    1.687033
N        12.823803   -8.853508    3.831746
H        13.543792   -8.780637    4.513470
C        11.439923   -9.003939    4.287582
H        11.434955   -9.002678    5.369415
H        11.007985   -9.927468    3.934527
C        10.555333   -7.857365    3.781156
O         9.411987   -8.054592    3.466079
N        11.138297   -6.648006    3.747752
C        10.483268   -5.475422    3.224682
C        10.288886   -5.521776    1.706617
O         9.246362   -5.114637    1.250554
C        11.233710   -4.218143    3.694152
O        12.622337   -4.375018    3.688899
H        12.094883   -6.558886    4.000622
H         9.478963   -5.419564    3.611836
H        10.927740   -3.370907    3.090634
H        10.962070   -4.017001    4.720055
H        12.956067   -4.542453    2.815940
N        11.272040   -5.999729    0.924796
C        11.019500   -6.186421   -0.501178
C         9.887295   -7.194729   -0.709151
O         9.129440   -7.089064   -1.645806
C        12.314587   -6.542583   -1.293264
C        12.539535   -8.040921   -1.534188
C        12.330415   -5.795703   -2.628388
H        12.107479   -6.366904    1.326214
H        10.631264   -5.255776   -0.883326
H        13.140419   -6.168813   -0.698406
H        13.494484   -8.180374   -2.030027
H        12.550322   -8.621334   -0.622573
H        11.777751   -8.450558   -2.190463
H        13.230445   -6.034757   -3.183603
H        11.476128   -6.073732   -3.235985
H        12.310978   -4.722842   -2.478401
N         9.788710   -8.165126    0.206857
H        10.539996   -8.293077    0.846176
C         8.835229   -9.266292    0.109763
H         9.200195  -10.065763    0.741948
H         8.777316   -9.634660   -0.902912
C         7.418290   -8.932314    0.527070
O         6.500071   -9.395116   -0.109264
N         7.226687   -8.139272    1.587863
C         5.882809   -7.694146    1.875354
C         5.430755   -6.672073    0.838112
O         4.246624   -6.579518    0.590536
C         5.624650   -7.223470    3.320930
C         5.808935   -8.380505    4.307954
C         6.432240   -5.996707    3.750322
H         7.999855   -7.869353    2.156693
H         5.219878   -8.527573    1.700843
H         4.574467   -6.947693    3.327346
H         5.515744   -8.069759    5.305100
H         5.195064   -9.232363    4.034166
```

| | | | |
|---|---:|---:|---:|
| H | 6.841852 | -8.706821 | 4.350452 |
| H | 6.081997 | -5.646366 | 4.715318 |
| H | 7.481813 | -6.238691 | 3.853137 |
| H | 6.338921 | -5.175118 | 3.047714 |
| N | 6.332409 | -5.929333 | 0.187184 |
| C | 5.917325 | -5.109999 | -0.938888 |
| C | 5.511828 | -5.988836 | -2.130498 |
| O | 4.475192 | -5.760275 | -2.710110 |
| C | 6.979118 | -4.052606 | -1.301895 |
| C | 7.141671 | -3.044796 | -0.150508 |
| C | 6.593064 | -3.339891 | -2.602583 |
| C | 8.404008 | -2.188784 | -0.252130 |
| H | 7.298771 | -5.986744 | 0.426790 |
| H | 5.001661 | -4.606165 | -0.668615 |
| H | 7.926365 | -4.561664 | -1.451504 |
| H | 6.262684 | -2.399390 | -0.125873 |
| H | 7.169523 | -3.570765 | 0.794343 |
| H | 7.273842 | -2.526363 | -2.812027 |
| H | 6.616133 | -4.005315 | -3.455528 |
| H | 5.590032 | -2.929145 | -2.537488 |
| H | 8.479564 | -1.524355 | 0.601729 |
| H | 9.287975 | -2.815117 | -0.259960 |
| H | 8.418562 | -1.576189 | -1.146370 |
| N | 6.314753 | -6.997868 | -2.488269 |
| H | 7.214742 | -7.099775 | -2.067851 |
| C | 5.992153 | -7.833873 | -3.625676 |
| H | 6.835682 | -8.490424 | -3.798344 |
| H | 5.829762 | -7.243062 | -4.514822 |
| C | 4.737286 | -8.685277 | -3.433793 |
| O | 3.900165 | -8.746640 | -4.302432 |
| N | 4.633346 | -9.349564 | -2.275688 |
| H | 5.374251 | -9.311311 | -1.608343 |
| C | 3.460930 | -10.132167 | -1.962235 |
| H | 3.235980 | -10.837889 | -2.747737 |
| H | 3.659961 | -10.684354 | -1.052330 |
| C | 2.211504 | -9.277530 | -1.762558 |
| O | 1.131135 | -9.685085 | -2.118699 |
| N | 2.399834 | -8.079171 | -1.212257 |
| C | 1.254614 | -7.181101 | -1.186809 |
| C | 0.805230 | -6.719593 | -2.577704 |
| O | -0.377467 | -6.578061 | -2.788679 |
| C | 1.471818 | -5.909998 | -0.383131 |
| O | 1.595595 | -6.104561 | 0.992005 |
| H | 3.299032 | -7.762847 | -0.929532 |
| H | 0.411373 | -7.709916 | -0.773814 |
| H | 2.326145 | -5.380639 | -0.779937 |
| H | 0.587009 | -5.308497 | -0.526448 |
| H | 2.453898 | -6.471101 | 1.165690 |
| N | 1.731688 | -6.421031 | -3.497467 |
| C | 1.335507 | -6.013212 | -4.834388 |
| C | 0.587086 | -7.143482 | -5.533176 |
| O | -0.401914 | -6.900495 | -6.189430 |
| C | 2.542103 | -5.493068 | -5.649126 |
| C | 2.938013 | -4.101277 | -5.128172 |

```
C         2.244081    -5.459976    -7.153712
C         4.263484    -3.581415    -5.683498
H         2.701772    -6.488555    -3.277380
H         0.603521    -5.222098    -4.755227
H         3.369159    -6.177090    -5.483584
H         2.143335    -3.398946    -5.375170
H         3.007483    -4.133507    -4.048018
H         3.095244    -5.071321    -7.698654
H         2.040082    -6.447024    -7.549921
H         1.388355    -4.829772    -7.373765
H         4.550256    -2.666147    -5.175650
H         5.058965    -4.303771    -5.530250
H         4.210709    -3.362955    -6.743846
N         1.045345    -8.388206    -5.372742
H         1.896238    -8.553137    -4.877427
C         0.277122    -9.503383    -5.870967
H         0.166377    -9.464198    -6.944263
H         0.797334   -10.415548    -5.607540
C        -1.129420    -9.546180    -5.280784
O        -2.081357    -9.778179    -5.988272
N        -1.238847    -9.341562    -3.969335
C        -2.540865    -9.396277    -3.348814
C        -3.472815    -8.276566    -3.808433
O        -4.665696    -8.504971    -3.829727
C        -2.444399    -9.368922    -1.816910
O        -3.686544    -9.637257    -1.228460
H        -0.423441    -9.262708    -3.399295
H        -3.043477   -10.308833    -3.637559
H        -1.756559   -10.137507    -1.491127
H        -2.059769    -8.408378    -1.489722
H        -4.383556    -9.297046    -1.777012
N        -2.940828    -7.090989    -4.104650
C        -3.714396    -5.980530    -4.640072
C        -4.260359    -6.323574    -6.025316
O        -5.433722    -6.187528    -6.274239
C        -2.819733    -4.725051    -4.658333
C        -3.299232    -3.550229    -5.497915
C        -4.640055    -3.242571    -5.698945
C        -2.341094    -2.736927    -6.098929
C        -5.007848    -2.166723    -6.495492
C        -2.702859    -1.654288    -6.880787
C        -4.044417    -1.369303    -7.087692
H        -1.961402    -6.964638    -3.963165
H        -4.579928    -5.811204    -4.019772
H        -2.677814    -4.416749    -3.626000
H        -1.846694    -5.016256    -5.029495
H        -5.416477    -3.851185    -5.279418
H        -1.296572    -2.963066    -5.966893
H        -6.053452    -1.979860    -6.660766
H        -1.941004    -1.046095    -7.335689
H        -4.332357    -0.544644    -7.715810
N        -3.380102    -6.770872    -6.929975
C        -3.843581    -7.123947    -8.252678
C        -4.957008    -8.168519    -8.197382
```

```
O   -5.858123   -8.123772   -8.990262
C   -2.704207   -7.650790   -9.127293
S   -1.493044   -6.397397   -9.657048
H   -2.431188   -6.931889   -6.667367
H   -4.289167   -6.261459   -8.725814
H   -2.192611   -8.469511   -8.639062
H   -3.132051   -8.027777  -10.046417
H   -0.898939   -6.204438   -8.489634
N   -4.856366   -9.133447   -7.257267
H   -4.014838   -9.250838   -6.735925
C   -5.837667  -10.206109   -7.284835
H   -6.000986  -10.550626   -8.294471
H   -5.448208  -11.027997   -6.694970
C   -7.220768   -9.847502   -6.717447
O   -8.231163  -10.243145   -7.230182
N   -7.192887   -9.152143   -5.575263
C   -8.409975   -8.810074   -4.867786
C   -9.298337   -7.820014   -5.608869
O  -10.492149   -7.842398   -5.455414
C   -8.117941   -8.262012   -3.460051
C   -7.225540   -7.008977   -3.366201
C   -7.834828   -5.644811   -3.179516
O   -9.119418   -5.541677   -3.330617
H   -9.393987   -4.628418   -3.171123
O   -7.137181   -4.697290   -2.892675
H   -6.311813   -8.876639   -5.198532
H   -9.015615   -9.699082   -4.760630
H   -9.072668   -8.085126   -2.986868
H   -7.627586   -9.055070   -2.905695
H   -6.540452   -7.114976   -2.536623
H   -6.597092   -6.905914   -4.236447
N   -8.663754   -6.893808   -6.343234
C   -9.410469   -5.872015   -7.054845
C  -10.082072   -6.487822   -8.279182
H   -9.656814   -7.433588   -8.623464
O  -11.014472   -5.988574   -8.822358
C   -8.482302   -4.674510   -7.369569
O   -8.222315   -3.967633   -6.180946
C   -9.021022   -3.730005   -8.440284
H   -7.678771   -6.954787   -6.482382
H  -10.212834   -5.520999   -6.419807
H   -7.523425   -5.057113   -7.692313
H   -9.015061   -3.503038   -5.926155
H   -8.361343   -2.871487   -8.500318
H   -9.062824   -4.202027   -9.413328
H  -10.018428   -3.382571   -8.195534
N  -12.707331   -9.374677    3.087214
H  -12.961416   -9.890726    2.270111
C  -11.554139   -8.495243    2.823605
C  -10.560347   -8.453213    3.965134
O   -9.431357   -8.043321    3.768192
C  -11.952603   -7.030873    2.542213
O  -12.477533   -6.543697    3.751232
C  -12.962595   -6.888529    1.409664
```

```
H    -13.502160    -8.835944     3.372267
H    -11.005503    -8.874816     1.973550
H    -11.047999    -6.494762     2.287430
H    -12.570336    -5.594205     3.703863
H    -13.136608    -5.836769     1.209891
H    -12.610590    -7.348401     0.493772
H    -13.914460    -7.331888     1.676366
N    -10.959379    -8.869883     5.156823
H    -11.881972    -9.244434     5.198159
C    -10.169178    -8.859743     6.362248
H    -10.113527    -7.867438     6.782144
H    -10.670944    -9.489590     7.087456
C     -8.720771    -9.337977     6.269503
O     -7.892065    -8.749900     6.920582
N     -8.383634   -10.408175     5.527540
H     -9.084947   -10.906602     5.027655
C     -6.985352   -10.804352     5.420052
H     -6.944463   -11.779079     4.950853
H     -6.540479   -10.880020     6.399382
C     -6.128134    -9.832217     4.586845
O     -4.961200    -9.668882     4.839695
N     -6.800928    -9.245762     3.603038
C     -6.090334    -8.226808     2.842068
C     -6.093301    -6.891908     3.582426
O     -5.139004    -6.152923     3.478294
C     -6.415666    -8.138171     1.337567
C     -7.638304    -7.376063     0.870465
C     -7.691516    -5.987608     0.962372
C     -8.678817    -8.034599     0.229581
C     -8.763412    -5.280797     0.445033
C     -9.744863    -7.328753    -0.306015
C     -9.794327    -5.949818    -0.195035
H     -7.794876    -9.281905     3.571067
H     -5.056907    -8.524532     2.878385
H     -5.542374    -7.685685     0.881078
H     -6.449420    -9.156302     0.968646
H     -6.877152    -5.450234     1.414336
H     -8.641753    -9.104427     0.114927
H     -8.779184    -4.207587     0.511925
H    -10.521078    -7.852672    -0.833803
H    -10.611047    -5.401513    -0.627039
N     -7.120372    -6.618423     4.387702
C     -7.099953    -5.492358     5.288833
C     -6.001429    -5.673597     6.335778
O     -5.396487    -4.702226     6.739483
C     -8.496341    -5.340532     5.930760
C     -8.783651    -3.995753     6.563403
C     -9.021403    -2.882782     5.761623
C     -8.878699    -3.855329     7.941075
C     -9.336589    -1.657239     6.321662
C     -9.199684    -2.630887     8.508925
C     -9.426461    -1.527889     7.702130
H     -7.956170    -7.156971     4.313647
H     -6.839838    -4.589493     4.755734
```

| | | | |
|---|---|---|---|
| H | -9.223845 | -5.513726 | 5.145265 |
| H | -8.629149 | -6.126275 | 6.663128 |
| H | -8.964860 | -2.973881 | 4.690196 |
| H | -8.705486 | -4.704491 | 8.578875 |
| H | -9.511737 | -0.810363 | 5.683576 |
| H | -9.263271 | -2.539941 | 9.578492 |
| H | -9.652464 | -0.572306 | 8.141448 |
| N | -5.733338 | -6.905810 | 6.761546 |
| H | -6.366956 | -7.641667 | 6.541979 |
| C | -4.674460 | -7.188124 | 7.704098 |
| H | -4.724131 | -6.544219 | 8.569045 |
| H | -4.795729 | -8.212369 | 8.032781 |
| C | -3.274263 | -7.021960 | 7.118866 |
| O | -2.414130 | -6.477402 | 7.769403 |
| N | -3.051545 | -7.499423 | 5.887964 |
| C | -1.771673 | -7.296142 | 5.230288 |
| C | -1.558255 | -5.817629 | 4.889137 |
| O | -0.470967 | -5.310220 | 5.057000 |
| C | -1.652434 | -8.213589 | 3.993226 |
| C | -0.472271 | -7.826738 | 3.097230 |
| C | -1.530743 | -9.677824 | 4.430739 |
| H | -3.756856 | -8.038518 | 5.431583 |
| H | -0.977341 | -7.529926 | 5.924696 |
| H | -2.563716 | -8.099265 | 3.411197 |
| H | -0.389636 | -8.536597 | 2.280938 |
| H | -0.577409 | -6.842107 | 2.661920 |
| H | 0.461582 | -7.845121 | 3.651236 |
| H | -1.515855 | -10.326594 | 3.561060 |
| H | -0.603760 | -9.833441 | 4.976718 |
| H | -2.353122 | -9.992052 | 5.059412 |
| N | -2.606641 | -5.139149 | 4.406276 |
| C | -2.476392 | -3.758144 | 3.991828 |
| C | -2.373594 | -2.814761 | 5.192496 |
| O | -1.534044 | -1.934854 | 5.184244 |
| C | -3.627917 | -3.323176 | 3.074659 |
| C | -3.664731 | -3.998342 | 1.689166 |
| C | -2.526095 | -3.561839 | 0.779737 |
| O | -1.389182 | -3.928959 | 1.094795 |
| O | -2.800786 | -2.848964 | -0.215574 |
| H | -3.465136 | -5.615203 | 4.229390 |
| H | -1.546839 | -3.648974 | 3.462350 |
| H | -4.571649 | -3.513100 | 3.572012 |
| H | -3.549356 | -2.247624 | 2.952222 |
| H | -3.619010 | -5.071918 | 1.805132 |
| H | -4.608523 | -3.754022 | 1.216162 |
| N | -3.216197 | -2.974453 | 6.220116 |
| C | -3.001161 | -2.209779 | 7.432520 |
| C | -1.665453 | -2.581887 | 8.083996 |
| O | -0.994042 | -1.721926 | 8.611699 |
| C | -4.131536 | -2.230539 | 8.471851 |
| O | -4.329744 | -3.498299 | 9.023340 |
| C | -5.419151 | -1.602619 | 7.949978 |
| H | -3.870173 | -3.726513 | 6.215925 |
| H | -2.871138 | -1.177907 | 7.145660 |

```
H                              -3.760853    -1.618561     9.284969
H                              -4.812374    -4.035623     8.406501
H                              -6.154436    -1.586222     8.745143
H                              -5.243654    -0.581117     7.627224
H                              -5.832843    -2.158657     7.118914
N                              -1.245431    -3.851978     7.992367
H                              -1.886022    -4.590845     7.798294
C                               0.022408    -4.207208     8.579889
H                               0.068058    -3.970492     9.633551
H                               0.157039    -5.273606     8.453191
C                               1.187882    -3.485863     7.916474
O                               2.131932    -3.090473     8.560245
N                               1.112798    -3.335904     6.581417
C                               2.200902    -2.720667     5.867715
C                               2.277470    -1.211163     6.071535
O                               3.365520    -0.674521     6.036483
C                               2.167463    -3.064623     4.367231
S                               1.284557    -1.901321     3.276137
H                               0.386647    -3.804321     6.078606
H                               3.132176    -3.093647     6.270541
H                               3.185837    -3.050380     4.000657
H                               1.775270    -4.062870     4.234717
H                               0.083332    -1.946482     3.831986
N                               1.141717    -0.533419     6.259064
C                               1.224178     0.894464     6.488985
C                               1.814391     1.172850     7.858856
O                               2.448694     2.178473     8.072705
C                              -0.083805     1.682477     6.252683
C                              -1.138974     1.509818     7.356999
C                              -0.621692     1.410279     4.843697
C                              -2.413709     2.324013     7.123725
H                               0.266124    -1.012706     6.259950
H                               1.949688     1.284890     5.792119
H                               0.233669     2.718672     6.291001
H                              -1.392371     0.465867     7.481818
H                              -0.706112     1.826133     8.300854
H                              -1.356755     2.154469     4.566249
H                               0.177314     1.451075     4.109524
H                              -1.090430     0.437989     4.769131
H                              -3.043228     2.296830     8.007638
H                              -2.190185     3.365930     6.909660
H                              -2.997648     1.934040     6.296943
N                               1.596921     0.227042     8.778624
H                               0.958952    -0.513704     8.586695
C                               2.027131     0.407520    10.146872
H                               1.519451    -0.330606    10.755064
H                               1.783021     1.394163    10.511190
C                               3.530073     0.236940    10.304051
O                               4.189668     0.998775    10.957270
N                               4.070432    -0.815904     9.651635
H                               3.483569    -1.469320     9.177609
C                               5.498961    -0.968315     9.618336
H                               5.733566    -1.923198     9.162705
H                               5.923124    -0.948524    10.610937
```

```
C         6.207498    0.124017    8.821372
O         7.334795    0.441287    9.109516
N         5.539586    0.622924    7.782318
C         6.110635    1.780825    7.141654
C         5.505747    3.082034    7.661052
H         5.020630    3.026358    8.633927
O         5.597272    4.105763    7.056670
C         6.011506    1.708438    5.615982
C         6.793488    0.563402    5.007504
C         8.170800    0.439909    5.196847
C         6.150476   -0.387785    4.222006
C         8.885989   -0.586393    4.595872
C         6.864120   -1.412339    3.620337
C         8.234441   -1.512067    3.795306
H         4.610664    0.332883    7.569095
H         7.146983    1.811663    7.442942
H         4.970101    1.619465    5.332029
H         6.364725    2.654620    5.221072
H         8.685956    1.129693    5.843681
H         5.083335   -0.340551    4.114330
H         9.945472   -0.667129    4.767515
H         6.347487   -2.142088    3.022343
H         8.776468   -2.310872    3.323360
N        12.432461    9.850267   -1.180400
H        11.683400   10.159274   -1.771337
C        12.154309    8.506610   -0.665063
C        11.124043    8.659084    0.446648
O        10.130917    7.964982    0.472927
C        13.428759    7.844072   -0.123256
C        14.513692    7.575274   -1.171701
S        13.905159    6.501747   -2.504494
C        15.385146    6.402339   -3.539842
H        13.250644    9.851724   -1.756688
H        11.695136    7.856534   -1.398432
H        13.856321    8.468817    0.655135
H        13.143606    6.904929    0.336810
H        14.891503    8.501134   -1.590565
H        15.351950    7.096504   -0.678260
H        15.144400    5.753264   -4.367958
H        15.658862    7.378458   -3.918884
H        16.216849    5.977550   -2.993092
N        11.336346    9.617449    1.364819
C        10.425605    9.783018    2.480426
C         9.091458   10.420251    2.090041
O         8.101787   10.166145    2.730321
C        11.092666   10.553153    3.626878
C        12.283586    9.771912    4.151371
O        13.410970   10.058344    3.806981
N        12.005028    8.760617    4.983754
H        12.147132   10.191504    1.276085
H        10.151391    8.803835    2.837813
H        11.451046   11.520749    3.298643
H        10.360837   10.700596    4.411042
H        12.762270    8.180350    5.273957
```

```
H        11.091610   8.356537   5.024316
N         9.056620  11.224004   1.010668
H         9.898897  11.416780   0.518722
C         7.784613  11.651662   0.466670
H         7.962926  12.407626  -0.287308
H         7.162697  12.077954   1.236798
C         6.991346  10.507697  -0.167263
O         5.789240  10.563730  -0.196848
N         7.690398   9.479933  -0.692832
C         7.025942   8.281700  -1.176302
C         6.515754   7.408112  -0.019250
O         5.430039   6.884882  -0.105383
C         7.921063   7.516038  -2.179927
C         8.189644   8.392494  -3.418236
C         7.284873   6.179229  -2.579321
C         9.284746   7.860027  -4.345650
H         8.655275   9.399379  -0.473128
H         6.125549   8.592056  -1.683688
H         8.867280   7.307314  -1.688737
H         7.262758   8.502900  -3.976628
H         8.467895   9.391187  -3.098592
H         7.913774   5.664583  -3.294552
H         7.153221   5.523120  -1.727485
H         6.309772   6.323765  -3.032753
H         9.459514   8.556868  -5.158850
H        10.225088   7.729401  -3.816925
H         9.026342   6.905899  -4.789926
N         7.273355   7.291384   1.089405
C         6.784678   6.553910   2.242785
C         5.422152   7.101866   2.688328
O         4.485126   6.373184   2.909552
C         7.739800   6.653189   3.442788
C         9.174172   6.140854   3.277471
O        10.048797   6.785811   3.848148
O         9.374968   5.094057   2.611306
H         8.225031   7.589342   1.070693
H         6.609803   5.523179   1.976145
H         7.803919   7.676317   3.782096
H         7.296862   6.081723   4.255975
N         5.356324   8.437143   2.847322
H         6.175864   8.990969   2.714907
C         4.153102   9.076505   3.319512
H         4.368428  10.126166   3.476316
H         3.826054   8.661236   4.259670
C         2.987574   8.991187   2.344007
O         1.844747   9.074010   2.731463
N         3.318448   8.851230   1.057095
H         4.277311   8.895193   0.801559
C         2.356256   8.924057   0.001326
H         2.887901   9.135156  -0.917826
H         1.664737   9.738736   0.166945
C         1.474817   7.688649  -0.242168
O         0.419497   7.802375  -0.799649
N         1.987524   6.540536   0.243995
```

```
C         1.207990    5.344252    0.442761
C         0.388113    5.402852    1.736002
O        -0.794382    5.152766    1.710863
C         2.111689    4.111290    0.428077
H         2.898449    6.586256    0.649526
H         0.480977    5.273652   -0.349169
H         1.529707    3.215015    0.613784
H         2.588281    4.027167   -0.541823
H         2.885992    4.180191    1.185039
N         1.012122    5.704134    2.889945
C         0.293999    5.510539    4.139960
C        -0.844668    6.512324    4.316133
O        -1.867855    6.171712    4.872385
C         1.219622    5.440272    5.379634
C         1.799347    6.789392    5.830818
C         2.316973    4.388714    5.156703
C         2.632401    6.696065    7.111646
H         1.989968    5.899749    2.902518
H        -0.211612    4.558260    4.076522
H         0.570524    5.088648    6.176007
H         2.405648    7.205775    5.033650
H         0.990046    7.491931    6.001303
H         2.705332    4.036459    6.101308
H         1.927569    3.529373    4.619833
H         3.146277    4.791805    4.585939
H         2.929382    7.688549    7.437404
H         2.058908    6.245863    7.916570
H         3.532234    6.108483    6.979376
N        -0.667461    7.757480    3.856319
H         0.208312    8.028156    3.458704
C        -1.661659    8.770938    4.101037
H        -1.276603    9.715956    3.739406
H        -1.866050    8.874956    5.156765
C        -3.013036    8.481456    3.452225
O        -4.017501    8.671381    4.095633
N        -3.040071    7.984605    2.201643
C        -4.377475    7.864543    1.619795
C        -5.076093    6.570816    1.998475
O        -6.285289    6.525021    1.980859
C        -4.175290    8.106342    0.118376
C        -2.939573    9.012631    0.115514
C        -2.059239    8.372621    1.177050
H        -5.012048    8.640182    2.020302
H        -3.979773    7.178233   -0.402487
H        -5.052745    8.559682   -0.324206
H        -2.436160    9.082959   -0.830534
H        -3.221574   10.016087    0.425066
H        -1.552338    7.500307    0.804888
H        -1.317223    9.047601    1.571270
N        -4.311176    5.530145    2.366455
C        -4.908585    4.358661    2.979578
C        -5.588387    4.695419    4.313341
O        -6.656107    4.195752    4.587004
C        -3.867379    3.248582    3.226064
```

```
C         -3.543430    2.339464    2.056078
C         -4.033420    1.040398    2.031960
C         -2.690142    2.725574    1.026735
C         -3.669438    0.143966    1.037200
C         -2.309130    1.839302    0.034699
C         -2.779776    0.536092    0.047441
O         -2.367062   -0.318455   -0.922378
H         -3.325388    5.668922    2.437657
H         -5.697859    3.991103    2.342732
H         -2.960325    3.707632    3.603875
H         -4.249690    2.628082    4.027837
H         -4.708143    0.710420    2.802834
H         -2.280634    3.714476    1.012997
H         -4.046949   -0.860861    1.041189
H         -1.628168    2.146094   -0.739046
H         -2.482289   -1.234323   -0.628364
N         -4.931885    5.514728    5.151398
H         -4.009062    5.824226    4.926208
C         -5.483308    5.890676    6.435056
H         -4.679846    6.297772    7.036352
H         -5.892534    5.032313    6.944270
C         -6.603337    6.928281    6.353072
O         -7.557164    6.875164    7.091530
N         -6.471637    7.856266    5.400665
H         -5.611186    7.913850    4.900940
C         -7.412171    8.967382    5.186505
H         -6.951673    9.632509    4.467419
H         -7.578956    9.513875    6.103094
C         -8.778397    8.567800    4.671702
O         -9.769758    9.135252    5.056262
N         -8.814356    7.579701    3.760072
C        -10.068047    7.166286    3.181611
C        -10.925144    6.317316    4.113058
O        -12.014004    5.947305    3.717038
C         -9.826353    6.401761    1.859989
O        -11.028794    6.311522    1.126185
C         -9.217530    5.019415    2.067130
H         -7.960177    7.171661    3.440955
H        -10.664767    8.042055    2.963730
H         -9.150550    7.003106    1.266504
H        -11.745540    6.195234    1.745611
H         -8.996424    4.570236    1.105423
H         -8.294448    5.061492    2.624509
H         -9.908295    4.369621    2.589612
N        -10.443212    6.006375    5.316303
C        -11.283305    5.363149    6.311042
C        -12.364924    6.289334    6.875335
O        -13.352990    5.807973    7.364530
C        -10.444337    4.814942    7.473298
C         -9.416925    3.734829    7.102446
C         -8.617569    3.350986    8.351458
C        -10.059115    2.495914    6.471647
H         -9.535934    6.315628    5.585994
H        -11.829136    4.565571    5.833449
```

| | | | |
|---|---:|---:|---:|
| H | -9.925337 | 5.642717 | 7.946671 |
| H | -11.141819 | 4.417280 | 8.203792 |
| H | -8.717353 | 4.146488 | 6.384172 |
| H | -7.853414 | 2.616627 | 8.112795 |
| H | -8.125922 | 4.218806 | 8.777975 |
| H | -9.262511 | 2.923105 | 9.116135 |
| H | -9.303006 | 1.741695 | 6.275367 |
| H | -10.806885 | 2.060554 | 7.132654 |
| H | -10.540038 | 2.717785 | 5.525181 |
| N | -12.110284 | 7.614422 | 6.867441 |
| C | -13.108370 | 8.554178 | 7.284742 |
| C | -14.020591 | 9.014772 | 6.167902 |
| H | -14.816354 | 9.695908 | 6.490131 |
| O | -13.909004 | 8.703506 | 5.029004 |
| H | -11.353844 | 7.967692 | 6.322838 |
| H | -13.714522 | 8.113847 | 8.065458 |
| H | -12.630948 | 9.435271 | 7.705316 |
| N | -12.161858 | 10.388978 | -3.085274 |
| H | -11.967931 | 10.787747 | -2.187885 |
| C | -11.979893 | 8.944091 | -3.062133 |
| C | -11.506812 | 8.612037 | -4.467330 |
| O | -10.428452 | 8.120371 | -4.704778 |
| C | -13.202308 | 8.127857 | -2.583182 |
| C | -13.535532 | 8.490491 | -1.124370 |
| C | -12.968373 | 6.623874 | -2.752129 |
| C | -14.805753 | 7.834922 | -0.577912 |
| H | -13.106994 | 10.643171 | -3.302434 |
| H | -11.139733 | 8.723351 | -2.417407 |
| H | -14.056536 | 8.412204 | -3.201515 |
| H | -12.693392 | 8.216893 | -0.493330 |
| H | -13.650226 | 9.567347 | -1.042935 |
| H | -13.861980 | 6.063534 | -2.509027 |
| H | -12.694946 | 6.362878 | -3.769606 |
| H | -12.175478 | 6.285962 | -2.093451 |
| H | -15.044376 | 8.243237 | 0.397657 |
| H | -15.656699 | 8.016513 | -1.229554 |
| H | -14.694674 | 6.764392 | -0.454814 |
| N | -12.336366 | 8.999238 | -5.455141 |
| H | -13.244716 | 9.315260 | -5.205349 |
| C | -12.143360 | 8.617686 | -6.852869 |
| H | -12.210506 | 7.547235 | -6.987249 |
| H | -12.921278 | 9.093303 | -7.436114 |
| C | -10.781884 | 9.042941 | -7.408967 |
| O | -10.130205 | 8.308099 | -8.093667 |
| N | -10.412699 | 10.311633 | -7.120649 |
| H | -10.965282 | 10.818989 | -6.466970 |
| C | -9.115058 | 10.872417 | -7.512661 |
| H | -8.988560 | 10.774598 | -8.579688 |
| H | -9.123505 | 11.924288 | -7.257797 |
| C | -7.877489 | 10.227644 | -6.873597 |
| O | -6.951812 | 9.831645 | -7.527551 |
| N | -7.861536 | 10.228083 | -5.524705 |
| H | -8.697245 | 10.462119 | -5.039105 |
| C | -6.687246 | 9.832494 | -4.724095 |

```
H                           -5.799554       10.270902       -5.149481
H                           -6.825559       10.221439       -3.723011
C                           -6.426191        8.340134       -4.619217
O                           -5.276578        7.965215       -4.645082
N                           -7.441228        7.480704       -4.424721
C                           -7.097549        6.118499       -4.038204
C                           -6.484779        5.325872       -5.187396
O                           -5.617327        4.515067       -4.952652
C                           -8.217357        5.327050       -3.320966
C                           -9.407127        4.929314       -4.210986
C                           -8.641247        6.072121       -2.050113
C                          -10.340402        3.902631       -3.564284
H                           -8.392419        7.790843       -4.432547
H                           -6.282595        6.188435       -3.333693
H                           -7.723571        4.409202       -3.012279
H                           -9.968284        5.809987       -4.495853
H                           -9.030240        4.495836       -5.131832
H                           -9.292375        5.460837       -1.439967
H                           -7.779078        6.330258       -1.442748
H                           -9.176064        6.985327       -2.284980
H                          -11.137686        3.638950       -4.252409
H                           -9.814497        2.988868       -3.315179
H                          -10.806475        4.281831       -2.662776
N                           -6.927231        5.533487       -6.439802
C                           -6.291206        4.847411       -7.549519
C                           -4.886078        5.387179       -7.812344
O                           -3.996511        4.640271       -8.146438
C                           -7.165640        4.895114       -8.811479
C                           -8.402430        3.990593       -8.715334
C                           -8.073668        2.499457       -8.866993
N                           -9.201887        1.635242       -8.563022
C                           -9.349340        0.907288       -7.410615
N                           -8.589236        1.329473       -6.347354
N                          -10.114974       -0.102382       -7.281569
H                           -7.627776        6.222473       -6.608868
H                           -6.125672        3.823679       -7.255705
H                           -7.471665        5.920086       -9.001457
H                           -6.554869        4.592966       -9.656973
H                           -8.913559        4.157919       -7.772384
H                           -9.109357        4.272018       -9.489036
H                           -7.766899        2.304967       -9.888774
H                           -7.240380        2.213762       -8.240578
H                           -9.706735        1.281797       -9.343556
H                           -8.466137        2.314326       -6.265955
H                           -8.854044        0.912038       -5.475554
H                          -10.526490       -0.380120       -8.152303
N                           -4.699324        6.703471       -7.652177
H                           -5.465762        7.298274       -7.432028
C                           -3.399771        7.330340       -7.937278
H                           -3.065857        7.092060       -8.935476
H                           -3.524419        8.401253       -7.848533
C                           -2.330344        6.864044       -6.961771
O                           -1.196512        6.654699       -7.317939
N                           -2.750277        6.734795       -5.697894
```

```
C         -1.860851    6.438570   -4.603194
C         -1.554928    4.957905   -4.475086
O         -0.421089    4.596296   -4.259988
C         -2.451923    7.020764   -3.312637
C         -1.852497    6.324447   -2.092156
C         -2.234301    8.540287   -3.341651
H         -3.683627    7.007686   -5.476049
H         -0.899518    6.894322   -4.792310
H         -3.517501    6.826389   -3.325427
H         -2.073781    6.858988   -1.187963
H         -2.249793    5.319466   -1.990738
H         -0.779507    6.266601   -2.165366
H         -2.973807    9.056175   -2.745738
H         -1.246763    8.795434   -2.971579
H         -2.330840    8.925911   -4.349751
N         -2.563946    4.076375   -4.597407
C         -2.199558    2.681622   -4.714544
C         -1.251699    2.509566   -5.907124
O         -0.307897    1.752474   -5.823936
C         -3.412015    1.741739   -4.778493
C         -4.272094    1.802762   -3.506275
C         -5.373755    0.736748   -3.429929
N         -4.869207   -0.521891   -2.873207
C         -5.572423   -1.644133   -2.775570
N         -6.789719   -1.741127   -3.290225
N         -5.085396   -2.691696   -2.129741
H         -3.485193    4.378942   -4.837116
H         -1.601935    2.407689   -3.857455
H         -4.026902    1.983907   -5.640470
H         -3.033542    0.737397   -4.932222
H         -3.643411    1.728750   -2.625030
H         -4.757726    2.766559   -3.455560
H         -6.180706    1.096945   -2.801846
H         -5.774798    0.554612   -4.419929
H         -3.980033   -0.502802   -2.416929
H         -7.185423   -1.028559   -3.860868
H         -7.237624   -2.639172   -3.296120
H         -4.246906   -2.667523   -1.577969
H         -5.604803   -3.546857   -2.157154
N         -1.508449    3.201604   -7.009865
H         -2.352352    3.721578   -7.125013
C         -0.640520    3.033431   -8.160167
H         -1.003955    3.680635   -8.947683
H         -0.637312    2.012664   -8.514466
C          0.803608    3.402973   -7.835577
O          1.746185    2.780735   -8.257846
N          0.928428    4.468792   -7.029027
H          0.128048    5.030218   -6.839960
C          2.196342    4.966707   -6.603683
H          2.825903    5.249180   -7.438277
H          2.024542    5.851665   -6.004440
C          3.026265    4.023811   -5.755368
O          4.228953    4.173350   -5.669953
N          2.387109    3.056479   -5.086398
```

```
C         3.105172    2.236259   -4.153448
C         4.302152    1.488900   -4.744529
O         5.247232    1.231213   -4.024088
C         2.110556    1.263258   -3.518575
C         2.707384    0.417387   -2.404633
C         1.646050   -0.478603   -1.796688
N         2.276012   -1.365813   -0.831927
C         1.623376   -2.351178   -0.214719
N         0.341282   -2.500019   -0.353291
N         2.356535   -3.210258    0.507071
H         1.397925    2.949027   -5.181022
H         3.548134    2.856517   -3.383339
H         1.293209    1.859651   -3.126899
H         1.684095    0.632983   -4.290767
H         3.513137   -0.197913   -2.790113
H         3.135082    1.051154   -1.632936
H         0.884767    0.109284   -1.298694
H         1.165002   -1.059452   -2.578234
H         3.261937   -1.478195   -0.906100
H        -0.178292   -1.889300   -0.943822
H        -0.234978   -3.127785    0.230264
H         3.162198   -2.833286    0.953631
H         1.870691   -3.892269    1.057001
N         4.284694    1.074097   -6.018924
H         3.530255    1.335531   -6.620648
C         5.381998    0.270655   -6.525685
H         5.162761    0.014028   -7.554569
H         5.479094   -0.645928   -5.965279
C         6.740289    0.940462   -6.490451
O         7.756195    0.294324   -6.383123
N         6.740730    2.270788   -6.610537
H         5.876524    2.772640   -6.577470
C         7.964896    3.035091   -6.553211
H         8.673155    2.673037   -7.284971
H         7.735910    4.063336   -6.799298
C         8.715664    3.042362   -5.219487
O         9.849690    3.470937   -5.171368
N         8.085253    2.535910   -4.151861
C         8.677432    2.374958   -2.835077
C         9.956362    1.527105   -2.853928
O        10.777126    1.618302   -1.955425
C         7.590096    1.684184   -1.981879
C         7.801646    1.545100   -0.473935
C         7.435288    2.794408    0.336353
C         7.345308    2.503885    1.830941
N         8.691026    2.501957    2.476229
O        11.065708    1.832380    0.868015
H        10.908969    2.051225   -0.050862
H        11.561458    1.016547    0.806249
H         7.143858    2.217567   -4.262091
H         8.962528    3.334884   -2.425162
H         6.665502    2.219178   -2.152745
H         7.432494    0.699762   -2.407648
H         7.143183    0.745302   -0.148238
```

| | | | |
|---|---:|---:|---:|
| H | 8.806619 | 1.212398 | -0.250589 |
| H | 8.126508 | 3.615059 | 0.166641 |
| H | 6.458596 | 3.147106 | 0.020854 |
| H | 6.766668 | 3.252861 | 2.348724 |
| H | 6.892952 | 1.541665 | 2.017448 |
| H | 9.410269 | 2.064327 | 1.908594 |
| H | 8.984370 | 3.480386 | 2.638138 |
| H | 8.658226 | 2.013938 | 3.359077 |
| N | 10.078196 | 0.676847 | -3.863911 |
| C | 11.145854 | -0.283332 | -4.107104 |
| C | 12.508845 | 0.310932 | -4.465630 |
| O | 13.481609 | -0.399605 | -4.351992 |
| C | 10.751419 | -1.192043 | -5.292703 |
| C | 9.634222 | -2.213651 | -5.036637 |
| C | 9.209184 | -2.841906 | -6.367578 |
| C | 10.087092 | -3.298937 | -4.059541 |
| H | 9.340050 | 0.669095 | -4.537138 |
| H | 11.318508 | -0.872428 | -3.218305 |
| H | 10.468346 | -0.548645 | -6.120398 |
| H | 11.643780 | -1.725453 | -5.597426 |
| H | 8.769261 | -1.700521 | -4.627179 |
| H | 8.422838 | -3.574593 | -6.212831 |
| H | 8.834356 | -2.089660 | -7.049868 |
| H | 10.044342 | -3.351751 | -6.841196 |
| H | 9.314155 | -4.047431 | -3.921855 |
| H | 10.967742 | -3.804211 | -4.441840 |
| H | 10.333426 | -2.900909 | -3.079748 |
| N | 12.568936 | 1.557330 | -4.941369 |
| C | 13.796405 | 2.170640 | -5.412581 |
| C | 14.938523 | 2.372348 | -4.398228 |
| O | 15.984784 | 2.776812 | -4.829317 |
| C | 13.450361 | 3.497514 | -6.091453 |
| O | 12.852289 | 4.391416 | -5.196046 |
| H | 11.739836 | 2.112175 | -4.932298 |
| H | 14.242299 | 1.520168 | -6.155611 |
| H | 14.364089 | 3.939872 | -6.456333 |
| H | 12.803043 | 3.299137 | -6.939727 |
| H | 11.905655 | 4.368447 | -5.266600 |
| N | 14.780380 | 2.046613 | -3.095631 |
| C | 15.918938 | 1.987237 | -2.200266 |
| C | 16.820840 | 0.770297 | -2.437684 |
| O | 17.949465 | 0.773078 | -2.019875 |
| C | 15.489617 | 1.897278 | -0.730873 |
| C | 14.617745 | 3.037363 | -0.208976 |
| C | 14.279127 | 2.777428 | 1.259320 |
| N | 13.239062 | 3.648723 | 1.791853 |
| C | 13.396224 | 4.793060 | 2.437559 |
| N | 14.604632 | 5.287994 | 2.680451 |
| O | 14.911328 | 7.732274 | 4.263083 |
| H | 14.608243 | 8.613443 | 4.028128 |
| H | 15.627207 | 7.851275 | 4.874341 |
| N | 12.333773 | 5.454915 | 2.846496 |
| H | 13.894362 | 1.728384 | -2.780199 |
| H | 16.534868 | 2.858312 | -2.363270 |

```
H     14.963097    0.959896   -0.582916
H     16.403597    1.835944   -0.150446
H     15.119238    3.993955   -0.323366
H     13.693561    3.107148   -0.770836
H     13.904598    1.769429    1.365349
H     15.161667    2.836068    1.885174
H     12.304392    3.338533    1.617793
H     15.421683    4.818326    2.370552
H     14.723189    6.139288    3.202539
H     11.383787    5.158906    2.685672
H     12.411861    6.291970    3.378690
N     16.240472   -0.294073   -3.017081
C     16.937951   -1.539361   -3.233691
C     17.731282   -1.485034   -4.535925
H     18.110329   -0.495580   -4.801288
O     17.948043   -2.434899   -5.214886
C     15.960839   -2.719747   -3.224928
C     15.429514   -3.076805   -1.869606
N     14.492808   -2.334111   -1.178259
O     12.537203   -0.202120   -0.505441
H     13.197458   -0.778007   -0.895441
H     12.059293    0.199250   -1.223030
C     15.752770   -4.168244   -1.140496
C     14.280416   -2.975859   -0.076091
N     15.017391   -4.090125    0.015658
H     15.324980   -0.209539   -3.403699
H     17.668167   -1.650936   -2.439877
H     15.136347   -2.502136   -3.895038
H     16.479039   -3.577927   -3.631081
H     16.427256   -4.973840   -1.337048
H     13.597529   -2.670829    0.690978
H     14.936151   -4.782437    0.727651
O    -14.844693   -2.510762    4.924427
H    -15.120967   -1.834495    4.310246
H    -14.467057   -2.054233    5.668822
O    -12.997310   -0.985125    6.839079
H    -12.268212   -1.481828    7.193669
H    -13.262480   -0.378144    7.519124
O    -12.215584   -0.405678    4.046063
H    -12.423184   -0.414763    4.975687
H    -13.060907   -0.432644    3.595117
O    -13.351473   -2.418727    0.883140
H    -12.619013   -1.797940    0.870805
H    -13.169481   -3.016298    1.606711
O     -7.050858    0.717354    0.256460
H     -7.303607    1.153106    1.072298
H     -6.104872    0.667627    0.287605
O     -7.745231    1.842191    2.836540
H     -8.696650    1.768397    2.877485
H     -7.500924    2.570317    3.400943
O     -9.119952   -0.719767   -1.002267
H     -8.309912   -0.416984   -0.587949
H     -8.974926   -1.553209   -1.442894
O    -12.566418   -2.709654   -1.860314
```

```
H                              -13.024195    -2.809930    -1.025857
H                              -13.154744    -2.948864    -2.570114
O                              -13.199470     1.397229    -0.601929
H                              -12.659533     0.711273    -0.203015
H                              -12.732367     2.203983    -0.382233
O                              -14.163434     4.582854     2.140962
H                              -13.875648     4.979485     2.957793
H                              -14.730472     3.848459     2.348198
O                              -15.293822     1.805519     1.315464
H                              -16.132296     1.924317     0.888185
H                              -14.652556     1.685272     0.609654
O                              -12.376580    -0.049991    -3.037487
H                              -12.746739     0.620936    -2.467339
H                              -12.303059    -0.826125    -2.485046
O                              -13.369210    -2.175239    -4.646563
H                              -13.248804    -1.271745    -4.357738
H                              -14.087794    -2.179811    -5.264922
O                               -9.868946    -2.924458    -2.750987
H                               -9.997361    -2.536470    -3.621761
H                              -10.741583    -2.944045    -2.358577
O                              -11.378976    -0.439350     0.600649
H                              -10.635609    -0.593676     0.011713
H                              -11.029627     0.097386     1.313382
O                              -10.633308     1.471188     2.582947
H                              -11.196285     1.049971     3.232880
H                              -11.123855     2.174830     2.159717
O                              -11.980470     3.469156     0.850831
H                              -12.753852     3.844334     1.284183
H                              -11.441872     4.224146     0.636107
O                               -9.452234     0.407786    -3.608494
H                              -10.394331     0.283217    -3.698195
H                               -9.282858     0.208631    -2.689596
O                              -14.736743    -0.640512     2.716858
H                              -15.061580     0.175172     2.335681
H                              -14.514563    -1.215401     1.985130
O                              -12.734980    -3.622064     3.473648
H                              -13.519911    -3.381055     3.978309
H                              -12.103354    -2.945444     3.695847
O                              -10.500123    -2.268870    -5.392021
H                              -11.443761    -2.408179    -5.383079
H                              -10.340253    -1.491119    -5.940211
```

Case 12
Voltage = -70mV; R297 neutral; E183 neutral; Y266 neutral; R300 positive; S4 Free
```
N         15.580943    -6.451467     5.391409
H         14.775824    -6.208984     5.935849
C         15.508704    -5.858249     4.055130
```

```
C         14.783014    -6.782958     3.060463
O         14.071890    -6.337178     2.176035
C         16.919244    -5.582430     3.521476
H         16.383855    -6.119906     5.888776
H         14.946219    -4.934166     4.050116
H         16.888271    -5.204600     2.506195
H         17.404380    -4.833302     4.137075
H         17.526193    -6.480861     3.535426
N         15.032376    -8.085603     3.186390
H         15.602804    -8.362472     3.956235
C         14.398657    -9.092387     2.333239
H         14.489322    -8.827225     1.292915
H         14.886017   -10.044104     2.498546
C         12.911238    -9.211259     2.668498
O         12.078650    -9.399302     1.820210
N         12.582803    -9.055885     3.966868
H         13.303757    -8.999744     4.648971
C         11.195100    -9.170815     4.421404
H         11.187948    -9.155989     5.503111
H         10.743471   -10.089010     4.078858
C         10.336960    -8.010602     3.900707
O          9.186594    -8.179854     3.594958
N         10.954712    -6.818653     3.846617
C         10.320195    -5.626766     3.343729
C         10.079161    -5.661894     1.832707
O          9.021939    -5.250608     1.416879
C         11.115127    -4.390984     3.796737
O         12.498741    -4.584496     3.767241
H         11.915813    -6.754250     4.089064
H          9.328768    -5.543824     3.758060
H         10.819665    -3.536276     3.199041
H         10.867820    -4.180148     4.826719
H         12.816356    -4.772263     2.891905
N         11.033749    -6.128715     1.005716
C         10.698338    -6.292206    -0.405570
C          9.618144    -7.359928    -0.580786
O          8.830228    -7.283540    -1.496061
C         11.933868    -6.517159    -1.312066
C         12.449017    -7.959717    -1.340913
C         11.639875    -6.028949    -2.735124
H         11.877792    -6.518814     1.367597
H         10.226871    -5.376048    -0.720899
H         12.712178    -5.885209    -0.901571
H         13.383955    -7.998199    -1.889882
H         12.623002    -8.365664    -0.353406
H         11.750964    -8.617945    -1.849243
H         12.513343    -6.166568    -3.362215
H         10.815703    -6.575293    -3.178335
H         11.391860    -4.972984    -2.743220
N          9.570411    -8.340325     0.328332
H         10.335698    -8.473019     0.950415
C          8.593033    -9.421196     0.241217
H          8.919390   -10.207489     0.909852
H          8.545550    -9.820499    -0.760035
```

```
C         7.181474    -9.020575     0.615778
O         6.258048    -9.462883    -0.027657
N         7.003382    -8.190949     1.651152
C         5.672519    -7.688529     1.904122
C         5.270082    -6.679227     0.833467
O         4.096307    -6.585679     0.540222
C         5.406736    -7.176732     3.334464
C         5.530154    -8.319273     4.347718
C         6.251456    -5.972497     3.754607
H         7.776948    -7.936260     2.226731
H         4.983880    -8.502598     1.738503
H         4.367661    -6.861661     3.316868
H         5.230159    -7.977415     5.332609
H         4.890593    -9.153897     4.079519
H         6.549583    -8.681760     4.416835
H         5.899913    -5.594803     4.708904
H         7.290102    -6.249713     3.876115
H         6.197546    -5.160882     3.037200
N         6.192829    -5.951919     0.198976
C         5.803410    -5.131258    -0.936877
C         5.337640    -6.008168    -2.107747
O         4.321904    -5.725360    -2.700022
C         6.923153    -4.144557    -1.329460
C         7.115366    -3.097073    -0.218143
C         6.605822    -3.470187    -2.668495
C         8.411265    -2.296141    -0.339521
H         7.150411    -5.998822     0.473280
H         4.921623    -4.569358    -0.666279
H         7.842852    -4.710459    -1.441421
H         6.261177    -2.421101    -0.230217
H         7.110632    -3.584739     0.747813
H         7.353832    -2.724854    -2.906941
H         6.589768    -4.178277    -3.486385
H         5.637874    -2.980127    -2.638535
H         8.495991    -1.586665     0.477031
H         9.276053    -2.947356    -0.293774
H         8.471213    -1.732612    -1.262813
N         6.082160    -7.069871    -2.440333
H         6.961511    -7.222514    -1.993684
C         5.730993    -7.903712    -3.568923
H         6.563450    -8.570857    -3.755162
H         5.560697    -7.311142    -4.455171
C         4.471204    -8.741208    -3.355241
O         3.615448    -8.790034    -4.207294
N         4.386069    -9.411249    -2.199621
H         5.132519    -9.367357    -1.538680
C         3.209943   -10.176911    -1.860584
H         2.963009   -10.884127    -2.638119
H         3.418711   -10.726341    -0.951220
C         1.974189    -9.306843    -1.644831
O         0.883525    -9.704407    -1.983931
N         2.187874    -8.107007    -1.107150
C         1.056059    -7.192773    -1.015226
C         0.594937    -6.669828    -2.378361
```

```
O          -0.590438    -6.517825    -2.575534
C           1.269528    -6.000587    -0.078308
O           1.469961    -6.396631     1.237601
H           3.092981    -7.814907    -0.815330
H           0.214198    -7.747566    -0.634619
H           2.083237    -5.379107    -0.433947
H           0.359633    -5.418642    -0.105302
H           2.383184    -6.630798     1.333970
N           1.510446    -6.368844    -3.310027
C           1.095814    -5.952377    -4.637436
C           0.334347    -7.074880    -5.341754
O          -0.664699    -6.814786    -5.978474
C           2.289782    -5.412297    -5.462411
C           2.644558    -4.000922    -4.960387
C           1.995712    -5.410909    -6.968071
C           3.934301    -3.426016    -5.547371
H           2.484115    -6.438300    -3.104536
H           0.362105    -5.165406    -4.540323
H           3.135600    -6.069122    -5.285477
H           1.816170    -3.333849    -5.195878
H           2.736413    -4.022595    -3.881542
H           2.844857    -5.024797    -7.517766
H           1.803484    -6.407630    -7.345887
H           1.134948    -4.793506    -7.204417
H           4.178444    -2.479424    -5.073090
H           4.768167    -4.100650    -5.383302
H           3.856103    -3.241918    -6.612897
N           0.789708    -8.324210    -5.215040
H           1.650629    -8.501087    -4.740214
C           0.037620    -9.430425    -5.760663
H          -0.080151    -9.342728    -6.830532
H           0.580824   -10.342105    -5.546376
C          -1.365280    -9.540117    -5.166936
O          -2.309835    -9.812220    -5.869895
N          -1.472630    -9.318869    -3.858022
C          -2.759693    -9.428921    -3.216028
C          -3.714897    -8.302887    -3.610713
O          -4.907902    -8.528722    -3.621373
C          -2.616082    -9.468992    -1.689635
O          -3.816267    -9.851132    -1.082580
H          -0.652057    -9.230561    -3.295651
H          -3.250357   -10.338474    -3.532538
H          -1.871577   -10.205816    -1.423186
H          -2.281394    -8.503282    -1.323253
H          -4.555685    -9.476746    -1.547054
N          -3.185508    -7.107524    -3.877287
C          -3.962126    -5.989841    -4.384222
C          -4.491209    -6.284871    -5.787103
O          -5.660609    -6.115657    -6.042419
C          -3.091499    -4.708328    -4.328939
C          -3.386845    -3.649081    -5.377513
C          -4.677517    -3.215315    -5.665268
C          -2.333299    -3.100326    -6.100650
C          -4.902876    -2.284780    -6.666090
```

```
C          -2.552986   -2.156850   -7.091474
C          -3.843937   -1.753888   -7.385240
H          -2.203191   -6.987941   -3.751879
H          -4.843013   -5.859879   -3.776414
H          -3.189856   -4.297414   -3.329024
H          -2.057821   -4.998582   -4.439448
H          -5.516258   -3.638094   -5.147652
H          -1.329052   -3.435685   -5.908390
H          -5.912069   -1.990257   -6.894184
H          -1.721191   -1.746577   -7.633191
H          -4.023242   -1.039956   -8.169462
N          -3.613200   -6.724612   -6.695506
C          -4.067654   -7.011991   -8.037957
C          -5.191963   -8.045450   -8.028222
O          -6.113075   -7.941648   -8.794220
C          -2.928463   -7.516207   -8.926451
S          -1.713949   -6.252233   -9.420637
H          -2.659489   -6.871161   -6.440251
H          -4.500420   -6.125104   -8.476724
H          -2.420328   -8.350762   -8.462204
H          -3.356188   -7.865600   -9.856655
H          -1.126418   -6.087345   -8.245910
N          -5.078569   -9.066921   -7.155614
H          -4.239510   -9.208649   -6.635682
C          -6.088709  -10.110661   -7.192012
H          -6.267711  -10.436410   -8.205371
H          -5.717864  -10.951420   -6.617375
C          -7.459078   -9.722657   -6.613127
O          -8.477351  -10.107175   -7.119688
N          -7.418079   -9.015934   -5.478385
C          -8.645358   -8.676643   -4.784162
C          -9.483813   -7.601509   -5.472627
O         -10.662346   -7.534646   -5.233008
C          -8.395344   -8.283153   -3.319476
C          -7.563433   -7.007005   -3.099045
C          -8.243266   -5.735719   -2.658193
O          -9.494941   -5.817003   -2.304751
H          -9.820908   -4.940921   -2.054204
O          -7.625405   -4.702198   -2.583710
H          -6.537961   -8.757623   -5.087086
H          -9.283599   -9.548356   -4.784547
H          -9.361313   -8.190500   -2.843998
H          -7.883962   -9.113279   -2.848612
H          -6.816727   -7.189264   -2.334990
H          -7.000554   -6.733773   -3.978495
N          -8.837281   -6.734017   -6.273824
C          -9.555173   -5.702658   -7.000838
C          -9.765983   -6.149571   -8.445475
H          -9.144232   -6.983780   -8.777287
O         -10.566790   -5.644658   -9.165827
C          -8.814750   -4.352440   -6.862768
O          -8.860083   -3.924275   -5.521975
C          -9.335846   -3.256887   -7.787103
H          -7.858102   -6.841916   -6.430965
```

```
H           -10.536873    -5.610389    -6.558827
H            -7.765690    -4.519869    -7.069824
H            -9.757510    -3.714463    -5.272729
H            -8.826798    -2.330332    -7.544182
H            -9.160415    -3.487487    -8.829062
H           -10.401063    -3.105547    -7.653143
N           -12.952182    -9.022387     3.174334
H           -13.191626    -9.579349     2.380085
C           -11.773294    -8.180298     2.898413
C           -10.800342    -8.106032     4.054309
O            -9.680650    -7.670944     3.859705
C           -12.137342    -6.732056     2.522637
O           -12.857846    -6.258265     3.625185
C           -12.931330    -6.628053     1.226691
H           -13.739161    -8.447262     3.406325
H           -11.209631    -8.626773     2.090493
H           -11.210031    -6.178641     2.408509
H           -12.984139    -5.311310     3.578610
H           -13.162273    -5.587754     1.022766
H           -12.364055    -7.011522     0.383900
H           -13.868663    -7.168383     1.292359
N           -11.197404    -8.533614     5.242057
H           -12.115208    -8.922145     5.276423
C           -10.404553    -8.508671     6.445489
H           -10.328525    -7.507538     6.841626
H           -10.914405    -9.113083     7.186278
C            -8.964806    -9.011974     6.351351
O            -8.119110    -8.415561     6.971054
N            -8.655485   -10.111799     5.640730
H            -9.371582   -10.621808     5.174403
C            -7.266602   -10.543981     5.541598
H            -7.249939   -11.527575     5.090030
H            -6.821863   -10.610691     6.521841
C            -6.387450    -9.608355     4.692672
O            -5.217748    -9.452979     4.936212
N            -7.045371    -9.018002     3.701142
C            -6.417642    -8.068355     2.809858
C            -6.265634    -6.715861     3.502761
O            -5.250443    -6.072899     3.374677
C            -7.191487    -7.926298     1.487240
C            -7.490660    -9.245923     0.799749
C            -8.779166    -9.521377     0.360987
C            -6.497073   -10.193394     0.572296
C            -9.078331   -10.716621    -0.277732
C            -6.790250   -11.384437    -0.069635
C            -8.084285   -11.654727    -0.490830
H            -8.029353    -9.136122     3.631334
H            -5.412439    -8.404249     2.612714
H            -8.122091    -7.404274     1.671987
H            -6.595920    -7.288616     0.840726
H            -9.551307    -8.784170     0.491634
H            -5.483076   -10.003418     0.870706
H           -10.080667   -10.904649    -0.619193
H            -6.002725   -12.093405    -0.249001
```

```
H          -8.309476   -12.579140   -0.991204
N          -7.287557    -6.315997    4.264987
C          -7.216114    -5.164115    5.128445
C          -6.176572    -5.367282    6.228792
O          -5.552133    -4.408813    6.633583
C          -8.624182    -4.893477    5.699191
C          -8.813352    -3.555646    6.381519
C          -8.797235    -2.376679    5.639822
C          -9.062653    -3.479830    7.744112
C          -9.012985    -1.151422    6.243748
C          -9.287389    -2.253431    8.357130
C          -9.262502    -1.087598    7.609272
H          -8.127549    -6.852050    4.266965
H          -6.874485    -4.303855    4.572196
H          -9.315861    -4.964624    4.866833
H          -8.877219    -5.692258    6.385394
H          -8.618919    -2.413327    4.578869
H          -9.078677    -4.377940    8.336872
H          -8.977376    -0.254916    5.651920
H          -9.460232    -2.212351    9.418409
H          -9.418179    -0.135922    8.084701
N          -5.974366    -6.595367    6.695698
H          -6.626245    -7.314472    6.477457
C          -4.947093    -6.895932    7.667921
H          -5.002976    -6.248412    8.529746
H          -5.099755    -7.915726    7.996525
C          -3.532365    -6.764350    7.109760
O          -2.662604    -6.269925    7.786720
N          -3.318173    -7.215338    5.868943
C          -2.036060    -7.043859    5.210380
C          -1.795413    -5.574129    4.842070
O          -0.702845    -5.082250    5.022068
C          -1.932959    -7.998352    4.001088
C          -0.758014    -7.646805    3.088829
C          -1.816577    -9.448702    4.484112
H          -4.036633    -7.725951    5.401917
H          -1.247357    -7.272922    5.912423
H          -2.848990    -7.895510    3.424784
H          -0.686161    -8.372686    2.286351
H          -0.856494    -6.670161    2.635519
H           0.181479    -7.658345    3.631037
H          -1.815713   -10.124817    3.635116
H          -0.883760    -9.593480    5.023230
H          -2.633668    -9.736577    5.131722
N          -2.821974    -4.877665    4.336683
C          -2.662153    -3.488125    3.955914
C          -2.474242    -2.596829    5.190244
O          -1.588162    -1.766139    5.196428
C          -3.842614    -2.971435    3.117312
C          -4.041190    -3.592904    1.720444
C          -2.983123    -3.221859    0.691983
O          -1.809013    -3.442578    0.981021
O          -3.365084    -2.734364   -0.409478
H          -3.685029    -5.339173    4.146052
```

```
H        -1.755777    -3.387600     3.386547
H        -4.759631    -3.116206     3.678714
H        -3.713178    -1.898226     3.015107
H        -4.050773    -4.672609     1.794721
H        -5.013210    -3.290595     1.349491
N        -3.311574    -2.757148     6.225089
C        -3.093976    -2.024914     7.456999
C        -1.764288    -2.419979     8.108651
O        -1.069209    -1.568227     8.620576
C        -4.228811    -2.057114     8.490904
O        -4.472056    -3.340999     8.983524
C        -5.493097    -1.367072     7.993832
H        -4.003350    -3.472937     6.191842
H        -2.951656    -0.987104     7.197000
H        -3.843591    -1.493669     9.332094
H        -4.965209    -3.832051     8.337203
H        -6.232367    -1.368032     8.785303
H        -5.286539    -0.336704     7.721771
H        -5.917092    -1.869701     7.134802
N        -1.380399    -3.700440     8.043327
H        -2.033704    -4.424984     7.837062
C        -0.107950    -4.080313     8.602592
H        -0.035230    -3.840160     9.653899
H        -0.001044    -5.150498     8.480107
C         1.069485    -3.393499     7.919839
O         2.024381    -3.017881     8.560809
N         1.003101    -3.256861     6.582649
C         2.114819    -2.675295     5.872936
C         2.213653    -1.159757     6.042928
O         3.312595    -0.645119     5.994248
C         2.154289    -3.097326     4.388516
S         1.406297    -1.967607     3.169492
H         0.250082    -3.680371     6.079034
H         3.024084    -3.041228     6.327873
H         3.189641    -3.152121     4.081793
H         1.720033    -4.079946     4.277590
H         0.149072    -1.997426     3.584876
N         1.095202    -0.458169     6.248588
C         1.204069     0.968491     6.491348
C         1.792551     1.248724     7.860096
O         2.448003     2.243725     8.061461
C        -0.085128     1.781567     6.235237
C        -1.159681     1.629519     7.323095
C        -0.604654     1.516833     4.817363
C        -2.416345     2.465371     7.069673
H         0.209262    -0.919502     6.244308
H         1.942336     1.345074     5.800379
H         0.251206     2.811554     6.277610
H        -1.432102     0.590503     7.444455
H        -0.735031     1.939017     8.272999
H        -1.327989     2.269286     4.529434
H         0.206960     1.551748     4.096385
H        -1.079787     0.548743     4.733949
H        -3.058625     2.452531     7.944744
```

```
H          -2.172276    3.502546    6.854939
H          -2.995447    2.082253    6.236373
N           1.549785    0.323262    8.791262
H           0.902226   -0.410608    8.602189
C           1.983444    0.505383   10.157729
H           1.455165   -0.212407   10.772714
H           1.765069    1.501851   10.511342
C           3.481167    0.297266   10.320295
O           4.156399    1.045581   10.973283
N           3.998391   -0.771002    9.674217
H           3.397963   -1.413526    9.201999
C           5.423861   -0.955558    9.644624
H           5.637510   -1.917209    9.192950
H           5.845932   -0.941470   10.638168
C           6.156885    0.119120    8.844832
O           7.285818    0.422442    9.139165
N           5.502078    0.621275    7.799023
C           6.098615    1.759278    7.146320
C           5.552900    3.073975    7.701812
H           5.091935    3.016697    8.686605
O           5.660471    4.105804    7.114288
C           5.950565    1.716204    5.621254
C           6.805375    0.661738    4.952119
C           8.198392    0.757436    4.967410
C           6.225592   -0.414321    4.288244
C           8.986816   -0.184349    4.322551
C           7.012814   -1.352858    3.636119
C           8.393174   -1.238590    3.642844
H           4.568894    0.347499    7.584612
H           7.145264    1.750077    7.413175
H           4.909955    1.552525    5.370098
H           6.221081    2.695702    5.245248
H           8.670148    1.565078    5.500887
H           5.157727   -0.527079    4.309898
H          10.058956   -0.096555    4.358129
H           6.546112   -2.178724    3.129128
H           8.994816   -1.967559    3.131574
N          12.613547    9.596234   -1.248800
H          11.907032    9.888338   -1.897448
C          12.304853    8.264738   -0.719488
C          11.235779    8.430981    0.350142
O          10.214476    7.778609    0.315099
C          13.557412    7.603139   -0.124697
C          14.647984    7.251345   -1.143320
S          14.037621    6.086024   -2.396004
C          15.523954    5.882159   -3.408828
H          13.472113    9.589713   -1.762483
H          11.869220    7.606402   -1.459847
H          13.990600    8.260130    0.623231
H          13.249282    6.697165    0.385310
H          15.034126    8.140071   -1.629638
H          15.479729    6.804647   -0.610629
H          15.303053    5.133114   -4.156121
H          15.788888    6.809193   -3.900355
```

```
H       16.357506    5.539062   -2.809030
N       11.445089    9.351567    1.305501
C       10.510690    9.490032    2.403370
C        9.221362   10.219114    2.016349
O        8.210084    9.998273    2.635568
C       11.189684   10.124005    3.624557
C       12.310879    9.215069    4.098104
O       13.450563    9.389648    3.719800
N       11.956109    8.220042    4.922018
H       12.295591    9.871648    1.294229
H       10.173154    8.503241    2.679196
H       11.620421   11.087693    3.382554
H       10.447810   10.258332    4.401072
H       12.651902    7.536591    5.129214
H       11.014238    7.886516    4.937483
N        9.245158   11.046686    0.954092
H       10.104625   11.193036    0.475628
C        8.003499   11.517476    0.368851
H        8.233842   12.245491   -0.398096
H        7.380657   11.985559    1.113407
C        7.186388   10.384433   -0.254315
O        5.984528   10.460481   -0.273692
N        7.858716    9.347055   -0.792967
C        7.158458    8.183882   -1.309546
C        6.580338    7.316990   -0.180398
O        5.475291    6.843383   -0.299571
C        8.048701    7.391074   -2.297278
C        8.424995    8.280375   -3.497695
C        7.350032    6.107701   -2.761024
C        9.464780    7.668668   -4.440333
H        8.821669    9.233910   -0.579484
H        6.285621    8.537349   -1.836874
H        8.958319    7.110577   -1.775382
H        7.522075    8.516276   -4.057035
H        8.807313    9.228859   -3.135467
H        7.996004    5.552114   -3.428344
H        7.100654    5.459052   -1.930699
H        6.426239    6.331081   -3.286181
H        9.746824    8.386832   -5.203424
H       10.367915    7.382177   -3.908599
H        9.095229    6.786950   -4.950527
N        7.306031    7.141714    0.942705
C        6.757447    6.385234    2.056070
C        5.437936    6.999645    2.532623
O        4.454302    6.321726    2.719891
C        7.705629    6.356175    3.260796
C        9.096743    5.763669    3.042377
O       10.028031    6.355256    3.580880
O        9.208001    4.713431    2.362190
H        8.268175    7.405938    0.954016
H        6.514845    5.383353    1.736455
H        7.836621    7.353289    3.654879
H        7.225434    5.769527    4.040715
N        5.462068    8.323317    2.763706
```

```
H         6.310204    8.834478    2.637504
C         4.294516    9.019609    3.245967
H         4.559931   10.058548    3.396380
H         3.954729    8.622508    4.189479
C         3.127584    8.972889    2.270517
O         1.984476    9.069312    2.652896
N         3.465880    8.836296    0.985811
H         4.425864    8.879780    0.734753
C         2.517175    8.905269   -0.078291
H         3.062528    9.080259   -0.996629
H         1.843276    9.740952    0.056928
C         1.609148    7.687008   -0.310212
O         0.580801    7.792050   -0.922381
N         2.070896    6.543125    0.236327
C         1.268998    5.355621    0.385241
C         0.453543    5.356056    1.683101
O        -0.706194    5.011852    1.660307
C         2.143014    4.102510    0.310887
H         2.973224    6.568009    0.662202
H         0.537950    5.341976   -0.407121
H         1.539293    3.213040    0.454825
H         2.620701    4.052633   -0.661254
H         2.915478    4.120542    1.072663
N         1.061914    5.711507    2.828308
C         0.350911    5.549018    4.086549
C        -0.762236    6.581487    4.258057
O        -1.793612    6.273841    4.817226
C         1.283926    5.473506    5.321416
C         1.882307    6.817295    5.764438
C         2.366575    4.407861    5.096810
C         2.721477    6.719295    7.041005
H         2.022672    5.979342    2.827367
H        -0.174866    4.608095    4.035879
H         0.635811    5.134435    6.123812
H         2.488023    7.225319    4.962249
H         1.082342    7.529713    5.937805
H         2.764455    4.063038    6.040390
H         1.960872    3.546650    4.575863
H         3.190020    4.795854    4.507485
H         3.030575    7.709707    7.361806
H         2.147403    6.277865    7.850224
H         3.614280    6.121438    6.907585
N        -0.551356    7.816635    3.783853
H         0.331949    8.058294    3.385077
C        -1.510324    8.867276    4.015453
H        -1.092032    9.791988    3.637926
H        -1.710912    8.994204    5.069561
C        -2.866997    8.618770    3.365897
O        -3.874757    8.854361    3.989208
N        -2.902661    8.108961    2.120752
C        -4.219192    7.953691    1.505310
C        -4.990380    6.750548    2.019754
O        -6.198153    6.782003    2.058497
C        -3.923114    7.884144   -0.000152
```

```
C           -2.658240      8.733626     -0.127444
C           -1.858567      8.307990      1.097398
H           -4.837595      8.810845      1.729678
H           -3.719461      6.860326     -0.294718
H           -4.760763      8.243324     -0.583873
H           -2.103931      8.570653     -1.039682
H           -2.911544      9.788273     -0.064044
H           -1.342079      7.382480      0.911508
H           -1.137243      9.048033      1.406469
N           -4.281032      5.669995      2.386893
C           -4.938004      4.536011      3.006957
C           -5.566814      4.898081      4.352168
O           -6.646804      4.446246      4.656335
C           -3.962331      3.358635      3.195857
C           -3.824354      2.456297      1.989218
C           -4.552117      1.274224      1.927421
C           -2.968215      2.755035      0.935738
C           -4.420021      0.401166      0.860646
C           -2.825422      1.887834     -0.134659
C           -3.537256      0.702518     -0.163954
O           -3.358882     -0.153367     -1.209634
H           -3.286675      5.739890      2.425770
H           -5.765422      4.229402      2.386278
H           -2.995846      3.747472      3.496189
H           -4.324697      2.762094      4.024448
H           -5.230451      1.020363      2.723384
H           -2.377801      3.649881      0.960581
H           -4.985908     -0.511781      0.831793
H           -2.144101      2.112663     -0.935644
H           -3.326550     -1.058491     -0.872589
N           -4.850710      5.692627      5.162236
H           -3.927980      5.974691      4.905094
C           -5.382442      6.145489      6.424388
H           -4.572585      6.588258      6.990784
H           -5.792305      5.323834      6.990481
C           -6.496070      7.176820      6.275922
O           -7.467107      7.148035      6.995008
N           -6.342178      8.091224      5.313945
H           -5.477873      8.129980      4.819138
C           -7.257639      9.220197      5.085232
H           -6.781172      9.866934      4.360185
H           -7.419134      9.777596      5.996001
C           -8.629733      8.833900      4.568455
O           -9.610533      9.425487      4.951773
N           -8.687634      7.836221      3.670770
C           -9.953945      7.451874      3.088784
C          -10.834758      6.647929      4.034690
O          -11.946660      6.334810      3.659724
C           -9.758683      6.696479      1.752696
O          -10.933372      6.760988      0.992044
C           -9.294807      5.251801      1.919389
H           -7.841288      7.415130      3.345398
H          -10.523449      8.344910      2.871788
H           -9.011466      7.245471      1.194744
```

```
H            -11.656711        6.469741        1.538307
H             -9.082933        4.831936        0.942461
H             -8.399358        5.179140        2.523228
H            -10.073292        4.649155        2.375583
N            -10.353349        6.327969        5.240897
C            -11.224038        5.807928        6.279070
C            -12.254021        6.835177        6.758153
O            -13.309619        6.455067        7.210769
C            -10.415138        5.337086        7.497163
C             -9.453589        4.166567        7.246326
C             -8.623276        3.916499        8.508994
C            -10.182748        2.889536        6.821939
H             -9.426056        6.593022        5.489170
H            -11.805335        4.995699        5.874011
H             -9.849648        6.179892        7.880787
H            -11.131002        5.056830        8.264227
H             -8.765055        4.441513        6.455014
H             -7.910803        3.111467        8.351008
H             -8.068278        4.805039        8.788543
H             -9.257520        3.637830        9.347880
H             -9.468877        2.086845        6.666452
H            -10.888167        2.571172        7.587187
H            -10.732979        3.010558        5.896212
N            -11.902320        8.130033        6.727768
C            -12.841298        9.147185        7.102042
C            -13.869578        9.457386        6.034415
H            -14.608231       10.210943        6.330022
O            -13.901950        8.960084        4.959067
H            -11.097506        8.416866        6.212027
H            -13.366551        8.855677        8.003334
H            -12.302801       10.063301        7.323999
N            -14.581147       10.608317       -6.352208
H            -14.295356       11.220747       -7.091454
C            -13.475396       10.280376       -5.460569
C            -12.692412        9.111561       -6.051821
O            -11.503439        8.971224       -5.854062
C            -14.022540        9.944525       -4.044285
C            -14.426700       11.249226       -3.325750
C            -13.041046        9.137403       -3.186197
C            -15.385132       11.047886       -2.150530
H            -15.326986       11.062973       -5.866319
H            -12.751965       11.085582       -5.363213
H            -14.916237        9.342044       -4.197919
H            -13.523466       11.746784       -2.980620
H            -14.894008       11.936706       -4.025616
H            -13.437627        8.999787       -2.188859
H            -12.856621        8.150330       -3.593153
H            -12.085568        9.643257       -3.099584
H            -15.649834       12.004327       -1.712964
H            -16.304187       10.567509       -2.473415
H            -14.951367       10.441870       -1.364069
N            -13.390870        8.244218       -6.808026
H            -14.362723        8.434773       -6.922087
C            -12.852166        6.952482       -7.164268
```

```
H            -12.635985       6.353102      -6.291458
H            -13.593309       6.428249      -7.755635
C            -11.565887       7.012656      -7.972230
O            -10.700467       6.187470      -7.816853
N            -11.449841       7.997479      -8.880941
H            -12.190258       8.651105      -8.992037
C            -10.253228       8.104314      -9.681364
H            -10.078688       7.200585     -10.245379
H            -10.378016       8.921979     -10.380114
C             -8.993890       8.369869      -8.864864
O             -7.931439       7.925898      -9.223768
N             -9.128637       9.157069      -7.783768
H            -10.037220       9.334507      -7.414954
C             -7.985548       9.465857      -6.958267
H             -7.182911       9.860319      -7.560928
H             -8.283978      10.218734      -6.239373
C             -7.421989       8.266802      -6.206608
O             -6.232855       8.056535      -6.190785
N             -8.300632       7.480643      -5.559258
C             -7.815324       6.347198      -4.795857
C             -7.270710       5.242168      -5.704842
O             -6.321412       4.587469      -5.332619
C             -8.817868       5.837101      -3.733323
C            -10.185477       5.424089      -4.303704
C             -8.949970       6.878583      -2.613646
C            -11.044808       4.614464      -3.328227
H             -9.274241       7.691932      -5.595133
H             -6.933801       6.671402      -4.264150
H             -8.348183       4.951899      -3.312209
H            -10.732769       6.314174      -4.603349
H            -10.045686       4.836672      -5.205517
H             -9.564761       6.513547      -1.801179
H             -7.977712       7.127661      -2.199959
H             -9.401538       7.795981      -2.978971
H            -11.988724       4.349082      -3.795276
H            -10.559568       3.687856      -3.041645
H            -11.274543       5.169983      -2.426318
N             -7.835218       5.039549      -6.906882
C             -7.252795       4.084336      -7.830951
C             -5.870688       4.521658      -8.319274
O             -4.985303       3.705131      -8.437338
C             -8.192693       3.818503      -9.015999
C             -9.429231       2.993799      -8.637536
C             -9.125891       1.503526      -8.448148
N            -10.265180       0.759013      -7.945346
C            -10.448003       0.328180      -6.660967
N             -9.590082       0.852316      -5.719010
N            -11.329161      -0.520222      -6.303534
H             -8.625469       5.580814      -7.188844
H             -7.058837       3.165715      -7.299054
H             -8.496854       4.768327      -9.442714
H             -7.627403       3.298629      -9.784250
H             -9.886892       3.394259      -7.739627
H            -10.175493       3.094674      -9.419607
```

```
H           -8.836692     1.071029    -9.399733
H           -8.287639     1.350864    -7.782466
H          -10.829090     0.301602    -8.624744
H           -9.331353     1.803943    -5.867926
H           -9.922004     0.716746    -4.777606
H          -11.831411    -0.890940    -7.087954
N           -5.692306     5.817958    -8.607097
H           -6.451496     6.463535    -8.563095
C           -4.407416     6.295513    -9.054259
H           -4.090233     5.797755    -9.958941
H           -4.496617     7.355582    -9.254453
C           -3.297124     6.082283    -8.032290
O           -2.196306     5.731887    -8.394771
N           -3.594393     6.309946    -6.746653
C           -2.604578     6.108067    -5.706068
C           -2.210117     4.637014    -5.578518
O           -1.053706     4.335323    -5.377720
C           -3.106863     6.678294    -4.363112
C           -2.222954     6.236263    -3.192500
C           -3.176358     8.208506    -4.428134
H           -4.501360     6.653149    -6.505902
H           -1.688042     6.610349    -5.983671
H           -4.108148     6.287912    -4.198800
H           -2.578071     6.692843    -2.280121
H           -2.241058     5.162955    -3.045994
H           -1.190847     6.538085    -3.331988
H           -3.589934     8.603127    -3.505756
H           -2.182154     8.629421    -4.549944
H           -3.799289     8.559588    -5.240223
N           -3.178077     3.713230    -5.662454
C           -2.843936     2.309537    -5.538401
C           -1.913158     1.832861    -6.663126
O           -1.096448     0.974546    -6.430797
C           -4.091222     1.426201    -5.409555
C           -4.922794     1.745540    -4.158840
C           -6.073816     0.760631    -3.921215
N           -5.614794    -0.422010    -3.197517
C           -6.273136    -1.571572    -3.087673
N           -7.382347    -1.802308    -3.773697
N           -5.821659    -2.515152    -2.279225
H           -4.118413     3.988737    -5.852751
H           -2.244344     2.181481    -4.647617
H           -4.707080     1.539508    -6.297466
H           -3.756185     0.396770    -5.384537
H           -4.283467     1.766745    -3.280394
H           -5.354771     2.731365    -4.248608
H           -6.863067     1.245647    -3.356262
H           -6.493336     0.460356    -4.872629
H           -4.842557    -0.293248    -2.572766
H           -7.748802    -1.140662    -4.419135
H           -7.767590    -2.727035    -3.817418
H           -5.003633    -2.415412    -1.711478
H           -6.327274    -3.377731    -2.218136
N           -2.025180     2.412283    -7.872125
```

```
H        -2.761919     3.062463    -8.044602
C        -1.071231     2.132980    -8.921969
H        -1.488471     2.481098    -9.858822
H        -0.885336     1.073592    -8.998337
C         0.293193     2.802448    -8.726706
O         1.320526     2.199370    -8.942737
N         0.283254     4.080183    -8.316167
H        -0.584876     4.560133    -8.196921
C         1.503101     4.814063    -8.091204
H         2.124095     4.837471    -8.976501
H         1.243017     5.833613    -7.838385
C         2.381749     4.249743    -6.979470
O         3.564835     4.508539    -6.937327
N         1.773194     3.430713    -6.115426
C         2.445654     2.645234    -5.109844
C         3.614600     1.838666    -5.662888
O         4.507185     1.505968    -4.904096
C         1.382826     1.709190    -4.515195
C         1.823519     0.784223    -3.383505
C         0.679321    -0.174726    -3.075960
N         1.009726    -1.021712    -1.936619
C         0.152319    -1.864861    -1.359861
N        -0.988246    -2.164497    -1.952967
N         0.452573    -2.439162    -0.213529
H         0.774726     3.426116    -6.109447
H         2.879938     3.274673    -4.341841
H         0.574033     2.338736    -4.160926
H         0.971576     1.121302    -5.324652
H         2.705927     0.217788    -3.655339
H         2.070334     1.362815    -2.495843
H        -0.224022     0.384276    -2.856338
H         0.486040    -0.790860    -3.946121
H         1.859515    -0.842143    -1.452674
H        -1.260530    -1.695501    -2.785905
H        -1.730688    -2.600775    -1.421829
H         1.224438    -2.129272     0.332189
H        -0.287082    -2.930529     0.281571
N         3.632132     1.506366    -6.954313
H         2.872573     1.757518    -7.557758
C         4.776520     0.830232    -7.517875
H         4.625121     0.747308    -8.586449
H         4.890149    -0.166409    -7.115719
C         6.108478     1.534225    -7.276207
O         7.132424     0.888716    -7.228616
N         6.100487     2.865543    -7.113350
H         5.240244     3.378052    -7.141998
C         7.317772     3.576304    -6.819849
H         8.063142     3.415997    -7.586485
H         7.095597     4.635072    -6.790005
C         7.988615     3.207932    -5.500698
O         9.144205     3.530843    -5.302013
N         7.273589     2.525184    -4.603261
C         7.833749     2.045056    -3.358600
C         8.973948     1.042313    -3.566492
```

```
O         9.764761    0.843965   -2.657975
C         6.692113    1.405606   -2.536903
C         6.933396    1.225401   -1.033366
C         6.694480    2.495965   -0.205222
C         6.731829    2.241168    1.301620
N         8.121925    2.263365    1.845660
O        10.351239    1.416045    0.137490
H        10.161417    1.581687   -0.784959
H        10.782528    0.562212    0.101269
H         6.327799    2.274743   -4.814781
H         8.279380    2.871693   -2.821389
H         5.809807    2.016287   -2.680160
H         6.455423    0.445246   -2.982241
H         6.231263    0.472007   -0.686504
H         7.920808    0.823629   -0.851213
H         7.402745    3.280691   -0.457224
H         5.711889    2.892064   -0.439671
H         6.182507    2.996066    1.844034
H         6.306693    1.279105    1.547421
H         8.799800    1.808289    1.239443
H         8.452113    3.229636    2.002696
H         8.159349    1.795565    2.741174
N         9.052144    0.411148   -4.737158
C        10.109811   -0.528529   -5.073992
C        11.496337    0.104621   -5.221163
O        12.456270   -0.630855   -5.278907
C         9.772955   -1.245437   -6.396231
C         8.661213   -2.304155   -6.306304
C         8.150792   -2.634963   -7.711960
C         9.150522   -3.574763   -5.603485
H         8.371398    0.609116   -5.443805
H        10.213875   -1.249328   -4.275544
H         9.498749   -0.489314   -7.125014
H        10.679496   -1.718360   -6.753246
H         7.827286   -1.892971   -5.741718
H         7.373716   -3.392265   -7.673313
H         7.739722   -1.754057   -8.191067
H         8.952972   -3.018787   -8.336911
H         8.354535   -4.309515   -5.539571
H         9.971623   -4.025180   -6.153901
H         9.497630   -3.384250   -4.592540
N        11.597863    1.436928   -5.303047
C        12.804764    2.106172   -5.758472
C        13.999757    2.101975   -4.801524
O        15.034370    2.574588   -5.193632
C        12.468995    3.543756   -6.162109
O        11.995393    4.285742   -5.077767
H        10.766293    1.984313   -5.252994
H        13.179820    1.588428   -6.634231
H        13.372573    4.007836   -6.525854
H        11.744761    3.522453   -6.970503
H        11.046239    4.285747   -5.071391
N        13.879427    1.532017   -3.584956
C        15.046758    1.275499   -2.765867
```

```
C         15.826655    0.012725   -3.152845
O         16.920204   -0.165237   -2.682810
C         14.678605    1.108939   -1.288152
C         13.950774    2.296058   -0.667458
C         13.614372    1.989519    0.789459
N         12.719873    2.964997    1.396685
C         13.064216    4.058685    2.059761
N         14.340023    4.359411    2.287989
O         14.699562    6.843580    3.875635
H         14.464902    7.761122    3.715252
H         15.381568    6.864004    4.535321
N         12.127338    4.856899    2.517218
H         13.000545    1.153155   -3.317038
H         15.729861    2.101286   -2.889733
H         14.064160    0.221296   -1.180804
H         15.603245    0.915381   -0.757173
H         14.548778    3.199246   -0.740763
H         13.027732    2.499594   -1.197178
H         13.107329    1.037292    0.850717
H         14.507487    1.894109    1.395569
H         11.751513    2.844244    1.180518
H         15.074037    3.849077    1.858043
H         14.572018    5.202211    2.781807
H         11.139577    4.677660    2.413344
H         12.349937    5.651363    3.073564
N         15.192286   -0.875650   -3.939264
C         15.723626   -2.185717   -4.228808
C         16.094228   -2.276339   -5.703614
H         16.478092   -1.346116   -6.130911
O         16.004815   -3.266288   -6.352463
C         14.760582   -3.311677   -3.823987
C         14.622770   -3.520614   -2.343631
N         13.770249   -2.785174   -1.544837
O         11.590261   -0.821194   -1.087942
H         12.262691   -1.428852   -1.400573
H         11.062958   -0.570939   -1.838905
C         15.253569   -4.443047   -1.581071
C         13.904320   -3.260869   -0.349636
N         14.794733   -4.260735   -0.298878
H         14.291802   -0.654346   -4.308757
H         16.652161   -2.273477   -3.677018
H         13.784802   -3.109157   -4.252395
H         15.117622   -4.227642   -4.276184
H         15.975871   -5.191179   -1.827736
H         13.373945   -2.917009    0.515023
H         14.965315   -4.846205    0.488377
O        -15.115763   -2.662351    5.685802
H        -15.417285   -1.868926    5.250243
H        -14.563173   -2.371195    6.403365
O        -12.837431   -1.415749    7.418892
H        -12.038795   -1.825873    7.732607
H        -12.999078   -0.679335    7.995808
O        -12.415814   -0.875889    4.579013
H        -12.422590   -0.950959    5.530098
```

```
H      -13.282808    -0.536174     4.352382
O      -14.089092    -1.912697     1.408550
H      -13.317698    -1.339412     1.419233
H      -13.920859    -2.586260     2.071081
O       -7.796929     0.236653     1.186745
H       -7.989144     0.702379     2.003245
H       -7.176842     0.792225     0.734053
O       -8.278779     1.778388     3.567600
H       -9.230850     1.868170     3.577846
H       -7.908963     2.595813     3.887865
O       -9.895709    -0.645541    -0.534303
H       -9.144414    -0.463476     0.033430
H       -9.857187    -1.558948    -0.815724
O      -13.232671    -2.771375    -1.162621
H      -13.715714    -2.669803    -0.343607
H      -13.828024    -3.074929    -1.841841
O      -14.637132     0.744311    -0.044901
H      -14.982627    -0.113350     0.178957
H      -13.735487     0.722317     0.287733
O      -14.425580     5.176820     4.764327
H      -13.806742     5.733972     4.299730
H      -14.404039     5.502364     5.659530
O      -15.745295     1.975199     2.294639
H      -16.314438     2.733396     2.264790
H      -15.486197     1.771566     1.396057
O      -13.326677    -0.200567    -2.532907
H      -13.879631     0.336670    -1.971319
H      -13.066989    -0.937728    -1.983160
O      -14.178417    -2.576311    -3.887433
H      -14.073341    -1.636742    -3.742313
H      -14.849936    -2.685739    -4.547499
O      -10.487568    -3.239180    -1.749667
H      -10.591922    -2.950445    -2.660085
H      -11.381619    -3.241502    -1.408964
O      -12.130483     0.089727     1.040618
H      -11.396943    -0.151164     0.469173
H      -11.754170     0.596553     1.764905
O      -11.138180     1.397934     3.370232
H      -11.330827     0.616836     3.888018
H      -11.851819     2.001688     3.601770
O      -13.511639     2.603332     4.114384
H      -13.751528     3.465244     4.461833
H      -14.183991     2.413684     3.467670
O      -10.559281     0.842216    -2.940410
H      -11.488598     0.626775    -2.960702
H      -10.227754     0.421837    -2.149567
O      -15.169795    -0.461095     3.774261
H      -15.520706     0.401126     3.564054
H      -15.097250    -0.914641     2.938895
O      -13.340656    -3.458008     3.665409
H      -14.036770    -3.386249     4.326630
H      -12.725467    -2.772912     3.926772
O      -11.256698    -2.694227    -4.422693
H      -12.178498    -2.933847    -4.373172
```

| | | | |
|---|---|---|---|
| H | -11.237700 | -1.869943 | -4.923428 |

Case 13
Voltage=-70 mV; R297 neutral; E183 neutral; Y266 negative; E226 neutral
| | | | |
|---|---|---|---|
| N | 15.580944 | -6.451469 | 5.391416 |
| H | 14.782258 | -6.195286 | 5.938893 |
| C | 15.513293 | -5.857897 | 4.053313 |
| C | 14.783016 | -6.782962 | 3.060452 |
| O | 14.050703 | -6.345513 | 2.187864 |
| C | 16.924932 | -5.585446 | 3.519928 |
| H | 16.392015 | -6.132800 | 5.883906 |
| H | 14.953632 | -4.931994 | 4.048190 |
| H | 16.896483 | -5.200743 | 2.507438 |
| H | 17.411847 | -4.841563 | 4.140259 |
| H | 17.528376 | -6.486334 | 3.530551 |
| N | 15.032378 | -8.085605 | 3.186405 |
| H | 15.613965 | -8.361720 | 3.948318 |
| C | 14.398660 | -9.092385 | 2.333233 |
| H | 14.482886 | -8.824286 | 1.292746 |
| H | 14.890877 | -10.042435 | 2.493202 |
| C | 12.912584 | -9.219873 | 2.671112 |
| O | 12.080408 | -9.421647 | 1.825842 |
| N | 12.582804 | -9.055884 | 3.966877 |
| H | 13.302856 | -8.989037 | 4.649121 |
| C | 11.195108 | -9.170813 | 4.421378 |
| H | 11.188191 | -9.157686 | 5.503131 |
| H | 10.743677 | -10.088408 | 4.077368 |
| C | 10.336945 | -8.010597 | 3.900733 |
| O | 9.189852 | -8.186673 | 3.588443 |
| N | 10.947586 | -6.815721 | 3.854811 |
| C | 10.323857 | -5.634238 | 3.313850 |
| C | 10.128990 | -5.698369 | 1.796602 |
| O | 9.093505 | -5.280941 | 1.335366 |
| C | 11.107697 | -4.390700 | 3.766310 |
| O | 12.490854 | -4.588165 | 3.779546 |
| H | 11.906222 | -6.747364 | 4.106147 |
| H | 9.320859 | -5.546165 | 3.698536 |
| H | 10.834791 | -3.547485 | 3.142135 |
| H | 10.831523 | -4.160165 | 4.784809 |
| H | 12.829246 | -4.775918 | 2.912422 |
| N | 11.104026 | -6.203151 | 1.021110 |
| C | 10.848458 | -6.400460 | -0.403001 |
| C | 9.693854 | -7.384933 | -0.600853 |
| O | 8.942940 | -7.276331 | -1.542508 |
| C | 12.135991 | -6.796348 | -1.188418 |
| C | 12.327545 | -8.302560 | -1.408204 |
| C | 12.170602 | -6.068588 | -2.533667 |
| H | 11.933522 | -6.579024 | 1.426591 |
| H | 10.482048 | -5.465767 | -0.796838 |

```
H    12.969441    -6.433029    -0.597611
H    13.280067    -8.470424    -1.899941
H    12.323292    -8.870536    -0.488710
H    11.557922    -8.703810    -2.060460
H    13.065536    -6.336110    -3.084173
H    11.310758    -6.335017    -3.138559
H    12.175769    -4.993659    -2.398369
N     9.570429    -8.340341     0.328317
H    10.316795    -8.476606     0.971499
C     8.593019    -9.421181     0.241224
H     8.947632   -10.228769     0.869119
H     8.519100    -9.787715    -0.770990
C     7.183199    -9.065843     0.671329
O     6.254302    -9.527222     0.048853
N     7.009970    -8.263832     1.727919
C     5.674184    -7.804613     2.035982
C     5.216742    -6.774614     1.008402
O     4.030420    -6.675544     0.773174
C     5.444705    -7.333963     3.486245
C     5.629745    -8.495931     4.467364
C     6.274254    -6.117887     3.903182
H     7.792685    -7.996208     2.284122
H     5.000813    -8.631206     1.869208
H     4.400166    -7.040434     3.509765
H     5.355297    -8.184496     5.469868
H     5.001235    -9.339695     4.201028
H     6.658774    -8.835687     4.494060
H     5.953293    -5.775289     4.881046
H     7.324661    -6.367396     3.974008
H     6.161269    -5.289070     3.212941
N     6.125014    -6.042988     0.352524
C     5.729917    -5.224945    -0.781382
C     5.314062    -6.111496    -1.961982
O     4.268641    -5.892880    -2.529567
C     6.816049    -4.199217    -1.164499
C     7.013337    -3.179924    -0.029816
C     6.436853    -3.495744    -2.473012
C     8.291599    -2.352174    -0.164770
H     7.090660    -6.116625     0.590196
H     4.828165    -4.693501    -0.517533
H     7.748689    -4.735149    -1.313844
H     6.147501    -2.519768    -0.000323
H     7.046713    -3.695847     0.920387
H     7.143566    -2.710870    -2.705249
H     6.427955    -4.176164    -3.314706
H     5.451225    -3.046816    -2.401681
H     8.397000    -1.678364     0.678666
H     9.163048    -2.995976    -0.181770
H     8.301518    -1.750335    -1.066295
N     6.116407    -7.119355    -2.325827
H     7.023765    -7.215122    -1.919826
C     5.791145    -7.951479    -3.465341
H     6.627307    -8.619214    -3.630679
H     5.643763    -7.359921    -4.356808
```

```
C         4.523563    -8.785344    -3.283161
O         3.692687    -8.837576    -4.158708
N         4.398180    -9.444433    -2.124088
H         5.133790    -9.418186    -1.449807
C         3.204024   -10.195110    -1.816850
H         2.970831   -10.904744    -2.596438
H         3.377370   -10.740384    -0.897609
C         1.974188    -9.306857    -1.644803
O         0.886054    -9.690432    -2.004750
N         2.187871    -8.107001    -1.107183
C         1.099009    -7.141924    -1.113516
C         0.671449    -6.662579    -2.506787
O        -0.507121    -6.458721    -2.698990
C         1.368182    -5.913524    -0.248960
O         1.327873    -6.233229     1.115623
H         3.098600    -7.819351    -0.830948
H         0.219316    -7.627632    -0.722083
H         2.309864    -5.453969    -0.507568
H         0.577357    -5.201437    -0.428406
H         2.188745    -6.531103     1.390630
N         1.598001    -6.446197    -3.443903
C         1.215128    -6.059631    -4.791333
C         0.419860    -7.180439    -5.453367
O        -0.567526    -6.923075    -6.106268
C         2.443083    -5.624609    -5.624203
C         2.932101    -4.252160    -5.131149
C         2.129022    -5.596631    -7.125711
C         4.288408    -3.834840    -5.699710
H         2.564695    -6.551801    -3.224298
H         0.520338    -5.233846    -4.735313
H         3.227467    -6.357332    -5.456912
H         2.186059    -3.502956    -5.390169
H         3.002816    -4.264636    -4.050931
H         2.993650    -5.263786    -7.685958
H         1.864640    -6.575716    -7.505863
H         1.307704    -4.921661    -7.344519
H         4.637043    -2.930281    -5.212682
H         5.034999    -4.605368    -5.531480
H         4.248852    -3.638173    -6.764935
N         0.839383    -8.435997    -5.266785
H         1.688068    -8.616226    -4.772873
C         0.043862    -9.537680    -5.752611
H        -0.063209    -9.509114    -6.826670
H         0.540598   -10.459220    -5.476654
C        -1.365271    -9.540097    -5.166933
O        -2.321059    -9.756638    -5.873445
N        -1.472642    -9.318874    -3.858022
C        -2.774953    -9.325137    -3.238886
C        -3.672520    -8.187381    -3.720344
O        -4.870158    -8.386870    -3.746706
C        -2.676022    -9.268296    -1.708400
O        -3.928201    -9.462597    -1.112184
H        -0.654948    -9.259304    -3.287989
H        -3.304668   -10.227763    -3.510256
```

```
H         -2.021732   -10.058312   -1.364920
H         -2.247959    -8.318237   -1.404155
H         -4.613838    -9.146097   -1.688635
N         -3.109362    -7.017878   -4.024287
C         -3.852890    -5.897374   -4.585261
C         -4.407592    -6.253204   -5.964039
O         -5.577640    -6.096672   -6.214665
C         -2.926432    -4.666384   -4.627807
C         -3.385827    -3.489314   -5.474755
C         -4.720577    -3.146516   -5.655565
C         -2.414438    -2.710278   -6.098947
C         -5.070703    -2.068044   -6.456276
C         -2.758562    -1.624617   -6.884014
C         -4.094870    -1.304022   -7.071264
H         -2.125261    -6.917052   -3.894321
H         -4.714143    -5.693481   -3.969577
H         -2.766318    -4.349016   -3.600773
H         -1.963548    -4.985933   -5.001992
H         -5.507028    -3.727562   -5.215805
H         -1.374629    -2.962927   -5.978954
H         -6.112738    -1.851435   -6.605010
H         -1.987603    -1.040420   -7.354523
H         -4.369890    -0.475313   -7.699777
N         -3.539319    -6.738441   -6.861046
C         -4.012453    -7.098346   -8.178525
C         -5.153924    -8.111358   -8.108136
O         -6.054191    -8.053876   -8.901125
C         -2.887651    -7.669890   -9.044085
S         -1.644448    -6.458299   -9.596126
H         -2.591734    -6.908847   -6.600021
H         -4.434199    -6.231343   -8.665129
H         -2.398008    -8.493570   -8.541862
H         -3.325489    -8.050942   -9.956838
H         -1.052025    -6.250315   -8.430378
N         -5.078559    -9.066911   -7.155611
H         -4.240363    -9.198113   -6.632336
C         -6.083538   -10.117055   -7.173947
H         -6.253358   -10.468157   -8.180236
H         -5.713995   -10.941622   -6.575042
C         -7.459135    -9.722603   -6.613169
O         -8.476990   -10.103914   -7.122104
N         -7.417947    -9.016183   -5.478396
C         -8.618022    -8.624224   -4.764913
C         -9.489234    -7.626915   -5.516418
O        -10.681635    -7.614068   -5.349181
C         -8.296487    -8.044566   -3.372303
C         -7.364204    -6.815089   -3.317196
C         -7.914963    -5.429728   -3.094340
O         -9.199690    -5.275657   -3.225379
H         -9.436452    -4.353045   -3.062815
O         -7.176197    -4.515851   -2.807077
H         -6.529060    -8.747980   -5.114647
H         -9.245476    -9.493710   -4.626910
H         -9.244126    -7.827778   -2.898811
```

```
H            -7.826537    -8.837634    -2.800199
H            -6.640519    -6.947843    -2.526368
H            -6.778335    -6.725027    -4.218444
N            -8.837268    -6.733933    -6.273831
C            -9.555215    -5.702643    -7.000854
C           -10.263653    -6.318770    -8.203092
H            -9.869186    -7.280003    -8.541182
O           -11.191715    -5.802253    -8.737777
C            -8.582769    -4.552075    -7.353891
O            -8.261329    -3.847240    -6.181000
C            -9.106375    -3.598577    -8.424069
H            -7.854499    -6.821383    -6.413901
H           -10.335034    -5.304862    -6.365010
H            -7.650970    -4.981889    -7.696388
H            -9.018726    -3.331656    -5.917615
H            -8.413378    -2.769193    -8.510116
H            -9.190292    -4.080354    -9.389649
H           -10.082787    -3.206763    -8.161812
N           -12.952093    -9.022577     3.174130
H           -13.209305    -9.557156     2.370071
C           -11.773585    -8.180046     2.898864
C           -10.786158    -8.128810     4.049551
O            -9.650766    -7.733077     3.855245
C           -12.147130    -6.724119     2.564961
O           -12.731841    -6.216988     3.735169
C           -13.096093    -6.611749     1.377094
H           -13.733274    -8.451299     3.432382
H           -11.222535    -8.599365     2.067521
H           -11.230898    -6.192414     2.343847
H           -12.878689    -5.277214     3.634355
H           -13.262488    -5.565947     1.144746
H           -12.693244    -7.092369     0.492929
H           -14.059127    -7.054220     1.602274
N           -11.197212    -8.533632     5.241858
H           -12.123922    -8.900460     5.271995
C           -10.417194    -8.551389     6.453287
H           -10.329935    -7.561964     6.873475
H           -10.950875    -9.158681     7.174893
C            -8.986258    -9.085943     6.389387
O            -8.163151    -8.556324     7.096443
N            -8.660237   -10.136668     5.616396
H            -9.362372   -10.590156     5.076280
C            -7.270059   -10.566454     5.517962
H            -7.248882   -11.540348     5.045793
H            -6.837027   -10.655322     6.501102
C            -6.387487    -9.608344     4.692717
O            -5.221731    -9.457124     4.956767
N            -7.045383    -9.018035     3.701132
C            -6.304018    -7.971553     3.000774
C            -6.252610    -6.695931     3.842941
O            -5.241561    -6.025106     3.859181
C            -6.641575    -7.770101     1.507372
C            -7.773889    -6.855706     1.083343
C            -7.715235    -5.483224     1.309188
```

```
C          -8.832151    -7.354167     0.334995
C          -8.700289    -4.638684     0.825997
C          -9.805884    -6.508935    -0.172931
C          -9.750104    -5.149068     0.080524
H          -8.040488    -9.008242     3.682980
H          -5.282357    -8.308549     3.006537
H          -5.729326    -7.398393     1.053174
H          -6.791255    -8.759601     1.092617
H          -6.881993    -5.061122     1.842172
H          -8.879844    -8.406861     0.115772
H          -8.632293    -3.580433     1.005691
H         -10.591436    -6.907218    -0.787736
H         -10.500236    -4.493289    -0.321598
N          -7.306122    -6.389535     4.599884
C          -7.275342    -5.260375     5.501130
C          -6.193861    -5.461413     6.560450
O          -5.505188    -4.523925     6.906484
C          -8.676380    -5.088768     6.124612
C          -8.924605    -3.776231     6.836942
C          -9.190305    -2.621698     6.106695
C          -8.954180    -3.708416     8.223698
C          -9.470370    -1.426563     6.746546
C          -9.240370    -2.514985     8.870145
C          -9.497978    -1.370340     8.133585
H          -8.175705    -6.853901     4.441387
H          -6.993925    -4.361386     4.970865
H          -9.394036    -5.187405     5.317231
H          -8.850716    -5.910713     6.806291
H          -9.191399    -2.655918     5.030781
H          -8.761345    -4.592133     8.807095
H          -9.668157    -0.545350     6.164182
H          -9.257257    -2.481225     9.944721
H          -9.705716    -0.440190     8.632264
N          -6.035717    -6.678210     7.076344
H          -6.702797    -7.389989     6.871070
C          -5.050053    -6.949406     8.097757
H          -5.138040    -6.267670     8.930411
H          -5.225029    -7.954762     8.459563
C          -3.608985    -6.850710     7.608382
O          -2.784348    -6.275826     8.277331
N          -3.301267    -7.438329     6.441477
C          -1.939056    -7.387468     5.941513
C          -1.592433    -5.993207     5.401762
O          -0.455206    -5.583439     5.479424
C          -1.680959    -8.522741     4.923241
C          -0.264778    -8.463822     4.339515
C          -1.894323    -9.888940     5.588614
H          -3.993853    -7.954479     5.939296
H          -1.267301    -7.520150     6.777263
H          -2.393100    -8.416601     4.108739
H          -0.124455    -9.293120     3.654237
H          -0.072873    -7.548860     3.799413
H           0.480441    -8.547975     5.125040
H          -1.708898   -10.680411     4.870122
```

```
H          -1.200412    -10.021855     6.414638
H          -2.899831    -10.019778     5.961587
N          -2.587028     -5.268983     4.859835
C          -2.342165     -3.960753     4.289793
C          -2.340741     -2.843326     5.336207
O          -1.428583     -2.048498     5.325386
C          -3.275650     -3.652381     3.113252
C          -3.173476     -4.696953     1.987099
C          -1.738031     -4.962984     1.634751
O          -1.355492     -6.199918     1.784128
H          -0.406890     -6.281210     1.616404
O          -0.984575     -4.094009     1.277522
H          -3.497516     -5.667958     4.773246
H          -1.325482     -3.947952     3.934925
H          -4.305348     -3.607291     3.442721
H          -3.009343     -2.672510     2.734613
H          -3.656120     -5.615289     2.268114
H          -3.663174     -4.308277     1.099841
N          -3.322579     -2.798407     6.252695
C          -3.093984     -2.024902     7.456980
C          -1.773315     -2.415512     8.124154
O          -1.086384     -1.569520     8.652552
C          -4.224546     -1.987927     8.496259
O          -4.418852     -3.221777     9.119069
C          -5.510377     -1.382301     7.945469
H          -4.005490     -3.525343     6.273006
H          -2.936508     -0.999790     7.156283
H          -3.848305     -1.333418     9.272809
H          -4.895379     -3.798326     8.533853
H          -6.240890     -1.316689     8.742138
H          -5.329505     -0.382137     7.564727
H          -5.932641     -1.980380     7.149215
N          -1.380401     -3.700454     8.043332
H          -2.078321     -4.405566     7.930792
C          -0.212529     -4.103727     8.792167
H          -0.304957     -3.884299     9.846633
H          -0.095464     -5.172083     8.668090
C           1.063424     -3.428464     8.320672
O           1.937063     -3.150756     9.107913
N           1.171243     -3.202528     6.993040
C           2.350381     -2.592712     6.447752
C           2.211731     -1.136649     5.995320
O           3.202693     -0.603047     5.541562
C           3.121649     -3.480041     5.471615
S           2.325308     -3.918651     3.897423
H           0.428073     -3.500921     6.396899
H           3.019388     -2.478093     7.288726
H           4.033666     -2.972085     5.197121
H           3.387072     -4.393171     5.986233
H           1.358641     -4.699442     4.359103
N           1.089820     -0.460023     6.279345
C           1.165658      0.974756     6.464040
C           1.750502      1.263787     7.845333
O           2.405883      2.256692     8.043874
```

```
C         -0.138312    1.752135    6.181127
C         -1.188978    1.651129    7.298065
C         -0.670334    1.376165    4.794167
C         -2.458647    2.460419    7.024797
H          0.272654   -0.966285    6.545130
H          1.903589    1.338123    5.769792
H          0.175342    2.789690    6.150059
H         -1.450522    0.617816    7.483828
H         -0.748145    2.019221    8.220297
H         -1.433585    2.070729    4.469753
H          0.127433    1.404971    4.056887
H         -1.092374    0.379872    4.780617
H         -3.085415    2.486118    7.910666
H         -2.227449    3.486986    6.753877
H         -3.047083    2.030771    6.221759
N          1.549781    0.323269    8.791262
H          0.846842   -0.369273    8.651559
C          1.974277    0.559832   10.155277
H          1.470349   -0.159586   10.788562
H          1.729004    1.558593   10.484319
C          3.477800    0.394703   10.298817
O          4.165565    1.206355   10.851368
N          3.993393   -0.714806    9.712404
H          3.384965   -1.466949    9.464875
C          5.418329   -0.903195    9.703203
H          5.635361   -1.888609    9.307974
H          5.846147   -0.831065   10.692604
C          6.156883    0.119117    8.844832
O          7.306979    0.379424    9.092469
N          5.502084    0.621277    7.799023
C          6.041179    1.755482    7.087275
C          5.439801    3.076040    7.559369
H          4.954743    3.061552    8.533488
O          5.544286    4.074881    6.918145
C          5.932796    1.606590    5.569708
C          6.756464    0.469120    5.001563
C          8.124488    0.355033    5.253267
C          6.158199   -0.480907    4.179820
C          8.872667   -0.664571    4.681287
C          6.904174   -1.499176    3.607435
C          8.264973   -1.590748    3.847956
H          4.537202    0.408032    7.686128
H          7.080960    1.813598    7.377564
H          4.894705    1.462580    5.298548
H          6.245386    2.548079    5.131554
H          8.608124    1.044329    5.923520
H          5.098325   -0.436889    4.018512
H          9.923701   -0.738787    4.900654
H          6.419994   -2.225815    2.980032
H          8.833689   -2.382699    3.397820
N         12.613550    9.596234   -1.248801
H         11.872989    9.914821   -1.845350
C         12.304852    8.264740   -0.719487
C         11.277015    8.451101    0.389051
```

```
O            10.270616    7.776939     0.423178
C            13.563604    7.580101    -0.168365
C            14.643852    7.275864    -1.211959
S            14.013688    6.201003    -2.533521
C            15.492117    6.058709    -3.566075
H            13.433082    9.573505    -1.822707
H            11.832657    7.617079    -1.446686
H            14.003694    8.203778     0.603903
H            13.257022    6.652567     0.301388
H            15.042421    8.188481    -1.640552
H            15.470792    6.784525    -0.711791
H            15.238305    5.405805    -4.387241
H            15.786865    7.024540    -3.955562
H            16.314121    5.622325    -3.013817
N            11.508085    9.415962     1.295961
C            10.599703    9.613726     2.408077
C             9.280410   10.276147     2.008890
O             8.284925   10.051557     2.651134
C            11.282414   10.380699     3.547382
C            12.455273    9.578840     4.081395
O            13.589502    9.839260     3.739142
N            12.153127    8.579402     4.920157
H            12.329490    9.973497     1.199801
H            10.303236    8.644626     2.775075
H            11.662393   11.336718     3.209680
H            10.553299   10.552297     4.329166
H            12.897657    7.985869     5.216365
H            11.231782    8.193434     4.959373
N             9.264806   11.067915     0.920263
H            10.111749   11.235624     0.427160
C             8.003502   11.517473     0.368853
H             8.199916   12.261192    -0.392762
H             7.389705   11.965603     1.133065
C             7.186385   10.384437    -0.254317
O             5.985814   10.465788    -0.285411
N             7.863796    9.336504    -0.768201
C             7.175283    8.147357    -1.240715
C             6.645838    7.295871    -0.075902
O             5.550913    6.792207    -0.158913
C             8.056499    7.353954    -2.235168
C             8.341520    8.211534    -3.482998
C             7.396990    6.024233    -2.620110
C             9.428246    7.649626    -4.402848
H             8.826874    9.238703    -0.547809
H             6.282137    8.470931    -1.752648
H             8.998537    7.133658    -1.740894
H             7.417307    8.331381    -4.043786
H             8.636611    9.208913    -3.174416
H             8.017926    5.490415    -3.328228
H             7.252840    5.380428    -1.760923
H             6.425228    6.180510    -3.076713
H             9.615709    8.333867    -5.223884
H            10.365704    7.509038    -3.871492
H             9.153919    6.694967    -4.836240
```

```
N         7.398620    7.176633    1.036244
C         6.892639    6.459521    2.194638
C         5.540550    7.038422    2.632201
O         4.589285    6.330498    2.859443
C         7.845591    6.550811    3.396866
C         9.268269    6.003649    3.244148
O        10.155185    6.635456    3.810803
O         9.447625    4.943995    2.592562
H         8.356579    7.453529    1.016173
H         6.698009    5.430043    1.936988
H         7.931637    7.576049    3.724781
H         7.386256    5.998835    4.214449
N         5.501638    8.376114    2.779311
H         6.331619    8.912444    2.640408
C         4.310465    9.043470    3.242939
H         4.546549   10.089750    3.392305
H         3.972411    8.641883    4.185246
C         3.146084    8.973450    2.264905
O         2.004237    9.076925    2.650054
N         3.476933    8.821680    0.979455
H         4.436812    8.847316    0.725212
C         2.517189    8.905262   -0.078310
H         3.053840    9.095369   -0.999016
H         1.842837    9.735785    0.078730
C         1.609134    7.687018   -0.310189
O         0.561925    7.822151   -0.877635
N         2.083317    6.531179    0.195676
C         1.269434    5.354944    0.378754
C         0.438395    5.417821    1.664778
O        -0.739856    5.155103    1.635783
C         2.141475    4.099683    0.366160
H         2.994616    6.551259    0.602332
H         0.550350    5.306636   -0.421736
H         1.529375    3.220593    0.537942
H         2.627991    4.009370   -0.597993
H         2.906797    4.143153    1.134410
N         1.061046    5.731000    2.818139
C         0.343185    5.564331    4.070620
C        -0.760201    6.602897    4.253472
O        -1.782618    6.302126    4.831682
C         1.272804    5.472446    5.305287
C         1.887548    6.808294    5.750477
C         2.342583    4.394495    5.077873
C         2.718782    6.700222    7.031138
H         2.034905    5.943797    2.826157
H        -0.192555    4.628274    4.011598
H         0.619964    5.139308    6.106575
H         2.503654    7.206526    4.951509
H         1.096174    7.531186    5.918548
H         2.731224    4.039529    6.021035
H         1.927291    3.542363    4.548220
H         3.176089    4.774421    4.497793
H         3.038293    7.686840    7.353473
H         2.135166    6.266294    7.837920
```

```
H         3.604948      6.092225      6.899984
N        -0.547729      7.835912      3.775232
H         0.330492      8.072649      3.362624
C        -1.509120      8.882224      4.010498
H        -1.096929      9.810679      3.636244
H        -1.708377      9.004459      5.065233
C        -2.866972      8.618759      3.365892
O        -3.867089      8.837238      4.006798
N        -2.902685      8.108971      2.120754
C        -4.243641      7.994344      1.542560
C        -4.978530      6.726590      1.950900
O        -6.189515      6.733165      1.965971
C        -4.034500      8.192755      0.036110
C        -2.790454      9.086984      0.013407
C        -1.917296      8.462246      1.089926
H        -4.862860      8.793238      1.920778
H        -3.845213      7.248701     -0.457883
H        -4.906191      8.641710     -0.422163
H        -2.285921      9.131415     -0.933204
H        -3.062761     10.099760      0.300740
H        -1.419663      7.576440      0.736666
H        -1.168542      9.138700      1.469405
N        -4.254084      5.653843      2.304308
C        -4.898051      4.503066      2.917031
C        -5.562182      4.877911      4.249309
O        -6.642530      4.415260      4.539589
C        -3.906553      3.339641      3.154792
C        -3.668177      2.387246      1.996895
C        -4.240451      1.119574      1.995468
C        -2.824307      2.694193      0.930706
C        -3.991722      0.197424      0.988139
C        -2.567225      1.785982     -0.077300
C        -3.133531      0.494417     -0.088552
O        -2.867443     -0.362641     -1.032949
H        -3.259321      5.730937      2.329042
H        -5.704340      4.171695      2.281765
H        -2.968240      3.761942      3.501444
H        -4.300284      2.759828      3.982010
H        -4.902027      0.838547      2.798719
H        -2.341537      3.652215      0.893054
H        -4.448447     -0.777904      1.027146
H        -1.913492      2.061428     -0.887871
N        -4.879639      5.692560      5.074044
H        -3.950893      5.974520      4.838669
C        -5.411793      6.100245      6.355679
H        -4.591569      6.484155      6.949827
H        -5.848124      5.261129      6.874383
C        -6.496076      7.176796      6.275894
O        -7.442167      7.164689      7.027007
N        -6.342148      8.091257      5.313951
H        -5.489223      8.109658      4.798787
C        -7.257646      9.220181      5.085234
H        -6.781555      9.869407      4.362036
H        -7.421362      9.776354      5.996194
```

```
C         -8.628519      8.830978      4.566477
O         -9.611269      9.418888      4.948642
N         -8.684984      7.832910      3.666612
C         -9.953905      7.451896      3.088789
C        -10.835684      6.652036      4.037345
O        -11.950065      6.343802      3.664075
C         -9.771772      6.693348      1.752723
O        -10.957745      6.766070      1.003856
C         -9.316390      5.247634      1.921598
H         -7.840572      7.407490      3.341190
H        -10.520466      8.347604      2.875073
H         -9.027049      7.236737      1.186211
H        -11.672400      6.470761      1.559733
H         -9.112219      4.815575      0.949501
H         -8.415503      5.174917      2.515957
H        -10.093051      4.653140      2.391239
N        -10.353416      6.327971      5.240916
C        -11.207501      5.752449      6.261697
C        -12.260835      6.739481      6.771564
O        -13.309099      6.321692      7.206736
C        -10.383684      5.258995      7.460660
C         -9.403464      4.111747      7.171381
C         -8.568177      3.837772      8.425955
C        -10.109853      2.834782      6.707506
H         -9.422782      6.583060      5.485911
H        -11.768775      4.939807      5.830333
H         -9.830051      6.099235      7.866905
H        -11.090168      4.944023      8.223047
H         -8.720815      4.420375      6.387387
H         -7.841373      3.051267      8.242840
H         -8.029848      4.727313      8.733852
H         -9.196973      3.521208      9.255506
H         -9.383509      2.042349      6.552040
H        -10.826268      2.493380      7.452860
H        -10.640825      2.969003      5.772118
N        -11.934195      8.042045      6.784641
C        -12.893399      9.030338      7.184777
C        -13.893609      9.389093      6.105702
H        -14.650671     10.115872      6.420816
O        -13.886178      8.954266      5.003103
H        -11.136625      8.356324      6.273853
H        -13.439870      8.684536      8.053324
H        -12.371994      9.938308      7.472194
N        -11.959308     10.656091     -3.211148
H        -11.754655     11.059495     -2.318251
C        -11.814410      9.207262     -3.172602
C        -11.340185      8.853023     -4.571788
O        -10.269603      8.342018     -4.801539
C        -13.063492      8.427528     -2.702861
C        -13.413722      8.814832     -1.254653
C        -12.860541      6.916178     -2.848745
C        -14.734352      8.233070     -0.746703
H        -12.897273     10.932365     -3.432120
H        -10.987111      8.969554     -2.517496
```

```
H        -13.899936     8.724012    -3.339773
H        -12.605959     8.498580    -0.600010
H        -13.473936     9.896438    -1.177819
H        -13.773170     6.379076    -2.623653
H        -12.568492     6.637239    -3.856228
H        -12.092728     6.568045    -2.166078
H        -14.958248     8.621457     0.240520
H        -15.561209     8.495915    -1.401452
H        -14.700563     7.152733    -0.667319
N        -12.160013     9.244786    -5.566149
H        -13.063461     9.578314    -5.321661
C        -11.972873     8.843878    -6.959263
H        -12.063308     7.773744    -7.082453
H        -12.738923     9.330337    -7.549267
C        -10.601341     9.233126    -7.517250
O         -9.965740     8.476739    -8.193822
N        -10.204772    10.496260    -7.241744
H        -10.745194    11.020525    -6.591206
C         -8.894336    11.023974    -7.637100
H         -8.768065    10.910105    -8.702541
H         -8.881414    12.079078    -7.395954
C         -7.670672    10.363396    -6.988186
O         -6.749925     9.948210    -7.637144
N         -7.659042    10.373849    -5.639645
H         -8.492749    10.623923    -5.158619
C         -6.495251     9.961280    -4.832300
H         -5.597571    10.379323    -5.257229
H         -6.629460    10.358613    -3.833997
C         -6.269002     8.464691    -4.721383
O         -5.128663     8.062103    -4.738373
N         -7.307710     7.632841    -4.536525
C         -7.003250     6.259161    -4.160411
C         -6.400078     5.469433    -5.315853
O         -5.524300     4.665918    -5.092671
C         -8.148178     5.489039    -3.458846
C         -9.338492     5.123085    -4.361796
C         -8.566040     6.235321    -2.186375
C        -10.308474     4.126483    -3.722378
H         -8.249836     7.968557    -4.545907
H         -6.192181     6.300521    -3.449366
H         -7.678581     4.558075    -3.151207
H         -9.871042     6.018437    -4.655851
H         -8.964019     4.675013    -5.276336
H         -9.233057     5.635246    -1.582374
H         -7.702103     6.472451    -1.573100
H         -9.079794     7.160790    -2.419850
H        -11.100177     3.876386    -4.422320
H         -9.807472     3.203216    -3.457042
H        -10.778088     4.524836    -2.831434
N         -6.854163     5.666196    -6.566634
C         -6.223953     4.957354    -7.665951
C         -4.798146     5.450610    -7.906209
O         -3.931564     4.680466    -8.246371
C         -7.077697     5.030462    -8.940688
```

```
C         -8.335364    4.152872   -8.862352
C         -8.035748    2.656741   -9.031394
N         -9.162774    1.805504   -8.685738
C         -9.283708    1.102235   -7.515560
N         -8.492869    1.542479   -6.484380
N        -10.051216    0.098944   -7.344792
H         -7.554407    6.353882   -6.742009
H         -6.095515    3.928789   -7.369254
H         -7.357104    6.062786   -9.131415
H         -6.462367    4.716177   -9.778337
H         -8.845766    4.319165   -7.918890
H         -9.034262    4.459298   -9.633821
H         -7.773862    2.461211  -10.065103
H         -7.181142    2.357642   -8.440818
H         -9.686060    1.437597   -9.447094
H         -8.353311    2.526400   -6.427039
H         -8.723277    1.140575   -5.596974
H        -10.485559   -0.192867   -8.199775
N         -4.571416    6.757547   -7.721978
H         -5.322515    7.371420   -7.502093
C         -3.257793    7.352338   -8.011316
H         -2.930553    7.097344   -9.007626
H         -3.359396    8.426629   -7.933485
C         -2.195555    6.876798   -7.034384
O         -1.068888    6.633338   -7.392581
N         -2.610902    6.784275   -5.767066
C         -1.724804    6.480955   -4.671929
C         -1.452332    4.995135   -4.530429
O         -0.329173    4.606875   -4.303963
C         -2.304133    7.089248   -3.388079
C         -1.729461    6.384819   -2.160607
C         -2.045914    8.602479   -3.431311
H         -3.540837    7.070587   -5.547593
H         -0.754800    6.915340   -4.867353
H         -3.374571    6.923730   -3.403540
H         -1.937303    6.931486   -1.260749
H         -2.155995    5.392748   -2.053494
H         -0.658644    6.295066   -2.231364
H         -2.778323    9.145731   -2.851048
H         -1.056362    8.836279   -3.052650
H         -2.120245    8.977973   -4.445290
N         -2.483319    4.143744   -4.655743
C         -2.174787    2.733412   -4.745815
C         -1.220413    2.507858   -5.924341
O         -0.296267    1.729744   -5.829171
C         -3.434918    1.861689   -4.821392
C         -4.348001    2.052429   -3.601345
C         -5.476740    1.022912   -3.498125
N         -4.975159   -0.201823   -2.888624
C         -5.648194   -1.324880   -2.745240
N         -6.860672   -1.510241   -3.241285
N         -5.089111   -2.307374   -2.041311
H         -3.393026    4.474721   -4.900485
H         -1.610021    2.443671   -3.871862
```

```
H        -3.990669    2.087943   -5.727200
H        -3.111625    0.829660   -4.896835
H        -3.764347    2.020204   -2.688150
H        -4.801465    3.029819   -3.644086
H        -6.280293    1.426877   -2.890745
H        -5.878706    0.805066   -4.481320
H        -4.115860   -0.149910   -2.352313
H        -7.333226   -0.812722   -3.773252
H        -7.276086   -2.421087   -3.189573
H        -4.212219   -2.127138   -1.596727
H        -5.521493   -3.206311   -2.010306
N        -1.459765    3.193317   -7.035089
H        -2.291003    3.731135   -7.156740
C        -0.595017    2.993452   -8.182625
H        -0.943783    3.639053   -8.978094
H        -0.613177    1.968835   -8.525352
C         0.857465    3.335366   -7.862690
O         1.786148    2.690572   -8.282168
N         1.005838    4.404207   -7.064171
H         0.216488    4.981385   -6.877237
C         2.282317    4.880664   -6.640287
H         2.920781    5.138664   -7.476075
H         2.127074    5.776103   -6.052556
C         3.094719    3.928904   -5.773657
O         4.299243    4.059771   -5.680474
N         2.428048    2.983354   -5.101947
C         3.084473    2.129958   -4.148713
C         4.257212    1.311988   -4.705090
O         5.155603    0.991578   -3.952567
C         1.992396    1.252818   -3.520154
C         2.313152    0.639922   -2.160709
C         0.998456    0.242231   -1.496511
N         1.234744   -0.348893   -0.189031
C         0.472622   -1.287589    0.367510
N        -0.646637   -1.682048   -0.192287
N         0.892780   -1.849665    1.494035
H         1.440818    2.898021   -5.228118
H         3.546908    2.738377   -3.380807
H         1.135915    1.903690   -3.398470
H         1.681519    0.488111   -4.222874
H         2.964728   -0.222754   -2.249982
H         2.829468    1.366925   -1.537696
H         0.349866    1.104242   -1.402685
H         0.483627   -0.488611   -2.103503
H         1.935247    0.062573    0.385857
H        -1.280422   -1.061951   -0.703155
H        -1.099776   -2.459002    0.241947
H         1.830860   -1.756606    1.806892
H         0.358194   -2.582829    1.906578
N         4.279521    0.942029   -5.996895
H         3.549462    1.240358   -6.611063
C         5.363430    0.116655   -6.504622
H         5.137241   -0.134737   -7.533322
H         5.440406   -0.801518   -5.943797
```

```
C         6.736304      0.756125     -6.473520
O         7.738688      0.090201     -6.358880
N         6.766110      2.085325     -6.608285
H         5.911590      2.604519     -6.573395
C         8.004948      2.824348     -6.552897
H         8.709102      2.440133     -7.277304
H         7.798899      3.854763     -6.809856
C         8.747892      2.828093     -5.215313
O         9.888715      3.237125     -5.162014
N         8.102821      2.339213     -4.147628
C         8.684908      2.176464     -2.826890
C         9.949180      1.306678     -2.833735
O        10.766720      1.389818     -1.931306
C         7.581314      1.511095     -1.974759
C         7.783360      1.387887     -0.464573
C         7.454094      2.660345      0.325309
C         7.354118      2.396032      1.824143
N         8.697992      2.367874      2.473107
O        11.057133      1.630559      0.885061
H        10.903870      1.839504     -0.036888
H        11.536486      0.804316      0.834427
H         7.157191      2.037176     -4.260658
H         8.984756      3.134284     -2.422700
H         6.665432      2.057845     -2.156542
H         7.409607      0.524374     -2.389221
H         7.099812      0.613233     -0.130497
H         8.777432      1.028441     -0.233096
H         8.169694      3.457292      0.143920
H         6.488922      3.037553      0.002705
H         6.794784      3.168177      2.329042
H         6.876791      1.448718      2.023374
H         9.407724      1.906335      1.912107
H         9.015329      3.340391      2.624962
H         8.652485      1.891639      3.361760
N        10.062466      0.448216     -3.837536
C        11.109720     -0.538022     -4.063206
C        12.485299      0.020955     -4.432700
O        13.441980     -0.710360     -4.314048
C        10.692455     -1.461528     -5.229098
C         9.544570     -2.443378     -4.952056
C         9.102064     -3.087922     -6.269421
C         9.963253     -3.519847     -3.950369
H         9.326728      0.446670     -4.513524
H        11.270579     -1.113440     -3.163301
H        10.429686     -0.829138     -6.071930
H        11.569745     -2.027136     -5.519164
H         8.694680     -1.896267     -4.555163
H         8.294310     -3.793313     -6.099345
H         8.749793     -2.340097     -6.968395
H         9.922180     -3.632361     -6.730806
H         9.166309     -4.239089     -3.794944
H        10.826427     -4.061952     -4.321745
H        10.223410     -3.107547     -2.980044
N        12.574050      1.261828     -4.919242
```

```
C    13.814962    1.843533   -5.394464
C    14.959287    2.031588   -4.379939
O    16.014496    2.410098   -4.812841
C    13.498192    3.170319   -6.087675
O    12.917038    4.085428   -5.202472
H    11.758211    1.836002   -4.912410
H    14.248078    1.175918   -6.129767
H    14.421612    3.589901   -6.455061
H    12.848531    2.976743   -6.935239
H    11.970345    4.083354   -5.276140
N    14.792294    1.722326   -3.074279
C    15.928331    1.647847   -2.176900
C    16.803566    0.408980   -2.399662
O    17.931630    0.391505   -1.980629
C    15.495699    1.582971   -0.707225
C    14.645271    2.745455   -0.199925
C    14.295226    2.506090    1.269049
N    13.274145    3.406232    1.790191
C    13.457177    4.554372    2.421978
N    14.676647    5.024715    2.658861
O    15.037977    7.482719    4.210924
H    14.753691    8.368211    3.968853
H    15.758311    7.591744    4.818793
N    12.410210    5.244341    2.824117
H    13.899807    1.424087   -2.757490
H    16.563129    2.503530   -2.348660
H    14.950808    0.657851   -0.549519
H    16.407570    1.510575   -0.124742
H    15.166720    3.690497   -0.321237
H    13.724989    2.828749   -0.766262
H    13.896450    1.508225    1.381699
H    15.176401    2.549395    1.898170
H    12.332778    3.116011    1.617460
H    15.482561    4.536147    2.348997
H    14.814485    5.880154    3.169195
H    11.453751    4.967408    2.667417
H    12.507605    6.087410    3.343396
N    16.200238   -0.648980   -2.967242
C    16.870723   -1.911312   -3.169655
C    17.669346   -1.888048   -4.469628
H    18.065758   -0.908896   -4.747533
O    17.872772   -2.850980   -5.134143
C    15.867550   -3.069577   -3.152405
C    15.324452   -3.400827   -1.795222
N    14.401645   -2.630976   -1.114956
O    12.503485   -0.438849   -0.464063
H    13.145419   -1.038331   -0.848943
H    12.026984   -0.040834   -1.184531
C    15.622174   -4.490961   -1.053460
C    14.172042   -3.256220   -0.006839
N    14.885002   -4.384830    0.099358
H    15.287835   -0.548364   -3.357348
H    17.595863   -2.031418   -2.372442
H    15.050049   -2.840464   -3.827217
```

```
H          16.367856   -3.943090   -3.548161
H          16.279868   -5.312830   -1.239245
H          13.493126   -2.928648    0.754427
H          14.787152   -5.067586    0.818346
O         -15.222082   -2.397519    4.863101
H         -15.361843   -1.621889    4.324705
H         -14.798702   -2.088381    5.656519
O         -13.219930   -1.101582    6.936753
H         -12.596257   -1.542990    7.501452
H         -13.469882   -0.304315    7.387867
O         -12.200149   -0.861924    4.208203
H         -12.410926   -0.868805    5.138838
H         -12.978974   -0.493170    3.787186
O         -13.417866   -1.950300    0.759984
H         -12.666441   -1.360786    0.861251
H         -13.343133   -2.605457    1.454946
O          -7.237915    1.128633    0.289282
H          -7.464103    1.507390    1.141138
H          -6.298597    0.991713    0.319550
O          -7.874253    2.047801    2.945889
H          -8.809770    1.933676    3.099028
H          -7.593828    2.798470    3.461898
O          -9.272610   -0.246742   -0.994812
H          -8.478391    0.082579   -0.562431
H          -9.106451   -1.114433   -1.353302
O         -12.617772   -2.268728   -1.948294
H         -13.047611   -2.388420   -1.101496
H         -13.179282   -2.602570   -2.641464
O         -13.724606    1.033897   -0.778680
H         -14.149956    0.185193   -0.818763
H         -12.905355    0.864242   -0.302113
O         -14.301902    4.959760    4.736376
H         -13.725804    5.585401    4.304864
H         -14.320232    5.249655    5.643604
O         -15.059769    2.145928    1.477030
H         -15.726791    2.810154    1.363943
H         -14.689503    1.954954    0.614177
O         -12.318300    0.260893   -3.325775
H         -12.763545    0.921930   -2.804857
H         -12.236272   -0.483909   -2.733424
O         -13.341573   -1.970004   -4.776980
H         -13.218382   -1.044783   -4.571967
H         -14.006754   -2.025399   -5.450402
O          -9.888188   -2.609922   -2.710778
H          -9.964377   -2.252581   -3.599884
H         -10.779437   -2.593582   -2.364780
O         -11.504675    0.129613    0.657972
H         -10.732986   -0.009257    0.101874
H         -11.203938    0.612982    1.431982
O         -10.743689    1.373410    3.105047
H         -10.970572    0.593529    3.608273
H         -11.457947    1.985321    3.305964
O         -13.168505    2.592614    3.699243
H         -13.497321    3.346933    4.191720
```

```
H            -13.734116     2.537363     2.935261
O             -9.328020     0.613271    -3.721136
H            -10.258680     0.501758    -3.898192
H             -9.264296     0.528992    -2.770846
O            -14.731438    -0.361184     2.878967
H            -14.999971     0.514156     2.607902
H            -14.567599    -0.845349     2.073639
O            -13.154495    -3.422539     3.255631
H            -13.963091    -3.252307     3.748676
H            -12.556592    -2.754155     3.589030
O            -10.438575    -2.026779    -5.399976
H            -11.378720    -2.184859    -5.419447
H            -10.271665    -1.269990    -5.975172
```

Case 14:
Voltage = -70mV; R300 positive; E183 negative; Y266 neutral (all amino acids in "native" charge state); S4 FREE
```
N                               15.626239    -6.300447     5.440039
H                               14.801636    -6.080905     5.965082
C                               15.563757    -5.706958     4.100715
C                               14.842600    -6.633509     3.104470
O                               14.098210    -6.198846     2.248509
C                               16.976689    -5.429177     3.578286
H                               16.405311    -5.938400     5.954169
H                               14.999330    -4.783803     4.089911
H                               16.944508    -5.061277     2.559231
H                               17.456294    -4.673528     4.190664
H                               17.587495    -6.324910     3.601684
N                               15.104918    -7.933853     3.227881
H                               15.699509    -8.208703     3.979209
C                               14.485631    -8.944699     2.368945
H                               14.576431    -8.676165     1.329231
H                               14.980794    -9.892315     2.534730
C                               12.998086    -9.079931     2.696787
O                               12.171623    -9.280474     1.844934
N                               12.661952    -8.931733     3.994189
H                               13.379481    -8.848043     4.676928
C                               11.273425    -9.062347     4.441917
H                               11.261027    -9.051043     5.523615
H                               10.833976    -9.984085     4.093406
C                               10.405699    -7.909618     3.920501
O                                9.265378    -8.095898     3.584889
N                               10.996877    -6.703509     3.904055
C                               10.342988    -5.520782     3.401296
C                               10.113097    -5.562129     1.889384
O                                9.053978    -5.169951     1.458384
C                               11.111601    -4.271040     3.862354
O                               12.499190    -4.434498     3.830596
H                               11.958333    -6.623840     4.142055
H                                9.348711    -5.457161     3.811816
H                               10.791498    -3.416495     3.275525
H                               10.861500    -4.077567     4.895151
```

```
H    12.825748    -4.585328     2.951059
N    11.083267    -6.022093     1.079501
C    10.790925    -6.169294    -0.340811
C     9.740988    -7.263227    -0.570194
O     9.000316    -7.201671    -1.526017
C    12.069193    -6.330643    -1.204710
C    12.565421    -7.774574    -1.332213
C    11.853431    -5.710442    -2.589944
H    11.942928    -6.367786     1.450874
H    10.299311    -5.262700    -0.653031
H    12.836081    -5.750791    -0.698790
H    13.524641    -7.786236    -1.838932
H    12.685200    -8.263128    -0.374031
H    11.880350    -8.372053    -1.926233
H    12.752209    -5.812725    -3.188381
H    11.040894    -6.200307    -3.114643
H    11.625076    -4.652148    -2.517950
N     9.659180    -8.237032     0.343698
H    10.394229    -8.345349     1.005441
C     8.693500    -9.327721     0.248938
H     9.024046   -10.111100     0.918768
H     8.658890    -9.726602    -0.752944
C     7.274203    -8.944539     0.612562
O     6.361599    -9.408373    -0.032091
N     7.076795    -8.104985     1.635999
C     5.741855    -7.598699     1.857195
C     5.358662    -6.609118     0.760253
O     4.187990    -6.510925     0.453116
C     5.450443    -7.064997     3.274303
C     5.560012    -8.192071     4.306329
C     6.286225    -5.853520     3.690321
H     7.843952    -7.827938     2.210140
H     5.057006    -8.416271     1.692669
H     4.411093    -6.752620     3.234804
H     5.243874    -7.836293     5.281186
H     4.927181    -9.032713     4.040680
H     6.579264    -8.550599     4.396706
H     5.918667    -5.462333     4.633058
H     7.322543    -6.128479     3.833109
H     6.246023    -5.051804     2.961210
N     6.295952    -5.912661     0.114862
C     5.938338    -5.122965    -1.053183
C     5.483305    -6.028435    -2.204157
O     4.467210    -5.765331    -2.805965
C     7.083947    -4.166288    -1.443887
C     7.206700    -3.057412    -0.384994
C     6.866195    -3.576638    -2.841415
C     8.508598    -2.265211    -0.468219
H     7.245299    -5.943953     0.418990
H     5.058916    -4.541270    -0.819104
H     8.004276    -4.739753    -1.462340
H     6.357458    -2.383649    -0.492681
H     7.134154    -3.489457     0.605325
H     7.642355    -2.858689    -3.074338
```

```
H         6.896126    -4.337888    -3.611381
H         5.907387    -3.072914    -2.910325
H         8.532754    -1.497148     0.296405
H         9.363918    -2.911750    -0.308682
H         8.642697    -1.774867    -1.423358
N         6.232132    -7.094769    -2.511282
H         7.117292    -7.228276    -2.068490
C         5.873336    -7.953628    -3.618325
H         6.693255    -8.642243    -3.780270
H         5.718421    -7.384429    -4.522685
C         4.596836    -8.759488    -3.383934
O         3.741767    -8.820718    -4.236258
N         4.498211    -9.390964    -2.207997
H         5.244921    -9.339525    -1.547589
C         3.320523   -10.148249    -1.857922
H         3.077754   -10.873001    -2.620489
H         3.524873   -10.676326    -0.934962
C         2.082623    -9.276650    -1.667580
O         0.997466    -9.677243    -2.022398
N         2.281347    -8.076193    -1.125563
C         1.147511    -7.165033    -1.055726
C         0.701281    -6.650325    -2.426905
O        -0.481426    -6.491475    -2.635056
C         1.352205    -5.965069    -0.126870
O         1.568050    -6.349263     1.189639
H         3.183951    -7.773823    -0.835020
H         0.301897    -7.718228    -0.680982
H         2.156008    -5.336208    -0.491936
H         0.435581    -5.394135    -0.153082
H         2.486910    -6.564479     1.279442
N         1.626632    -6.358953    -3.352812
C         1.222926    -5.955335    -4.687843
C         0.463970    -7.083026    -5.386003
O        -0.533453    -6.828729    -6.028060
C         2.422591    -5.426962    -5.512092
C         2.777672    -4.010854    -5.024426
C         2.134077    -5.440758    -7.018851
C         4.069658    -3.442176    -5.612221
H         2.598865    -6.440377    -3.143072
H         0.489773    -5.166159    -4.604046
H         3.266537    -6.082600    -5.324052
H         1.949863    -3.346351    -5.270913
H         2.864912    -4.021218    -3.944857
H         2.985082    -5.060357    -7.569523
H         1.943280    -6.441225    -7.387225
H         1.273573    -4.826229    -7.264291
H         4.305791    -2.484505    -5.156676
H         4.904088    -4.109850    -5.426645
H         3.999774    -3.279073    -6.681670
N         0.920735    -8.330283    -5.248958
H         1.779280    -8.504841    -4.768664
C         0.170248    -9.440101    -5.788523
H         0.062709    -9.365078    -6.860475
H         0.707601   -10.351030    -5.557202
```

```
C    -1.237907   -9.534456   -5.205893
O    -2.180497   -9.792998   -5.917294
N    -1.353570   -9.318049   -3.896874
C    -2.656869   -9.424660   -3.278082
C    -3.606556   -8.303293   -3.699323
O    -4.798368   -8.532391   -3.738402
C    -2.566012   -9.449949   -1.745578
O    -3.802746   -9.779297   -1.177218
H    -0.539548   -9.239903   -3.322772
H    -3.139081  -10.336723   -3.600136
H    -1.857820  -10.208868   -1.443182
H    -2.210997   -8.491078   -1.380227
H    -4.507750   -9.403595   -1.691698
N    -3.076845   -7.104689   -3.946764
C    -3.849866   -5.980868   -4.445985
C    -4.371132   -6.258987   -5.855944
O    -5.533496   -6.064515   -6.126236
C    -2.961107   -4.718176   -4.365894
C    -3.314812   -3.566884   -5.290693
C    -4.623478   -3.152252   -5.515839
C    -2.290098   -2.899692   -5.955832
C    -4.894362   -2.120822   -6.399825
C    -2.556753   -1.850567   -6.821496
C    -3.865064   -1.463261   -7.054158
H    -2.097881   -6.981043   -3.798008
H    -4.734276   -5.851449   -3.842407
H    -2.971943   -4.393025   -3.329851
H    -1.945289   -5.011918   -4.584754
H    -5.439280   -3.660909   -5.041116
H    -1.272859   -3.227453   -5.822791
H    -5.917357   -1.852648   -6.592488
H    -1.748180   -1.355200   -7.328875
H    -4.082518   -0.669334   -7.746089
N    -3.493910   -6.719411   -6.755261
C    -3.948180   -7.028468   -8.092895
C    -5.071670   -8.064948   -8.072827
O    -6.004924   -7.956291   -8.823832
C    -2.805404   -7.536932   -8.974176
S    -1.589982   -6.272211   -9.464694
H    -2.549927   -6.900998   -6.486249
H    -4.383213   -6.148575   -8.542668
H    -2.297754   -8.368678   -8.504521
H    -3.228187   -7.890795   -9.905009
H    -0.994598   -6.121893   -8.291874
N    -4.946583   -9.094105   -7.210551
H    -4.106658   -9.231307   -6.690559
C    -5.945603  -10.149582   -7.253493
H    -6.116083  -10.475161   -8.268535
H    -5.566907  -10.987013   -6.679212
C    -7.322686   -9.775960   -6.680993
O    -8.337923  -10.152588   -7.200341
N    -7.294287   -9.072267   -5.543993
C    -8.547814   -8.746439   -4.883372
C    -9.399011   -7.740430   -5.653175
```

```
O        -10.605202   -7.765590   -5.559613
C         -8.351873   -8.246603   -3.441764
C         -7.574228   -6.935987   -3.255431
C         -8.353536   -5.625243   -3.143471
O         -9.549554   -5.625532   -2.830590
O         -7.664928   -4.595112   -3.344260
H         -6.425460   -8.820720   -5.122655
H         -9.154477   -9.639687   -4.844861
H         -9.334854   -8.145791   -3.008981
H         -7.844140   -9.039365   -2.903211
H         -7.011656   -7.000214   -2.327782
H         -6.833365   -6.798830   -4.025233
N         -8.733494   -6.802863   -6.339756
C         -9.394631   -5.751211   -7.105786
C        -10.047341   -6.278709   -8.387040
H         -9.648640   -7.218824   -8.776369
O        -10.939606   -5.706397   -8.923011
C         -8.378130   -4.608953   -7.364243
O         -8.227598   -3.834017   -6.195788
C         -8.790099   -3.643063   -8.465115
H         -7.750856   -6.925459   -6.467983
H        -10.194362   -5.334102   -6.511516
H         -7.424132   -5.049268   -7.630098
H         -7.885736   -4.358360   -5.476914
H         -8.082864   -2.820786   -8.477902
H         -8.777894   -4.113346   -9.439211
H         -9.787801   -3.253849   -8.296981
N        -12.868062   -9.160769    3.082702
H        -13.050827   -9.718593    2.274597
C        -11.736051   -8.245302    2.852567
C        -10.657664   -8.272362    3.966784
O         -9.521625   -7.923612    3.760735
C        -12.173713   -6.777182    2.685559
O        -12.814424   -6.406862    3.888348
C        -13.090879   -6.566062    1.488235
H        -13.697035   -8.652697    3.319068
H        -11.211195   -8.539661    1.958796
H        -11.277465   -6.180574    2.544998
H        -13.345719   -5.620271    3.755041
H        -13.350522   -5.516875    1.410268
H        -12.601307   -6.854364    0.567156
H        -14.011799   -7.129140    1.587719
N        -11.128070   -8.659873    5.159896
H        -12.059233   -9.015568    5.143994
C        -10.376600   -8.790469    6.387711
H        -10.364832   -7.862952    6.941144
H        -10.877718   -9.528467    7.003783
C         -8.909435   -9.200119    6.269395
O         -8.085192   -8.557279    6.871794
N         -8.558678  -10.284507    5.556924
H         -9.255110  -10.824763    5.095113
C         -7.155760  -10.660794    5.456762
H         -7.099806  -11.636226    4.990537
H         -6.712902  -10.727791    6.437978
```

```
C      -6.304709   -9.683070    4.629961
O      -5.141842   -9.514063    4.897407
N      -6.964106   -9.096727    3.637051
C      -6.251040   -8.139177    2.810364
C      -6.289345   -6.726858    3.400591
O      -5.346422   -5.992317    3.209009
C      -6.512922   -8.234072    1.287792
C      -7.901737   -8.567735    0.784183
C      -8.736902   -7.588049    0.262593
C      -8.314406   -9.896667    0.714187
C      -9.938363   -7.925157   -0.341002
C      -9.524395  -10.237728    0.131929
C     -10.335155   -9.251511   -0.408990
H      -7.948347   -9.194652    3.542365
H      -5.209672   -8.384434    2.922631
H      -6.180697   -7.297657    0.854935
H      -5.836251   -8.991377    0.910851
H      -8.435951   -6.555496    0.283658
H      -7.664915  -10.673576    1.079186
H     -10.527132   -7.154265   -0.799769
H      -9.813342  -11.271837    0.065381
H     -11.252160   -9.516622   -0.905226
N      -7.306955   -6.396400    4.195184
C      -7.275203   -5.269277    5.094196
C      -6.173406   -5.441887    6.138754
O      -5.524216   -4.477462    6.481034
C      -8.672655   -5.148070    5.731414
C      -8.860268   -4.045219    6.750307
C      -9.110436   -2.736444    6.349026
C      -8.841744   -4.334642    8.110395
C      -9.343716   -1.743417    7.284836
C      -9.070383   -3.339720    9.050773
C      -9.327250   -2.043503    8.639381
H      -8.095954   -6.999981    4.231202
H      -7.027080   -4.361962    4.563760
H      -9.374063   -5.013356    4.914732
H      -8.905116   -6.097866    6.195410
H      -9.136788   -2.492148    5.301154
H      -8.643720   -5.340932    8.439474
H      -9.539068   -0.738648    6.958242
H      -9.047070   -3.579556   10.098308
H      -9.509787   -1.271791    9.364915
N      -5.952669   -6.659839    6.629662
H      -6.601890   -7.390416    6.442156
C      -4.890955   -6.928952    7.573780
H      -4.928756   -6.260177    8.420542
H      -5.019634   -7.942953    7.929789
C      -3.493857   -6.788792    6.977563
O      -2.613101   -6.288174    7.634085
N      -3.300958   -7.230373    5.728663
C      -2.028304   -7.034311    5.057801
C      -1.805975   -5.550954    4.739456
O      -0.725592   -5.045443    4.953402
C      -1.932844   -7.941750    3.813412
```

```
C         -0.773393    -7.541737     2.901195
C         -1.790941    -9.408130     4.238200
H         -4.005958    -7.782574     5.289833
H         -1.227056    -7.281743     5.739359
H         -2.856032    -7.826049     3.250831
H         -0.698246    -8.240272     2.075301
H         -0.894246    -6.552116     2.481316
H          0.171716    -7.554915     3.432101
H         -1.801387   -10.051719     3.364339
H         -0.844304    -9.561069     4.749999
H         -2.588484    -9.730760     4.893631
N         -2.834245    -4.852651     4.239994
C         -2.673117    -3.451936     3.898803
C         -2.449997    -2.596165     5.151334
O         -1.573166    -1.754211     5.152857
C         -3.876638    -2.903588     3.115875
C         -4.133243    -3.491861     1.716773
C         -3.093157    -3.153053     0.663954
O         -1.911647    -3.359422     0.932824
O         -3.495895    -2.714192    -0.450441
H         -3.694767    -5.312108     4.029926
H         -1.783720    -3.339114     3.306078
H         -4.773959    -3.053179     3.707173
H         -3.738635    -1.829967     3.031238
H         -4.185756    -4.570803     1.771403
H         -5.102568    -3.147928     1.376483
N         -3.264788    -2.793897     6.194195
C         -3.094699    -2.057349     7.429141
C         -1.906247    -2.536557     8.254759
O         -1.382254    -1.741316     9.005653
C         -4.343981    -1.945843     8.315630
O         -4.717187    -3.180118     8.858300
C         -5.500152    -1.239094     7.619264
H         -3.950372    -3.515035     6.143439
H         -2.820081    -1.051050     7.159059
H         -4.027459    -1.345906     9.157936
H         -5.164841    -3.692529     8.196467
H         -6.320160    -1.123631     8.316880
H         -5.200960    -0.253524     7.276679
H         -5.859995    -1.802345     6.767106
N         -1.374881    -3.733186     8.020608
H         -1.922119    -4.462031     7.619526
C         -0.122208    -4.094723     8.627565
H         -0.089815    -3.840789     9.675792
H          0.006859    -5.164132     8.518562
C          1.053778    -3.394191     7.958285
O          1.995723    -2.994785     8.603398
N          0.993358    -3.264319     6.619258
C          2.101037    -2.670158     5.914335
C          2.185843    -1.157347     6.110199
O          3.282503    -0.644361     6.155553
C          2.150537    -3.085595     4.427655
S          1.414175    -1.956958     3.199562
H          0.236855    -3.676073     6.112529
```

| | | | |
|---|---:|---:|---:|
| H | 3.011650 | -3.031826 | 6.369425 |
| H | 3.188717 | -3.139751 | 4.130547 |
| H | 1.717959 | -4.068348 | 4.312208 |
| H | 0.155378 | -1.977193 | 3.611434 |
| N | 1.059360 | -0.447759 | 6.248343 |
| C | 1.174099 | 0.981702 | 6.484322 |
| C | 1.739204 | 1.259173 | 7.878431 |
| O | 2.359988 | 2.270931 | 8.097394 |
| C | -0.122814 | 1.779079 | 6.233695 |
| C | -1.195537 | 1.571898 | 7.315525 |
| C | -0.628549 | 1.541492 | 4.805944 |
| C | -2.491547 | 2.344202 | 7.062703 |
| H | 0.170694 | -0.886781 | 6.120002 |
| H | 1.920087 | 1.364794 | 5.805068 |
| H | 0.191600 | 2.814662 | 6.304607 |
| H | -1.418918 | 0.519080 | 7.433090 |
| H | -0.790277 | 1.895990 | 8.268395 |
| H | -1.359860 | 2.290699 | 4.532081 |
| H | 0.188052 | 1.606727 | 4.092648 |
| H | -1.091049 | 0.570184 | 4.692329 |
| H | -3.137446 | 2.288918 | 7.933057 |
| H | -2.297573 | 3.394556 | 6.861838 |
| H | -3.044330 | 1.942120 | 6.220601 |
| N | 1.509856 | 0.318561 | 8.793538 |
| H | 0.898053 | -0.437125 | 8.584616 |
| C | 1.927600 | 0.490036 | 10.167474 |
| H | 1.399266 | -0.239089 | 10.768699 |
| H | 1.692876 | 1.480579 | 10.526309 |
| C | 3.426437 | 0.298209 | 10.349034 |
| O | 4.078250 | 1.040680 | 11.032137 |
| N | 3.967395 | -0.751006 | 9.692807 |
| H | 3.382556 | -1.382726 | 9.188876 |
| C | 5.395638 | -0.917153 | 9.670842 |
| H | 5.623459 | -1.879095 | 9.226729 |
| H | 5.809407 | -0.892916 | 10.667579 |
| C | 6.118528 | 0.162072 | 8.868004 |
| O | 7.241662 | 0.485842 | 9.169033 |
| N | 5.463393 | 0.660407 | 7.820574 |
| C | 6.134674 | 1.786409 | 7.201386 |
| C | 6.062620 | 3.015846 | 8.112630 |
| H | 5.662745 | 2.846595 | 9.111550 |
| O | 6.449346 | 4.085994 | 7.760693 |
| C | 5.742213 | 2.017768 | 5.737581 |
| C | 6.423302 | 0.999978 | 4.836017 |
| C | 7.806402 | 1.049196 | 4.655236 |
| C | 5.719433 | -0.028982 | 4.225035 |
| C | 8.467169 | 0.072740 | 3.928566 |
| C | 6.377861 | -1.002466 | 3.482808 |
| C | 7.753953 | -0.965660 | 3.341184 |
| H | 4.529036 | 0.405335 | 7.592323 |
| H | 7.188435 | 1.548329 | 7.211627 |
| H | 4.666589 | 1.971640 | 5.620414 |
| H | 6.062923 | 3.017528 | 5.469086 |
| H | 8.370035 | 1.853617 | 5.091097 |

```
H            4.659323    -0.097882     4.366716
H            9.540160     0.112692     3.838282
H            5.812605    -1.802903     3.037406
H            8.264109    -1.739953     2.795041
N           12.523095     9.734609    -1.168596
H           11.759041    10.063661    -1.729479
C           12.225776     8.398446    -0.644570
C           11.259680     8.562228     0.522674
O           10.335218     7.796529     0.696409
C           13.503310     7.693910    -0.154437
C           14.597130     7.503512    -1.212473
S           13.976873     6.635237    -2.682854
C           15.515537     6.405986    -3.609335
H           13.310239     9.703122    -1.786439
H           11.719913     7.764176    -1.362019
H           13.929734     8.258820     0.668934
H           13.216932     6.724632     0.236073
H           15.030202     8.455159    -1.499771
H           15.395647     6.921446    -0.767628
H           15.277868     5.789438    -4.464614
H           15.915109     7.354231    -3.944171
H           16.252034     5.890211    -3.006830
N           11.471921     9.607637     1.341280
C           10.599025     9.855842     2.470144
C            9.217364    10.358213     2.040832
O            8.235682    10.028104     2.651361
C           11.239307    10.814671     3.472384
C           12.366921    10.154383     4.253920
O           12.404626     8.960616     4.420456
N           13.287824    10.987797     4.756074
H           12.198268    10.242226     1.087977
H           10.409332     8.923360     2.972939
H           11.585281    11.723423     2.988994
H           10.485937    11.108662     4.197408
H           13.990586    10.619280     5.358893
H           13.208871    11.976425     4.686545
N            9.161600    11.174191     0.967599
H            9.999330    11.416767     0.491604
C            7.885888    11.603299     0.432828
H            8.062737    12.352691    -0.327810
H            7.271909    12.036472     1.206428
C            7.083483    10.463499    -0.197110
O            5.882443    10.502249    -0.194635
N            7.788300     9.486765    -0.804468
C            7.128241     8.367366    -1.452348
C            6.478368     7.430855    -0.430356
O            5.332840     7.082748    -0.580137
C            8.089418     7.652120    -2.425823
C            8.482925     8.611025    -3.565151
C            7.454350     6.364546    -2.959405
C            9.520517     8.056391    -4.543088
H            8.751789     9.392101    -0.585934
H            6.293933     8.761296    -2.012857
H            8.987470     7.375677    -1.875226
```

```
H         7.584054    8.888839   -4.111429
H         8.865770    9.532923   -3.137292
H         8.124819    5.864497   -3.643828
H         7.230358    5.671238   -2.157193
H         6.525512    6.575471   -3.481523
H         9.837989    8.833492   -5.230894
H        10.404747    7.686088   -4.033481
H         9.128224    7.242615   -5.139692
N         7.196125    7.066021    0.651318
C         6.594317    6.181138    1.629795
C         5.389089    6.848398    2.294134
O         4.378332    6.217722    2.512813
C         7.605840    5.745362    2.697665
C         8.554526    4.638635    2.231040
O         8.892379    3.793886    3.046437
O         8.918751    4.663216    1.020999
H         8.192238    7.118539    0.623263
H         6.183587    5.310347    1.138573
H         8.192571    6.603304    3.017867
H         7.082432    5.380948    3.571784
N         5.508933    8.143878    2.621292
H         6.369068    8.621824    2.459779
C         4.396371    8.870606    3.187721
H         4.716210    9.889635    3.364629
H         4.087111    8.447928    4.130717
C         3.173440    8.910396    2.274817
O         2.058791    9.010459    2.732590
N         3.432459    8.838834    0.966969
H         4.376345    8.854191    0.659825
C         2.429071    8.935620   -0.047232
H         2.924088    9.162265   -0.981924
H         1.747064    9.749145    0.161289
C         1.534925    7.708571   -0.286710
O         0.506261    7.812835   -0.898341
N         1.995659    6.563289    0.257722
C         1.191330    5.377167    0.403725
C         0.395002    5.360207    1.713544
O        -0.755826    4.984479    1.705729
C         2.055567    4.119049    0.297892
H         2.897653    6.581652    0.683170
H         0.447535    5.378989   -0.376711
H         1.447309    3.231849    0.436687
H         2.517665    4.080815   -0.682443
H         2.839464    4.120247    1.047503
N         1.005750    5.737165    2.849737
C         0.305022    5.580143    4.114474
C        -0.820461    6.600792    4.281383
O        -1.844656    6.283118    4.846274
C         1.242431    5.524833    5.347296
C         1.808638    6.883138    5.790095
C         2.344308    4.479003    5.124168
C         2.700410    6.797330    7.031364
H         1.959027    6.029309    2.834863
H        -0.210978    4.633570    4.074961
```

```
H         0.603420     5.172155     6.151131
H         2.361599     7.335894     4.973998
H         0.987356     7.558044     6.008988
H         2.745024     4.141630     6.070221
H         1.954046     3.608894     4.606724
H         3.159853     4.877200     4.530044
H         2.973012     7.793309     7.366602
H         2.185813     6.304248     7.850273
H         3.617696     6.251132     6.845296
N        -0.631068     7.835514     3.793601
H         0.250751     8.087278     3.397456
C        -1.613472     8.868873     4.013944
H        -1.217037     9.797612     3.622853
H        -1.813562     9.004808     5.066990
C        -2.967688     8.583286     3.371066
O        -3.978936     8.791606     3.997620
N        -2.992285     8.076816     2.124302
C        -4.294721     7.852451     1.501237
C        -5.035720     6.651432     2.059304
O        -6.241950     6.660359     2.119339
C        -3.974306     7.702488     0.004563
C        -2.726448     8.567218    -0.162778
C        -1.938810     8.260223     1.106451
H        -4.942347     8.700903     1.671184
H        -3.744035     6.668350    -0.230928
H        -4.810489     8.008956    -0.609944
H        -2.153989     8.337683    -1.050084
H        -3.000752     9.617857    -0.191479
H        -1.374481     7.352097     0.990977
H        -1.261528     9.054973     1.379430
N        -4.304380     5.586030     2.430895
C        -4.946799     4.445906     3.050550
C        -5.561791     4.789525     4.407149
O        -6.606387     4.274013     4.738839
C        -3.971199     3.264204     3.199389
C        -3.855938     2.393695     1.966277
C        -4.658633     1.265279     1.842169
C        -2.951046     2.670703     0.949055
C        -4.549697     0.421201     0.750111
C        -2.830152     1.829671    -0.145702
C        -3.613624     0.694474    -0.235526
O        -3.453282    -0.142041    -1.303146
H        -3.309751     5.660263     2.446515
H        -5.784718     4.147111     2.440086
H        -2.999324     3.645108     3.490849
H        -4.322119     2.648700     4.018921
H        -5.375984     1.031944     2.609579
H        -2.309721     3.527410     1.020206
H        -5.172184    -0.451605     0.674848
H        -2.111448     2.033694    -0.919027
H        -3.404984    -1.050646    -0.979064
N        -4.889451     5.640812     5.191357
H        -3.989745     5.974394     4.916110
C        -5.447358     6.104927     6.439235
```

```
H         -4.657366    6.588754    7.000087
H         -5.831476    5.282232    7.020782
C         -6.595046    7.095409    6.259999
O         -7.575692    7.037395    6.961827
N         -6.446173    8.014179    5.301271
H         -5.576337    8.070477    4.818835
C         -7.372225    9.134357    5.071783
H         -6.895875    9.792308    4.356821
H         -7.547925    9.682903    5.985568
C         -8.736903    8.738819    4.537806
O         -9.725353    9.332981    4.894767
N         -8.770409    7.728210    3.657174
C        -10.013810    7.301650    3.045943
C        -10.951279    6.612206    4.038682
O        -12.101143    6.399105    3.721493
C         -9.686029    6.423297    1.811056
O        -10.780250    6.385124    0.943231
C         -9.197873    5.027259    2.172429
H         -7.908428    7.331559    3.344855
H        -10.560174    8.163638    2.690001
H         -8.886078    6.945490    1.298169
H        -11.103917    5.497871    0.842297
H         -8.906679    4.499903    1.270047
H         -8.345465    5.057645    2.834950
H         -9.985947    4.459948    2.651551
N        -10.438532    6.209954    5.204804
C        -11.262177    5.653265    6.264473
C        -12.010161    6.694567    7.105414
O        -12.834608    6.315165    7.898827
C        -10.417135    4.768872    7.195183
C         -9.956176    3.441668    6.572725
C         -8.812581    2.840389    7.394187
C        -11.107404    2.436585    6.450383
H         -9.491421    6.425680    5.425617
H        -12.043092    5.068978    5.805321
H         -9.551818    5.341854    7.513382
H        -11.007063    4.565538    8.080956
H         -9.570217    3.645494    5.578238
H         -8.461941    1.912044    6.951997
H         -7.970266    3.520644    7.440079
H         -9.132684    2.629778    8.412279
H        -10.771159    1.524314    5.964322
H        -11.488576    2.171345    7.433654
H        -11.938465    2.824929    5.872176
N        -11.678907    7.987875    6.960201
C        -12.383313    8.998877    7.689554
C        -13.741796    9.352047    7.121300
H        -14.296900   10.080466    7.725281
O        -14.198747    8.920450    6.117421
H        -11.073401    8.281338    6.225599
H        -12.522574    8.688020    8.718256
H        -11.785720    9.905579    7.704330
N        -18.243788    5.829572   -8.845791
H        -18.065544    5.953896   -9.823566
```

```
C    -17.193113   6.417302   -8.023766
C    -16.047593   5.417258   -7.893750
O    -14.897949   5.775413   -7.748753
C    -17.769671   6.810440   -6.634746
C    -18.608193   8.099069   -6.770154
C    -16.694840   6.977361   -5.553780
C    -19.591098   8.332864   -5.621479
H    -19.136165   6.238068   -8.656359
H    -16.746792   7.302491   -8.469266
H    -18.429062   5.997052   -6.336781
H    -17.930043   8.945313   -6.850289
H    -19.174190   8.085388   -7.697448
H    -17.142205   7.325019   -4.631457
H    -16.189173   6.045185   -5.331356
H    -15.941966   7.697626   -5.855029
H    -20.164853   9.236681   -5.795153
H    -20.292894   7.508677   -5.533288
H    -19.090160   8.446541   -4.667248
N    -16.381019   4.114238   -7.949810
H    -17.347097   3.903542   -8.075973
C    -15.449295   3.083596   -7.557193
H    -15.148220   3.187266   -6.524284
H    -15.937754   2.123134   -7.669731
C    -14.168641   3.057253   -8.375693
O    -13.115550   2.765295   -7.865389
N    -14.265881   3.345650   -9.685431
H    -15.154818   3.558977  -10.075192
C    -13.088685   3.325110  -10.519729
H    -12.614921   2.355081  -10.505386
H    -13.384083   3.542861  -11.538405
C    -12.031344   4.341995  -10.108650
O    -10.860596   4.094616  -10.266611
N    -12.466404   5.517413   -9.626438
H    -13.416363   5.614584   -9.340519
C    -11.522545   6.539859   -9.239911
H    -10.846506   6.755232  -10.052286
H    -12.079022   7.438204   -9.003524
C    -10.660833   6.169869   -8.038688
O     -9.473131   6.381786   -8.048563
N    -11.291552   5.624838   -6.981944
C    -10.547687   5.293857   -5.782501
C     -9.683746   4.039360   -5.951167
O     -8.667500   3.934419   -5.307216
C    -11.417408   5.269876   -4.505401
C    -12.587793   4.274825   -4.571088
C    -11.883515   6.691268   -4.165458
C    -13.284000   4.053343   -3.225709
H    -12.273718   5.464378   -7.028181
H     -9.809526   6.068540   -5.643046
H    -10.743367   4.945275   -3.717917
H    -13.324063   4.620484   -5.292371
H    -12.233148   3.318390   -4.942588
H    -12.349470   6.723823   -3.188892
H    -11.045742   7.380091   -4.144118
```

```
H            -12.602019     7.059462    -4.891872
H            -14.021397     3.260675    -3.309064
H            -12.578045     3.766062    -2.454059
H            -13.799956     4.943327    -2.884808
N            -10.061167     3.093071    -6.839169
C             -9.127392     2.048093    -7.208646
C             -7.929400     2.580367    -7.999873
O             -6.823130     2.138485    -7.781420
C             -9.810870     0.906737    -7.976991
C            -10.799032     0.079666    -7.144682
C            -10.123296    -0.694837    -6.008213
N            -11.054222    -1.440384    -5.178187
C            -11.764084    -0.930593    -4.182580
N            -11.773810     0.382034    -3.932383
N            -12.494455    -1.726297    -3.434464
H            -10.906388     3.205570    -7.357403
H             -8.676274     1.670844    -6.306507
H            -10.326687     1.317867    -8.838703
H             -9.030977     0.257051    -8.363030
H            -11.584374     0.718180    -6.757285
H            -11.285623    -0.637186    -7.797816
H             -9.430804    -1.420835    -6.409321
H             -9.550709    -0.027361    -5.373378
H            -11.081830    -2.442961    -5.260143
H            -11.188652     0.987699    -4.457559
H            -12.020441     0.695894    -3.007946
H            -12.412968    -2.722591    -3.547412
H            -13.132124    -1.351888    -2.765154
N             -8.151543     3.529318    -8.916852
H             -9.079631     3.819615    -9.138079
C             -7.070273     4.050169    -9.717722
H             -6.586476     3.271066   -10.289003
H             -7.486778     4.774789   -10.405527
C             -5.972491     4.728662    -8.908681
O             -4.814666     4.621174    -9.245786
N             -6.341051     5.432539    -7.828596
C             -5.357899     6.118134    -7.005937
C             -4.337152     5.139456    -6.422592
O             -3.190090     5.485684    -6.246862
C             -6.075891     6.945843    -5.912086
C             -5.123953     7.434648    -4.814452
C             -6.791321     8.145740    -6.546988
H             -7.313836     5.565976    -7.646755
H             -4.765629     6.780809    -7.623481
H             -6.816032     6.296085    -5.451345
H             -5.670499     8.065635    -4.121653
H             -4.692529     6.619376    -4.246877
H             -4.307567     8.016769    -5.228358
H             -7.351211     8.686757    -5.790929
H             -6.067771     8.833690    -6.975769
H             -7.486545     7.857247    -7.324189
N             -4.761289     3.904216    -6.103826
C             -3.844303     2.916959    -5.576020
C             -2.725910     2.550725    -6.555684
```

```
O        -1.704084    2.051783   -6.127507
C        -4.592719    1.655673   -5.114551
C        -5.520588    1.916714   -3.919936
C        -6.504337    0.779166   -3.643630
N        -5.869480   -0.397924   -3.054166
C        -6.464367   -1.580527   -2.932745
N        -7.616137   -1.832833   -3.553002
N        -5.959285   -2.534785   -2.174027
H        -5.672066    3.610308   -6.385367
H        -3.318712    3.350650   -4.735359
H        -5.170048    1.273038   -5.948175
H        -3.855402    0.904084   -4.863080
H        -4.938954    2.130978   -3.027145
H        -6.128721    2.793387   -4.110961
H        -7.289930    1.133206   -2.984007
H        -6.964670    0.505974   -4.584217
H        -5.020659   -0.259880   -2.538155
H        -7.842893   -1.316823   -4.368863
H        -7.931179   -2.795354   -3.537979
H        -5.142761   -2.434130   -1.600620
H        -6.367096   -3.447894   -2.263105
N        -2.892207    2.814901   -7.854831
H        -3.716983    3.272669   -8.181763
C        -1.827885    2.603943   -8.803980
H        -2.200962    2.865082   -9.785709
H        -1.516778    1.569629   -8.825842
C        -0.564806    3.417710   -8.548124
O         0.501911    3.023994   -8.965777
N        -0.680922    4.553820   -7.843252
H        -1.570123    4.846170   -7.490875
C         0.488089    5.319223   -7.487525
H         1.033209    5.640347   -8.363833
H         0.167614    6.200969   -6.948155
C         1.485400    4.566179   -6.614434
O         2.635506    4.935674   -6.530363
N         1.021606    3.484269   -5.974291
C         1.862169    2.605003   -5.201679
C         3.030840    2.029935   -5.991773
O         4.002494    1.616684   -5.386966
C         0.972050    1.475553   -4.667289
C         1.586215    0.572016   -3.602336
C         0.574947   -0.520630   -3.279525
N         0.960700   -1.272106   -2.094198
C         0.111709   -2.032429   -1.401098
N        -1.056499   -2.352663   -1.917405
N         0.445778   -2.491238   -0.211161
H         0.065548    3.218600   -6.087379
H         2.318740    3.146738   -4.382342
H         0.084556    1.937938   -4.255388
H         0.628656    0.881999   -5.507751
H         2.519831    0.136061   -3.936420
H         1.796056    1.151109   -2.706123
H        -0.392055   -0.067831   -3.096123
H         0.474873   -1.187225   -4.129493
```

```
H        1.870475   -1.135834   -1.716854
H       -1.315316   -2.042486   -2.825259
H       -1.799111   -2.714517   -1.334451
H        1.219490   -2.115095    0.288296
H       -0.280724   -2.937513    0.335112
N        2.963054    1.983575   -7.323976
H        2.154138    2.319341   -7.809353
C        4.097071    1.518697   -8.087987
H        3.866583    1.636071   -9.139058
H        4.303150    0.474786   -7.900403
C        5.392275    2.267279   -7.793274
O        6.459921    1.711378   -7.911739
N        5.298221    3.549997   -7.403798
H        4.405093    3.981034   -7.271296
C        6.483671    4.295432   -7.071899
H        7.178904    4.317339   -7.899048
H        6.195858    5.315146   -6.850324
C        7.265297    3.769382   -5.874877
O        8.424752    4.103751   -5.720642
N        6.636416    2.954214   -5.026072
C        7.291384    2.371401   -3.873681
C        8.481401    1.491935   -4.250510
O        9.362102    1.286994   -3.437122
C        6.247798    1.554446   -3.082531
C        6.649721    1.147430   -1.660901
C        6.513823    2.281300   -0.638595
C        6.905887    1.879568    0.780212
N        8.381522    1.900144    0.972883
O       10.223220    0.671089   -0.838540
H        9.950995    1.038002   -1.676990
H       10.694030   -0.122141   -1.096400
H        5.698879    2.666273   -5.222568
H        7.706640    3.158555   -3.260166
H        5.338594    2.142745   -3.040812
H        6.001211    0.671220   -3.661050
H        5.991283    0.339531   -1.352221
H        7.650727    0.738433   -1.655791
H        7.093293    3.152881   -0.928177
H        5.477720    2.604791   -0.609032
H        6.505809    2.569230    1.507776
H        6.557487    0.890030    1.037285
H        8.894125    1.248405    0.384552
H        8.735984    2.845880    0.803604
H        8.600843    1.716070    1.941712
N        8.516717    0.952761   -5.470455
C        9.620688    0.106716   -5.885694
C       10.973041    0.820908   -5.851452
O       11.981862    0.155776   -5.751091
C        9.377524   -0.428549   -7.306909
C        8.233744   -1.450026   -7.437969
C        7.894410   -1.648864   -8.918149
C        8.580149   -2.789562   -6.780337
H        7.795219    1.159854   -6.130957
H        9.721571   -0.712733   -5.189420
```

```
H            9.180642     0.418623    -7.957684
H           10.300514    -0.884589    -7.648830
H            7.347954    -1.048832    -6.952338
H            7.096968    -2.375758    -9.037537
H            7.568876    -0.718272    -9.369488
H            8.756317    -2.012557    -9.471498
H            7.756146    -3.488447    -6.884210
H            9.453289    -3.236386    -7.247914
H            8.787275    -2.690533    -5.719603
N           10.980805     2.154151    -5.970650
C           12.177847     2.936828    -6.213364
C           13.220283     2.960253    -5.095517
O           14.305097     3.417636    -5.354781
C           11.790866     4.376284    -6.565161
O           11.190734     5.032574    -5.490232
H           10.104481     2.627945    -6.007231
H           12.710636     2.513460    -7.056876
H           12.693707     4.910152    -6.818580
H           11.142992     4.366974    -7.437307
H           10.253738     4.879890    -5.493363
N           12.931097     2.440419    -3.887703
C           13.989118     2.276823    -2.909724
C           15.022240     1.207005    -3.292504
O           16.092390     1.212397    -2.742524
C           13.456379     1.964392    -1.507884
C           12.676911     3.116097    -0.868165
C           12.509548     2.891878     0.633692
N           11.791635     4.004735     1.249508
C           12.116925     4.632490     2.363202
N           13.091807     4.204997     3.167475
O           12.726091     6.265786     5.325563
H           12.839524     7.206943     5.210688
H           12.286013     6.146706     6.158031
N           11.517602     5.770323     2.663974
H           12.018500     2.084216    -3.712780
H           14.556974     3.193789    -2.872008
H           12.851336     1.065174    -1.532317
H           14.329356     1.741810    -0.904194
H           13.202899     4.050215    -1.025739
H           11.700862     3.222643    -1.329609
H           11.982543     1.964985     0.823067
H           13.489572     2.824257     1.086243
H           10.879903     4.240383     0.888954
H           13.356216     3.247919     3.159552
H           13.262919     4.724192     4.006467
H           10.842449     6.156401     2.038589
H           11.615254     6.145713     3.584761
N           14.650436     0.288787    -4.206103
C           15.570841    -0.673269    -4.767453
C           15.785690    -0.354187    -6.242437
H           15.763333     0.712953    -6.480548
O           15.985885    -1.171989    -7.079020
C           15.136736    -2.131165    -4.559449
C           15.313485    -2.615323    -3.151410
```

```
N         14.368709   -2.432603   -2.164569
O         11.654001   -1.291696   -2.389795
H         12.536143   -1.661839   -2.300246
H         11.601279   -0.989134   -3.288373
C         16.380084   -3.261496   -2.630222
C         14.868678   -2.949482   -1.090516
N         16.085515   -3.469547   -1.305619
H         13.754084    0.369810   -4.638226
H         16.523850   -0.506855   -4.279638
H         14.100724   -2.242885   -4.861674
H         15.724991   -2.748952   -5.225147
H         17.301015   -3.582144   -3.067186
H         14.391518   -2.979707   -0.132592
H         16.667679   -3.910460   -0.632387
O        -15.167871   -3.479925    6.399732
H        -15.135576   -2.533964    6.513297
H        -14.382182   -3.822480    6.812295
O        -12.285624   -4.616023    6.421987
H        -12.308145   -5.327629    5.787298
H        -11.582808   -4.804486    7.031288
O        -12.275630   -2.090722    4.942717
H        -11.964488   -2.755331    5.552508
H        -12.876882   -1.527887    5.433193
O        -14.648005   -2.431468    1.489409
H        -13.797366   -2.005174    1.498563
H        -14.646419   -3.014386    2.249632
O         -8.053240    0.463699    1.106869
H         -8.027644    0.889498    1.966505
H         -7.480805    0.983627    0.559612
O         -8.092280    1.876274    3.578060
H         -9.031422    2.035427    3.607906
H         -7.664700    2.643767    3.951534
O        -10.399246   -0.894035    0.123274
H         -9.571545   -0.503391    0.410044
H        -10.909619   -1.018396    0.920033
O        -13.570098   -3.483245   -1.047756
H        -13.987680   -3.693188   -0.221402
H        -12.641623   -3.717293   -0.963390
O        -14.968067   -0.038896    3.128347
H        -15.348057   -0.703957    2.560544
H        -14.039516   -0.045980    2.897523
O        -13.088040    3.757688    3.242109
H        -12.897327    4.676175    3.442964
H        -13.810258    3.481836    3.802331
O        -15.129727    2.317272    4.761319
H        -15.922391    2.774327    5.011448
H        -15.369969    1.692786    4.081235
O        -14.330373   -0.820835   -1.170164
H        -15.266068   -0.750168   -1.034299
H        -14.109900   -1.743159   -1.003861
O        -13.020103   -6.316751   -5.495081
H        -12.260172   -6.854747   -5.718011
H        -13.544393   -6.855784   -4.918779
O        -10.789241   -3.942691   -0.763094
```

```
H    -10.407401   -3.073738   -0.686063
H    -10.324557   -4.401149   -1.465787
O    -12.149293   -0.617307    2.514699
H    -11.723333    0.237766    2.641352
H    -11.947032   -1.129330    3.296882
O    -10.973471    1.909730    2.872233
H    -11.629971    2.481314    3.275777
H    -10.885844    2.293191    2.000879
O    -11.852192    3.446641    0.608956
H    -12.576208    3.541006    1.223680
H    -12.094888    2.774437   -0.024788
O    -12.183422    1.018942   -1.069978
H    -12.997691    0.538988   -0.923101
H    -11.498494    0.453316   -0.704043
O    -14.496612   -0.607185    5.898738
H    -14.484801    0.268261    6.270749
H    -14.828152   -0.482718    5.008867
O    -14.165895   -3.924299    3.831458
H    -14.770478   -3.902870    4.578332
H    -13.469602   -3.325225    4.104774
O    -11.528188   -4.211255   -4.390310
H    -12.106161   -4.892927   -4.743922
H    -10.844160   -4.672767   -3.899416
```